\documentclass[prd,preprint,aps,floats,superscriptaddress,floatfix,nofootinbib]{revtex4}
\usepackage{url}
\usepackage{graphicx}
\begin{document}

\title{TASI 2004 Lecture Notes on Higgs Boson Physics\footnote{These lectures are
    dedicated to Filippo, who listened to them before he
    was born and behaved really well while I was writing
    these proceedings.}}

\author{Laura Reina}

\affiliation{Physics Department, Florida State University,\\
315 Keen Building, \\ 
Tallahassee, FL 32306-4350, USA\\ 
E-mail: reina@hep.fsu.edu}

\begin{abstract}
In these lectures I briefly review the Higgs mechanism of spontaneous
symmetry breaking and focus on the most relevant aspects of the
phenomenology of the Standard Model and of the Minimal Supersymmetric
Standard Model Higgs bosons at both hadron (Tevatron, Large Hadron
Collider) and lepton (International Linear Collider) colliders. Some
emphasis is put on the perturbative calculation of both Higgs boson
branching ratios and production cross sections, including the most
important radiative corrections.
\end{abstract}

\maketitle
\newpage
\tableofcontents

\section{Introduction}
\label{sec:intro}
The origin of the electroweak symmetry breaking is among the most
important open questions of contemporary particle physics.  Since it
was first proposed in 1964 by Higgs, Kibble, Guralnik, Hagen, Englert,
and Brout \cite{Englert:1964et,Higgs:1964pj,Guralnik:1964eu}, the
mechanism of spontaneous symmetry breaking known as
\emph{Higgs mechanism} has become part of the Standard Model (SM) of
particle physics and of one of its most thoroughly studied extensions,
the Minimal Supersymmetric Standard Model (MSSM). A substantial
theoretical and experimental effort has been devoted to the study of
the physics of the single scalar Higgs boson predicted by the
realization of the Higgs mechanism in the SM, and of the multiple
scalar and pseudoscalar Higgs bosons arising in the MSSM.  Indeed, the
discovery of one or more Higgs bosons is among the most important
goals of both the Tevatron and the Large Hadron Collider (LHC), while
the precise determination of their physical properties strongly
support the need for a future high energy International Linear
Collider (ILC).

In these lectures I would like to present a self contained
introduction to the physics of the Higgs boson(s). Given the huge
amount of work that has been done in this field, I will not even come
close to being exhaustively complete. This is not actually my aim.
For this series of lectures, I would like to present the reader with
some important background of informations that could prepare her or
him to explore further topical issues in Higgs physics.  Also,
alternative theoretical approaches to the electroweak symmetry
breaking and the generation of both boson and fermion masses will not
be considered here. They have been covered in several other series of
lectures at this school, and to them I refer.

In Section \ref{sec:theory_framework}, after a brief glance at the
essence of the Higgs mechanism, I will review how it is embedded in
the Standard Model and what constraints are directly and indirectly
imposed on the mass of the single Higgs boson that is predicted in
this context. Among the extensions of the SM I will only consider the
case of the MSSM, and in this context I will mainly focus on those
aspects that could be more relevant to distinguish the MSSM Higgs
bosons. Section \ref{sec:pheno} will review the phenomenology of both
the SM and the MSSM Higgs bosons, at the Tevatron and the LHC, and
will then focus on the role that a high energy ILC could play in this
context. Finally, in Section
\ref{sec:theory}, I will briefly summarize the state of the art of existing
theoretical calculations for both decay rates and production cross
sections of a Higgs boson.

Let me conclude by pointing the reader to some selected references
available in the literature. The theoretical bases of the Higgs
mechanism are nowadays a matter for textbooks in Quantum Field
Theory. They are presented in depth in both Refs.~\cite{Peskin:1995ev}
and \cite{Weinberg:1995mt}. An excellent review of both SM and MSSM
Higgs physics, containing a very comprehensive discussion of both
theoretical and phenomenological aspects as well as an exhaustive
bibliography, has recently appeared
\cite{Djouadi:2005gi,Djouadi:2005gj}.
 The phenomenology of Higgs physics has also been thoroughly covered
 in a fairly recent review paper
\cite{Carena:2002es}, which can be complemented by several workshop
proceedings and reports
\cite{cms:1994tdr,atlas:1999tdr,Carena:2000yx,Cavalli:2002vs,Babukhadia:2003zu,Assamagan:2004mu,Abe:2001wn,Aguilar-Saavedra:2001rg}. Finally,
series of lectures given at previous summer schools
\cite{Dawson:1994ri,Dawson:1998yi} can provide excellent references.

\section{Theoretical framework: the Higgs mechanism and its consequences.}
\label{sec:theory_framework}
In Yang-Mills theories gauge invariance forbids to have an explicit
mass term for the gauge vector bosons in the Lagrangian. If this is
acceptable for theories like QED (Quantum Electrodynamics) and QCD
(Quantum Chromodynamics), where both photons and gluons are massless,
it is unacceptable for the gauge theory of weak interactions, since
both the charged ($W^\pm$) and neutral ($Z^0$) gauge bosons have very
heavy masses ($M_W\!\simeq\!80$~GeV, $M_Z\!\simeq\!91$~GeV). A
possible solution to this problem, inspired by similar phenomena
happening in the study of spin systems, was proposed by several
physicists in 1964 \cite{Englert:1964et,Higgs:1964pj,Guralnik:1964eu},
and it is known today simply as \emph{the Higgs mechanism}. We will
review the basic idea behind it in
Section~\ref{subsec:higgs_mechanism}. In Section~\ref{subsec:higgs_sm}
we will recall how the Higgs mechanism is implemented in the Standard
Model and we will discuss which kind of theoretical constraints are
imposed on \emph{the Higgs boson}, the only physical scalar particle
predicted by the model. Finally, in Section~\ref{subsec:higgs_mssm} we
will generalize our discussion to the case of the MSSM, and use its
extended Higgs sector to illustrate how differently the Higgs
mechanism can be implemented in extensions of the SM.

\subsection{A brief introduction to the Higgs mechanism}
\label{subsec:higgs_mechanism}
\begin{figure}
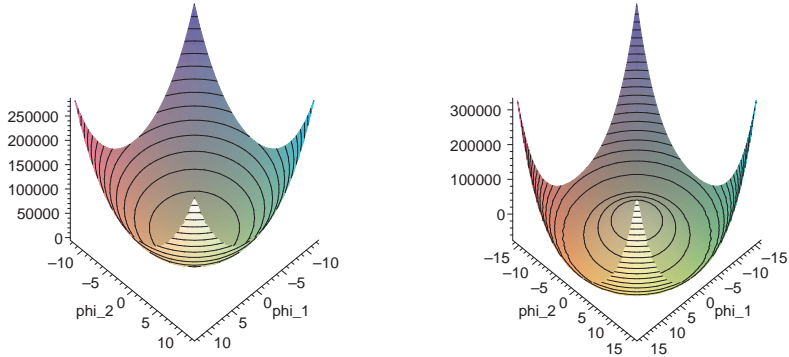

\centering
\includegraphics[scale=0.4]{higgs_1.ps}
\hspace{1.truecm}
\includegraphics[scale=0.4]{higgs_2.ps}
\caption[]{The potential $V(\phi)$ ($\phi\!=\phi_1+i\phi_2$) plotted
  for an arbitrary positive value of $\lambda$ and for an arbitrary
  positive (right) or negative (left) value of $\mu^2$. \label{fig:higgs_potential}
}
\end{figure}

The essence of the Higgs mechanism can be very easily illustrated
considering the case of a classical abelian Yang-Mills theory. In this
case, it is realized by adding to the Yang-Mills Lagrangian
\begin{equation}
\label{eq:L_ym_ab}
\mathcal{L}_A=-\frac{1}{4}F^{\mu\nu}F_{\mu\nu}\,\,\,\,\,\,\mbox{with}
\,\,\,\,\,\,
F^{\mu\nu}=(\partial^\mu A^\nu-\partial^\nu A^\mu)\,\,\,,
\end{equation}
a complex scalar field with Lagrangian
\begin{equation}
\label{eq:L_phi}
\mathcal{L}_\phi=(D^\mu\phi)^\ast D_\mu\phi -V(\phi)=
(D^\mu\phi)^\ast D_\mu\phi
-\mu^2\phi^\ast\phi-\lambda(\phi^\ast\phi)^2
\,\,\,,
\end{equation}
where $D^\mu\!=\!\partial^\mu +igA^\mu$, and $\lambda\!>\! 0$ for the
scalar potential to be bounded from below. The full Lagrangian
\begin{equation}
\label{eq:L_tot}
\mathcal{L}=\mathcal{L}_A+\mathcal{L}_\phi
\end{equation}
is invariant under a $U(1)$ \emph{gauge transformation} acting on
the fields as:
\begin{equation}
\label{eq:gauge_trans_ab}
\phi(x)\rightarrow e^{i\alpha(x)}\phi(x)\,\,\,,\,\,\,
A^\mu(x)\rightarrow A^\mu(x)+\frac{1}{g}\partial^\mu\alpha(x)\,\,\,,
\end{equation}
while a gauge field mass term (i.e., a term quadratic in the fields
$A^\mu$) would not be gauge invariant and cannot be added to
$\mathcal{L}$ if the $U(1)$ gauge symmetry has to be
preserved. Indeed, the Lagrangian in Eq.~(\ref{eq:L_tot}) can still
describe the physics of a massive gauge boson, provided the potential
$V(\phi)$ develops a non trivial minimum
($\phi^\ast\phi\!\neq\!0$). The occurrence of a non trivial minimum,
or, better, of a non trivial degeneracy of minima only depends on the
sign of the $\mu^2$ parameter in $V(\phi)$. For $\mu^2\!>\!0$ there is
a unique minimum at $\phi^\ast\phi\!=\!0$, while for $\mu^2\!<\!0$ the
potential develops a degeneracy of minima satisfying the equation
$\phi^\ast\phi\!=\!-\mu^2/(2\lambda)$. This is illustrated in
Fig.~\ref{fig:higgs_potential}, where the potential $V(\phi)$ is
plotted as a function of the real and imaginary parts of the field
$\phi\!=\!\phi_1+i\phi_2$.
In the case of a unique minimum at $\phi^\ast\phi\!=\!0$ the
Lagrangian in Eq.~(\ref{eq:L_tot}) describes the physics of a massless
vector boson (e.g. the photon, in electrodynamics, with $g\!=\!-e$)
interacting with a massive charged scalar particle. On the other hand,
something completely different takes place when $\mu^2\!<\!0$.
Choosing the ground state of the theory to be a particular $\phi$
among the many satisfying the equation of the minimum, and expanding
the potential in the vicinity of the chosen minimum, transforms the
Lagrangian in such a way that the original gauge symmetry is now
\emph{hidden} or \emph{spontaneously broken}, and new interesting
features emerge. To be more specific, let's pick the following
$\phi_0$ minimum (along the direction of the real part of $\phi$, as
traditional) and shift the $\phi$ field accordingly:
\begin{equation}
\label{eq:phi0}
\phi_0=\left(-\frac{\mu^2}{2\lambda}\right)^{1/2}=\frac{v}{\sqrt{2}}
\,\,\,\longrightarrow\,\,\,
\phi(x)=\phi_0+\frac{1}{\sqrt{2}}\left(\phi_1(x)+i\phi_2(x)\right)
\,\,\,. 
\end{equation}
The Lagrangian in Eq.~(\ref{eq:L_tot}) can then be rearranged as follows:
\begin{equation}
\label{eq:L_about_phi0}
{\mathcal L}=
\underbrace{
-\frac{1}{4}F^{\mu\nu}F_{\mu\nu}+\frac{1}{2}g^2v^2A^\mu A_\mu}_
{\mathrm{massive}\,\,\mathrm{vector}\,\,\mathrm{field}}
+
\underbrace{
\frac{1}{2}(\partial^\mu\phi_1)^2+\mu^2\phi_1^2}_
{\mathrm{massive}\,\,\mathrm{scalar}\,\,\mathrm{field}}+
\underbrace{
\frac{1}{2}(\partial^\mu\phi_2)^2+gvA_\mu\partial^\mu\phi_2}_{
\mathrm{Goldstone}\,\,\mathrm{boson}}+\ldots
\end{equation}
and now contains the correct terms to describe a massive vector field
$A^\mu$ with mass $m_A^2\!=\!g^2v^2$ (originating from the kinetic
term of $\mathcal{L}_\phi$), a massive real scalar field $\phi_1$ with
mass $m_{\phi_1}\!=\!-2\mu^2$, that will become a \emph{Higgs
boson}, and a massless scalar field $\phi_2$, a so called
\emph{Goldstone boson} which couples to the gauge vector boson
$A^\mu$. The terms omitted contain couplings between the $\phi_1$ and
$\phi_2$ fields irrelevant to this discussion. The gauge symmetry of
the theory allows us to make the particle content more transparent.
Indeed, if we parameterize the complex scalar field $\phi$ as:
\begin{equation}
\label{eq:phi_unit_gauge}
\phi(x)=\frac{e^{i\frac{\chi(x)}{v}}}{\sqrt{2}}(v+H(x))
\,\,\,\,\stackrel{U(1)}{\longrightarrow}\,\,\,\,
\frac{1}{\sqrt{2}}(v+H(x))\,\,\,,
\end{equation}
the $\chi$ degree of freedom can be \emph{rotated away}, as indicated
in Eq.~(\ref{eq:phi_unit_gauge}), by enforcing the $U(1)$ gauge
invariance of the original Lagrangian. With this gauge choice, known
as \emph{unitary gauge} or \emph{unitarity gauge}, the Lagrangian
becomes:
\begin{equation}
\label{eq:L_unit_gauge}
\mathcal{L}=\mathcal{L}_A+\frac{g^2v^2}{2}A^\mu A_\mu+
\frac{1}{2}\left(\partial^\mu H\partial_\mu H+2\mu^2 H^2
\right)+\ldots
\end{equation}
which unambiguously describes the dynamics of a massive vector boson
$A^\mu$ of mass $m_A^2\!=\!g^2v^2$, and a massive real scalar field of
mass $m_H^2\!=\!-2\mu^2=2\lambda v^2$, the \emph{Higgs field}. It is
interesting to note that the total counting of degrees of freedom
(d.o.f.)  before the original $U(1)$ symmetry is spontaneously broken
and after the breaking has occurred is the same. Indeed, one goes from
a theory with one massless vector field (two d.o.f.) and one complex
scalar field (two d.o.f.) to a theory with one massive vector field
(three d.o.f.) and one real scalar field (one d.o.f.), for a total of
four d.o.f. in both cases.  This is what is colorfully described by
saying that each gauge boson has \emph{eaten up} one scalar degree of
freedom, becoming massive.

We can now easily generalize the previous discussion to the case of a
non-abelian Yang-Mills theory. $\mathcal{L}_A$ in Eq.~(\ref{eq:L_tot})
now becomes:
\begin{equation}
\label{eq:L_ym_non_ab}
\mathcal{L}_A=\frac{1}{4}F^{a,\mu\nu}F^a_{\mu\nu}\,\,\,\,\,\mbox{with}
\,\,\,\,\, F^a_{\mu\nu}=\partial_\mu A^a_\nu-\partial_\nu A^a_\mu+
gf^{abc}A^b_\mu A^c_\nu\,\,\,,
\end{equation}
where the latin indices are group indices and $f^{abc}$ are the
structure constants of the Lie Algebra associated to the non abelian
gauge symmetry Lie group, defined by the commutation relations of the Lie
Algebra generators $t^a$: $[t^a,t^b]\!=\!if^{abc}t^c$. Let us also
generalize the scalar Lagrangian to include several scalar fields
$\phi_i$ which we will in full generality consider as real:
\begin{equation}
\label{eq:L_phi_multi}
\mathcal{L}_\phi=\frac{1}{2}(D^\mu\phi_i)^2-V(\phi)\,\,\,\,\,\mbox{where}\,\,\,\,\,
V(\phi)=\mu^2\phi_i^2+\frac{\lambda}{2}\phi_i^4\,\,\,,
\end {equation}
where the sum over the index $i$ is understood and
$D_\mu\!=\!\partial_\mu-igt^aA^a_\mu$. The Lagrangian of
Eq.~(\ref{eq:L_tot}) is invariant under a non-abelian gauge
transformation of the form:
\begin{eqnarray}
\label{eq:gauge_trans_nab}
\phi_i(x)&\rightarrow& (1+i\alpha^a(x)t^a)_{ij}\phi_j \,\,\,,\\
A^a_\mu(x)&\rightarrow& A^a_\mu(x)+\frac{1}{g}\partial_\mu\alpha^a(x)+
f^{abc}A^b_\mu(x)\alpha^c(x)\,\,\,.\nonumber
\end{eqnarray}
When $\mu^2\!<\!0$ the potential develops a degeneracy of minima
described by the minimum condition:
$\phi^2\!=\!\phi_{0}^2\!=\!-\mu^2/\lambda$, which only fixes the
magnitude of the vector $\phi_{0}$. By arbitrarily choosing the
direction of $\phi_{0}$, the degeneracy is removed. The Lagrangian can
be expanded in a neighborhood of the chosen minimum and mass terms for
the gauge vector bosons can be introduced as in the abelian case,
i.e.:
\begin{eqnarray}
\label{eq:gauge_boson_mass_nonab}
\frac{1}{2}(D_\mu\phi_i)^2&\longrightarrow& \ldots\,\,\,+
\frac{1}{2}g^2(t^a\phi)_i(t^b\phi)_i A_\mu^a A^{b\mu}+\ldots\\ 
&\stackrel{\phi_{min}\!=\!\phi_0}{\longrightarrow}&
\ldots\,\,\,+ \frac{1}{2}
\underbrace{g^2(t^a\phi_0)_i(t^b\phi_0)_i}_{m_{ab}^2}A_\mu^a
A^{b\mu}+\ldots \nonumber
\end{eqnarray}
Upon diagonalization of the mass matrix $m_{ab}^2$ in
Eq.~(\ref{eq:gauge_boson_mass_nonab}), all gauge vector bosons
$A_\mu^a$ for which $t^a\phi_0\ne 0$ become massive, and to each of
them corresponds a Goldstone particle, i.e. an unphysical massless
particle like the $\chi$ field of the abelian example. The remaining
scalar degrees of freedom become massive, and correspond to the Higgs
field $H$ of the abelian example.

The Higgs mechanism can be very elegantly generalized to the case of a
quantum field theory when the theory is quantized via the path
integral method\footnote{Here I assume some familiarity with path
integral quantization and the properties of various generating
functionals introduced in that context, as I did while giving these
lectures. The detailed explanation of the formalism used would take us
too far away from our main track}. In this context, the quantum analog
of the potential $V(\phi)$ is the
\emph{effective potential} $V_{eff}(\varphi_{cl})$, defined in term of
the \emph{effective action} $\Gamma[\phi_{cl}]$ (the generating
functional of the 1PI connected correlation functions) as:
\begin{equation}
\label{eq:v_eff}
V_{eff}(\varphi_{cl})=-\frac{1}{VT}\Gamma[\phi_{cl}]\,\,\,\,\,\mbox{for}
\,\,\,\,\,\phi_{cl}(x)=\mbox{constant}=\varphi_{cl}\,\,\,,
\end{equation}
where $VT$ is the space-time extent of the functional integration and
$\phi_{cl}(x)$ is the \emph{vacuum expectation value} of the field
configuration $\phi(x)$:
\begin{equation}
\label{eq:phi_cl}
\phi_{cl}(x)=\langle\Omega| \phi(x)|\Omega\rangle\,\,\,.
\end{equation}

The stable quantum states of the theory are defined by the variational
condition:
\begin{equation}
\label{eq:delta_v_eff}
\frac{\delta}{\delta\phi_{cl}}\Gamma[\phi_{cl}]\bigg|_{\phi_{cl}=\varphi_{cl}}=0
\,\,\,\,\,\,\,\,\longrightarrow\,\,\,\,\,\,\,\,
\frac{\partial}{\partial\varphi_{cl}}V_{eff}(\varphi_{cl})=0\,\,\,,
\end{equation}
which identifies in particular the states of minimum energy of the
theory, i.e. the stable vacuum states. A system with spontaneous
symmetry breaking has several minima, all with the same
energy. Specifying one of them, as in the classical case, breaks
the original symmetry on the vacuum. The relation between the
classical and quantum case is made even more transparent by the
perturbative form of the effective potential. Indeed,
$V_{eff}(\varphi_{cl})$ can be organized as a loop expansion and
calculated systematically order by order in $\hbar$:
\begin{equation}
\label{eq:veff_exp}
V_{eff}(\varphi_{cl})=V(\varphi_{cl})+\mbox{loop effects}\,\,\,,
\end{equation}
with the lowest order being the classical potential in
Eq.~(\ref{eq:L_phi}). Quantum corrections to $V_{eff}(\varphi_{cl})$
affect some of the properties of the potential and therefore
have to be taken into account in more sophisticated studies of the
Higgs mechanism for a spontaneously broken quantum gauge theory.  We
will see how this can be important in Section 
\ref{subsec:higgs_sm_constraints} when
we discuss how the mass of the SM Higgs boson is related to the
energy scale at which we expect new physics effect to become relevant
in the SM.

Finally, let us observe that at the quantum level the choice of gauge
becomes a delicate issue. For example, in the \emph{unitarity gauge}
of Eq.~(\ref{eq:phi_unit_gauge}) the particle content of the theory
becomes transparent but the propagator of a massive vector field
$A^\mu$ turns out to be:
\begin{equation}
\label{eq:prop_unit_gauge}
\Pi^{\mu\nu}(k)=-\frac{i}{k^2-m_A^2} \left(g^{\mu\nu}-\frac{k^\mu
k^\nu}{m_A^2}\right)\,\,\,,
\end{equation}
and has a problematic ultra-violet behavior, which makes more
difficult to consistently define and calculate ultraviolet-stable
scattering amplitudes and cross sections. Indeed, for the very purpose
of studying the renormalizability of quantum field theories with
spontaneous symmetry breaking, the so called \emph{renormalizable} or
\emph{renormalizability gauges} ($R_\xi$ \emph{gauges}) are
introduced. If we consider the abelian Yang-Mills theory of
Eqs.~(\ref{eq:L_ym_ab})-(\ref{eq:L_tot}), the \emph{renormalizable
gauge} choice is implemented by quantizing with a gauge condition $G$
of the form:
\begin{equation}
\label{eq:ren_gauge}
G=\frac{1}{\sqrt{\xi}}(\partial_\mu A^\mu+\xi gv\phi_2)\,\,\,,
\end{equation}
in the generating functional
\begin{equation}
\label{eq:Z_ren_gauge}
Z[J]=C\int DA\,D\phi_1\,D\phi_2\exp\left[i\int(\mathcal{L}-\frac{1}{2}G^2)\right]
\mbox{det}\left(\frac{\delta G}{\delta\alpha}\right)\,\,\,,
\end{equation}
where C is an overall factor independent of the fields, $\xi$ is an
arbitrary parameter, and $\alpha$ is the gauge transformation
parameter in Eq.~(\ref{eq:gauge_trans_ab}).  After having reduced the
determinant in Eq.~(\ref{eq:Z_ren_gauge}) to an integration over ghost
fields ($c$ and $\bar{c}$), the gauge plus scalar fields Lagrangian
looks like:
\begin{eqnarray}
\label{eq:L_ren_gauge}
\mathcal{L}-\frac{1}{2}G^2+\mathcal{L}_{ghost}&=&
-\frac{1}{2}A_\mu\left(-g^{\mu\nu}\partial^2+
\left(1-\frac{1}{\xi}\right)
\partial^\mu\partial^\nu-(gv)^2g^{\mu\nu}\right)A_\nu\nonumber\\
&+&\frac{1}{2}(\partial_\mu\phi_1)^2-\frac{1}{2}m_{\phi_1}^2\phi_1^2
+\frac{1}{2}(\partial_\mu\phi_2)^2
-\frac{\xi}{2}(gv)^2\phi_2^2+\cdots\nonumber\\
&+&\bar{c}\left[-\partial^2-
\xi(gv)^2\left(1+\frac{\phi_1}{v}\right)\right]c\,\,\,,
\end{eqnarray}
such that:
\begin{eqnarray}
\label{eq:prop_ren_gauge}
\langle A^\mu(k)A^\nu(-k)\rangle&=&
\frac{-i}{k^2-m_A^2}\left(g^{\mu\nu}-\frac{k^\mu k^\nu}{k^2}\right)+
\frac{-i\xi}{k^2-\xi m_A^2}
\left(\frac{k^\mu k^\nu}{k^2}\right)\,\,\,,
\nonumber\\
\langle \phi_1(k)\phi_1(-k)\rangle&=&\frac{-i}{k^2-m_{\phi_1}^2}\,\,\,,\\
\langle \phi_2(k)\phi_2(-k)\rangle&=&\langle c(k)\bar{c}(-k)\rangle=
\frac{-i}{k^2-\xi m_A^2}\,\,\,,
\nonumber
\end{eqnarray}
where the vector field propagator has now a safe ultraviolet behavior.
Moreover we notice that the $\phi_2$ propagator has the same
denominator of the longitudinal component of the gauge vector boson
propagator. This shows in a more formal way the relation between the
$\phi_2$ degree of freedom and the longitudinal component of the
massive vector field $A^\mu$, upon spontaneous symmetry breaking. 

\subsection{The Higgs sector of the Standard Model}
\label{subsec:higgs_sm}
The Standard Model is a spontaneously broken Yang-Mills theory based
on the $SU(2)_L\times U(1)_Y$ non-abelian symmetry
group\cite{Peskin:1995ev,Weinberg:1995mt}. The Higgs mechanism is
implemented in the Standard Model by introducing a complex scalar
field $\phi$, doublet of $SU(2)$ with hypercharge $Y_\phi=1/2$,
\begin{equation}
\label{eq:phi_sm}
\phi=
\left(
\begin{array}{c}
\phi^+\\
\phi^0
\end{array}
\right)\,\,\,,
\end{equation}
with Lagrangian
\begin{equation}
\label{eq:L_phi_sm}
\mathcal{L}_\phi=(D^\mu\phi)^\dagger D_\mu\phi-\mu^2\phi^\dagger\phi-
\lambda(\phi^\dagger\phi)^2\,\,\,,
\end{equation}
where $D_\mu\phi=(\partial_\mu-igA^a_\mu\tau^a-ig^\prime Y_\phi
B_\mu)$, and $\tau^a\!=\!\sigma^a/2$ 
(for $a\!=\!1,2,3$) are the $SU(2)$ Lie Algebra generators,
proportional to the Pauli matrix $\sigma^a$. The gauge symmetry 
of the Lagrangian is broken to $U(1)_{em}$ when a particular vacuum 
expectation value is chosen, e.g.:
\begin{equation}
\label{eq:phi_vev}
\langle\phi\rangle=\frac{1}{\sqrt{2}}
\left(
\begin{array}{c}
0\\ v
\end{array}
\right)\,\,\,\,\,\,\mbox{with}\,\,\,\,\,\,
v=\left(\frac{-\mu^2}{\lambda}\right)^{1/2}
\,\,\,\,\,(\mu^2<0,\,\lambda >0)\,\,\,.
\end{equation}
Upon spontaneous symmetry breaking the kinetic term in
Eq.~(\ref{eq:L_phi_sm}) gives origin to the SM gauge boson mass
terms. Indeed, specializing Eq.~(\ref{eq:gauge_boson_mass_nonab}) to
the present case, and using Eq.~(\ref{eq:phi_vev}), one gets:
\begin{eqnarray}
\label{eq:SM_gauge_boson_mass_terms}
(D^\mu\phi)^\dagger D_\mu\phi&\longrightarrow&\cdots +
\frac{1}{8}(0\,\,\,v)\left(gA_\mu^a\sigma^a+g^\prime B_\mu\right)
\left(gA^{b\mu}\sigma^b+g^\prime B^\mu\right)
\left(
\begin{array}{c}
0\\ v
\end{array}
\right)+\cdots
\nonumber\\
&\longrightarrow&
\cdots+\frac{1}{2}\frac{v^2}{4}\left[
g^2(A_\mu^1)^2+g^2(A_\mu^2)^2+(-gA_\mu^3+g^\prime B_\mu)^2\right] 
+\cdots\nonumber\\
\end{eqnarray}
One recognizes in Eq.~(\ref{eq:SM_gauge_boson_mass_terms}) the mass 
terms for the charged gauge bosons $W^\pm_\mu$:
\begin{equation}
\label{eq:W_mass}
W^\pm_\mu=\frac{1}{\sqrt{2}}(A_\mu^1\pm A_\mu^2)
\,\,\,\,\longrightarrow\,\,\,\,M_W=g\frac{v}{2}\,\,\,,
\end{equation}
and for the neutral gauge boson $Z^0_\mu$:
\begin{equation}
\label{eq:Z_mass}
Z^0_\mu=\frac{1}{\sqrt{g^2+g^{\prime 2}}}(gA_\mu^3-g^\prime B_\mu)
\,\,\,\,\longrightarrow\,\,\,\,
M_Z=\sqrt{g^2+g^{\prime 2}}\frac{v}{2}\,\,\,,
\end{equation} 
while the orthogonal linear combination of $A^3_\mu$ and $B_\mu$
remains massless and corresponds to the photon field ($A_\mu$):
\begin{equation}
\label{eq:photon_mass}
A_\mu=\frac{1}{\sqrt{g^2+g^{\prime 2}}}(g^\prime A_\mu^3+gB_\mu)
\,\,\,\,\longrightarrow\,\,\,\,M_A=0\,\,\,,
\end{equation}
the gauge boson of the residual $U(1)_{em}$ gauge symmetry.

The content of the scalar sector of the theory becomes more
transparent if one works in the unitary gauge and eliminate the
unphysical degrees of freedom using gauge invariance. In analogy to
what we wrote for the abelian case in Eq.~(\ref{eq:phi_unit_gauge}), 
this amounts to parametrize and rotate the $\phi(x)$ complex scalar 
field as follows:
\begin{equation}
\label{eq:phi_sm_unit_gauge}
\phi(x)=\frac{e^{\frac{i}{v}\vec\chi(x)\cdot\vec\tau}}{\sqrt{2}}
\left(
\begin{array}{c}
0\\ v+H(x)
\end{array}
\right)
\,\,\,\stackrel{SU(2)}{\longrightarrow}\,\,\,
\phi(x)=\frac{1}{\sqrt{2}}
\left(
\begin{array}{c}
0\\ v+H(x)
\end{array}
\right)\,\,\,,
\end{equation}
after which the scalar potential in Eq.~(\ref{eq:L_phi_sm}) becomes:
\begin{equation}
\label{eq:L_phi_sm_unit_gauge}
\mathcal{L}_\phi=\mu^2 H^2-\lambda v H^3-\frac{1}{4}H^4=
-\frac{1}{2}M_H^2 H^2-\sqrt{\frac{\lambda}{2}} M_H H^3
-\frac{1}{4}\lambda H^4\,\,\,.
\end{equation}
Three degrees of freedom, the $\chi^a(x)$ Goldstone bosons, have been
reabsorbed into the longitudinal components of the $W^\pm_\mu$ and
$Z^0_\mu$ weak gauge bosons. One real scalar field remains, the
\emph{Higgs boson} $H$, with mass $M_H^2\!=\!-2\mu^2=2\lambda v^2$ and
self-couplings:

\vspace{0.5truecm}
\begin{tabular}{cc}
\begin{minipage}{0.5\linewidth}
\includegraphics[scale=0.6]{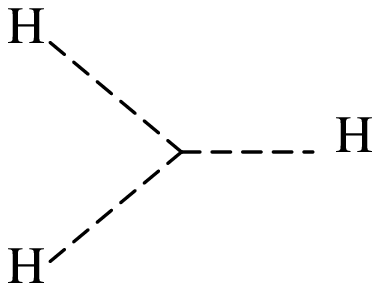}\parbox[b]{4.truecm}
{{\large $=-3i\frac{M_H^2}{v}$}\vspace{0.75truecm}}
\end{minipage}&
\begin{minipage}{0.5\linewidth}
\includegraphics[scale=0.6]{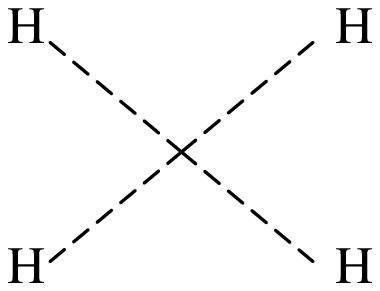}\parbox[b]{4.truecm}
{{\large $=-3i\frac{M_H^2}{v^2}$}\vspace{0.75truecm}}
\end{minipage}
\end{tabular}
\vspace{0.5truecm}

Furthermore, some of the terms that we omitted in
Eq.~(\ref{eq:SM_gauge_boson_mass_terms}), the terms linear in the
gauge bosons $W^\pm_\mu$ and $Z^0_\mu$, define the coupling
of the SM Higgs boson to the weak gauge fields:

\vspace{0.5truecm}
\begin{tabular}{cc}
\begin{minipage}{0.5\linewidth}
\includegraphics[scale=0.6]{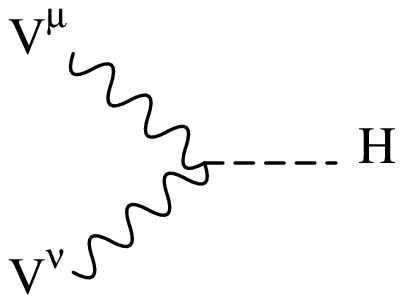}\parbox[b]{4.5truecm}
{{\large $=2i\frac{M_V^2}{v}g^{\mu\nu}$}\vspace{0.75truecm}}
\end{minipage}&
\begin{minipage}{0.5\linewidth}
\includegraphics[scale=0.6]{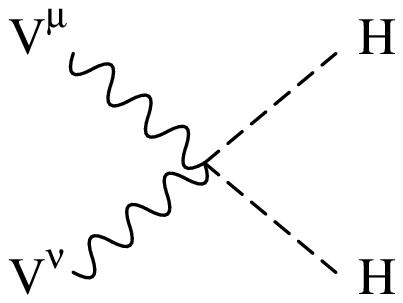}\parbox[b]{4.5truecm}
{{\large $=2i\frac{M_V^2}{v^2}g^{\mu\nu}$}\vspace{0.75truecm}}
\end{minipage}
\end{tabular}
\vspace{0.5truecm}

We notice that the couplings of the Higgs boson to the gauge fields
are proportional to their mass. Therefore $H$ does not couple to the
photon at tree level. It is important, however, to observe that
couplings that are absent at tree level may be induced at higher order
in the gauge couplings by loop corrections. Particularly relevant to
the SM Higgs boson phenomenology that will be discussed in
Section~\ref{sec:pheno} are the couplings of the SM Higgs boson to
pairs of photons, and to a photon and a $Z^0_\mu$ weak boson:

\begin{center}
\vspace{0.5truecm}
\begin{minipage}{0.5\linewidth}
\includegraphics[scale=0.6]{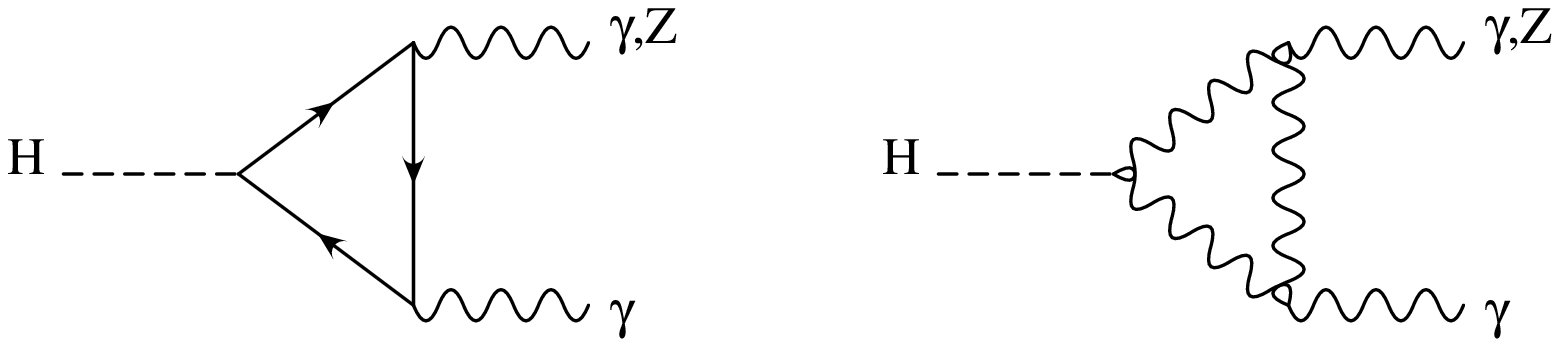}
\end{minipage}
\vspace{0.5truecm}
\end{center}

\noindent
as well as the coupling to pairs of gluons, when the SM Lagrangian is
extended through the QCD Lagrangian to include also the strong
interactions:

\begin{center}
\vspace{0.5truecm}
\begin{minipage}{0.5\linewidth}
\includegraphics[scale=0.6]{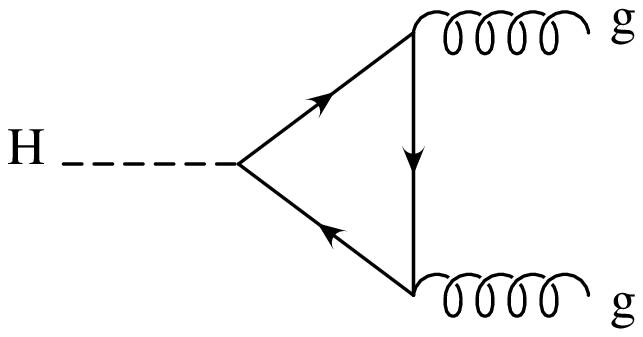}
\end{minipage}
\vspace{0.5truecm}
\end{center}

The analytical expressions for the $H\gamma\gamma$, $H\gamma Z$, and
$Hgg$ one-loop vertices are more involved and will be given in
Section~\ref{subsec:sm_higgs_branching_ratios}. As far as the Higgs
boson tree level couplings go, we observe that they are all expressed
in terms of just two parameters, either $\lambda$ and $\mu$ appearing
in the scalar potential of $\mathcal{L}_\phi$ (see Eq.~\ref{eq:L_phi_sm})) 
or, equivalently, $M_H$
and $v$, the Higgs boson mass and the scalar field vacuum expectation
value. Since $v$ is measured in muon decay to be
$v\!=\!(\sqrt{2}G_F)^{-1/2}\!=\!246$~GeV, the physics of the SM Higgs
boson is actually just function of its mass $M_H$.

The Standard Model gauge symmetry also forbids explicit mass terms for
the fermionic degrees of freedom of the Lagrangian. The fermion mass
terms are then generated via gauge invariant renormalizable Yukawa 
couplings to the scalar field $\phi$:
\begin{equation}
\label{eq:yukawa_lagrangian}
\mathcal{L}_\mathit{Yukawa}=-\Gamma_u^{ij}\bar{Q}^i_L\phi^c u^j_R
            -\Gamma_d^{ij}\bar{Q}^i_L\phi d^j_R
            -\Gamma_e^{ij}\bar{L}^i_L\phi l^j_R +h.c.
\end{equation}
where $\phi^c\!=\!-i\sigma^2\phi^\star$, and $\Gamma_f$ ($f=u,d,l$)
are matrices of couplings arbitrarily introduced to realize the Yukawa
coupling between the field $\phi$ and the fermionic fields of the SM.
$Q_L^i$ and $L_L^i$ (where $i=1,2,3$ is a generation index) represent
quark and lepton left handed doublets of $SU(2)_L$, while $u_R^i$,
$d_R^i$ and $l_R^i$ are the corresponding right handed singlets. When
the scalar fields $\phi$ acquires a non zero vacuum expectation value
through spontaneous symmetry breaking, each fermionic degree of
freedom coupled to $\phi$ develops a mass term with mass parameter
\begin{equation}
\label{eq:yukawa_coupling}
m_f=\Gamma_f\frac{v}{\sqrt{2}}\,\,\,,
\end{equation}
where the process of diagonalization from the current eigenstates in
Eq.~(\ref{eq:yukawa_lagrangian}) to the corresponding mass eigenstates
is understood, and $\Gamma_f$ are therefore the elements of the
diagonalized Yukawa matrices corresponding to a given fermion $f$. The
Yukawa couplings of the $f$ fermion to the Higgs boson ($y_f$) is
proportional to $\Gamma_f$:

\begin{center}
\vspace{0.5truecm}
\begin{minipage}{0.8\linewidth}
\includegraphics[scale=0.6]{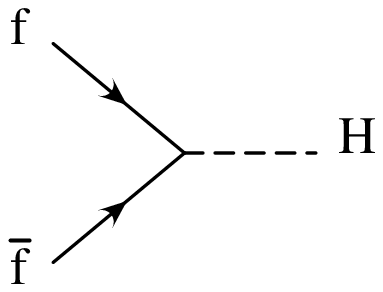}
\parbox[b]{6.truecm}
{{\large $=-i\frac{m_f}{v}=-i\frac{\Gamma_f}{\sqrt{2}}=-iy_f$}\vspace{0.75truecm}}
\end{minipage} 
\vspace{0.5truecm}
\end{center}

As long as the origin of fermion masses is not better understood in
some more general context beyond the Standard Model, the Yukawa
couplings $y_f$ represent free parameter of the SM Lagrangian. The
mechanism through which fermion masses are generated in the Standard
Model, although related to the mechanism of spontaneous symmetry
breaking, requires therefore further assumptions and involves a larger
degree of arbitrariness as compared to the gauge boson sector of the
theory.

\subsection{Theoretical constraints on the Standard Model Higgs boson mass}
\label{subsec:higgs_sm_constraints}

Several issues arising in the scalar sector of the Standard Model
link the mass of the Higgs boson to the energy scale where the
validity of the Standard Model is expected to fail. Below that scale,
the Standard Model is the extremely successful effective field theory
that emerges from the electroweak precision tests of the last
decades. Above that scale, the Standard Model has to be embedded into
some more general theory that gives origin to a wealth of new physics
phenomena. From this point of view, the Higgs sector of the Standard
Model contains actually two parameters, the Higgs mass ($M_H$) and the
scale of new physics ($\Lambda$).

In this Section we will review the most important theoretical
constraints that are imposed on the mass of the Standard Model Higgs
boson by the consistency of the theory up to a given energy scale
$\Lambda$. In particular we will touch on issues of unitarity,
triviality, vacuum stability, fine tuning and, finally, electroweak
precision measurements.

\subsubsection{Unitarity}
\label{subsubsec:unitarity}
The scattering amplitudes for longitudinal gauge bosons
($V_LV_L\rightarrow V_LV_L$, where $V=W^\pm,Z^0$) grow as the square
of the Higgs boson mass. This is easy to calculate using the
\emph{electroweak equivalence
theorem}~\cite{Peskin:1995ev,Weinberg:1995mt}, valid in the high
energy limit (i.e. for energies $s\!=\!Q^2\!\gg\! M_V^2$), according
to which the scattering amplitudes for longitudinal gauge bosons can
be expressed in terms of the scattering amplitudes for the
corresponding Goldstone bosons, i.e.:
\begin{equation}
\label{eq:equivalence_theorem}
\mathcal{A}(V_L^1\ldots V_L^n\rightarrow V_L^1\ldots V_L^m)=(i)^n(-i)^m
\mathcal{A}(\omega^1\dots\omega^n\rightarrow\omega^1\ldots\omega^m)+
O\left(\frac{M_V^2}{s}\right)\,\,\,,
\end{equation}
where we have indicated by $\omega^i$ the Goldstone boson associated
to the longitudinal component of the gauge boson $V^i$.
For instance, in the high energy limit, the scattering amplitude
for $W^+_LW^-_L\rightarrow W^+_LW^-_L$ satisfies:
\begin{equation}
\label{eq:ampl_wl}
\mathcal{A}(W_L^+ W_L^-\rightarrow W_L^+ W_L^-)=
\mathcal{A}(\omega^+\omega^-\rightarrow\omega^+\omega^-)+O\left(\frac{M_W^2}{s}\right)\,\,\,,
\end{equation}
where
\begin{equation}
\label{eq:ampl_omega}
\mathcal{A}(\omega^+\omega^-\rightarrow\omega^+\omega^-)=
-\frac{M_H^2}{v^2}\left(\frac{s}{s-M_H^2}+\frac{t}{t-M_H^2}\right)\,\,\,.
\end{equation}
Using a partial wave decomposition, we can also write $\mathcal A$ as:
\begin{equation}
\label{eq:ampl_partial_waves}
\mathcal{A}=16\pi\sum_{l=0}^{\infty}(2l+1)P_l(\cos\theta)a_l\,\,\,,
\end{equation}
where $a_l$ is the spin $l$ partial wave and $P_l(\cos\theta)$ are the
Legendre polynomials. In terms of partial wave amplitudes $a_l$, the
scattering cross section corresponding to $\mathcal{A}$ can be
calculated to be:
\begin{equation}
\label{eq:sigma_partial_waves}
\sigma=\frac{16\pi}{s}\sum_{l=0}^{\infty}(2l+1)|a_l|^2\,\,\,,
\end{equation}
where we have used the orthogonality of the Legendre polynomials.
Using the optical theorem, we can impose the unitarity constraint by
writing that:
\begin{equation}
\label{eq:optical_theorem}
\sigma=\frac{16\pi}{s}\sum_{l=0}^{\infty}(2l+1)|a_l|^2=
\frac{1}{s}\mathrm{Im}\left[\mathcal{A}(\theta=0)\right]\,\,\,,
\end{equation}
where $\mathcal{A}(\theta=0)$ indicates the scattering amplitude in
the forward direction. This implies that:
\begin{equation}
\label{eq:unitarity_bound}
|a_l|^2=\mathrm{Re}(a_l)^2+\mathrm{Im}(a_l)^2=\mathrm{Im}(a_l)\,\,\,
\longrightarrow\,\,\,|\mathrm{Re}(a_l)|\le\frac{1}{2}\,\,\,.
\end{equation}
Via Eq.~(\ref{eq:unitarity_bound}), different $a_l$ amplitudes can
than provide constraints on $M_H$. As an example, let us consider the
$J\!=\!0$ partial wave amplitude $a_0$ for the $W^+_LW^-_L\rightarrow
W^+_LW^-_L$ scattering we introduced above:
\begin{equation}
\label{eq:a0_wlwlwlwl}
a_0=\frac{1}{16\pi s}\int_{-s}^0\mathcal{A}\,dt = 
-\frac{M_H^2}{16\pi v^2}\left[2+\frac{M_H^2}{s-M_H^2}-
\frac{M_H^2}{s}\log\left(1+\frac{s}{M_H^2}\right)\right]\,\,\,.
\end{equation}
In the high energy limit ($M_H^2\ll s$), $a_0$ reduces to:
\begin{equation}
\label{eq:he_limit}
a_0\stackrel{M_H^2\ll s}{\longrightarrow}-\frac{M_H^2}{8\pi v^2}\,\,\,,
\end{equation}
from which, using Eq.~(\ref{eq:unitarity_bound}), one gets:
\begin{equation}
\label{eq:mh_bound_wlwlwlwl}
M_H< 870 \,\,\mbox{GeV}\,\,\,.
\end{equation}
Other more constraining relations can be obtained from different
longitudinal gauge boson scattering amplitudes. For instance,
considering the coupled channels like  $W^+_LW^-_L\rightarrow Z_LZ_L$, 
one can lower the bound to:
\begin{equation}
\label{eq:mh_bound_wlwlzlzl}
M_H<710\,\,\mbox{GeV}\,\,\,.
\end{equation}
Taking a different point of view, we can observe that if there is no Higgs
boson, or equivalently if $M_H^2\gg s$, Eq.~(\ref{eq:unitarity_bound})
gives indications on the critical scale $\sqrt{s_c}$ above which new physics 
should be expected. Indeed, considering again  
$W^+_LW^-_L\rightarrow W^+_LW^-_L$ scattering, we see that:
\begin{equation}
\label{eq:a0_no_higgs}
a_0(\omega^+\omega^-\rightarrow\omega^+\omega^-)
\stackrel{M_H^2\gg s}{\longrightarrow}-\frac{s}{32\pi v^2}\,\,\,,
\end{equation}
from which, using Eq.~(\ref{eq:unitarity_bound}), we get:
\begin{equation}
\label{eq:sc_upper_bound_wlwlwlwlw}
\sqrt{s_c}< 1.8\,\,\mbox{TeV}\,\,\,.
\end{equation}
Using more constraining channels the bound can be reduced to:
\begin{equation}
\label{eq:sc_upper_bound_best}
\sqrt{s_c}< 1.2\,\,\mbox{TeV}\,\,\,.
\end{equation}
This is very suggestive: it tells us that new physics ought to be
found around 1-2 TeV, i.e. exactly in the range of energies that will
be explored by the Tevatron and the Large Hadron Collider.

\subsubsection{Triviality and vacuum stability}
\label{subsubsec:triviality_vacuumstability}

The argument of triviality in a $\lambda\phi^4$ theory goes as
follows. The dependence of the quartic coupling $\lambda$ on the
energy scale ($Q$) is regulated by the renormalization group equation
\begin{equation}
\label{eq:phi4_lambda_rge}
\frac{d\lambda(Q)}{dQ^2}=\frac{3}{4\pi^2}\lambda^2(Q)\,\,\,.
\end{equation}
This equation states that the quartic coupling $\lambda$ decreases for
small energies and increases for large energies. Therefore, in the low
energy regime the coupling vanishes and the theory becomes trivial,
i.e. non-interactive. In the large energy regime, on the other hand,
the theory becomes non-perturbative, since $\lambda$ grows, and it can
remain perturbative only if $\lambda$ is set to zero, i.e. only if the
theory is made trivial.

The situation in the Standard Model is more complicated, since the
running of $\lambda$ is governed by more interactions. Including the
lowest orders in all the relevant couplings, we can write the equation
for the running of $\lambda(Q)$ with the energy scale as follows:
\begin{equation}
\label{eq:sm_lambda_rge}
32\pi^2\frac{d\lambda}{dt}=
24\lambda^2-(3g^{\prime 2}+9g^2-24y_t^2)\lambda
+\frac{3}{8}g^{\prime 4}+\frac{3}{4}g^{\prime 2}g^2+\frac{9}{8}g^4
-24y_t^4+\cdots
\end{equation}
where $t\!=\!\ln(Q^2/Q_0^2)$ is the logarithm of the ratio of the
energy scale and some reference scale $Q_0$ square, $y_t\!=\!m_t/v$ is
the top-quark Yukawa coupling, and the dots indicate the presence of
higher order terms that have been omitted. We see that when $M_H$
becomes large, $\lambda$ also increases (since $M_H^2\!=\!2\lambda
v^2$) and the first term in Eq.~(\ref{eq:sm_lambda_rge})
dominates. The evolution equation for $\lambda$ can then be easily
solved and gives:
\begin{equation}
\label{eq:lambda_sm_large_mh}
\lambda(Q)=\frac{\lambda(Q_0)}{1-\frac{3}{4\pi^2}\lambda(Q_0)
\ln\left(\frac{Q^2}{Q_0^2}\right)}\,\,\,.
\end{equation}
When the energy scale $Q$ grows, the
denominator in Eq.~(\ref{eq:lambda_sm_large_mh}) may vanish, in which
case $\lambda(Q)$ hits a pole, becomes infinite, and a triviality
condition needs to be imposed. This is avoided imposing that the
denominator in Eq.~(\ref{eq:lambda_sm_large_mh}) never vanishes, i.e. 
that $\lambda(Q)$ is always finite
and $1/\lambda(Q)>0$. This condition gives an explicit upper bound on
$M_H$:
\begin{equation}
M_H^2<\frac{8\pi^2v^2}{3\log\left(\frac{\Lambda^2}{v^2}\right)}\,\,\,,
\end{equation}
obtained from Eq.~(\ref{eq:lambda_sm_large_mh}) by setting
$Q\!=\!\Lambda$, the scale of new physics, and $Q_0\!=\!v$, the
electroweak scale.

On the other hand, for small $M_H$ , i.e. for small $\lambda$, the
last term in Eq.~(\ref{eq:sm_lambda_rge}) dominates and the
evolution of $\lambda(Q)$ looks like:
\begin{equation}
\lambda(\Lambda)=\lambda(v)-\frac{3}{4\pi^2}y_t^2
\log\left(\frac{\Lambda^2}{v^2}\right)\,\,\,.
\end{equation}
To assure the stability of the vacuum state of the theory we need to
require that $\lambda(\Lambda)\!>\!0$ and this gives a lower bound for
$M_H$:
\begin{equation}
\lambda(\Lambda)>0 \,\,\,\longrightarrow \,\,\,
M_H^2>\frac{3v^2}{2\pi^2}y_t^2\log\left(\frac{\Lambda^2}{v^2}\right)\,\,\,.
\end{equation}
More accurate analyses include higher order quantum correction in the
scalar potential and use a 2-loop renormalization group improved
effective potential, $V_{eff}$, whose nature and meaning has been
briefly sketched in Section~\ref{subsec:higgs_mechanism}.

\subsubsection{Indirect bounds from electroweak precision measurements}
\label{subsubsec:indirect_bound}

Once a Higgs field is introduced in the Standard Model, its virtual
excitations contribute to several physical observables, from the mass
of the $W$ boson, to various leptonic and hadronic asymmetries, to
many other electroweak observables that are usually considered in
precision tests of the Standard Model.  Since the Higgs boson mass is
the only parameter in the Standard Model that is not directly
determined either theoretically or experimentally, it can be extracted
indirectly from precision fits of all the measured electroweak
observables, within the fit uncertainty. This is actually one of the
most important results that can be obtained from precision tests of
the Standard Model and greatly illustrates the predictivity of the
Standard Model itself.  All available studies can be found on the LEP
Electroweak Working Group and on the LEP Higgs Working Group Web
pages~\cite{lepewwg:wp,lephwg:wp} as well as in their main
publications
\cite{lepewwg:2005di,ewreport:2005em,lepewwg:2004qh,lephwg:2001xx,Barate:2003sz}. 
An excellent recent series of lectures on the subject of
\emph{Precision Electroweak Physics} is also available from a previous
TASI school
\cite{Matchev:2004yw}.

\begin{figure}
\centering
\includegraphics[scale=0.4]{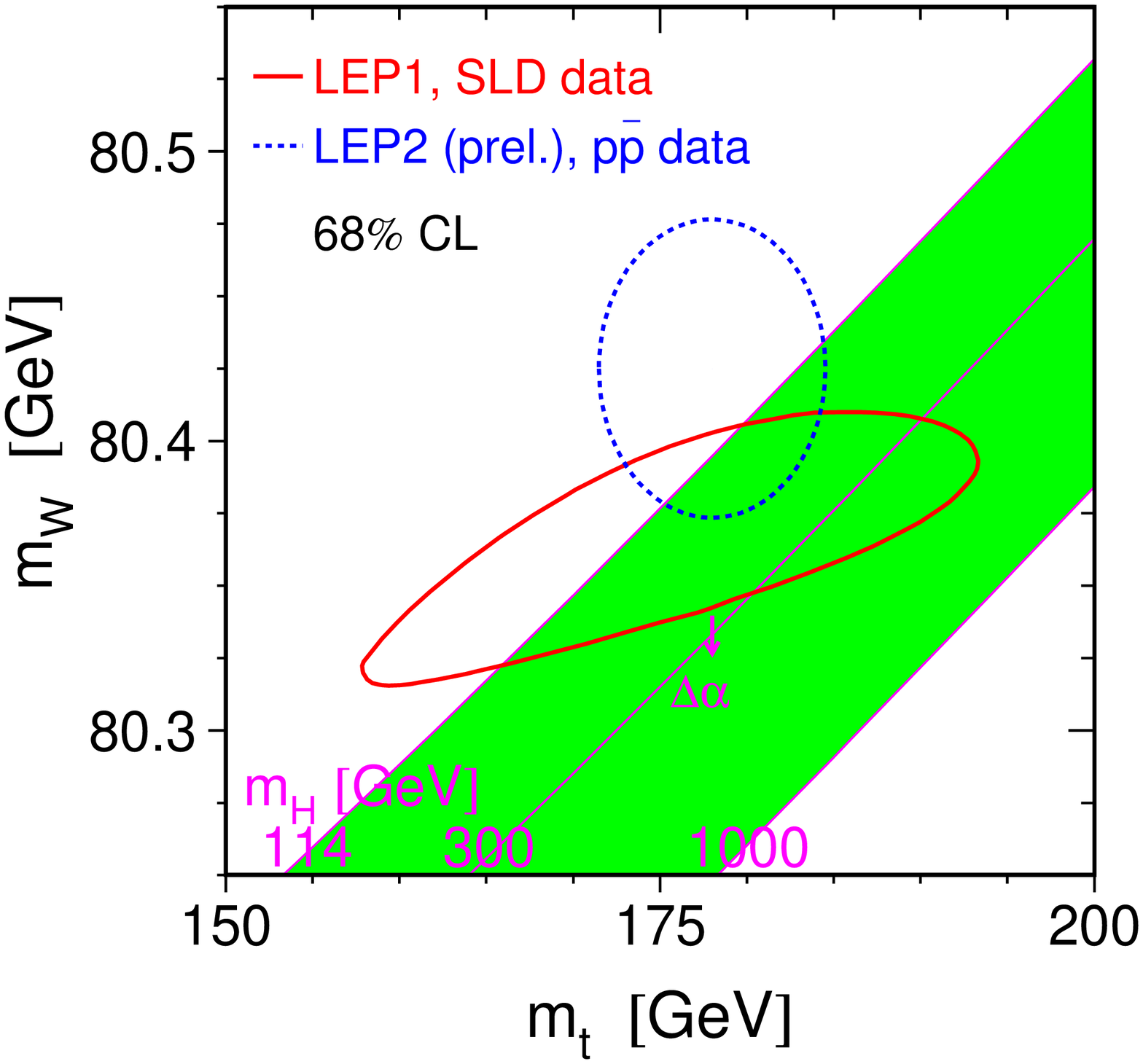}
\includegraphics[scale=0.4]{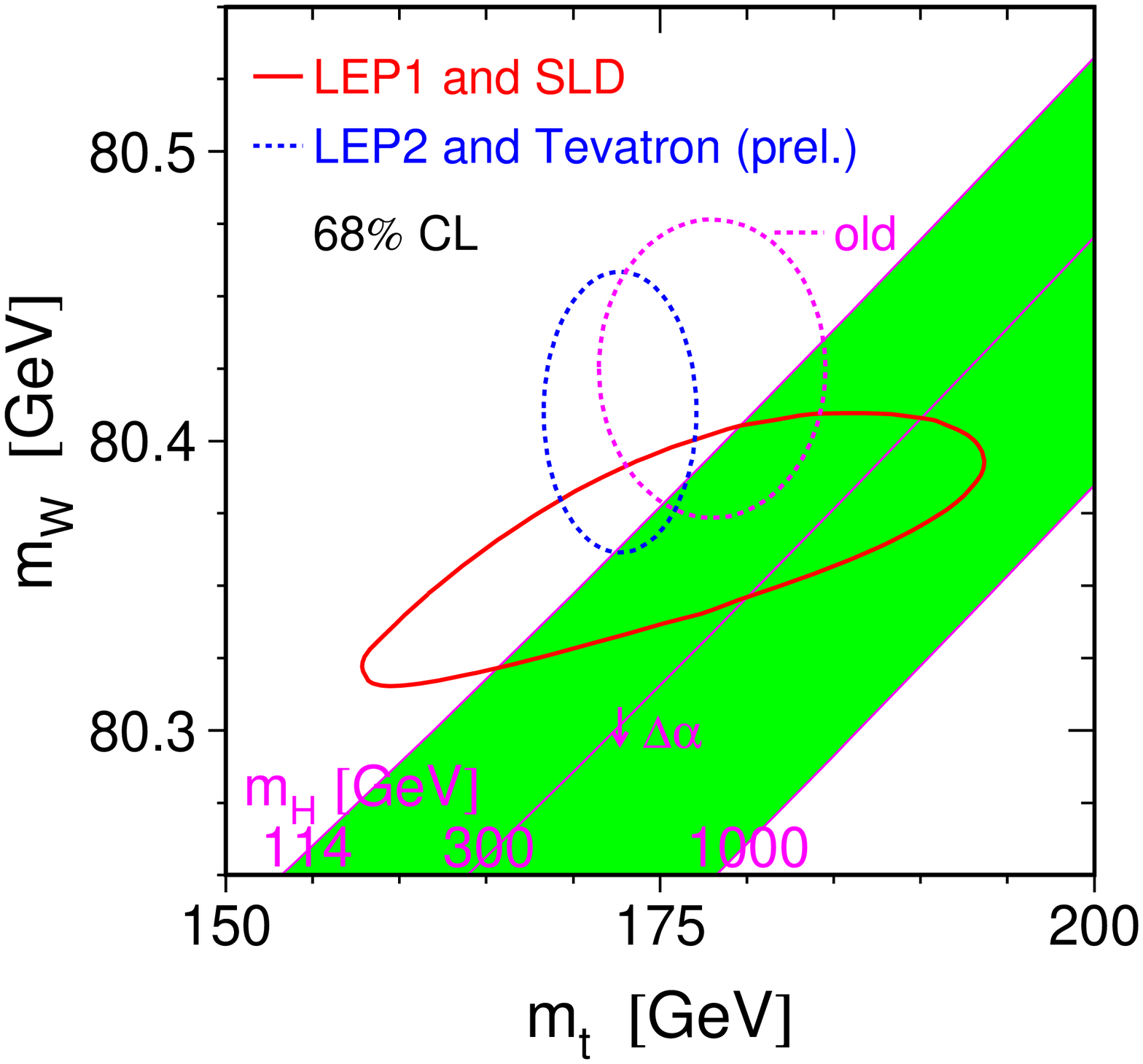}
\caption[]{Comparison of the indirect measurements of $M_W$ and $m_t$
(LEP I+SLD data) (solid contour) and the direct measurement
($p\bar{p}$ colliders and LEP II data) (dashed contour). In both cases
the 68\% CL contours are plotted. Also shown is the SM relationship
for these masses as a function of the Higgs boson mass, $m_H$. The
arrow labeled $\Delta\alpha$ shows the variation of this relation if
$\alpha(M_Z^2)$ is varied by one standard deviation. The left hand
side plot is from Ref.~\cite{ewreport:2005em} and corresponds to the
Winter 2005 situation (see Eq.~\ref{eq:mh_indirect_w05}), the right
hand side plot is from the Ref.~\cite{lepewwg:2005di} and corresponds
to the Summer 2005 situation (see Eq.~\ref{eq:mh_indirect_s05}). The
comparison between Winter and Summer 2005 is shown in the right hand
side plot.\label{fig:contour_mw_mt}}
\end{figure}
\begin{figure}
\centering
\includegraphics[scale=0.4]{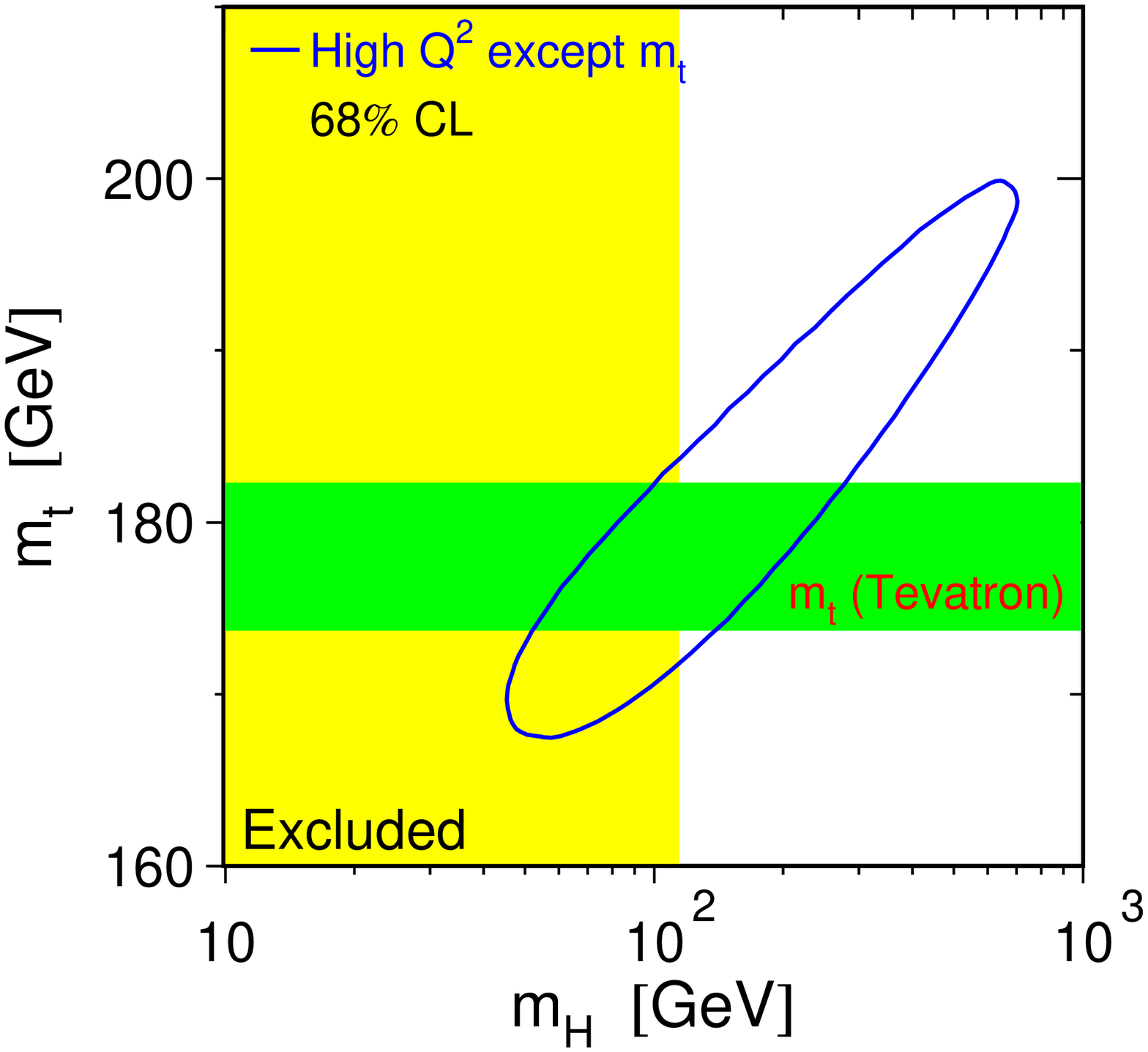}
\caption[]{The 68\% confidence level contour in $m_t$ and $M_H$ for
the fit to all data except the direct measurement of $m_t$, indicated
by the shaded horizontal band of $\pm 1\sigma$ width. The vertical
band shows the 95\% CL exclusion limit on $M_H$ from direct
searches. From Ref.~\cite{ewreport:2005em}.\label{fig:contour_mh_mt}}
\end{figure}
The correlation between the Higgs boson mass $M_H$, the $W$ boson mass
$M_W$, the top-quark mass $m_t$, and the precision data is illustrated
in Figs.~\ref{fig:contour_mw_mt} and \ref{fig:contour_mh_mt}. Apart
from the impressive agreement existing between the indirect
determination of $M_W$ and $m_t$ and their experimental measurements
we see in Fig.~\ref{fig:contour_mw_mt} that the 68\% CL contours from
LEP, SLD, and Tevatron measurements select a SM Higgs boson mass
region roughly below 200~GeV. Therefore, assuming no physics beyond
the Standard Model at the weak scale, all available electroweak
precision data are consistent with a light Higgs boson.

\begin{figure}
\centering
\includegraphics[scale=0.4]{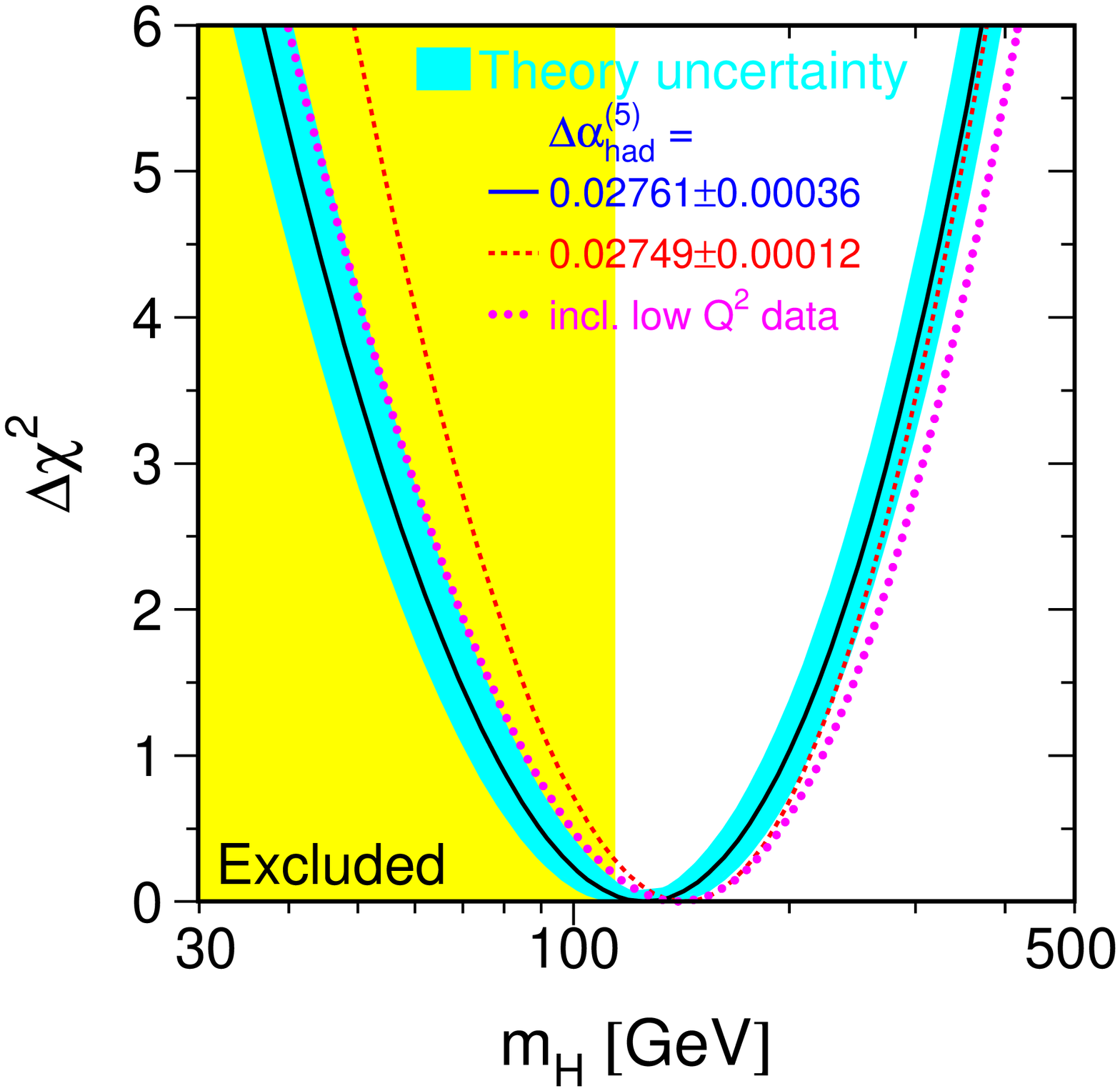}
\includegraphics[scale=0.4]{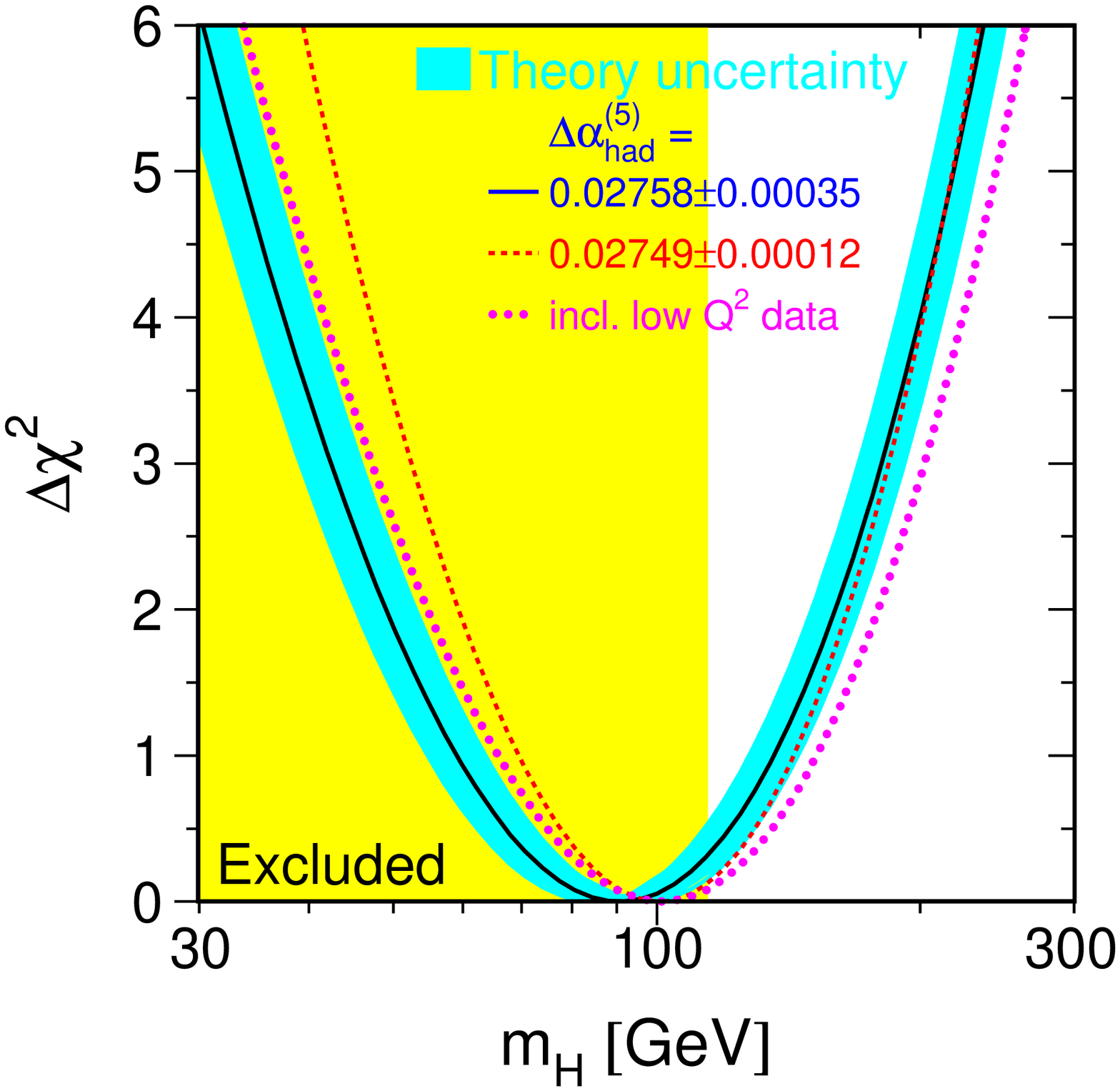}
\caption[]{$\Delta\chi^2=\chi^2-\chi^2_{min}$ \emph{vs.} $M_H$
curve. The line is the result of the fit using all electroweak data;
the band represents an estimate of the theoretical error due to
missing higher order corrections. The vertical band shows the $95\%$
CL exclusion limit on $M_H$ from direct searches. The solid and dashed
curves are derived using different evaluations of
$\Delta\alpha_{had}^{(5)}(M_Z^2)$. The dotted curve includes low $Q^2$
data. The left hand side plot is from Ref.~\cite{ewreport:2005em} and
corresponds to the Winter 2005 situation (see
Eq.~\ref{eq:mh_indirect_w05}), the right hand side plot is from the
Ref.~\cite{lepewwg:2005di} and corresponds to the Summer 2005
situation (see Eq.~\ref{eq:mh_indirect_s05}). Note the different
horizontal scale in the two plots.\label{fig:blue_band_mh}}
\end{figure}
The actual value of $M_H$ emerging from the electroweak precision fits
strongly depends on theoretical predictions of physical observables
that include different orders of strong and electroweak
corrections. As an example, in Fig.~\ref{fig:contour_mw_mt} the
magenta arrow shows how the yellow band would move for one standard
deviation variation in the QED fine-structure constant
$\alpha(m_Z^2)$. It also depends on the fit input parameters. As we
see in Fig.~\ref{fig:contour_mh_mt}, $M_H$ grows for larger $m_t$. The
sensitivity of the indirect bound on $M_H$ to $m_t$ is clearly visible
both in Fig.~\ref{fig:contour_mw_mt} and in
Fig.~\ref{fig:blue_band_mh},  where you can find the famous \emph{blue band} plot.
In both figures, we compare the results of the Winter 2005 and Summer 2005
electroweak precision fits. As far as the Higgs boson mass goes, the
main change between Winter and Summer 2005 has been the value of the
top-quark mass. We go from:
\begin{equation}
\label{eq:mh_indirect_w05}
\left\{
\begin{array}{l}
M_H=117^{+67}_{-45}\,\,\,\mbox{GeV}\\
M_H<251\,\,\,\mbox{GeV}\,\,\,(95\%\,\,\mbox{CL})
\end{array}
\right.
\,\,\,\,\,\mbox{for}\,\,\,\,\,
m_t\!=\!178.\pm 4.3\,\,\,\mbox{GeV}\,\,\,.
\end{equation}
in Winter 2005~\cite{ewreport:2005em}, to: 
\begin{equation}
\label{eq:mh_indirect_s05}
\left\{
\begin{array}{l}
M_H=91^{+ 45}_{-32}\,\,\,\mbox{GeV}\\
M_H<186-219\,\,\,\mbox{GeV}\,\,\,(95\%\,\,\mbox{CL})
\end{array}
\right.
\,\,\,\,\,\mbox{for}\,\,\,\,\,
m_t\!=\!172.7\pm 2.9\,\,\,\mbox{GeV}\,\,\,,
\end{equation}
in Summer 2005~\cite{lepewwg:2005di}. We see in Fig.~\ref{fig:contour_mw_mt} that the
overlap between the direct and indirect determination of $M_H$ is
greatly reduced when the value of $m_t$ decreases and at the same time
the minimum of the $\Delta\chi^2$ band in Fig.~\ref{fig:blue_band_mh}
considerably shifts. While in the first case the electroweak precision
fits are still largely compatible with the 
direct searches at LEP II that have placed a 95\% CL lower bound on
$M_H$ at:
\begin{equation}
\label{eq:mh_lower_bound}
M_H>114.4\,\,\,\mbox{GeV}\,\,\,,
\end{equation}
in the second case a large region of the $\Delta\chi^2$ band in
Fig.~\ref{fig:blue_band_mh}, in particular the region about the
minimum, is already excluded, and values of $M_H$ very close to the
experimental lower bound seem to be favored.

\begin{figure}
\centering
\includegraphics[scale=0.4]{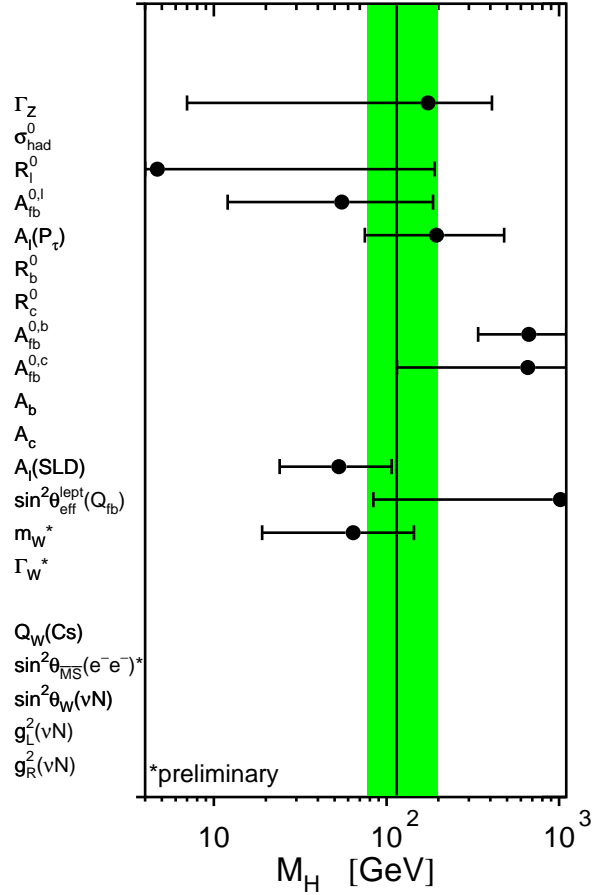}
\caption[]{Preferred range for the SM Higgs boson mass $M_H$ as determined
from various electroweak observables. The shaded band shows the
overall constraint on the mass of the Higgs boson as derived from the
full data set. From
Ref.~\cite{ewreport:2005em}.\label{fig:green_band_mh}}
\end{figure}
It is fair to conclude that the issue of constraining $M_H$ from
electroweak precision fits is open to controversies and, at a closer
look, emerges as a not clear cut statement. With this respect,
Fig.~\ref{fig:green_band_mh} illustrates the sensitivity of a few
selected electroweak observables to the Higgs boson mass as well as
the preferred range for the SM Higgs boson mass as determined from all
electroweak observables .  One can observe that $M_W$ and the leptonic
asymmetries prefer a lighter Higgs boson, while $A_{FB}^{b,c}$ and the
NuTeV determination of $\sin^2\theta_W$ prefer a heavier Higgs
boson. A certain
\emph{tension} is still present in the data.  We could just think that
things will progressively adjust and, after the discovery of a light
Higgs boson at either the Tevatron or the LHC, this will result in yet
another amazing success of the Standard Model. Or, one can interpret
the situation depicted in Fig.~\ref{fig:green_band_mh} as an
unavoidable indication of the presence of new physics beyond the
Standard Model. Indeed, since
the data compatible with a lighter Higgs boson are very solid, one
could either interpret the data compatible with a larger value of
$M_H$ as an indication of new physics beyond the Standard Model, or
one could drop them as wrong, and still, the Higgs boson mass would
turn out to be so small not to be compatible anymore with the Standard
Model, signaling once more the presence of new physics.

\subsubsection{Fine-tuning}
\label{subsubsec:fine_tuning}
One aspect of the Higgs sector of the Standard Model that is
traditionally perceived as problematic is that higher order
corrections to the Higgs boson mass parameter square contain quadratic
ultraviolet divergences. This is expected in a $\lambda\phi^4$ theory
and it does not pose a renormalizability problem, since a
$\lambda\phi^4$ theory is renormalizable. However, although per se
renormalizable, these quadratic divergences leave the \emph{inelegant}
feature that the Higgs boson renormalized mass square has to result
from the
\emph{adjusted} or \emph{fine-tuned} balance between a bare Higgs
boson mass square and a counterterm that is proportional to the
ultraviolet cutoff square. If the physical Higgs mass has to live at
the electroweak scale, this can cause a fine-tuning of several orders
of magnitude when the scale of new physics $\Lambda$ (the ultraviolet
cutoff of the Standard Model interpreted as an effective low energy
theory) is well above the electroweak scale. Ultimately this is
related to a symmetry principle, or better to the absence of a
symmetry principle. Indeed, setting to zero the mass of the scalar
fields in the Lagrangian of the Standard Model does not restore any
symmetry to the model. Hence, the mass of the scalar fields are not
protected against large corrections.

Models of new physics beyond the Standard Model should address this
fine-tuning problem and propose a more satisfactory mechanism to
obtain the mass of the Higgs particle(s) around the electroweak
scale. Supersymmetric models, for instance, have the remarkable
feature that fermionic and bosonic degrees of freedom conspire to
cancel the Higgs mass quadratic loop divergence, when the symmetry is
exact. Other non supersymmetric models, like little Higgs models,
address the problem differently, by interpreting the Higgs boson as a
Goldstone boson of some global approximate symmetry. In both cases the
Higgs mass turns out to be proportional to some small deviation from
an exact symmetry principle, and therefore intrinsically small.

\begin{figure}
\includegraphics[scale=.5]{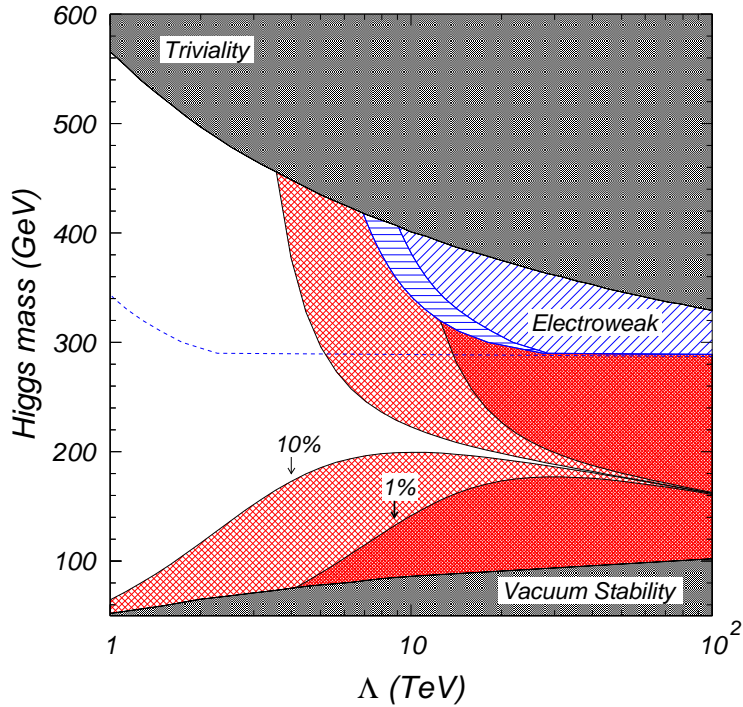}
\caption[]{The SM Higgs boson mass $M_H$ as a function of the scale of
new physics $\Lambda$, with all the constraints derived from
unitarity, triviality, vacuum stability, electroweak precision fits,
and the requirement of a limited fine-tuning. The empty region is
consistent with all the constraints and less than 1 part in 10
fine-tuning. From Ref.~\cite{Kolda:2000wi}.\label{fig:mh_vs_lambda}}
\end{figure}
As suggested in Ref.~\cite{Kolda:2000wi}, the \emph{no fine-tuning}
condition in the Standard Model can be softened and translated into a
\emph{maximum amount of allowed fine-tuning}, that can be directly
related to the scale of new physics. As derived in
Section~\ref{subsec:higgs_mechanism}, upon spontaneous breaking of the
electroweak symmetry, the SM Higgs boson mass at tree level is given
by $M_H^2\!=\!-2\mu^2$, where $\mu^2$ is the coefficient of the
quadratic term in the scalar potential. Higher order corrections to
$M_H^2$ can therefore be calculated as loop corrections to $\mu^2$,
i.e. by studying how the effective potential in
Eq.~(\ref{eq:veff_exp}) and its minimum condition are modified by loop
corrections.  If we interpret the Standard Model as the electroweak
scale effective limit of a more general theory living at a high scale
$\Lambda$, then the most general form of $\mu^2$ including all loop
corrections is:
\begin{equation}
\label{eq:mu_renormalized}
\bar\mu^2=\mu^2+\Lambda^2\sum_{n=0}^{\infty}c_n(\lambda_i)\log^n(\Lambda/Q)
\,\,\,,
\end{equation}
where $Q$ is the renormalization scale, $\lambda_i$ are a set of input
parameters (couplings) and the $c_n$ coefficients can be deduced from
the calculation of the effective potential at each loop order. As
noted originally by Veltman, there would be no fine-tuning problem if
the coefficient of $\Lambda^2$ in Eq.~(\ref{eq:mu_renormalized}) were
zero, i.e. if the loop corrections to $\mu^2$ had to vanish. This
condition, known as \emph{Veltman condition}, is usually over constraining,
since the number of independent $c_n$ (set to zero by the Veltman
condition) can be larger than the number of inputs
$\lambda_i$. However the Veltman condition can be relaxed, by
requiring that only the sum of a finite number of terms in the
coefficient of $\Lambda^2$ is zero, i.e. requiring that:
\begin{equation}
\label{eq:milder_veltman_condition}
\sum_0^{n_{max}}c_n(\lambda_i)\log^n(\Lambda/M_H)=0\,\,\,,
\end{equation}
where the renormalization scale $\mu$ has been arbitrarily set to
$M_H$ and the order $n$ has been set to $n_{max}$, fixed by the
required order of loop in the calculation of $V_{eff}$. This is based
on the fact that higher orders in $n$ come from higher loop effects
and are therefore suppressed by powers of $(16\pi^2)^{-1}$. Limiting
$n$ to $n_{max}$, Eq.~(\ref{eq:milder_veltman_condition}) can now have
a solution. Indeed, if the scale of new physics $\Lambda$ is not too
far from the electroweak scale, then the Veltman condition in
Eq.~(\ref{eq:milder_veltman_condition}) can be softened even more by
requiring that:
\begin{equation}
\label{eq:mildest_veltman_condition}
\sum_0^{n_{max}}c_n(\lambda_i)\log^n(\Lambda/M_H)<\frac{v^2}{\Lambda^2}
\,\,\,.
\end{equation}
This condition determines a value of $\Lambda_{max}$ such that for
$\Lambda\le\Lambda_{max}$ the stability of the electroweak scale does
not require any dramatic cancellation in $\bar{\mu}^2$. In other
words, for $\Lambda\le\Lambda_{max}$ the renormalization of the SM
Higgs boson mass does not require any fine-tuning. As an example, for
$n_{max}\!=\!0$, $c_0\!=\!(32\pi^2v^2)^{-1}
3(2M_W^2+M_Z^2+M_H^2-4m_t^2)$, and the stability of the electroweak
scale is assured up to $\Lambda$ of the order of $4\pi v\simeq 2$~TeV.
For $n_{max}\!=\!1$ the maximum $\Lambda$ is pushed up to
$\Lambda\simeq 15$~TeV and for $n_{max}\!=\!2$ up to $\Lambda\simeq
50$~TeV. So, just going up to 2-loops assures us that we can consider
the SM Higgs sector free of fine-tuning up to scales that are well
beyond where we would hope to soon discover new physics.

For each value of $n_{max}$, and for each corresponding
$\Lambda_{max}$, $M_H$ becomes a function of the cutoff $\Lambda$, and
the amount of fine-tuning allowed in the theory limits the region in
the $(\Lambda,M_H)$ plane allowed to $M_H(\Lambda)$. This is well
represented in Fig.~\ref{fig:mh_vs_lambda}, where also the constraint
from the conditions of unitarity (see
Section~\ref{subsubsec:unitarity}), triviality (see
Section~\ref{subsubsec:triviality_vacuumstability}), vacuum stability
(see Section~\ref{subsubsec:triviality_vacuumstability}) and
electroweak precision fits (see
Section~\ref{subsubsec:indirect_bound}) are summarized. Finally, the
main lesson we take away from this plot is that if a Higgs boson is
discovered new physics is just around the corner and should manifest
itself at the LHC.

\subsection{The Higgs sector of the Minimal Supersymmetric Standard Model}
\label{subsec:higgs_mssm}

In the supersymmetric extension of the Standard Model, the electroweak
symmetry is spontaneously broken via the Higgs mechanism introducing
two complex scalar $SU(2)_L$ doublets. The dynamics of the Higgs
mechanism goes pretty much unchanged with respect to the Standard
Model case, although the form of the scalar potential is more complex
and its minimization more involved. As a result, the $W^\pm$ and $Z^0$
weak gauge bosons acquire masses that depend on the parameterization
of the supersymmetric model at hand. At the same time, fermion masses
are generated by coupling the two scalar doublets to the fermions via
Yukawa interactions. A supersymmetric model is therefore a natural
reference to compare the Standard Model to, since it is a
theoretically sound extension of the Standard Model, still
fundamentally based on the same electroweak symmetry breaking
mechanism.

Far from being a simple generalization of the SM Higgs sector, the
scalar sector of a supersymmetric model can be theoretically more
satisfactory because: \emph{(i)} spontaneous symmetry breaking is
radiatively induced (\emph{i.e.} the sign of the quadratic term in the
Higgs potential is driven from positive to negative) mainly by the
evolution of the top-quark Yukawa coupling from the scale of
supersymmetry-breaking to the electroweak scale, and \emph{(ii)}
higher order corrections to the Higgs mass do not contain quadratic
divergences, since they cancel when the contribution of both scalars
and their super-partners is considered (see
Section~\ref{subsubsec:fine_tuning}).

At the same time, the fact of having a supersymmetric theory and two
scalar doublets modifies the phenomenological properties of the
supersymmetric physical scalar fields dramatically. In this Section we
will review only the most important properties of the Higgs sector of
the MSSM, so that in Section~\ref{sec:pheno} we can compare the
physics of the SM Higgs boson to that of the MSSM Higgs bosons.

I will start by recalling some general properties of a Two Higgs
Doublet Model in Section~\ref{subsubsec:2hdm}, and I will then specify
the discussion to the case of the MSSM in
Section~\ref{subsubsec:higgs_mssm}. In
Sections~\ref{subsubsec:mssm_higgs_couplings_bosons} and
\ref{subsubsec:mssm_higgs_couplings_fermions} I will review the form
of the couplings of the MSSM Higgs bosons to the SM gauge bosons and
fermions, including the impact of the most important supersymmetric
higher order corrections. A thorough introduction to Supersymmetry and
the Minimal Supersymmetric Standard Model has been given during this
school by Prof.~H.~Haber to whose lectures I
refer~\cite{Haber:tasi04}.

\subsubsection{About Two Higgs Doublet Models}
\label{subsubsec:2hdm}
The most popular and simplest extension of the Standard Model is
obtained by considering a scalar sector made of two instead of one
complex scalar doublets. These models, dubbed \emph{Two Higgs Doublet
Models} (2HDM), have a richer spectrum of physical scalar
fields. Indeed, after spontaneous symmetry breaking, only three of the
eight original scalar degrees of freedom (corresponding to two complex
doublet) are reabsorbed in transforming the originally massless vector
bosons into massive ones. The remaining five degrees of freedom
correspond to physical degrees of freedom in the form of: two neutral
scalar, one neutral pseudoscalar, and two charged scalar fields.

At the same time, having multiple scalar doublets in the Yukawa
Lagrangian (see Eq.~(\ref{eq:yukawa_lagrangian})) allows for scalar
flavor changing neutral current. Indeed, when generalized to the case 
of two scalar doublet $\phi^1$ and $\phi^2$,
Eq.~(\ref{eq:yukawa_lagrangian}) becomes (quark case only):
\begin{equation}
\label{eq:yukawa_lagrangian_2hdm}
\mathcal{L}_{Yukawa}=
-\sum_{k=1,2}\Gamma_{ij,k}^u\bar{Q}^i_L\Phi^{k,c} u^j_R
-\sum_{k=1,2}\Gamma_{ij,k}^d\bar{Q}^i_L\Phi^k d^j_R +
\mathrm{h.c.}\,\,\,,
\end{equation}
where each pair of fermions $(i,j)$ couple to a linear combination of
the scalar fields $\phi^1$ and $\phi^2$.
When, upon spontaneous symmetry
breaking, the fields $\phi^1$ and $\phi^2$ acquire vacuum
expectation values
\begin{equation}
\langle\Phi^k\rangle=\frac{v^k}{\sqrt{2}}\,\,\,\,\,\mbox{for}\,\,\,\,\,
k=1,2\,\,\,,
\end{equation}
the reparameterization of $\mathcal{L}_{Yukawa}$ of
Eq.~(\ref{eq:yukawa_lagrangian_2hdm}) in the vicinity of the minimum
of the scalar potential, with $
\Phi^k=\Phi^{\prime k}+v^k$ (for $k=1,2$), gives:
\begin{equation}
\mathcal{L}_{Yukawa}=
-\bar{u}^i_L
\underbrace{\sum_k\Gamma_{ij,k}^u\frac{v^k}{\sqrt{2}}}_{M^u_{ij}}u^j_R
-\bar{d}^i_L
\underbrace{\sum_k\Gamma_{ij,k}^d\frac{v^k}{\sqrt{2}}}_{M^d_{ij}}d^j_R+ 
\mathrm{h.c.} + \mbox{\small{FC couplings}}\,\,\,,
\end{equation}
where the fermion mass matrices $M^u_{ij}$ and $M^d_{ij}$ are now
proportional to a linear combination of the vacuum expectation values
of $\phi^1$ and $\phi^2$. The diagonalization of $M^u_{ij}$ and
$M^d_{ij}$ does not imply the diagonalization of the couplings of the
$\phi^{\prime k}$ fields to the fermions, and Flavor Changing (FC)
couplings arise.  This is perceived as a problem in view of the
absence of experimental evidence to support neutral flavor changing
effects. If present, these effects have to be tiny in most processes
involving in particular the first two generations of quarks, and a
safer way to build a 2HDM is to forbid them all together at the
Lagrangian level.  This is traditionally done by requiring either that
$u$-type and $d$-type quarks couple to the same doublet (Model I) or
that $u$-type quarks couple to one scalar doublet while $d$-type
quarks to the other (Model II). Indeed, these two different
realization of a 2HDM can be justified by enforcing on
$\mathcal{L}_{Yukawa}$ the following \emph{ad hoc} discrete symmetry:
\begin{equation}
\left\{
\begin{array}{c}
\Phi^1\rightarrow -\Phi^1\,\,\,\,\mathrm{and}\,\,\,\,\Phi^2\rightarrow\Phi^2 \\
d^i\rightarrow -d^i\,\,\,\,\mathrm{and}\,\,\,\, u^j\rightarrow\pm u^j
\end{array}
\right.
\end{equation}
The case in which FC scalar neutral current are not forbidden (Model
III) has also been studied in detail. In this case both up and
down-type quarks can couple to both scalar doublets, and strict
constraints have to be imposed on the FC scalar couplings in particular
between the first two generations of quarks.

2HDMs have indeed a very rich phenomenology that has been extensively
studied. In these lectures, however, we will only compare the SM Higgs
boson phenomenology to the phenomenology of the Higgs bosons of the
MSSM, a particular kind of 2HDM that we will illustrate in the
following Sections.

\subsubsection{The MSSM Higgs sector: introduction}
\label{subsubsec:higgs_mssm}
The Higgs sector of the MSSM is actually a Model II 2HDM. It contains
two complex $SU(2)_L$ scalar doublets:
\begin{equation}
\label{eq:phiu_phid}
\Phi_1=\left(\begin{array}{c}\phi_1^+\\\phi_1^0\end{array}\right)
\,\,\,\,\,\,,\,\,\,\,\,\,
\Phi_2=\left(\begin{array}{c}\phi_2^0\\\phi_2^-\end{array}\right)
\,\,\,,
\end{equation}
with opposite hypercharge ($Y\!=\!\pm 1$), as needed to make the
theory anomaly-free\footnote{Another reason for the choice of a 2HDM
is that in a supersymmetric model the superpotential should be
expressed just in terms of superfields, not their conjugates. So, one
needs to introduce two doublets to give mass to fermion fields of
opposite weak isospin. The second doublet plays the role of $\phi^c$
in the Standard Model (see Eq.~(\ref{eq:yukawa_lagrangian})), where
$\phi^c$ has opposite hypercharge and weak isospin with respect to
$\phi$.}. $\Phi_1$ couples to the up-type and $\Phi_2$ to the
down-type quarks respectively.  Correspondingly, the Higgs part of the
superpotential can be written as:
\begin{eqnarray}
\label{eq:higgs_superpotential}
V_H&=&(|\mu|^2+m_1^2)|\Phi_1|^2+(|\mu|^2+m_2^2)|\Phi_2|^2
-\mu B\epsilon_{ij}(\Phi_1^i\Phi_2^j+h.c.)\nonumber\\
&+&\frac{g^2+g^{\prime 2}}{8}\left(|\Phi_1|^2-|\Phi_2|^2\right)^2
+\frac{g^2}{2}|\Phi_1^\dagger\Phi_2|^2\,\,\,,
\end{eqnarray}
in which we can identify three different contributions
\cite{Haber:tasi04,Djouadi:2005gj}:
\begin{itemize}
\item[\emph{(i)}] the so
called $D$ terms, containing the quartic scalar interactions, which
for the Higgs fields $\Phi_1$ and $\Phi_2$ correspond to:
\begin{equation}
\label{eq:higgs_superpotential_d_terms}
\frac{g^2+g^{\prime 2}}{8}\left(|\Phi_1|^2-|\Phi_2|^2\right)^2
+\frac{g^2}{2}|\Phi_1^\dagger\Phi_2|^2\,\,\,,
\end{equation}
with $g$ and $g^\prime$ the gauge couplings of $SU(2)_L$ and $U(1)_Y$
respectively;
\item[\emph{(ii)}] the so called $F$ terms, corresponding to:
\begin{equation}
\label{eq:higgs_superpotential_f_terms}
|\mu|^2(|\Phi_1|^2+|\Phi_2|^2)\,\,\,;
\end{equation}
\item[\emph{(iii)}] the soft SUSY-breaking scalar Higgs mass and bilinear
terms, corresponding to:
\begin{equation}
\label{eq:higgs_superpotential_soft_terms}
m_1^2|\Phi_1|^2+m_2^2|\Phi_2|^2
-\mu B\epsilon_{ij}(\Phi_1^i\Phi_2^j+h.c.)\,\,\,.
\end{equation}
\end{itemize}
Overall, the scalar potential in Eq.~(\ref{eq:higgs_superpotential})
depends on three independent combinations of parameters,
$|\mu|^2+m_1^2$, $|\mu|^2+m_2^2$, and $\mu B$. One basic difference
with respect to the SM case is that the quartic coupling has been
replaced by gauge couplings. This reduced arbitrariness will play an
important role in the following.

Upon spontaneous symmetry breaking, the neutral components of
$\Phi_1$ and $\Phi_2$ acquire vacuum expectation values
\begin{equation}
\label{eq:phiu_phid_vev}
\langle\Phi_1\rangle=
\frac{1}{\sqrt{2}}\left(\begin{array}{c}0\\v_1\end{array}\right)
\,\,\,\,\,\,,\,\,\,\,\,\,
\langle\Phi_2\rangle=
\frac{1}{\sqrt{2}}\left(\begin{array}{c}v_2\\0\end{array}\right)\,\,\,,
\end{equation}
and the Higgs mechanism proceed as in the Standard Model except that
now one starts with eight degrees of freedom, corresponding to the
two complex doublets $\Phi_1$ and $\Phi_2$. Three degrees of freedom
are absorbed in making the $W^\pm$ and the $Z^0$ massive. The $W$ mass
is chosen to be: $M_W^2=g^2(v_1^2+v_2^2)/4=g^2v^2/4$, and this fixes
the normalization of $v_1$ and $v_2$, leaving only two independent
parameters to describe the entire MSSM Higgs sector.  The remaining
five degrees of freedom are physical and correspond to two neutral
scalar fields
\begin{eqnarray}
\label{eq:mssm_scalar_higgses}
h^0&=&-(\sqrt{2}\mbox{Re}\phi_2^0-v_2)\sin\alpha+
 (\sqrt{2}\mbox{Re}\phi_1^0-v_1)\cos\alpha\\
H^0&=&
(\sqrt{2}\mbox{Re}\phi_2^0-v_2)\cos\alpha+
(\sqrt{2}\mbox{Re}\phi_1^0-v_1)\sin\alpha\,\,\,,\nonumber
\end{eqnarray}
one neutral pseudoscalar field
\begin{equation}
\label{eq:mssm_pseudoscalar_higgs}
A^0=
\sqrt{2}\left(\mbox{Im}\phi_2^0\sin\beta+\mbox{Im}\phi_1^0\cos\beta\right)\,\,\,,
\end{equation}
and two charged scalar fields
\begin{equation}
\label{eq:mssm_charged_higgs}
H^\pm=\phi_2^\pm\sin\beta+\phi_1^\pm\cos\beta\,\,\,,
\end{equation}
where $\alpha$ and $\beta$ are mixing angles, and
$\tan\beta\!=\!v_1/v_2$. At tree level, the masses of the scalar and
pseudoscalar degrees of freedom satisfy the following relations:
\begin{eqnarray}
\label{eq:mssm_higgs_masses}
M_{H^\pm}^2&=&M_A^2+M_W^2\,\,\,,\\
M_{H,h}^2&=&\frac{1}{2}\left(
M_A^2+M_Z^2\pm((M_A^2+M_Z^2)^2-4M_Z^2M_A^2\cos^2 2\beta)^{1/2}\right)
\,\,\,,\nonumber
\end{eqnarray}
making it natural to pick $M_A$ and $\tan\beta$ as the two independent
parameters of the Higgs sector.

\begin{figure}
\centering
\includegraphics[scale=0.5]{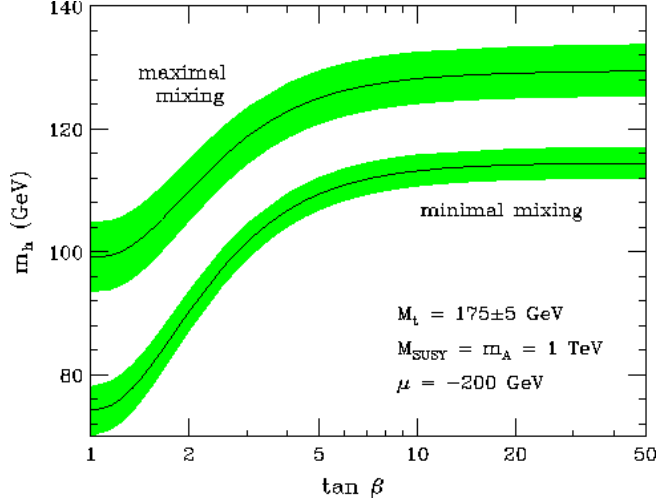}
\caption[]{The mass of the light neutral scalar Higgs boson, $h^0$, as
a function of $\tan\beta$, in the \emph{minimal mixing} and
\emph{maximal mixing} scenario. The green bands are obtained by varying
the top-quark mass in the $m_t\!=\!175\pm 5$~GeV range. 
The plot is built by fixing $M_A\!=\!1$~TeV and
$M_{SUSY}\!\equiv\!M_Q\!=\!\!M_U\!=\!M_D\!=\!1$~TeV. From
Ref.~\cite{Carena:2002es}.\label{fig:mh_upper_bound}}
\end{figure}

\begin{figure}
\centering
\includegraphics[scale=0.5]{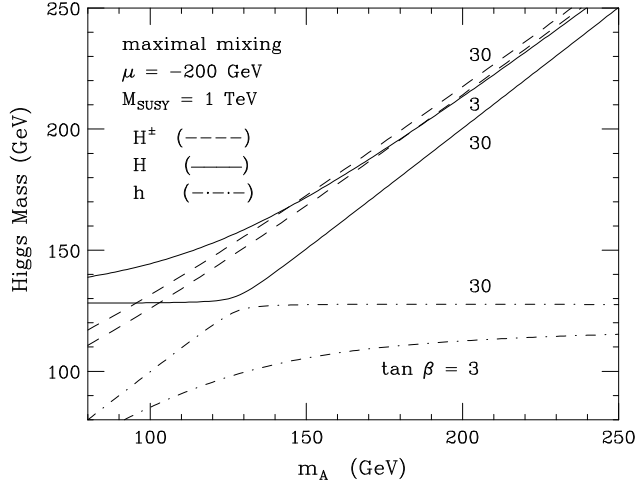}
\caption[]{The mass of the light ($h^0$) and heavy ($H^0$) neutral
scalar Higgs bosons, and of the charged scalar Higgs boson ($H^\pm$)
as a function of the neutral pseudoscalar mass $M_A$, for two
different values of $\tan\beta$ ($\tan\beta\!=\!3,30$). The top-quark
mass is fixed to $m_t\!=\!174.3$~GeV and
$M_{SUSY}\!\equiv\!M_Q\!=\!\!M_U\!=\!M_D\!=\!1$~TeV. The maximal
mixing scenario is chosen. From
Ref.~\cite{Carena:2002es}.\label{fig:mass_higgs_ma_tanb}}
\end{figure}

Eq.~(\ref{eq:mssm_higgs_masses}) provides the famous tree level upper
bound on the mass of one of the neutral scalar Higgs bosons, $h^0$:
\begin{equation}
\label{eq:mh0_upper_bound_tree_level}
M_{h}^2\le M_Z^2\cos 2\beta\le M_Z^2\,\,\,,
\end{equation}
which already contradicts the current experimental lower bound set by
LEP II: $M_{h}>93.0$~GeV~\cite{lephwg:2004mssm}. The contradiction is
lifted by including higher order radiative corrections to the Higgs
spectrum, in particular by calculating higher order corrections to the
neutral scalar mass matrix. Over the past few years a huge effort has
been dedicated to the calculation of the full one-loop corrections and
of several leading and sub-leading sets of two-loop corrections,
including resummation of leading and sub-leading logarithms via
appropriate renormalization group equation (RGE) methods. A detailed
discussion of this topic can be found in some recent
reviews~\cite{Carena:2002es,Heinemeyer:2004ms,Heinemeyer:2004gx} and
in the original literature referenced therein. For the purpose of
these lectures, let us just observe that, qualitatively, the impact of
radiative corrections on $M_{h}^{max}$ can be seen by just including
the leading two-loop corrections proportional to $y_t^2$, the square
of the top-quark Yukawa coupling, and applying RGE techniques to resum
the leading orders of logarithms. In this case, the upper bound on the
light neutral scalar in Eq.~(\ref{eq:mh0_upper_bound_tree_level}) is
modified as follows:
\begin{equation}
\label{eq:mh0_upper_bound_loop_level}
M_{h}^2\le M_Z^2+\frac{3g^2m_t^2}{8\pi^2M_W^2}
\left[\log\left(\frac{M_S^2}{m_t^2}\right)+
\frac{X_t^2}{M_S^2}\left(1-\frac{X_t^2}{12 M_S^2}\right)\right]\,\,\,,
\end{equation}
where $M_S^2=(M_{\tilde t_1}^2+M_{\tilde t_2}^2)/2$ is the average of
the two top-squark masses, $m_t$ is the running top-quark mass (to
account for the leading two-loop QCD corrections), and $X_t$ is the
top-squark mixing parameter defined by the top-squark mass matrix:
\begin{equation}
\label{eq:stop_mass_matrix}
\left(
\begin{array}{cc}
M_{Q_t}^{2}+m_t^2+D_L^t & m_t X_t\\
m_t X_t & M_{R_t}^2+m_t^2+D_R^t
\end{array}
\right)\,\,\,,
\end{equation}
with $X_t\equiv A_t-\mu\cot\beta$ ($A_t$ being one of the top-squark soft
SUSY breaking trilinear coupling),
$D_L^t=(1/2-2/3\sin\theta_W)M_Z^2\cos 2\beta$, and
$D_R^t=2/3\sin^2\theta_W M_Z^2\cos 2\beta$.
Fig.~\ref{fig:mh_upper_bound} illustrates the behavior of $M_h$
as a function of $\tan\beta$, in the case of minimal and maximal
mixing. For large $\tan\beta$ a plateau (i.e. an upper bound) is
clearly reached. The green bands represent the variation of $M_h$
as a function of $m_t$ when $m_t\!=\!175\pm 5$~GeV.
If top-squark mixing is maximal, the upper bound on $M_{h}$ is
approximately $M_h^{max}\!\simeq\! 135$~GeV\footnote{This limit is
obtained for $m_t\!=\!175$ GeV, and it can go up to
$M_h^{max}\!\simeq\!144$~GeV for $m_t\!=\!178$~GeV.}.  The behavior of
both $M_{h,H}$ and $M_{H^\pm}$ as a function of $M_A$ and $\tan\beta$
is summarized in Fig.~\ref{fig:mass_higgs_ma_tanb}, always for the
case of maximal mixing. It is interesting to notice that for all
values of $M_A$ and $\tan\beta$ the $M_H\!>\!M_h^{max}$. Also we
observe that, in the limit of large $\tan\beta$, \emph{i)} for
$M_A\!<\!M_h^{max}$: $M_h\simeq M_A$ and $M_H\simeq M_h^{max}$, while
\emph{ii)} for $M_A\!>\!M_h^{max}$: $M_H\simeq
M_A$ and $M_h\simeq M_h^{max}$.

\subsubsection{MSSM Higgs boson couplings to electroweak gauge bosons}
\label{subsubsec:mssm_higgs_couplings_bosons}
The Higgs boson couplings to the electroweak gauge bosons are obtained
from the kinetic term of the scalar Lagrangian, in strict analogy to
what we have explicitly seen in the case of the SM Higgs boson.
Here, we would like to recall the form of the $H_iVV$ and $H_iH_jV$
couplings (for $H_i\!=\!h^0,H^0,A^0,H^\pm$, and $V\!=\!W^\pm,Z^0$) that
are most important in order to understand the main features of the
MSSM plots that will be shown in Section~\ref{sec:pheno}.  

First of all, the couplings of the neutral scalar Higgs bosons to both
$W^\pm$ and $Z^0$ can be written as:
\begin{equation}
\label{eq:couplings_hVV}
g_{hVV}=g_V M_V\sin(\beta-\alpha)g^{\mu\nu}\,\,\,\,\,,\,\,\,\,\,
g_{HVV}=g_V M_V\cos(\beta-\alpha)g^{\mu\nu}\,\,\,,
\end{equation}
where $g_V\!=\!2M_V/v$, while the $A^0VV$ and $H^\pm VV$ couplings
vanish because of CP-invariance. As in the SM case, since the photon
is massless, there are no tree level $\gamma\gamma H_i$ and $\gamma
Z^0 H_i$ couplings.

Moreover, in the neutral Higgs sector, only the $h^0A^0Z^0$ and
$H^0A^0Z^0$ couplings are allowed and given by:
\begin{equation}
\label{eq:coupling_hAZ_HAZ}
g_{hAZ}=\frac{g\cos(\beta-\alpha)}{2\cos\theta_W}(p_h-p_A)^\mu
\,\,\,\,\,,\,\,\,\,\,
g_{HAZ}=-\frac{g\sin(\beta-\alpha)}{2\cos\theta_W}(p_H-p_A)^\mu\,\,\,,
\end{equation}
where all momenta are incoming. We also have several $H_iH_jV$
couplings involving the charge Higgs boson, namely:
\begin{eqnarray}
\label{eq:coupling_H+}
g_{H^+H^-Z}&=&
-\frac{g}{2\cos\theta_W}\cos 2\theta_W(p_{H^+}-p_{H^-})^\mu\,\,\,,\\
g_{H^+H^-\gamma}&=&-ie(p_{H^+}-p_{H^-})^\mu\,\,\,,\nonumber\\
g_{H^\mp hW^\pm}&=&
\mp i\frac{g}{2}\cos(\beta-\alpha)(p_{h}-p_{H^\mp})^\mu\,\,\,,
\nonumber\\
g_{H^\mp HW^\pm}&=&\pm i\frac{g}{2}\sin(\beta-\alpha)(p_{H}-p_{H^\mp})^\mu
\,\,\,,\nonumber\\
g_{H^\mp AW^\pm}&=&
\frac{g}{2}(p_{A}-p_{H^\pm})^\mu\,\,\,.\nonumber
\end{eqnarray}

At this stage it is interesting to introduce the so called
\emph{decoupling limit}, i.e. the limit of $M_A\gg M_Z$, and to
analyze how masses and couplings behave in this particular limit.
$M_{H^\pm}$ in Eq.~(\ref{eq:mssm_higgs_masses}) is unchanged, while
$M_{h,H}$ become:
\begin{equation}
\label{eq:M_H_decoupling_limit}
M_{h}\simeq M_{h}^{max}\,\,\,\,\,\mbox{and}\,\,\,\,\,
M_{H}\simeq M_A^2+M_Z^2\sin^2 2\beta\,\,\,.
\end{equation}
Moreover, as one can derive from the diagonalization of the
neutral scalar Higgs boson mass matrix:
\begin{equation}
\label{eq:cos_betamalpha_decoupling_limit}
\cos^2(\beta-\alpha)=\frac{M_h^2(M_Z^2-M_h^2)}
{M_A^2(M_H^2-M_h^2)}\,\,\,\stackrel{M_A^2\gg
M_Z^2}{\longrightarrow}\,\,\,
\frac{M_Z^4\sin^2 4\beta}{4M_A^4}\,\,\,.
\end{equation}
From the previous equations we then deduce that, in the decoupling
limit, the only light Higgs boson is $h^0$ with mass $M_{h}\simeq
M_{h}^{max}$, while $M_{H}\simeq M_{H^\pm}\simeq M_A\gg M_Z$, and
because $\cos(\beta-\alpha)\rightarrow 0$
($\sin(\beta-\alpha)\rightarrow 1)$), the couplings of $h^0$ to the
gauge bosons tend to the SM Higgs boson limit. This is to say that, in the
decoupling limit, the light MSSM Higgs boson will be hardly
distinguishable from the SM Higgs boson.

Finally, we need to remember that the tree level couplings may be
modified by radiative corrections involving both loops of SM and MSSM
particles, among which loops of third generation quarks and squarks
dominate. The very same radiative corrections that modify the Higgs
boson mass matrix, thereby changing the definition of the mass
eigenstates, also affect the couplings of the corrected mass
eigenstates to the gauge bosons. This can be reabsorbed into the
definition of a \emph{renormalized} mixing angle $\alpha$ or a
\emph{radiatively corrected} value for $\cos(\beta-\alpha)$
($\sin(\beta-\alpha)$). Using the notation of
Ref.~\cite{Carena:2002es}, the radiatively corrected
$\cos(\beta-\alpha)$ can be written as:
\begin{equation}
\label{eq:cos_betamalpha_rad_corrected}
\cos(\beta-\alpha)=K\left[\frac{M_Z^2\sin 4\beta}{2M_A^2}+
\mathcal{O}\left(\frac{M_Z^4}{M_A^4}\right)\right]\,\,\,,
\end{equation}
where
\begin{equation}
\label{eq:cos_betamalpha_K_factor}
K\equiv 1+
\frac{\delta\mathcal{M}_{11}^2-\delta\mathcal{M}_{22}^2}{2M_Z^2\cos 2\beta}-
\frac{\delta\mathcal{M}_{12}^2}{M_Z^2\sin 2\beta}\,\,\,,
\end{equation}
and $\delta\mathcal{M}_{ij}$ are the radiative corrections to the
corresponding elements of the CP-even Higgs squared-mass matrix (see
Ref.~\cite{Carena:2002es}).  It is interesting to notice that on top
of the traditional decoupling limit introduced above ($M_A\gg M_Z$),
there is now also the possibility that $\cos(\beta-\alpha)\rightarrow
0$ if $K\rightarrow 0$, and this happens independently of the value of
$M_A$.

\subsubsection{MSSM Higgs boson couplings to fermions}
\label{subsubsec:mssm_higgs_couplings_fermions}
As anticipated, $\Phi_1$ and $\Phi_2$ have Yukawa-type couplings to the
up-type and down-type components of all $SU(2)_L$ fermion
doublets. For example, the Yukawa Lagrangian corresponding to the
third generation of quarks reads:
\begin{equation}
\label{eq:yukawa_lagrangian_mssm}
\mathcal{L}_{Yukawa}=
-h_t\left[\bar{t}_R\phi_1^0 t_L-\bar{t}_R\phi_1^+ b_L\right]
-h_b\left[\bar{b}_R\phi_2^0 b_L-\bar{b}_R\phi_2^- t_L\right]+\mathrm{h.c.}
\end{equation}
Upon spontaneous symmetry breaking $\mathcal{L}_{Yukawa}$ provides both the
corresponding quark masses:
\begin{equation}
m_t=h_t\frac{v_1}{\sqrt{2}}=h_t\frac{v\sin\beta}{\sqrt{2}}
\,\,\,\,\,\mbox{and}\,\,\,\,\,
m_b=h_b\frac{v_2}{\sqrt{2}}=h_b\frac{v\cos\beta}{\sqrt{2}}\,\,\,,
\end{equation}
and the corresponding Higgs-quark couplings:
\begin{eqnarray}
\label{eq:yukawa_couplings_3rd_generation}
g_{ht\bar{t}}&=&\frac{\cos\alpha}{\sin\beta}y_t=
   \left[\sin(\beta-\alpha)+\cot\beta\cos(\beta-\alpha)\right]y_t\,\,\,,\\
g_{hb\bar{b}}&=&-\frac{\sin\alpha}{\cos\beta}y_b=
   \left[\sin(\beta-\alpha)-\tan\beta\cos(\beta-\alpha)\right]y_b\,\,\,,\nonumber\\
g_{Ht\bar{t}}&=&\frac{\sin\alpha}{\sin\beta}y_t=
   \left[\cos(\beta-\alpha)-\cot\beta\sin(\beta-\alpha)\right]y_t\,\,\,,\nonumber\\
g_{Hb\bar{b}}&=&\frac{\cos\alpha}{\cos\beta}y_b=
   \left[\cos(\beta-\alpha)+\tan\beta\sin(\beta-\alpha)\right]y_b\,\,\,,\nonumber\\
g_{At\bar{t}}&=&\cot\beta\, y_t\,\,\,\,,\,\,\,\,
g_{Ab\bar{b}}=\tan\beta\, y_b\,\,\,,\nonumber\\
g_{H^\pm t\bar{b}}&=&\frac{g}{2\sqrt{2}M_W}
\left[m_t\cot\beta(1-\gamma_5)+m_b\tan\beta(1+\gamma_5)\right]\,\,\,,\nonumber
\end{eqnarray}
where $y_q\!=\!m_q/v$ (for $q\!=\!t,b$) are the SM couplings.  It is
interesting to notice that in the $M_A\gg M_Z$ decoupling limit, as
expected, all the couplings in
Eq.~(\ref{eq:yukawa_couplings_3rd_generation}) reduce to the SM limit,
i.e. all $H^0$, $A^0$, and $H^\pm$ couplings vanish, while the
couplings of the light neutral Higgs boson, $h^0$, reduce to the
corresponding SM Higgs boson couplings.

The Higgs boson-fermion couplings are also modified directly by
one-loop radiative corrections (squarks-gluino loops for quarks
couplings and slepton-neutralino loops for lepton couplings). A
detailed discussion can be found in
Ref.~\cite{Carena:2002es,Djouadi:2005gj} and in the
literature referenced therein. Of particular relevance are the
corrections to the couplings of the third quark generation. These
can be parameterized at the Lagrangian level by writing the
radiatively corrected \emph{effective} Yukawa Lagrangian as:
\begin{eqnarray}
\label{eq:yukawa_lagrangian_mssm_rad_corrected}
\mathcal{L}_{Yukawa}^{eff}&=&
-\epsilon_{ij}\left[
(h_b+\delta h_b)\bar{b}_R Q^j_L\Phi^i_2+(h_t+\delta h_t)\bar{t}_R Q^i_L\Phi^j_1
\right]\\
&-&\Delta h_t\bar{t}_RQ^k_L\Phi^{k\ast}_2-\Delta h_b\bar{b}_RQ^k_L\Phi^{k\ast}_1+
\mathrm{h.c.}\,\,\,,\nonumber
\end{eqnarray}
where we notice that radiative corrections induce a small coupling
between $\Phi_1$ and down-type fields and between $\Phi_2$ and
up-type fields. Moreover the tree level relation between $h_b$, $h_t$,
$m_b$ and $m_t$ are modified as follows:
\begin{eqnarray}
\label{eq:mb_mt_rad_corrected}
m_b&=&\frac{h_b v}{\sqrt{2}}\cos\beta\left(
1+\frac{\delta h_b}{h_b}+\frac{\Delta h_b\tan\beta}{h_b}\right)\equiv
\frac{h_b v}{\sqrt{2}}\cos\beta(1+\Delta_b)\,\,\,,\\
m_t&=&\frac{h_t v}{\sqrt{2}}\sin\beta\left(
1+\frac{\delta h_t}{h_t}+\frac{\Delta h_t\tan\beta}{h_t}\right)\equiv
\frac{h_t v}{\sqrt{2}}\sin\beta(1+\Delta_t)\,\,\,,\nonumber
\end{eqnarray}
where the leading corrections are proportional to $\Delta h_b$ and
turn out to also be $\tan\beta$ enhanced.  On the other hand, the
couplings between Higgs mass eigenstates and third generation quarks given in
Eq.~(\ref{eq:yukawa_couplings_3rd_generation}) are corrected as
follows:
\begin{eqnarray}
\label{eq:yukawa_couplings_3rd_generation_rad_corrected}
g_{ht\bar{t}}&=&\frac{\cos\alpha}{\sin\beta}y_t\,
  \left[1-\frac{1}{1+\Delta_t}\frac{\Delta h_t}{h_t}\left(\cot\beta+\tan\alpha\right)
  \right]\,\,\,,\\
g_{hb\bar{b}}&=&-\frac{\sin\alpha}{\cos\beta}y_b\,
   \left[1+\frac{1}{1+\Delta_b}\left(\frac{\delta h_b}{h_b}-\Delta_b\right)
   \left(1+\cot\alpha\cot\beta\right)\right]\,\,\,,\nonumber\\
g_{Ht\bar{t}}&=&\frac{\sin\alpha}{\sin\beta}y_t\,
   \left[1-\frac{1}{1+\Delta_t}\frac{\Delta h_t}{h_t}\left(\cot\beta-\cot\alpha\right)
   \right]\nonumber\,\,\,,\\
g_{Hb\bar{b}}&=&\frac{\cos\alpha}{\cos\beta}y_b\,
   \left[1+\frac{1}{1+\Delta_b}\left(\frac{\delta h_b}{h_b}-\Delta_b\right)
   \left(1-\tan\alpha\cot\beta\right)\right]\,\,\,,\nonumber\\
g_{At\bar{t}}&=&\cot\beta\,y_t\,
   \left[1-\frac{1}{1+\Delta_t}\frac{\Delta h_t}{h_t}\left(\cot\beta+\tan\beta\right)
   \right]\nonumber\,\,\,,\\
g_{Ab\bar{b}}&=&\tan\beta\,y_b\,
   \left[1+\frac{1}{(1+\Delta_b)\sin^2\beta}\left(\frac{\delta h_b}{h_b}-\Delta_b\right)
   \right]\,\,\,,\nonumber\\
g_{H^\pm t\bar{b}}&\simeq&\frac{g}{2\sqrt{2}M_W}
 \left\{ m_t\cot\beta\left[1-\frac{1}{1+\Delta_t}\frac{\Delta h_t}{h_t}
        \left(\cot\beta+\tan\beta\right)\right](1+\gamma_5)\right.\nonumber\\
&+&\left.m_b\tan\beta\left[1+\frac{1}{(1+\Delta_b)\sin^2\beta}
        \left(\frac{\delta h_b}{h_b}-\Delta_b\right)\right](1-\gamma_5)
\right\}\,\,\,,\nonumber
\end{eqnarray}
where the last coupling is given in the approximation of small isospin
breaking effects, since interactions of this kind have been neglected
in the Lagrangian of
Eq.~(\ref{eq:yukawa_lagrangian_mssm_rad_corrected}).

\section{Phenomenology of the Higgs Boson}
\label{sec:pheno}

\subsection{Standard Model Higgs boson decay branching ratios}
\label{subsec:sm_higgs_branching_ratios}

\begin{figure}
\centering
\includegraphics[scale=0.6]{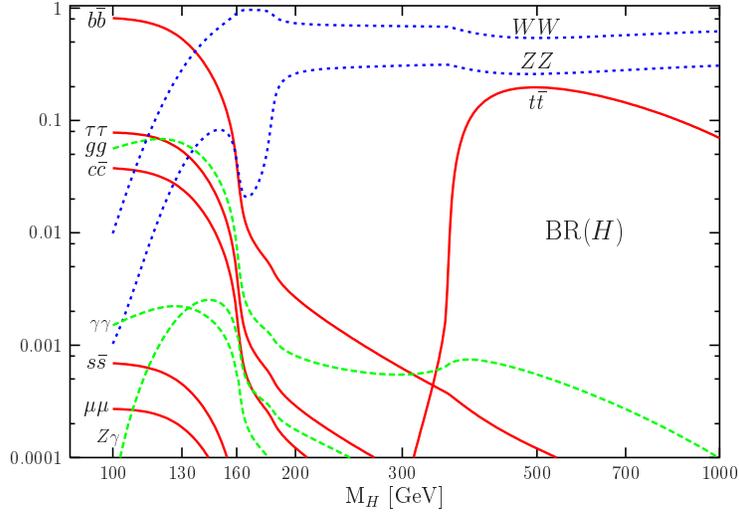}
\caption[]{SM Higgs decay branching ratios as a function of
$M_H$. The blue curves represent tree-level decays into electroweak
gauge bosons, the red curves tree level decays into quarks and
leptons, the green curves one-loop decays. From
Ref.~\cite{Djouadi:2005gi}.\label{fig:sm_higgs_br}}
\end{figure}
\begin{figure}
\centering
\includegraphics[scale=0.6]{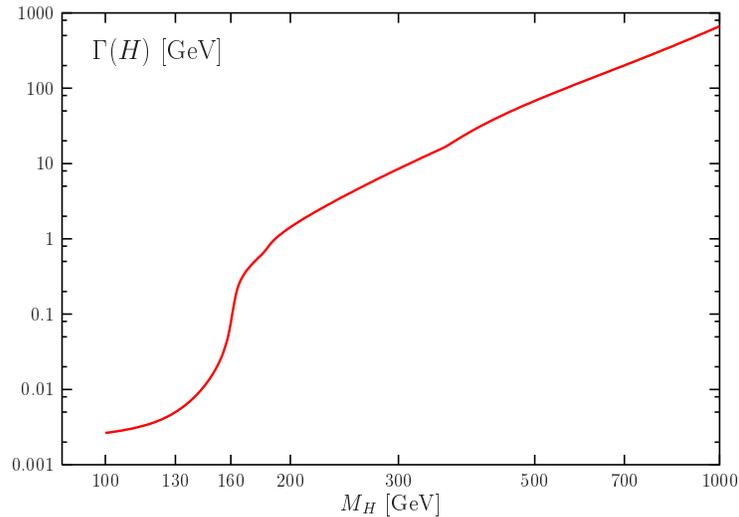}
\caption[]{SM Higgs total decay width as a function of
$M_H$. From Ref.~\cite{Djouadi:2005gi}.\label{fig:sm_higgs_width}}
\end{figure}
In this Section we approach the physics of the SM Higgs boson
by considering its branching ratios for various decay modes. In
Section~\ref{subsec:higgs_sm} we have derived the SM Higgs couplings
to gauge bosons and fermions. Therefore we know that, at the tree
level, the SM Higgs boson can decay into pairs of electroweak gauge
bosons ($H\rightarrow W^+W^-,ZZ$), and into pairs of quarks and
leptons ($H\rightarrow Q\bar{Q},l^+l^-$); while at one-loop it can
also decay into two photons ($H\rightarrow\gamma\gamma$), two gluons
($H\rightarrow gg$), or a $\gamma Z$ pair ($H\rightarrow\gamma Z$).
Fig.~\ref{fig:sm_higgs_br} represents all the decay branching ratios
of the SM Higgs boson as functions of its mass $M_H$. The SM Higgs
boson total width, sum of all the partial widths $\Gamma(H\rightarrow
XX)$, is represented in Fig.~\ref{fig:sm_higgs_width}.

Fig.~\ref{fig:sm_higgs_br} shows that a light Higgs boson ($M_H\le
130-140$~GeV) behaves very differently from a heavy Higgs boson
($M_H\ge 130-140$~GeV). Indeed, a light SM Higgs boson mainly decays
into a $b\bar{b}$ pair, followed hierarchically by all other pairs of
lighter fermions. Loop-induced decays also play a role in this
region. $H\rightarrow gg$ is dominant among them, and it is actually
larger than many tree level decays.  Unfortunately, this decay mode is
almost useless, in particular at hadron colliders, because of
background limitations. Among radiative decays,
$H\rightarrow\gamma\gamma$ is tiny, but it is actually
phenomenologically very important because the two photon signal can be
seen over large hadronic backgrounds. On the other hand, for larger
Higgs masses, the decays to $W^+W^-$ and $ZZ$ dominates. All decays
into fermions or loop-induced decays are suppressed, except
$H\rightarrow t\bar{t}$ for Higgs masses above the $t\bar{t}$
production threshold. There is an intermediate region, around
$M_H\simeq 160$~GeV, i.e. below the $W^+W^-$ and $ZZ$ threshold, where
the decays into $WW^*$ and $ZZ^*$ (when one of the two gauge bosons is
off-shell) become important. These are indeed three-body decays of the
Higgs boson that start to dominate over the $H\rightarrow b\bar{b}$
two-body decay mode when the largeness of the $HWW$ or $HZZ$ couplings
compensate for their phase space suppression\footnote{Actually, even
four-body decays, corresponding to $H\rightarrow W^*W^*,Z^*Z^*$ may
become important in the intermediate mass region and are indeed
accounted for in Fig.~\ref{fig:sm_higgs_br}.}.  The different decay
pattern of a light vs a heavy Higgs boson influences the role played,
in each mass region, by different Higgs production processes at hadron
and lepton colliders.

\begin{figure}
\centering
\includegraphics[scale=0.6]{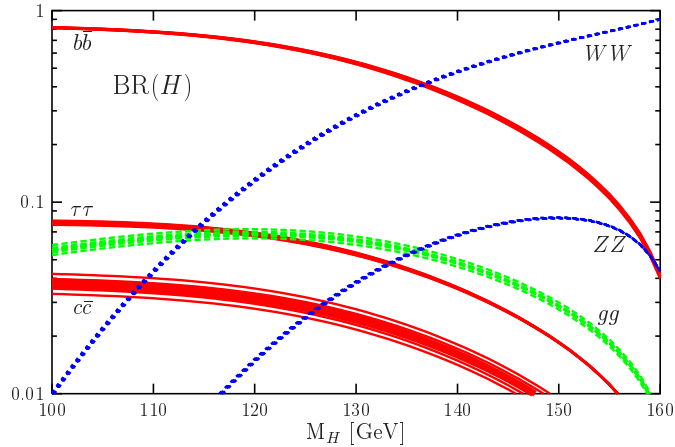}
\caption[]{SM Higgs boson decay branching ratios in the low and intermediate
Higgs boson mass range including the uncertainty from the quark masses
$m_t\!=\!178\pm4.3$~GeV, $m_b\!=\!4.88\pm 0.07$~GeV, and
$m_c\!=\!1.64\pm 0.07$~GeV, as well as from
$\alpha_s(M_Z)\!=\!0.1172\pm 0.002$. From
Ref.~\cite{Djouadi:2005gi}.\label{fig:sm_higgs_br_band} }
\end{figure}

The curves in Fig.~\ref{fig:sm_higgs_br} are obtained by including all
available QCD and electroweak (EW) radiative corrections. Indeed, the
problem of computing the relevant orders of QCD and EW corrections for
Higgs decays has been thoroughly explored and the results are nowadays
available in public codes like HDECAY~\cite{Djouadi:1997yw}, which has
been used to produce Fig.~\ref{fig:sm_higgs_br}. Indeed it would be
more accurate to represent each curve as a band, obtained by varying
the parameters that enters both at tree level and in particular
through loop corrections within their uncertainties. This is shown,
for a light and intermediate mass Higgs boson, in
Fig.~\ref{fig:sm_higgs_br_band} where each band has been obtained
including the uncertainty from the quark masses and from the strong
coupling constant.

In the following we will briefly review the various SM Higgs decay
channels. Giving a schematic but complete list of all available
radiative corrections goes beyond the purpose of these
lectures. Therefore we will only discuss those aspects that can be
useful as a general background. In particular I will comment on the
general structure of radiative corrections to Higgs decay and I will
add more details on QCD corrections to $H\rightarrow Q\bar{Q}$ ($Q=$
heavy quark).

For a detailed review of QCD correction in Higgs decays we refer the
reader to Ref.~\cite{Spira:1997dg}.  Ref.~\cite{Djouadi:2005gi} also
contain an excellent summary of both QCD and EW radiative corrections
to Higgs decays.

\subsubsection{General properties of radiative corrections to Higgs decays}
\label{subsubsection:sm_higgs_decays_rad_corr_general}
All Higgs boson decay rates are modified by both EW and QCD radiative
corrections. QCD corrections are particularly important for $H\rightarrow
Q\bar{Q}$ decays, where they mainly amount to a redefinition of the
Yukawa coupling by shifting the mass parameter in it from the pole
mass value to the running mass value, and for $H\rightarrow gg$.  EW
corrections can be further separated into: \emph{i)} corrections due
to fermion loops, \emph{ii)} corrections due to the Higgs boson
self-interaction, and \emph{iii)} other EW corrections. Both
corrections of type \emph{(ii)} and
\emph{(iii)} are in general very small if not for large Higgs boson
masses, i.e. for $M_H\gg M_W$. On the other hand, corrections of type
\emph{(i)} are very important over the entire Higgs mass range, and
are particularly relevant for $M_H\ll 2m_t$, where the top-quark loop
corrections play a leading role.  Indeed, for $M_H\ll 2m_t$, the
dominant corrections for both Higgs decays into fermion and gauge
bosons come from the top-quark contribution to the renormalization of
the Higgs wave function and vacuum expectation value.

Several higher order radiative corrections to Higgs decays have been
calculated in the large $m_t$ limit, specifically in the limit when
$M_H\ll 2m_t$. Results can then be derived applying some very powerful
\emph{low energy theorems}. The idea
is that, for an on-shell Higgs field ($p_H^2\!=\!M_H^2$), the limit of
small masses ($M_H\ll 2m_t$) is equivalent to a $p_H\rightarrow 0$
limit, in which case the Higgs couplings to the fermion fields can be
simply obtained by substituting
\begin{equation}
\label{eq:mi_subs_let}
m_i^0\rightarrow m_i^0\left(1+\frac{H^0}{v^0}\right)\,\,\,,
\end{equation}
in the (bare) Yukawa Lagrangian, for each massive particle $i$.  
In Eq.~(\ref{eq:mi_subs_let}) $H^0$
is a constant field and the upper zero indices indicate that all
formal manipulations are done on bare quantities. This induces a
simple relation between the bare matrix element for a process with
($X\rightarrow Y+H$) and without ($X\rightarrow Y$) a Higgs field,
namely
\begin{equation}
\label{eq:matrix_element_relation_let}
\lim_{p_H\rightarrow 0}\mathcal{A}(X\rightarrow Y+H)=
\frac{1}{v^0}\sum_im_i^0\frac{\partial}{\partial m_i^0}\mathcal{A}(X\rightarrow Y)\,\,\,.
\end{equation}
When the theory is renormalized, the only actual difference is that
the derivative operation in Eq.~(\ref{eq:matrix_element_relation_let})
needs to be modified as follows
\begin{equation}
\label{eq:derivative_renormalized_let}
m_i^0\frac{\partial}{\partial m_i^0}\longrightarrow
\frac{m_i}{1+\gamma_{m_i}}\frac{\partial}{\partial m_i}
\end{equation}
where $\gamma_{m_i}$ is the mass anomalous dimension of fermion $f_i$.
This accounts for the fact that the renormalized Higgs-fermion Yukawa
coupling is determined through the $Z_2$ and $Z_m$ counterterms, and
not via the $Hf\bar{f}$ vertex function at zero momentum transfer (as
used in the $p_H\to 0$ limit above).

The theorem summarized by Eq.~(\ref{eq:matrix_element_relation_let})
is valid also when higher order radiative corrections are included.
Therefore, outstanding applications of Eq.~(\ref{eq:matrix_element_relation_let})
include the determination of the one-loop $Hgg$ and $H\gamma\gamma$
vertices from the gluon or photon self-energies, as well as the
calculation of several orders of their QCD and EW radiative
corrections. Indeed, in the $m_t\rightarrow\infty$ limit, the loop-induced
$H\gamma\gamma$ and $Hgg$ interactions can be seen as effective
vertices derived from an effective Lagrangian of the form:
\begin{equation}
\label{eq:effective_lagrangian_let}
\mathcal{L}_{eff}=\frac{\alpha_s}{12\pi}F^{(a)\mu\nu}F^(a)_{\mu\nu}\frac{H}{v}
(1+O(\alpha_s))\,\,\,,
\end{equation}
where $F^(a)_{\mu\nu}$ is the field strength tensor of QED (for the
$H\gamma\gamma$ vertex) or QCD (for the $Hgg$ vertex).  The
calculation of higher order corrections to the
$H\rightarrow\gamma\gamma$ and $H\rightarrow gg$ decays is then
reduced by one order of loops! Since these vertices start as one-loop
effects, the calculation of the first order of corrections would
already be a strenuous task, and any higher order effect would be a
formidable challenge. Thanks to the low energy theorem results
sketched above, QCD NNLO corrections have indeed been calculated.

\subsubsection{Higgs boson decays to gauge bosons: 
$H\rightarrow W^+W^-,ZZ$}
\label{subsubsec:sm_higgs_to_gaugebosons}
The tree level decay rate for $H\rightarrow VV$ ($V\!=\!W^\pm,Z$) 
can be written as: 
\begin{equation}
\Gamma(H\rightarrow VV)=
\frac{G_FM_H^3}{16\sqrt{2}\pi}\delta_V\left(1-\tau_V+\frac{3}{4}\tau_V^2\right)
\beta_V\,\,\,,
\end{equation}
where $\beta_V=\sqrt{1-\tau_V}$, $\tau_V=4M_V^2/M_H^2$, and
$\delta_{W,Z}\!=\!2,1$. 

Below the $W^+W^-$ and $ZZ$ threshold, the SM Higgs boson can still
decay via three (or four) body decays mediated by $WW^*$ ($W^*W^*$)
or $ZZ^*$ ($Z^*Z^*$) intermediate states. As we can see from
Fig.~\ref{fig:sm_higgs_br}, the off-shell decays $H\rightarrow WW^*$
and $H\rightarrow ZZ^*$ are relevant in the intermediate mass region
around $M_H\simeq 160$~GeV, where they compete and overcome the
$H\rightarrow b\bar{b}$ decay mode. The decay rates for $H\rightarrow
VV^*\rightarrow Vf_i\bar{f_j}$ ($V\!=\!W^\pm,Z$) are given by:
\begin{eqnarray}
\Gamma(H\rightarrow WW^*)&=&\frac{3g^4M_H}{512\pi^3}
F\left(\frac{M_W}{M_H}\right)\,\,\,,\\
\Gamma(H\rightarrow ZZ^*)&=&\frac{g^4M_H}{2048(1-s_W^2)^2\pi^3}
\left(7-\frac{40}{3}s_W^2+\frac{160}{9}s_W^4\right)
F\left(\frac{M_Z}{M_H}\right)\,\,\,,\nonumber
\end{eqnarray}
where $s_W\!=\!\sin\theta_W$ is the sine of the Weinberg angle and the
function $F(x)$ is given by
\begin{eqnarray}
F(x)&=&-(1-x^2)\left(\frac{47}{2}x^2-\frac{13}{2}+\frac{1}{x^2}\right)
-3\left(1-6x^2+4x^4\right)\ln(x)\nonumber\\
&+&3\,\frac{1-8x^2+20x^4}{\sqrt{4x^2-1}}
\arccos\left(\frac{3x^2-1}{2x^3}\right)\,\,\,.
\end{eqnarray}

\subsubsection{Higgs boson decays to fermions:
$H\rightarrow Q\bar{Q},l^+l^-$}
\label{subsubsec:sm_higgs_to_fermions}
The tree level decay rate for $H\rightarrow f\bar{f}$ ($f\!=\!Q,l$,
$Q=$quark, $l=$lepton) can be written as: 
\begin{equation}
\Gamma(H\rightarrow f\bar f)=
\frac{G_FM_H}{4\sqrt{2}\pi}N_c^f m_f^2\beta_f^3\,\,\,,
\end{equation}
where $\beta_f=\sqrt{1-\tau_f}$, $\tau_f=4m_f^2/M_H^2$, and
$(N_c)^{l,Q}\!=\!1,3$.
QCD corrections dominate over other radiative corrections and they
modify the rate as follows:
\begin{equation}
\label{eq:gamma_hqq}
\Gamma(H\rightarrow Q\bar{Q})_{QCD}=
\frac{3G_FM_H}{4\sqrt{2}\pi}\bar{m}_Q^2(M_H)\beta_q^3
\left[\Delta_{QCD}+\Delta_t\right]\,\,\,,
\end{equation}
where $\Delta_t$ represents specifically QCD corrections
involving a top-quark loop. $\Delta_{QCD}$ and $\Delta_t$ have been
calculated up to three loops and are given by:
\begin{eqnarray}
\Delta_{QCD}&=&1+5.67\frac{\alpha_s(M_H)}{\pi}+(35.94-1.36N_F)
\left(\frac{\alpha_s(M_H)}{\pi}\right)^2+\\
&&(164.14-25.77N_F+0.26N_F^2)\left(\frac{\alpha_s(M_H)}{\pi}\right)^3
\,\,\,,\nonumber\\
\Delta_t&=&\left(\frac{\alpha_s(M_H)}{\pi}\right)^2
\left[1.57-\frac{2}{3}\ln\frac{M_H^2}{m_t^2}+
\frac{1}{9}\ln^2\frac{\bar{m}_Q^2(M_H)}{M_H^2}\right]\,\,\,,\nonumber
\end{eqnarray}
where $\alpha_s(M_H)$ and $\bar{m}_Q(M_H)$ are the renormalized
running QCD coupling and quark mass in the $\overline{MS}$ scheme.  It
is important to notice that using the $\overline{MS}$ running mass in
the overall Yukawa coupling square of Eq.~(\ref{eq:gamma_hqq})is very
important in Higgs decays, since it reabsorbs most of the QCD
corrections, including large logarithms of the form
$\ln(M_H^2/m_Q^2)$.  Indeed, for a generic scale $\mu$,
$\bar{m}_Q(\mu)$ is given at leading order by:
\begin{eqnarray}
\bar{m}_Q(\mu)_{LO}&=&\bar{m}_Q(m_Q)\left(\frac{\alpha_s(\mu)}{\alpha_s(m_Q)}\right)^
{\frac{2b_0}{\gamma_0}}\\
&=&\bar{m}_Q(m_Q)\left(1-\frac{\alpha_s(\mu)}{4\pi}\ln\left(\frac{\mu^2}{m_Q^2}\right)+
\cdots\right)\,\,\,,\nonumber
\end{eqnarray}
where $b_0$ and $\gamma_0$ are the first coefficients of the $\beta$
and $\gamma$ functions of QCD, while at higher orders it reads:
\begin{equation}
\label{eq:running_mass_lo}
\bar{m}_Q(\mu)=\bar{m}_Q(m_Q)\frac{f\left(\alpha_s(\mu)/\pi\right)}
{f\left(\alpha_s(m_Q)/\pi\right)}\,\,\,,
\end{equation}
where, from renormalization group techniques, the function $f(x)$
is of the form:
\begin{eqnarray}
\label{eq:running_mass_higher_order}
f(x)&=&\left(\frac{25}{6}x\right)^{\frac{12}{25}}\left[1+1.014x+\ldots\right]
\,\,\,\,\,\,\mbox{for}\,\,\,\,\,m_c\!<\!\mu\!<\!m_b\,\,\,,\\
f(x)&=&\left(\frac{23}{6}x\right)^{\frac{12}{23}}\left[1+1.175x+\ldots\right]
\,\,\,\,\,\,\mbox{for}\,\,\,\,\,m_b\!<\!\mu\!<\!m_t\,\,\,,\nonumber\\
f(x)&=&\left(\frac{7}{2}x\right)^{\frac{4}{7}}\left[1+1.398x+\ldots\right]
\,\,\,\,\,\,\mbox{for}\,\,\,\,\,\mu\!>\!m_t\,\,\,.\nonumber
\end{eqnarray}
As we can see from Eqs.~(\ref{eq:running_mass_lo}) and
(\ref{eq:running_mass_higher_order}), by using the $\overline{MS}$
running mass, leading and subleading logarithms up to the order of
the calculation are actually resummed at all orders in $\alpha_s$.

The overall mass factor coming from the quark Yukawa coupling square
is actually the only place where we want to employ a running mass. For
quarks like the $b$ quark this could indeed have a large impact, since,
in going from $\mu\simeq M_H$ to $\mu\simeq m_b$, $\bar{m}_n(\mu)$
varies by almost a factor of two, making therefore almost a factor of
four at the rate level.  All other mass corrections, in the matrix
element and phase space entering the calculation of the $H\rightarrow
Q\bar{Q}$ decay rate, can in first approximation be safely neglected.

\subsubsection{Loop induced Higgs boson decays: 
\label{subsubsec:sm_higgs_loop_decay}
$H\rightarrow \gamma\gamma,\gamma Z,gg$}
\label{subsubsec:sm_higgs_loop_decays}
As seen in Section~\ref{subsec:higgs_sm}, the $H\gamma\gamma$ and
$H\gamma Z$ couplings are induced at one loop via both a fermion loop
and a W-loop. At the lowest order the decay rate for
$H\rightarrow\gamma\gamma$ can be written as:
\begin{equation}
\label{eq:rate_Hgammagamma_lo}
\Gamma(H\rightarrow\gamma\gamma)=
\frac{G_F\alpha^2M_H^3}{128\sqrt{2}\pi^3}
\left|\sum_f N_c^fQ_f^2A_f^H(\tau_f)+A_W^H(\tau_W)\right|^2\,\,\,,
\end{equation}
where $N_c^f=1,3$ (for $f=l,q$ respectively), $Q_f$ is the charge of
the $f$ fermion species, $\tau_f=4m_f^2/M_H^2$, the function $f(\tau)$
is defined as:
\begin{equation}
\label{eq:f_tau}
f(\tau)=\left\{
\begin{array}{lr}
\arcsin^2\frac{1}{\sqrt{\tau}}&\tau\ge 1\\
-\frac{1}{4}\left[\ln\frac{1+\sqrt{1-\tau}}{1-\sqrt{1-\tau}}-i\pi\right]^2&
\tau<1\,\,\,,
\end{array}\right.
\end{equation}
and the form factors $A_f^H$ and $A_W^H$ are given by:
\begin{eqnarray}
\label{eq:rate_Hgammagamma_lo_form_factors}
A_f^H&=&2\tau\left[1+(1-\tau)f(\tau)\right]\,\,\,,\\
A_W^H(\tau)&=&-\left[2+3\tau+3\tau(2-\tau)f(\tau)\right]\,\,\,.\nonumber
\end{eqnarray}
On the other hand, the decay rate for $H\rightarrow\gamma Z$ is given by:
\begin{equation}
\label{eq:rate_HgammaZ_lo}
\Gamma(H\rightarrow \gamma Z)=
\frac{G_F^2M_W^2\alpha M_H^3}{64\pi^4}\left(1-\frac{M_Z^2}{M_H^2}\right)^3
\left|\sum_f A_f^H(\tau_f,\lambda_f)+A_W^H(\tau_W,\lambda_W)\right|^2\,\,\,,
\end{equation}
where $\tau_i\!=\!4M_i^2/M_H^2$ and $\lambda_i\!=\!4M_i^2/M_Z^2$
($i\!=\!f,W$), and the form factors $A_f^H(\tau,\lambda)$ and
$A_W^H(\tau,\lambda)$ are given by:
\begin{eqnarray}
\label{eq:rate_HgammaZ_lo_form_factors}
A_f^H(\tau,\lambda)&=&2N_c^f\frac{Q_f(I_{3f}-2Q_f\sin^2\theta_W)}{\cos\theta_W}
\left[I_1(\tau,\lambda)-I_2(\tau,\lambda)\right]\,\,\,,\\
A_W^H(\tau,\lambda)&=&\cos\theta_W\left\{
\left[\left(1+\frac{2}{\tau}\right)\tan^2\theta_W-\left(5+\frac{2}{\tau}\right)\right]
I_1(\tau,\lambda)\nonumber\right.\\
&&\left.\phantom{\frac{1}{2}}+4\left(3-\tan^2\theta_W\right)I_2(\tau,\lambda)
\right\}\,\,\,,
\end{eqnarray}
where $N_c^f$ and $Q_f$ are defined after
Eq.~(\ref{eq:rate_Hgammagamma_lo}), and $I_3^f$ is the weak isospin of the
$f$ fermion species. Moreover:
\begin{eqnarray}
\label{eq:rate_Hgammagamma_lo_I1I2}
I_1(\tau,\lambda)&=&\frac{\tau\lambda}{2(\tau-\lambda)}+
\frac{\tau^2\lambda^2}{2(\tau-\lambda)^2}[f(\tau)-f(\lambda)]+
\frac{\tau^2\lambda}{(\tau-\lambda)^2}[g(\tau)-g(\lambda)]\,\,\,,\nonumber\\
I_2(\tau,\lambda)&=&-\frac{\tau\lambda}{2(\tau-\lambda)}[f(\tau)-f(\lambda)]\,\,\,,
\end{eqnarray}
and
\begin{equation}
\label{eq:g_tau}
g(\tau)=\left\{
\begin{array}{lr}
\sqrt{\tau-1}\arcsin\frac{1}{\sqrt{\tau}}&\tau\ge 1\\
\frac{\sqrt{1-\tau}}{2}\left[\ln\frac{1+\sqrt{1-\tau}}{1-\sqrt{1-\tau}}-i\pi\right]&
\tau<1
\end{array}\right.
\end{equation}
while $f(\tau)$ is defined in Eq.~(\ref{eq:f_tau}). QCD and EW
corrections to both $\Gamma(H\rightarrow\gamma\gamma)$ and
$\Gamma(H\rightarrow\gamma Z)$ are pretty small and for their explicit
expression we refer the interested reader to the literature
\cite{Spira:1997dg,Djouadi:2005gi}.

\begin{figure}
\centering
\includegraphics[bb=125pt 470pt 490pt 760pt,scale=0.7]{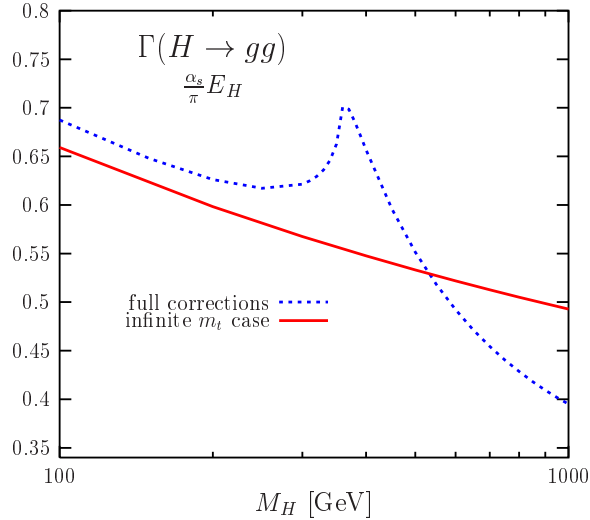}
\vspace{0.3truecm}
\caption[]{The QCD correction factor for the partial width
$\Gamma(H\rightarrow gg)$ as a function of the Higgs boson mass, in
the full massive case with $m_t\!=\!178$~GeV (dotted line) and in the
heavy-top-quark limit (solid line). The strong coupling constant is
normalized at $\alpha_s(M_Z)\!=\!0.118$. From
Ref.~\cite{Djouadi:2005gi}.\label{fig:hgg_nlo}}
\end{figure}

As far as $H\rightarrow gg$ is concerned, this decay can only be
induced by a fermion loop,
and therefore its rate, at the lowest order, can be written as:
\begin{equation}
\label{eq:rate_Hgg_lo}
\Gamma(H\rightarrow gg)=\frac{G_F\alpha_s^2M_H^3}{36\sqrt{2}\pi^3}
\left|\frac{3}{4}\sum_q A_q^H(\tau_q)\right|\,\,\,,
\end{equation}
where $\tau_q\!=\!4m_q^2/M_H^2$, $f(\tau)$ is defined in
Eq.(\ref{eq:f_tau}) and the form factor $A_q^H(\tau)$ is given in
Eq.~(\ref{eq:rate_HgammaZ_lo_form_factors}).  QCD corrections to
$H\rightarrow gg$ have been calculated up to NNLO in the
$m_t\rightarrow\infty$ limit, as explained in
Section~\ref{subsubsection:sm_higgs_decays_rad_corr_general}. At NLO
the expression of the corrected rate is remarkably simple
\begin{equation}
\Gamma(H\rightarrow gg(g),q\bar{q}g)=\Gamma_{LO}(H\rightarrow gg)
\left[1+E(\tau_Q)\frac{\alpha_s^{(N_L)}}{\pi}\right]\,\,\,,
\end{equation}
where
\begin{equation}
E(\tau_Q)\stackrel{M_H^2\ll 4m_q^2}{\longrightarrow} 
\frac{95}{4}-\frac{7}{6}N_L +
\frac{33-2N_F}{6}\log\left(\frac{\mu^2}{M_H^2}\right)\,\,\,.
\end{equation}
When compared with the fully massive NLO calculation (available in
this case), the two calculations display an impressive $10\%$
agreement, as illustrated in Fig.~\ref{fig:hgg_nlo}, even in regions
where the light Higgs approximation is not justified. This is actually
due to the presence of large constant factors in the first order of
QCD corrections.
We also observe that the first order of QCD corrections has quite a
large impact on the lowest order cross section, amounting to more than
50\% of $\Gamma_{LO}$ on average. This has been indeed the main reason
to prompt for a NNLO QCD calculation of $\Gamma(H\rightarrow gg)$. The
result, obtained in the heavy-top approximation, has shown that NNLO
QCD corrections amount to only 20\% of the NLO cross section,
therefore pointing to a convergence of the $\Gamma(H\rightarrow gg)$
perturbative series. We will refer to this discussion when dealing
with the $gg\rightarrow H$ production mode, since its cross section can be
easily related to $\Gamma(H\rightarrow gg)$.

\subsection{MSSM Higgs boson branching ratios}
\label{subsec:mssm_higgs_branching_ratios}
The decay patterns of the MSSM Higgs bosons are many and diverse,
depending on the specific choice of supersymmetric parameters. In
particular they depend on the choice of $M_A$ and $\tan\beta$, which
parameterize the MSSM Higgs sector, and they are clearly sensitive to
the choice of other supersymmetric masses (gluino masses, squark
masses, etc.) since this determines the possibility for the MSSM Higgs
bosons to decay into pairs of supersymmetric particles and for the
radiative induced decay channels ($h^0,H^0\rightarrow gg,\gamma\gamma,\gamma
Z$)  to receive supersymmetric loop contributions.

\begin{figure}
\begin{tabular}{cc}
\begin{minipage}{0.5\linewidth}
  {\includegraphics[scale=0.4]{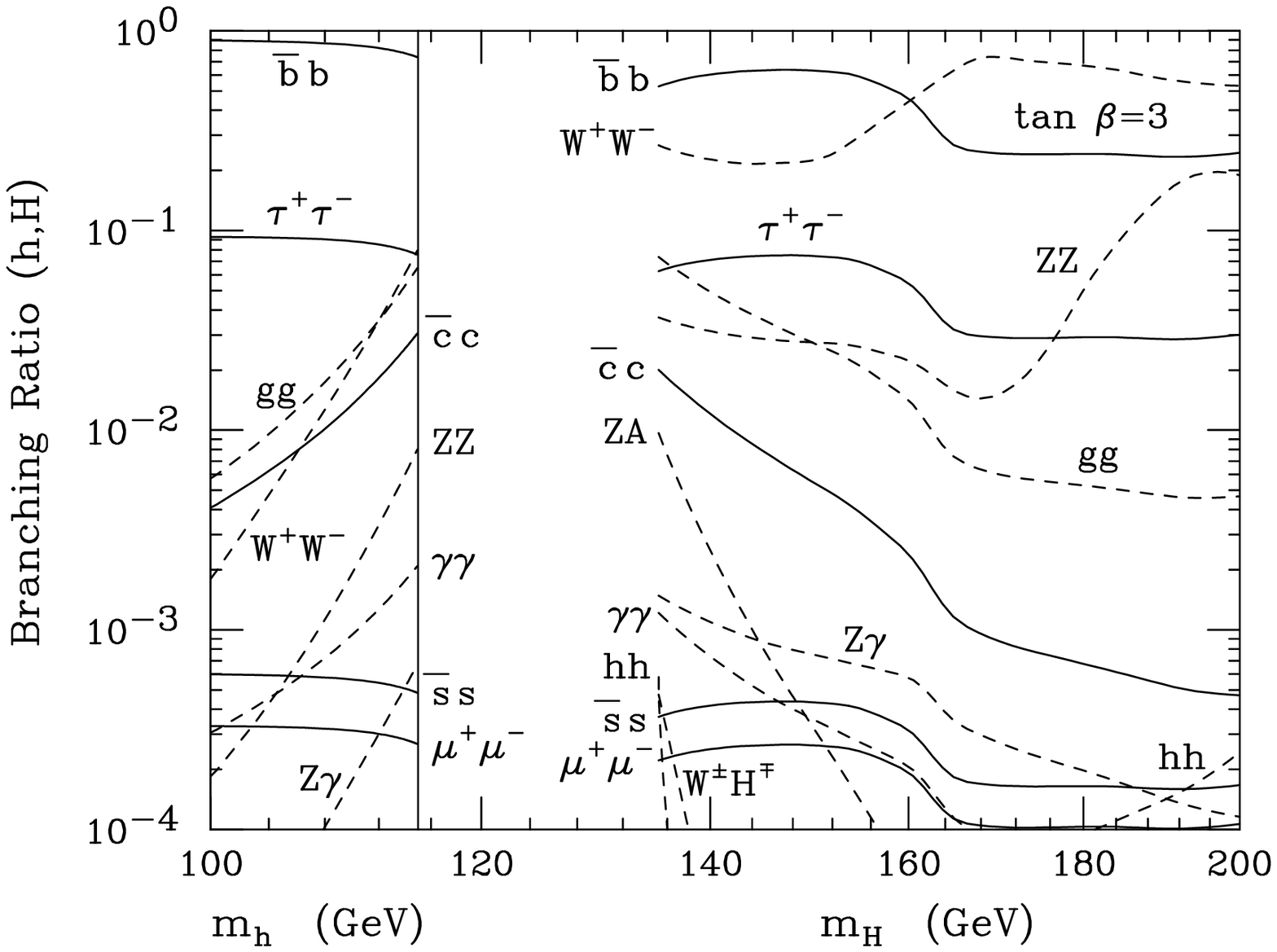}}
\end{minipage} &
\begin{minipage}{0.5\linewidth}
{\includegraphics[scale=0.4]{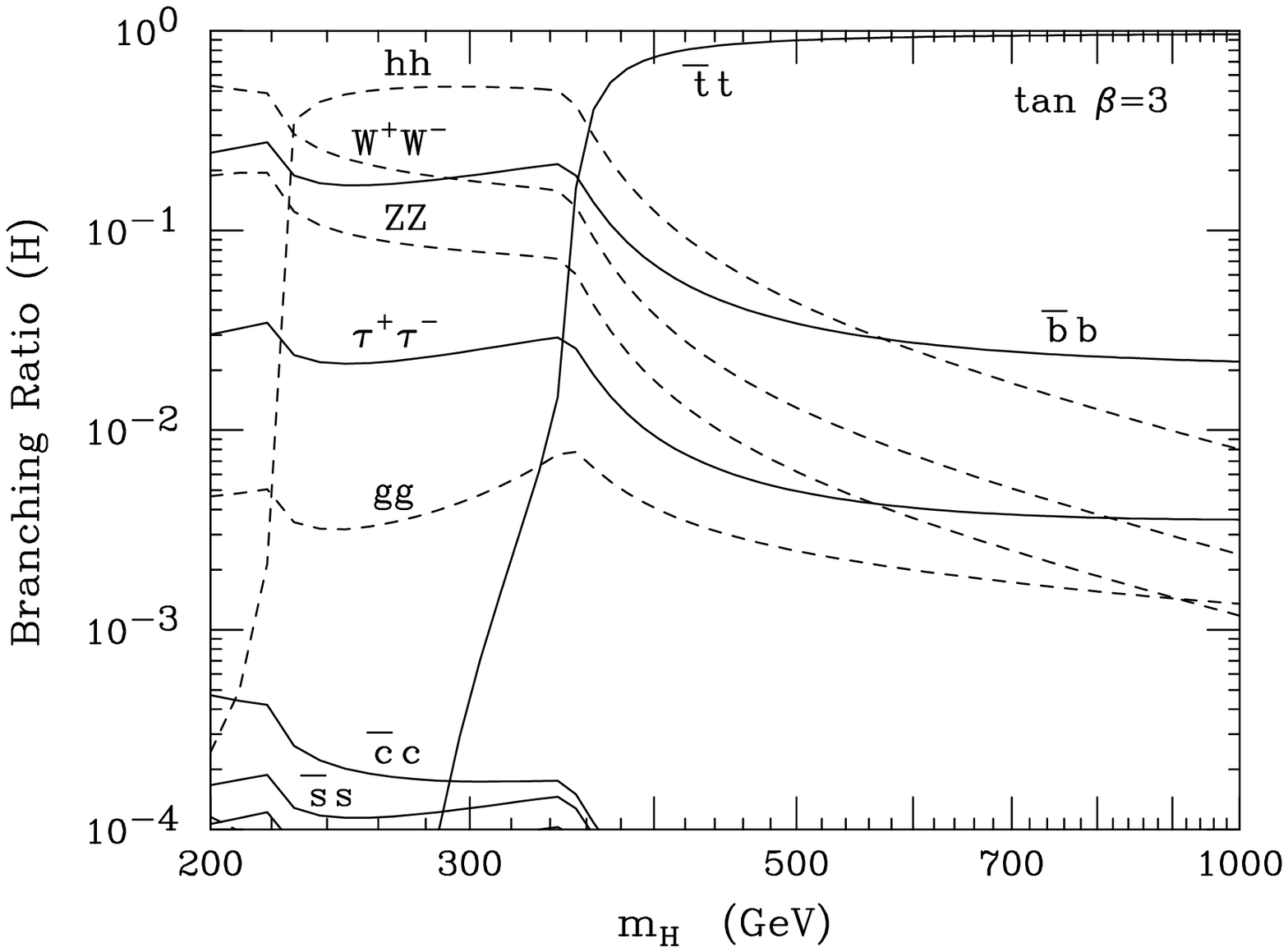}}
\end{minipage} 
\end{tabular}
\begin{tabular}{cc}
\begin{minipage}{0.5\linewidth}
{\includegraphics[scale=0.4]{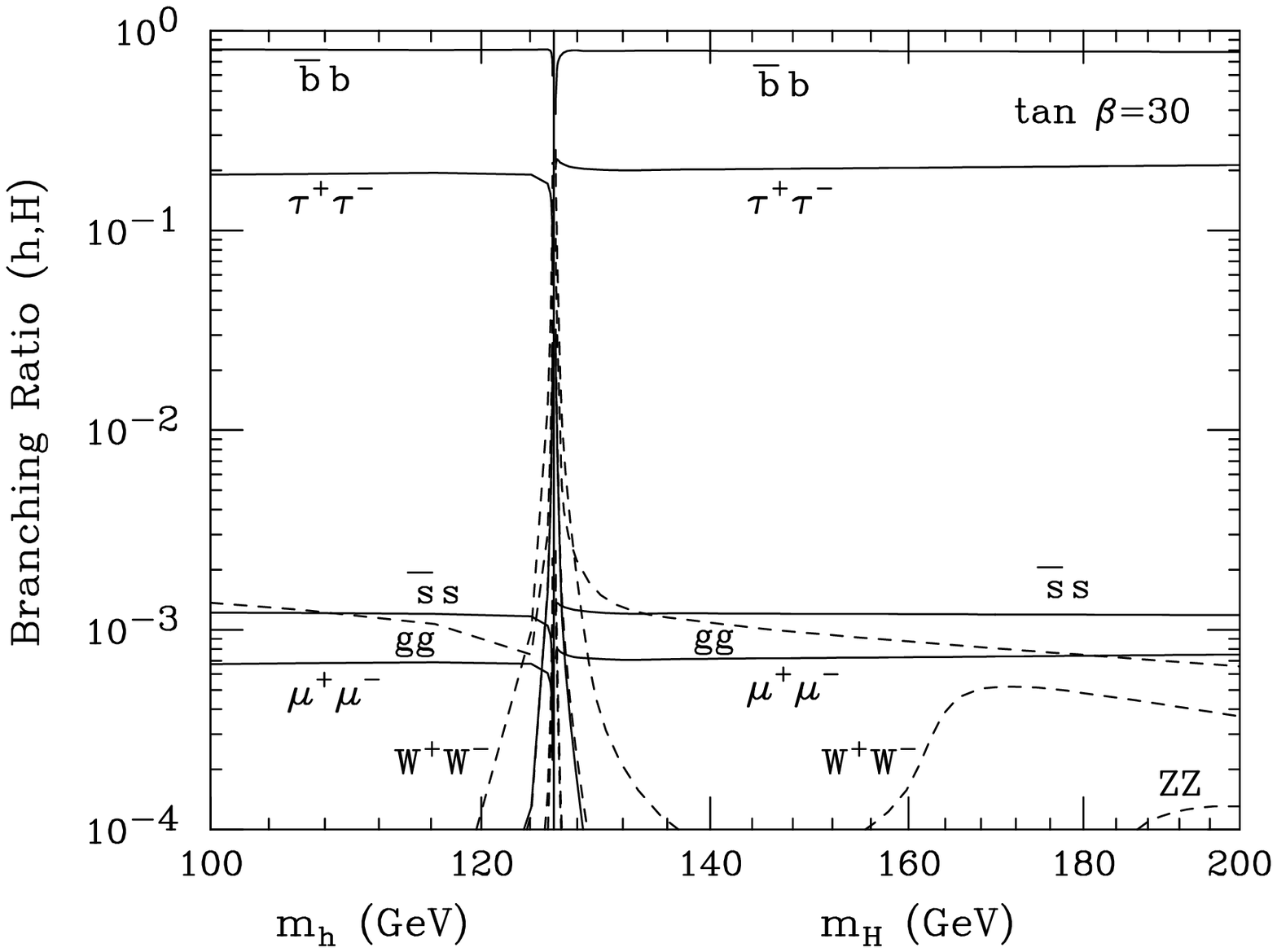}}
\end{minipage} &
\begin{minipage}{0.5\linewidth}
{\includegraphics[scale=0.4]{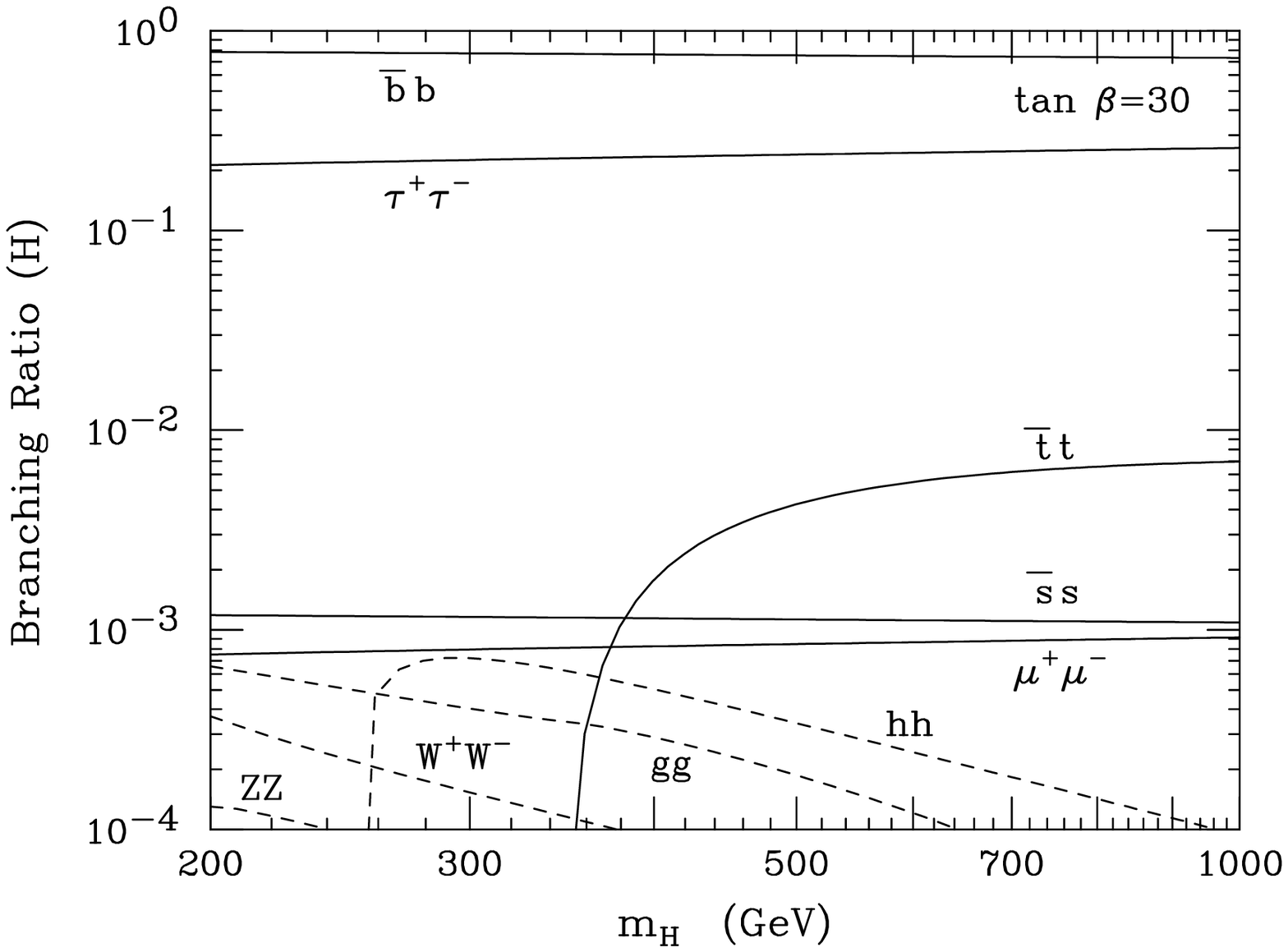}}
\end{minipage} 
\end{tabular}
\caption[]{Branching ratios for the $h^0$ and $H^0$ MSSM Higgs bosons,
for $\tan\beta\!=\!3,30$. The range of $M_H$ corresponds to
$M_A\!=\!90\,\mbox{GeV}-1\,\mbox{TeV}$, in the MSSM scenario discussed
in the text, with maximal top-squark mixing. The vertical line in the
left hand side plots indicates the upper bound on $M_h$, which, for
the given scenario is $M_h^{max}\!=\!115$~GeV ($\tan\beta=3$) or
$M_h^{max}\!=\!125.9$~GeV ($\tan\beta=30$). From
Ref.~\cite{Carena:2002es}.\label{fig:mssm_h_H_br_ratios_tanb3_30} }
\end{figure}
In order to be more specific, let us assume that all supersymmetric
masses are large enough to prevent the decay of the MSSM Higgs bosons
into pairs of supersymmetric particles (a good choice could be
$M_{\tilde g}\!=\!M_Q\!=\!=\!M_U\!=\!M_D\!=\!1$~TeV). Then, we only
need to examine the decays into SM particles and compare with the decay
patterns of a SM Higgs boson to identify any interesting difference.
From the study of the MSSM Higgs boson couplings in
Sections~\ref{subsubsec:mssm_higgs_couplings_bosons} and
\ref{subsubsec:mssm_higgs_couplings_fermions}, we expect that:
\emph{i)} in the decoupling regime, when $M_A\gg M_Z$, the properties
of the $h^0$ neutral Higgs boson are very much the same as the SM
Higgs boson; while away from the decoupling limit \emph{ii)} the
decay rates of $h^0$ and $H^0$ to electroweak gauge bosons are
suppressed with respect to the SM case, in particular for large Higgs
masses ($H^0$), \emph{iii)} the $A^0\rightarrow VV$ ($V=W^\pm,Z^0$)
decays are absent, \emph{iv)} the decay rates of $h^0$ and $H^0$ to
$\tau^+\tau^-$ and $b\bar{b}$ are enhanced for large $\tan\beta$,
\emph{v)} even for not too large values of $\tan\beta$, due to
\emph{ii)} above, the $h^0,H^0\rightarrow\tau^+\tau^-$ and 
$h^0,H^0\rightarrow b\bar{b}$ decay are large up
to the $t\bar{t}$ threshold, when the decay $H^0\rightarrow t\bar{t}$
becomes dominant, \emph{vi)} for the charged Higgs boson, the decay
$H^+\rightarrow\tau^+\nu_\tau$ dominates over $H^+\rightarrow
t\bar{b}$ below the $t\bar{b}$ threshold, and vice versa above it.

\begin{figure}
\begin{tabular}{cc}
\begin{minipage}{0.5\linewidth}
{\includegraphics[scale=0.4]{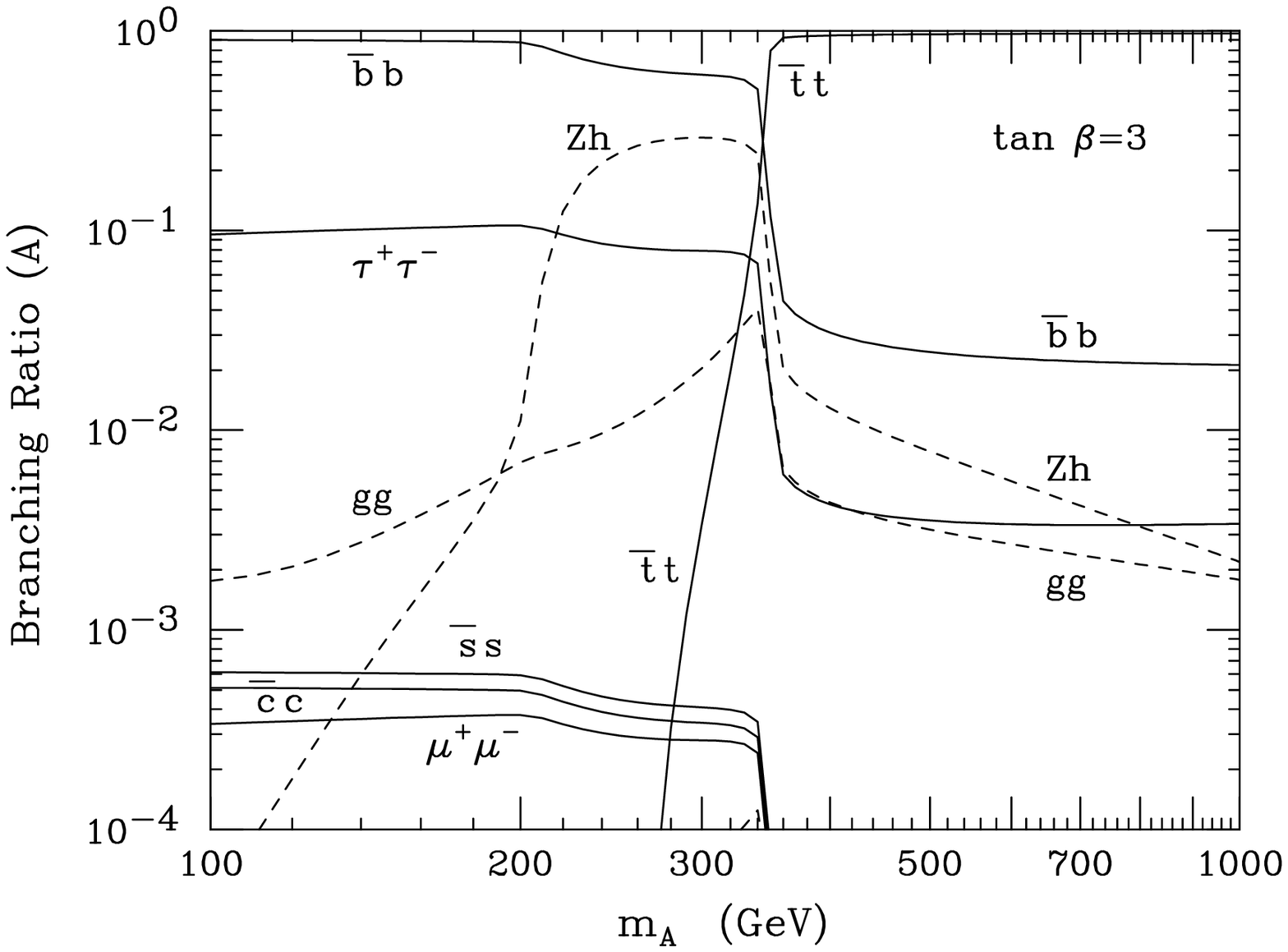}}
\end{minipage} &
\begin{minipage}{0.5\linewidth}
{\includegraphics[scale=0.4]{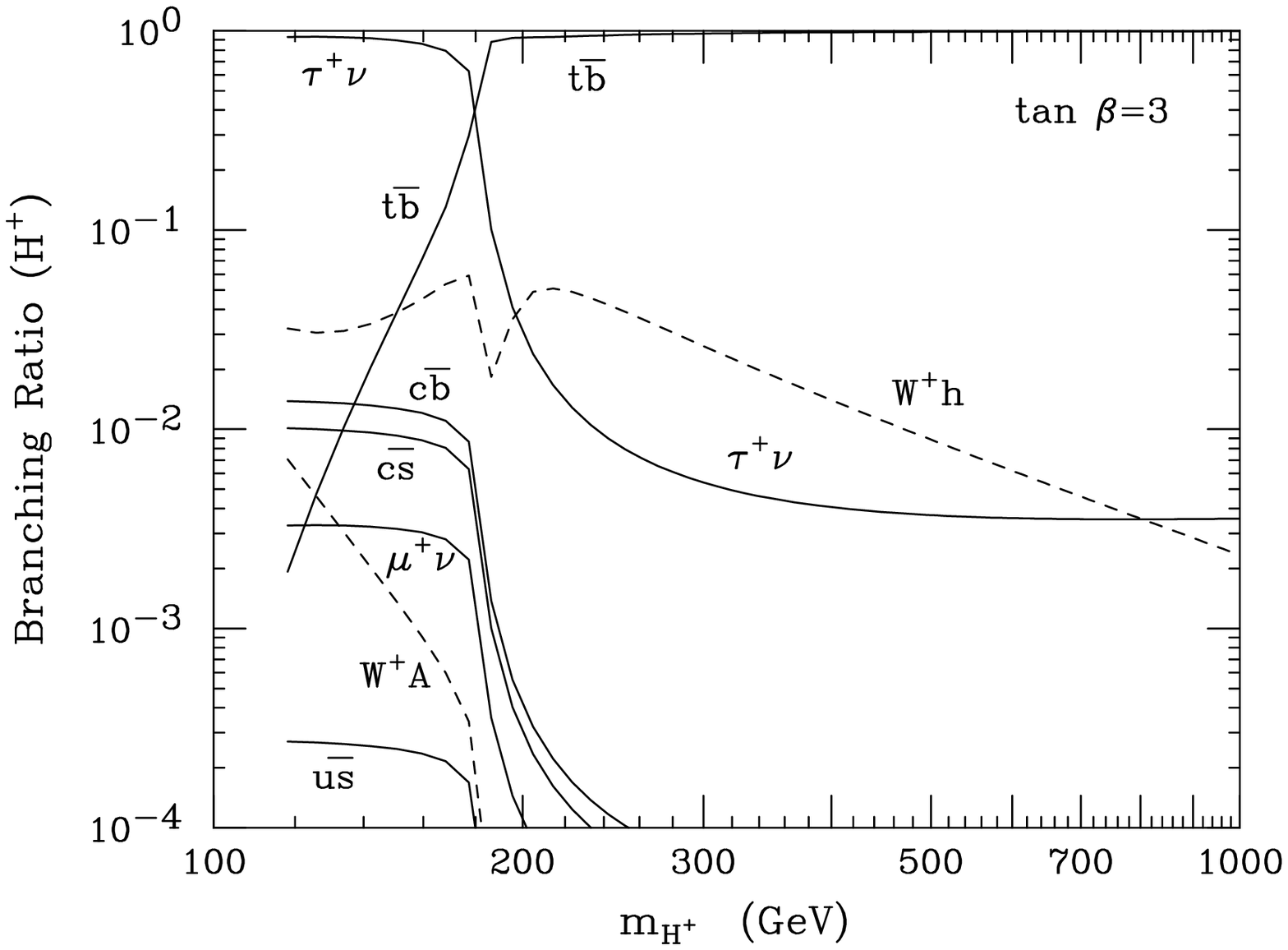}}
\end{minipage} \\
\begin{minipage}{0.5\linewidth}
{\includegraphics[scale=0.4]{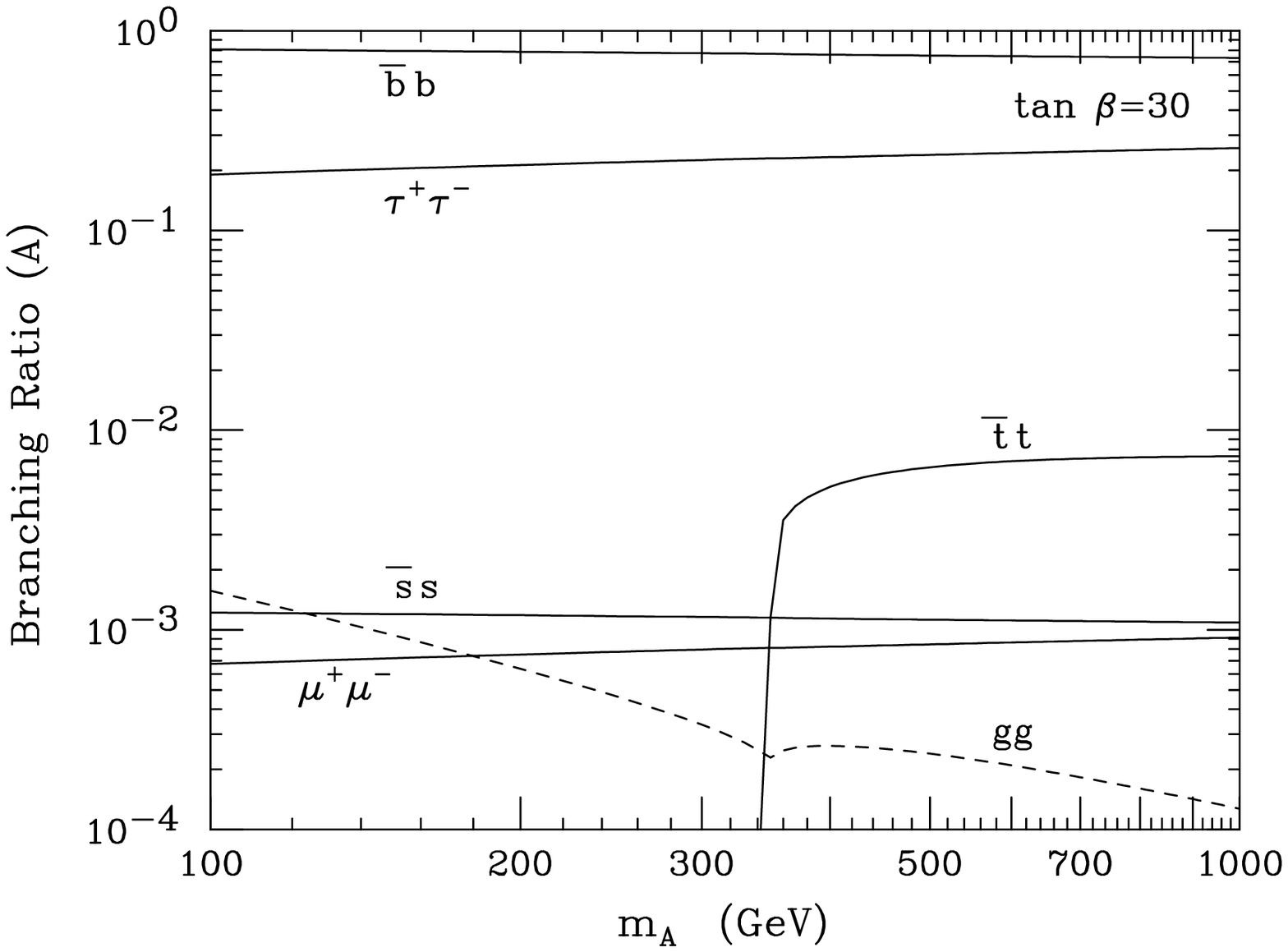}}
\end{minipage} &
\begin{minipage}{0.5\linewidth}
{\includegraphics[scale=0.4]{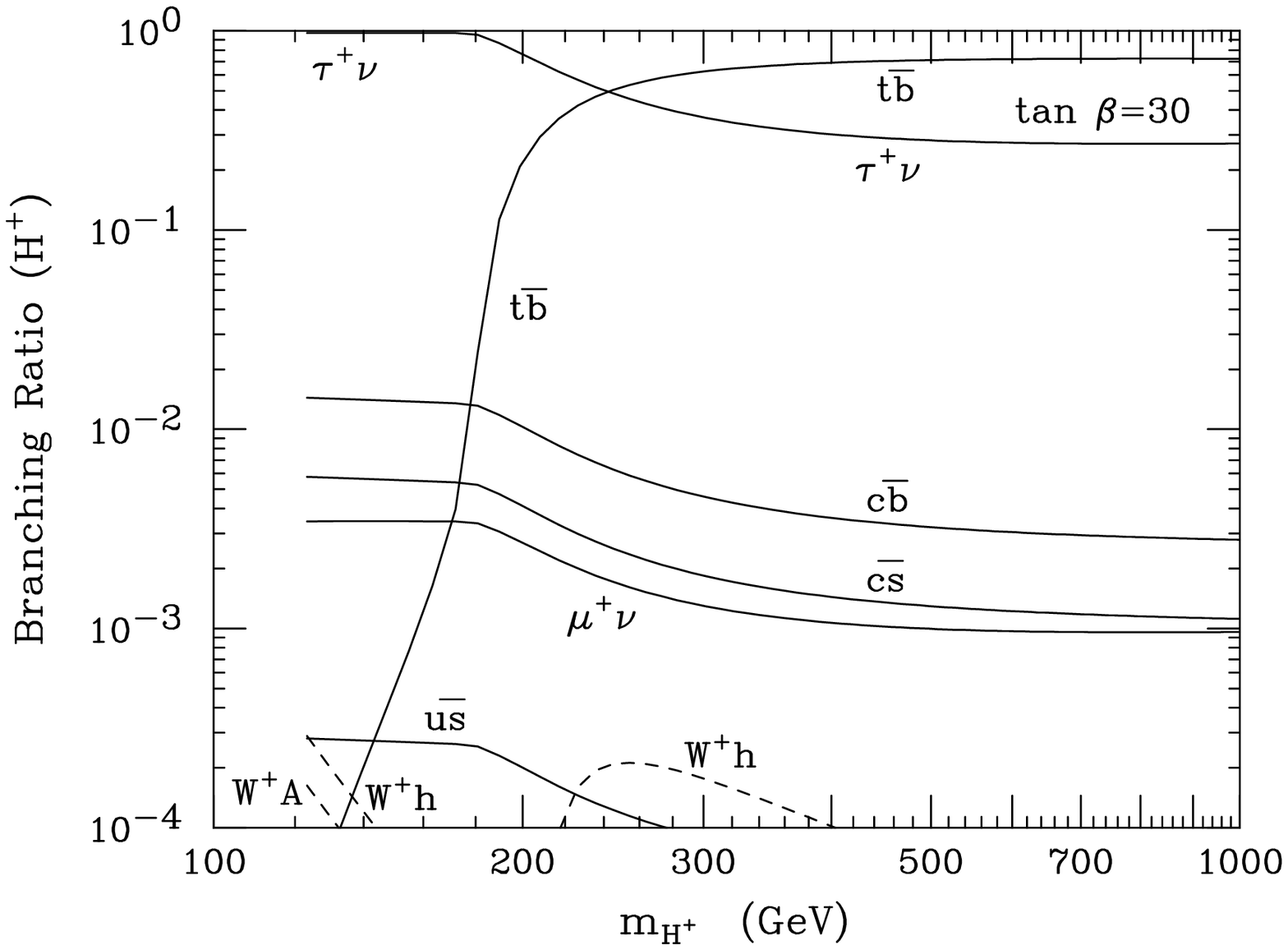}}
\end{minipage} 
\end{tabular}
\caption[]{Branching ratios for the $A^0$ and $H^+$ MSSM Higgs bosons,
for $\tan\beta\!=\!3,30$. The range of $M_{H^\pm}$ corresponds to
$M_A\!=\!90\,\mbox{GeV}-1\,\mbox{TeV}$, in the MSSM scenario discussed
in the text, with maximal top-squark mixing. From
Ref.~\cite{Carena:2002es}.\label{fig:mssm_A_H+_br_ratios_tanb3_30}}
\end{figure}

As far as QCD and EW radiative corrections go, what we have seen in
Sections~\ref{subsubsec:sm_higgs_to_gaugebosons}-\ref{subsubsec:sm_higgs_loop_decays}
for the SM case applies to the corresponding MSSM decays
too. Moreover, the truly MSSM corrections discussed in
Sections~\ref{subsubsec:mssm_higgs_couplings_bosons} and
\ref{subsubsec:mssm_higgs_couplings_fermions} need
to be taken into account and are included in
Figs.\ref{fig:mssm_h_H_br_ratios_tanb3_30} and \ref{fig:mssm_A_H+_br_ratios_tanb3_30}.

\subsection{Direct bounds on both SM and MSSM Higgs bosons}
\label{subsec:sm_mssm_higgs_direct_bounds}
\begin{figure}
\begin{tabular}{ccc}
\begin{minipage}{.3\linewidth}
\includegraphics[scale=0.6]{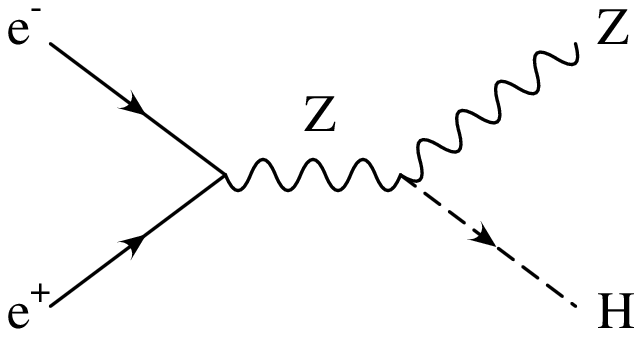}
\end{minipage} &
\begin{minipage}{.3\linewidth}
\includegraphics[scale=0.6]{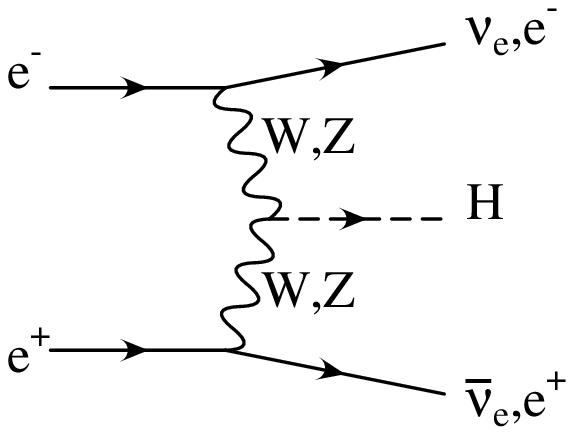}
\end{minipage} &
\begin{minipage}{.3\linewidth}
\includegraphics[scale=0.6]{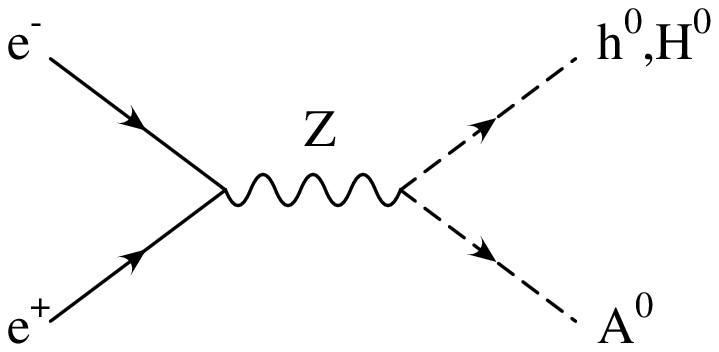}
\end{minipage}
\end{tabular}
\caption[]{SM and MSSM neutral Higgs boson production channels at LEP2. 
\label{fig:lep2_search_processes}}
\end{figure}

LEP2 has searched for a SM Higgs at center of mass energies between
189 and 209~GeV.  In this regime, a SM Higgs boson is produced mainly
through Higgs boson strahlung from $Z$ gauge bosons,
$e^+e^-\rightarrow Z^*\rightarrow HZ$, and to a lesser extent through
$WW$ and $ZZ$ gauge boson fusion, $e^+e^-\rightarrow WW,ZZ\rightarrow
H\nu_e\bar{\nu}_e, He^+e^-$ (see
Fig.~\ref{fig:lep2_search_processes}). Once produced, it decays mainly
into $b\bar{b}$ pairs, and more rarely into $\tau^+\tau^-$ pairs. The
four LEP2 experiments have been looking for: \emph{i)} a four jet
final state ($H\rightarrow b\bar{b}$, $Z\rightarrow q\bar{q}$), 
\emph{ii)} a missing
energy final state ($H\rightarrow b\bar{b}$,
$Z\rightarrow\nu\bar{\nu}$), 
\emph{iii)} a
leptonic final state ($H\rightarrow b\bar{b}$, $Z\rightarrow l^+l^-$) and
\emph{iv)} a specific  $\tau$-lepton final state ($H\rightarrow
b\bar{b}$, $Z\rightarrow\tau^+\tau^-$ plus $
H\rightarrow\tau^+\tau^-$, $Z\rightarrow q\bar{q}$).
The absence of any statistical significant signal has set a 95\% CL
lower bound on the SM Higgs boson at
\[
M_{H_{SM}}>114.4\,\,\mbox{GeV}\,\,\,. 
\]

\begin{figure}
\hspace{-1.truecm}
\begin{tabular}{lr}
\begin{minipage}{0.5\linewidth}
{\includegraphics[scale=0.45]{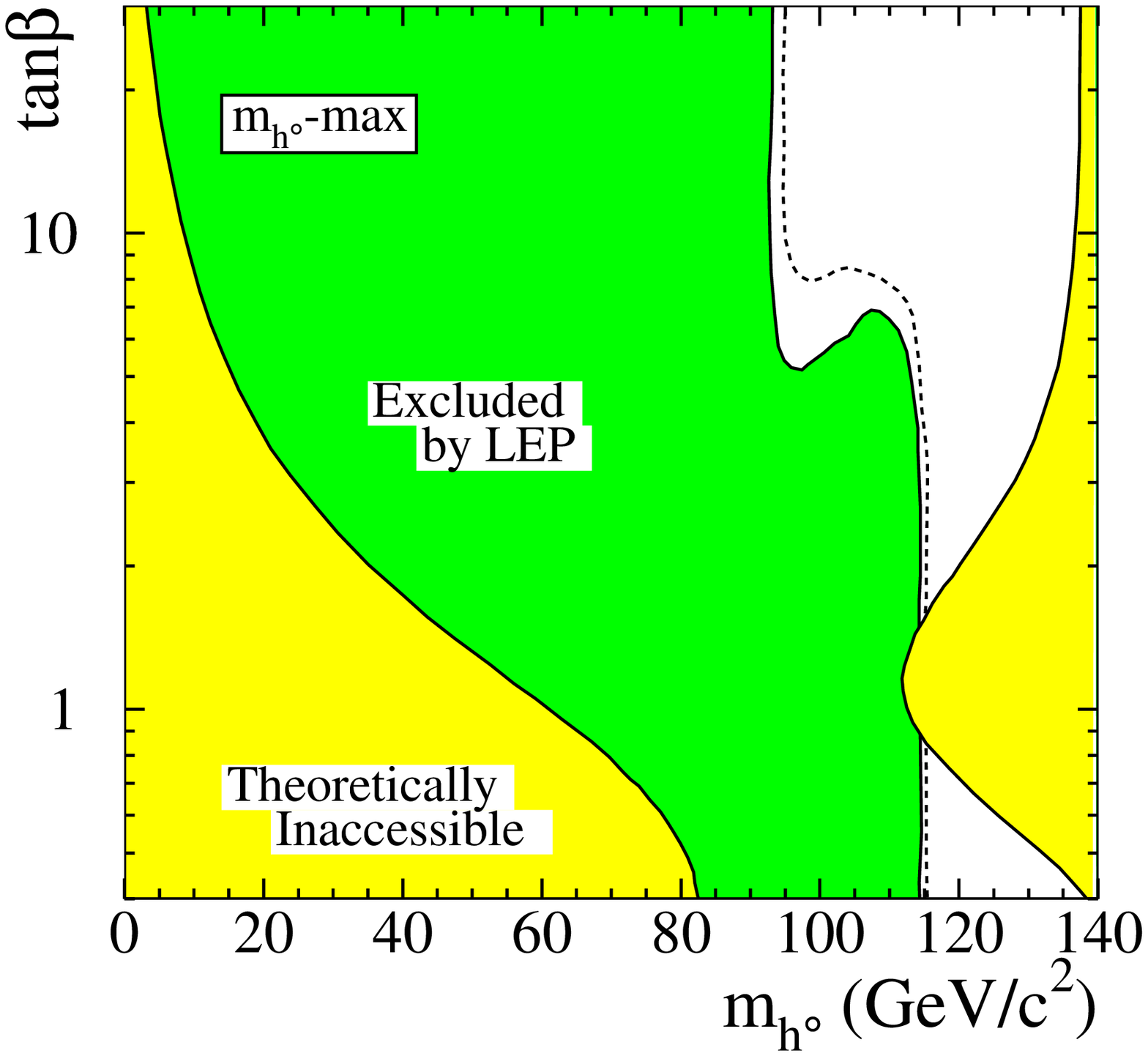}}
\end{minipage} &
\begin{minipage}{0.5\linewidth}
{\includegraphics[scale=0.45]{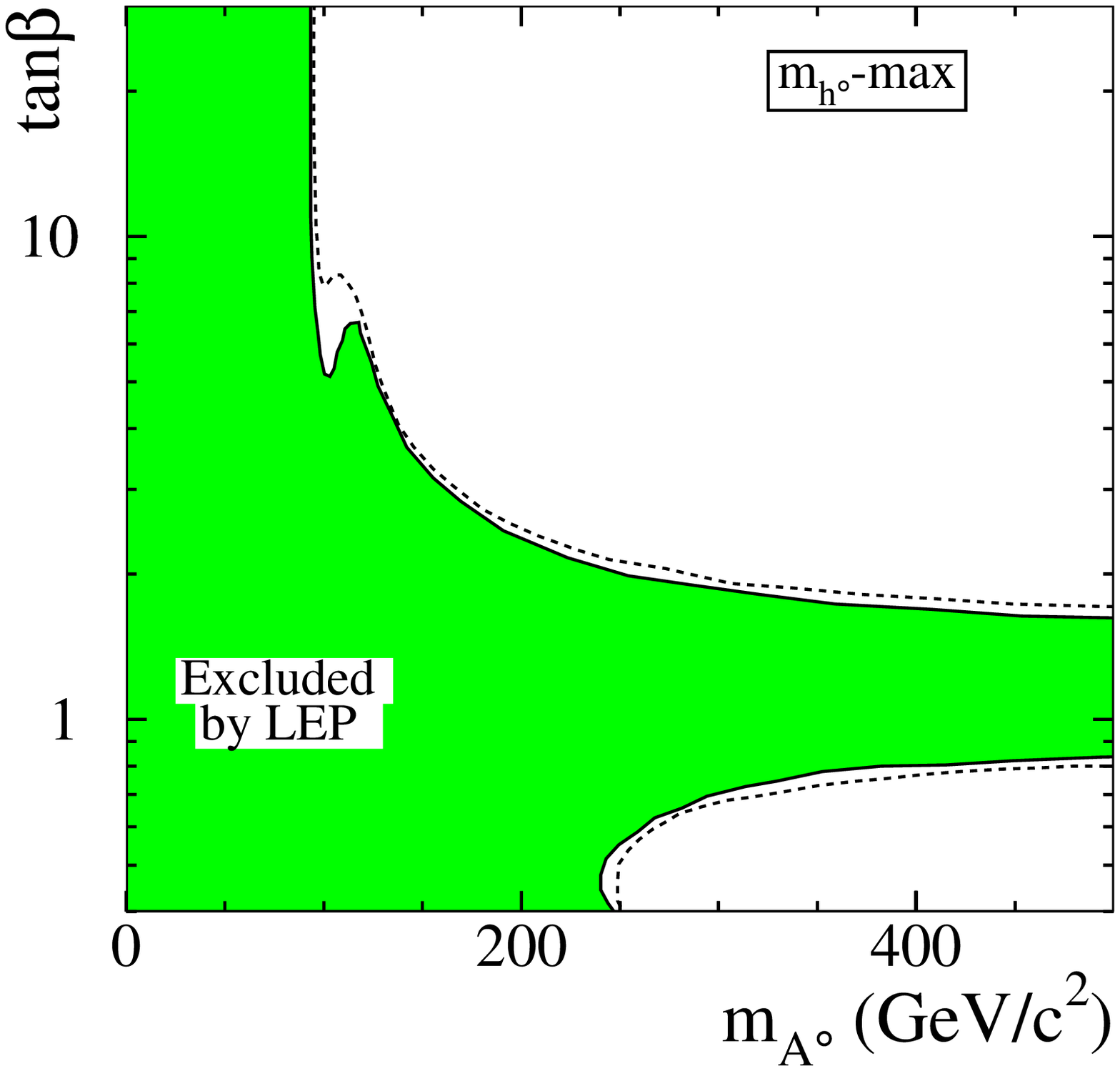}}
\end{minipage}\\
\begin{minipage}{0.5\linewidth}
{\includegraphics[scale=0.45]{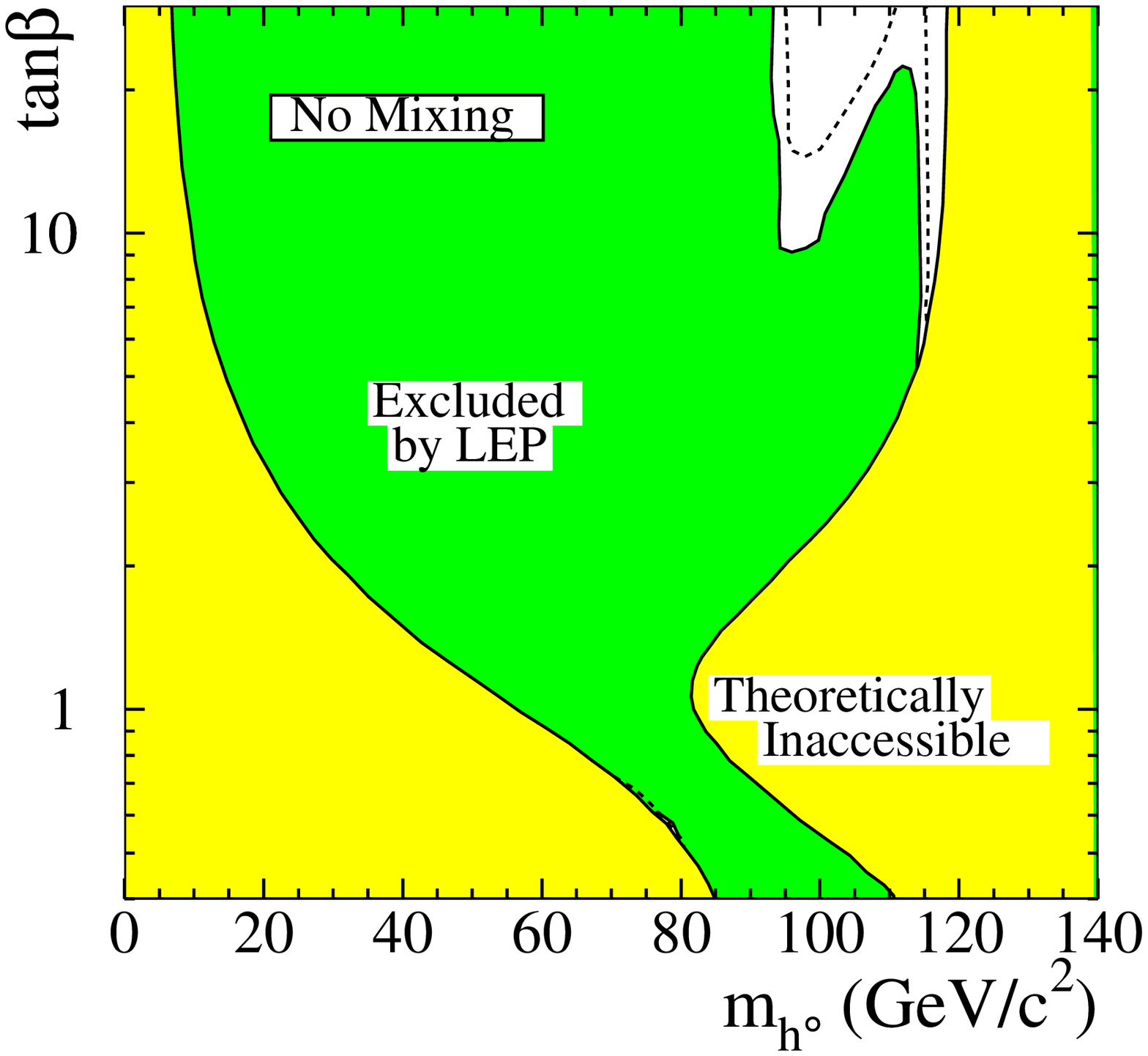}}
\end{minipage} &
\begin{minipage}{0.5\linewidth}
{\includegraphics[scale=0.45]{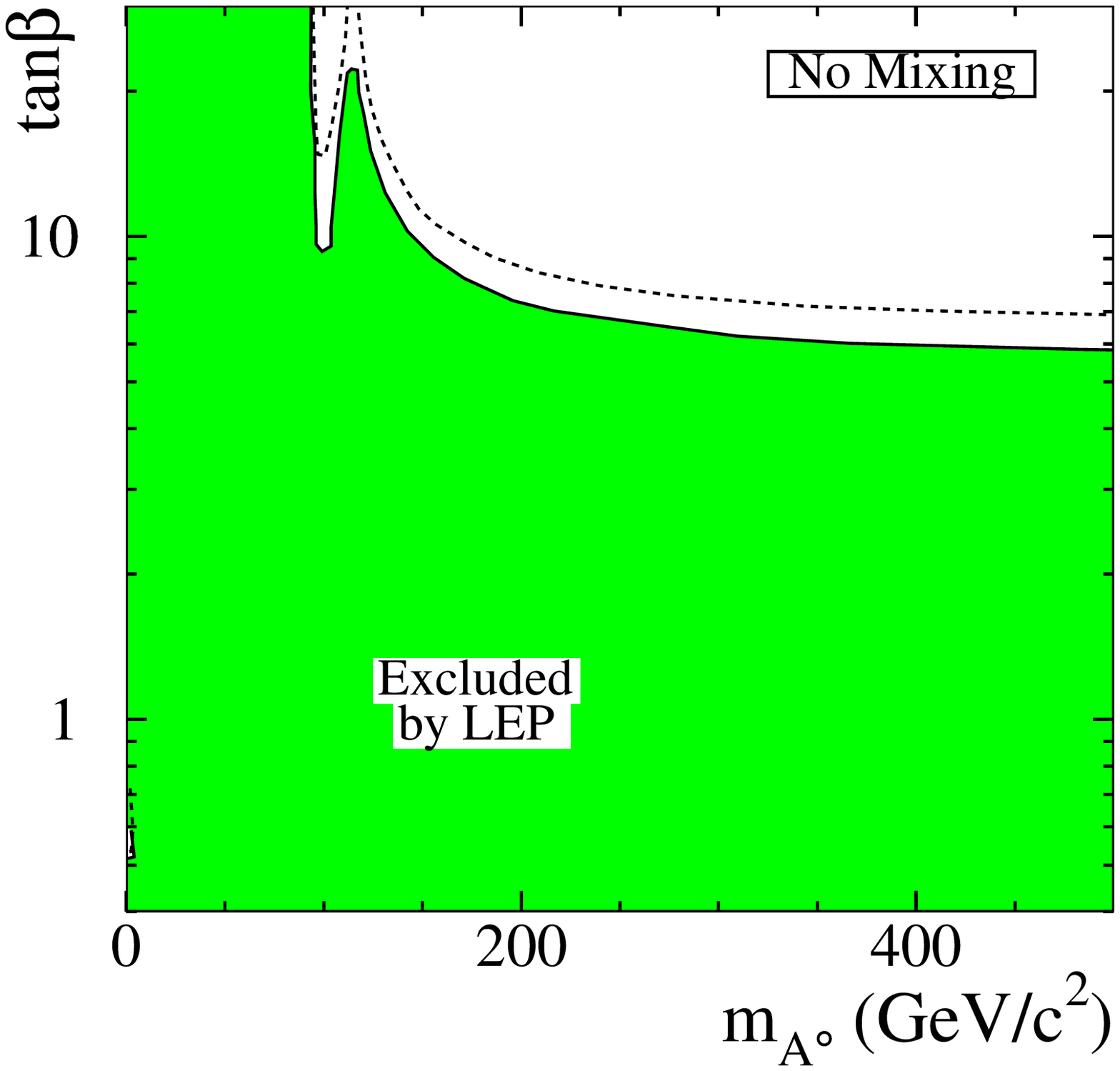}}
\end{minipage}
\end{tabular}
\caption[]{95\% CL exclusion limits for MSSM Higgs parameters from LEP2:
$(M_h,\tan\beta)$ (left) and $(M_A,\tan\beta)$ (right). Both the
maximal and no-mixing scenarios are illustrated, for $M_S\!=\!1$~TeV
and $m_t\!=\!179.3$~GeV. The dashed lines indicate the boundaries that
are excluded on the basis of Monte Carlo simulations in the absence of
a signal. From
Ref.~\cite{Djouadi:2005gj}.\label{fig:mh_mA_tanb_exclusion_from_searches}}
\end{figure}
LEP2 has also looked for the light scalar ($h^0$) and pseudoscalar
($A^0$) MSSM neutral Higgs bosons. In the decoupling regime, when
$A^0$ is very heavy and $h^0$ behaves like a SM Higgs bosons, only
$h^0$ can be observed and the same bounds established for the SM Higgs
boson apply. The bound can however be lowered when $m_A$ is
lighter. In that case, $h^0$ and $A^0$ can also be pair produced
through $e^+e^-\rightarrow Z\rightarrow h^0A^0$ (see
Fig.~\ref{fig:lep2_search_processes}). Combining the different
production channels one can derive plots like those shown in
Fig.~\ref{fig:mh_mA_tanb_exclusion_from_searches}, where the excluded
$(M_h,\tan\beta)$ and $(M_A,\tan\beta)$ regions of the MSSM parameter
space are shown.  The LEP2 collaborations~\cite{lephwg:2004mssm} have
been able to set the following bounds at 95\% CL:
\[
M_{h,A}>93.0\,\,\mbox{GeV}\,\,\,,
\]
obtained in the limit when $\cos(\beta-\alpha)\simeq 1$
(anti-decoupling regime) and for large $\tan\beta$. The plots in
Fig.~\ref{fig:mh_mA_tanb_exclusion_from_searches} have been obtained
in the maximal mixing scenario (explained in
Section~\ref{subsubsec:higgs_mssm}). For no-mixing, the corresponding
plots would exclude a much larger region of the MSSM parameter space.

Finally, the LEP collaborations have looked for the production of the
MSSM charged Higgs boson in the associated production channel:
$e^+e^-\rightarrow\gamma,Z^*\rightarrow H^+H^-$~\cite{lhwg:2001xy}.
An absolute lower bound of
\[
M_{H^\pm}>79.3\,\,\mbox{GeV}\,\,\,
\]
has been set by the ALEPH collaboration, and slightly lower values
have been obtained by the other LEP collaborations.

\subsection{Higgs boson studies at the Tevatron and at the LHC}
\label{subsec:higgs_tevatron_lhc}
\begin{figure}
\begin{tabular}{ccc}
\begin{minipage}{0.3\linewidth}
\includegraphics[scale=0.6]{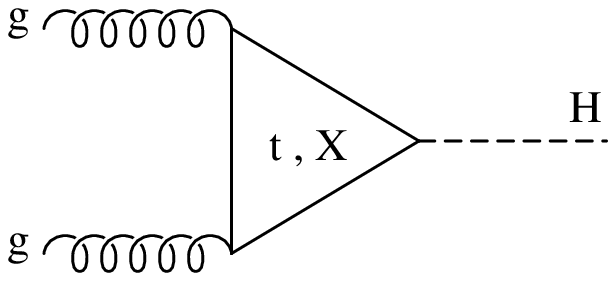}
\end{minipage}&
\begin{minipage}{0.3\linewidth}
\includegraphics[scale=0.6]{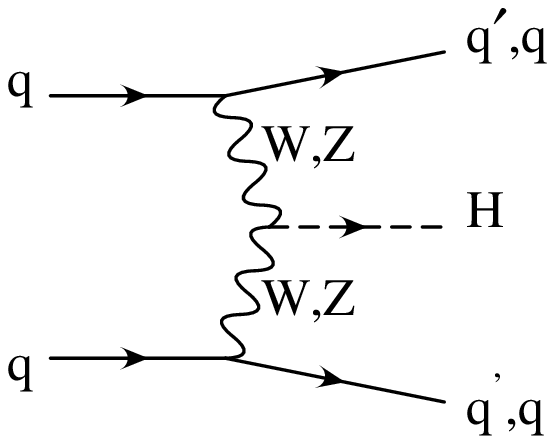}
\end{minipage}&
\begin{minipage}{0.3\linewidth}
\includegraphics[scale=0.6]{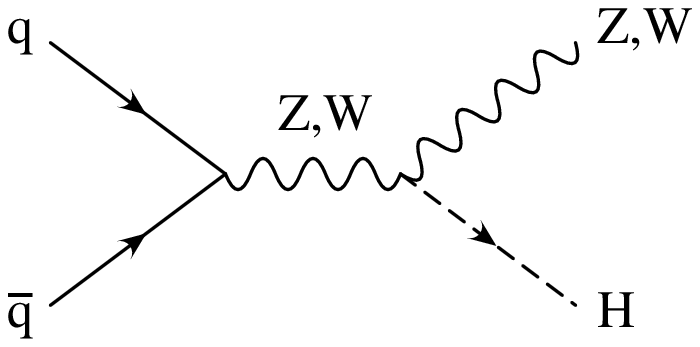}
\end{minipage}
\end{tabular}
\caption[]{Leading Higgs production processes at hadron colliders:
$gg\rightarrow H$, $qq\rightarrow qqH$, and $q\bar{q}\rightarrow
WH,ZH$.\label{fig:sm_higgs_production:ggh_qqHZ_qqHW_qqHqq}}
\end{figure}

\begin{figure}
\centering
\includegraphics[scale=0.6]{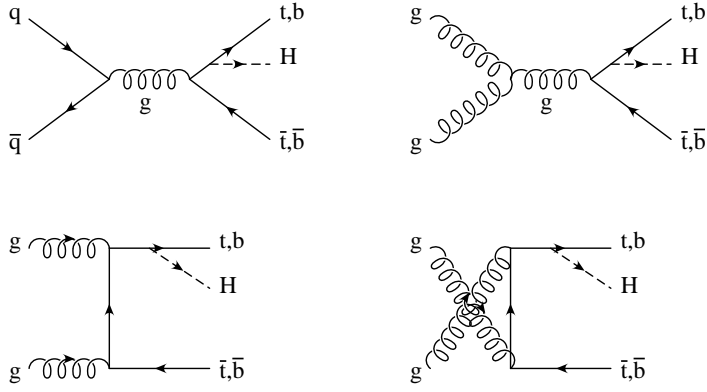}
\caption[]{Higgs production with heavy quarks: sample of Feynman
diagrams illustrating the two corresponding parton level processes
$q\bar{q},gg\rightarrow t\bar{t}H,b\bar{b}H$. Analogous diagrams with
the Higgs boson leg attached to the remaining top(bottom)-quark
legs are understood.\label{fig:sm_higgs_production:QQh}}
\end{figure}

The parton level processes through which a SM Higgs boson can be
produced at hadron colliders are illustrated in
Figs.~\ref{fig:sm_higgs_production:ggh_qqHZ_qqHW_qqHqq} and
\ref{fig:sm_higgs_production:QQh}.

\begin{figure}
\includegraphics[scale=0.45,angle=-90]{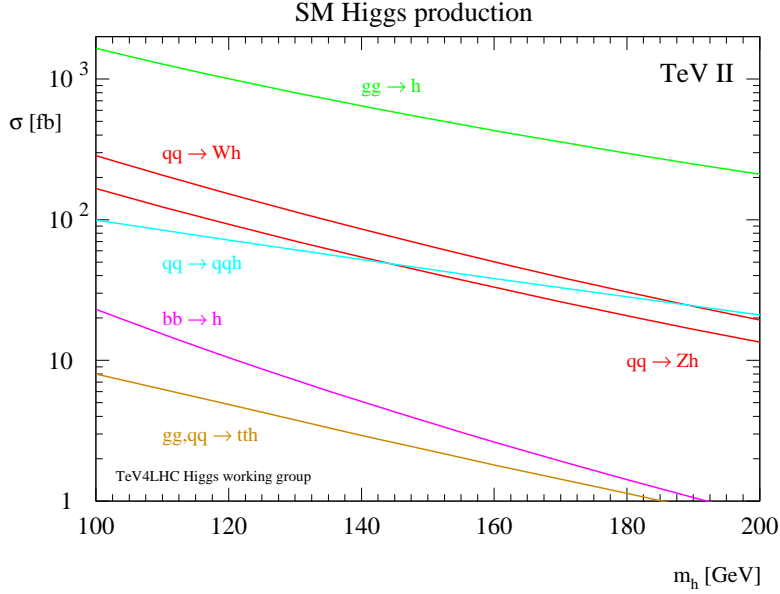}
\caption[]{Cross sections for SM Higgs boson production
processes at the Tevatron, Run II ($\sqrt{s}\!=\!1.96$~TeV). From
Ref.~\cite{Tev4lhc:hwg}.\label{fig:sm_higgs_tevatron}}
\end{figure}

\begin{figure}
\includegraphics[scale=0.45,angle=-90]{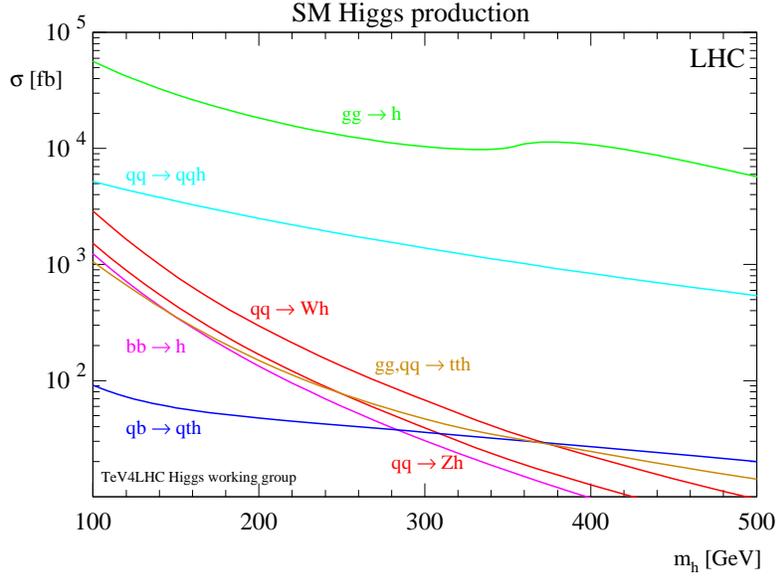}
\caption[]{Cross sections for SM Higgs boson production
processes at the LHC ($\sqrt{s}\!=\!14$~TeV). From
Ref.~\cite{Tev4lhc:hwg}.\label{fig:sm_higgs_lhc}}
\end{figure}

Figures~\ref{fig:sm_higgs_tevatron} and \ref{fig:sm_higgs_lhc}
summarize the cross sections for all these production modes as
functions of the SM Higgs boson mass, at the Tevatron (center of mass
energy: $\sqrt{s}\!=\!1.96$~TeV) and at the LHC (center of mass
energy: $\sqrt{s}\!=\!14$~TeV). These figures have been recently
produced during the TeV4LHC workshop~\cite{Tev4lhc:hwg}, and contain
all known orders of QCD corrections as well as the most up to date
input parameters. We postpone further details about QCD corrections
till Section~\ref{sec:theory}, while we comment here about some
general phenomenological aspects of hadronic Higgs production.

The leading production mode is gluon-gluon fusion, $gg\rightarrow H$
(see first diagram in
Fig.~\ref{fig:sm_higgs_production:ggh_qqHZ_qqHW_qqHqq}). In spite of
being a loop induced process, it is greatly enhanced by the top-quark
loop. For light and intermediate mass Higgs bosons, however, the very
large cross section of this process has to compete against a very
large hadronic background, since the Higgs boson mainly decays to
$b\bar{b}$ pairs, and there is no other non-hadronic probe that can
help distinguishing this mode from the overall hadronic activity in
the detector. To beat the background, one has often to employ
subleading if not rare Higgs decay modes, like
$H\rightarrow\gamma\gamma$, and this \emph{dilutes} the large cross
section. For larger Higgs masses, above the $ZZ$ threshold, on the
other hand, gluon-gluon fusion together with $H\rightarrow ZZ$
produces a very distinctive signal, and make this mode a
``\emph{gold-plated mode}'' for detection. For this reason,
$gg\rightarrow H$ plays a fundamental role at the LHC over the entire
Higgs boson mass range, but is of very limited use at the Tevatron,
where it can only be considered for Higgs boson masses very close to
the upper reach of the machine ($M_H\simeq 200$~GeV).

Weak boson fusion ($qq\rightarrow qqH$, see second diagram in
Fig.~\ref{fig:sm_higgs_production:ggh_qqHZ_qqHW_qqHqq}) and the
associated production with weak gauge bosons ($q\bar{q}\rightarrow
WH,ZH$, see third diagram in
Fig.~\ref{fig:sm_higgs_production:ggh_qqHZ_qqHW_qqHqq}) have also
fairly large cross sections, of different relative size at the
Tevatron and at the LHC. $q\bar{q}\rightarrow WH,ZH$ is particularly
important at the Tevatron, where only a relatively light Higgs boson
($M_H<200$~GeV) will be accessible. In this mass region, $gg\rightarrow H,
H\rightarrow \gamma\gamma$ is too small and $qq\rightarrow qqH$ is
suppressed (because the initial state is $p\bar{p}$). On the other
hand, $qq\rightarrow qqH$ becomes instrumental at the LHC ($pp$
initial state) for low and intermediate mass SM Higgs bosons, where
its characteristic final state configuration, with two very forward
jets, has been shown to greatly help in disentangling this signal from
the hadronic background, using different Higgs decay channels.

Finally, the production of a SM Higgs boson with heavy quarks, in the
two channels $q\bar{q},gg\rightarrow Q\bar{Q}H$ (with $Q\!=\!t,b$, see
Fig.~\ref{fig:sm_higgs_production:QQh}), is sub-leading at both the
Tevatron and the LHC, but has a great physics potential.  The
associated production with $t\bar{t}$ pairs is probably too small to
be seen at the Tevatron, given the expected luminosities, but will
play a very important role for a light SM Higgs boson
($M_H\!<\!130-140$~GeV) at the LHC, where enough statistics will be
available to fully exploit the spectacular signature of a $t\bar{t}H,
H\rightarrow b\bar{b}$ final state. Moreover, at the LHC, the
associated production of a Higgs boson with top quarks will offer a
direct handle on the top-quark Yukawa coupling (see
Section~\ref{subsubsec:sm_higgs_properties}). On the other hand, the
production of a SM Higgs boson with $b\bar{b}$ pairs is tiny, since
the SM bottom-quark Yukawa coupling is suppressed by the bottom-quark
mass. Therefore, the $b\bar{b}H,\,H\rightarrow b\bar{b}$ channel is
the ideal candidate to provide evidence of new physics, in particular
of extension of the SM, like supersymmetric models, where the
bottom-quark Yukawa coupling to one or more Higgs bosons is enhanced
(e.g., by large $\tan\beta$ in the MSSM). $b\bar{b}H$ production is
kinematically well within the reach of the Tevatron, RUN II. First
studies from both CDF~\cite{Affolder:2000rg} and
D$\emptyset$~\cite{Abazov:2005yr} have already translated the absence
of a $b\bar{b}h^0,H^0,A^0$ signal into an upper bound on the
$\tan\beta$ parameter of the MSSM. Were a signal observed, $b\bar{b}H$
could actually provide the first piece of evidence for new physics
from RUN II.

\subsubsection{Searching for a SM Higgs boson at the Tevatron and the LHC}
\label{subsubsec:sm_higgs_searches}
Discovering a Higgs boson during RUN II of the Tevatron is definitely
among the most important goal of this collider. It will be challenging
and mainly luminosity limited, but recent studies have confirmed that
RUN II can push the 95\% CL exclusion limit much farer than LEP2 and
also shoot for a $3\sigma$ or $5\sigma$ discovery, depending on the
integrated luminosity accumulated. 

\begin{figure}
\centering
\includegraphics[scale=0.7]{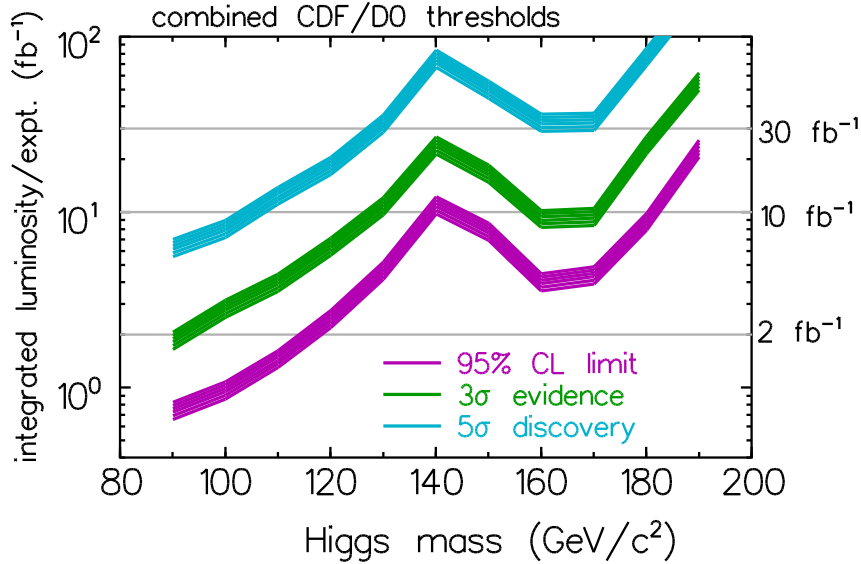}
\caption[]{Integrated luminosity required for each experiment at the
Tevatron, Run II, to exclude a SM Higgs boson at 95\% CL or to observe
it at the $3\sigma$ or $5\sigma$ level. Original study from
Ref.~\cite{Carena:2000yx}.\label{fig:tevatron_run2_reach_old}}
\end{figure}
The plot in Fig.~\ref{fig:tevatron_run2_reach_old} shows the 
integrated luminosity that was originally estimated to be necessary to
reach the 95\% CL exclusion limit, the $3\sigma$, and the $5\sigma$
discovery levels. It is given for a SM Higgs boson mass up to 200~GeV,
that is to be considered as the highest Higgs boson mass reachable by RUN
II. The curves have been obtained mainly by using the associated
production with weak gauge bosons, $q\bar{q}\rightarrow VH$
($V\!=\!W^\pm,Z^0$), with $H\rightarrow b\bar{b}$ and $H\rightarrow
W^+W^-$, over the entire Higgs boson mass range, and $gg\rightarrow H$
with $H\rightarrow ZZ$ in the upper mass region. As discussed in the
introduction to Section~\ref{subsec:higgs_tevatron_lhc}, this can be
understood in terms of production cross sections (see
Fig.~\ref{fig:sm_higgs_tevatron}) and decay branching ratios (see
Fig.~\ref{fig:sm_higgs_br}) over the $M_H\!=\!115-200$~GeV mass range.
From Fig.~\ref{fig:tevatron_run2_reach_old} we see that with, e.g.,
10~fb$^{-1}$ of integrated luminosity RUN II will be able to put a
95\% CL exclusion limit on a SM Higgs boson of mass up to $180$~GeV,
while it could claim a $3\sigma$ discovery of a SM Higgs boson with
mass up to $125$~GeV. A $5\sigma$ discovery of a SM Higgs boson up to
130~GeV, i.e. in the region immediately above the LEP2 lower bound,
seemed to require 30~fb$^{-1}$ of integrated luminosity, well beyond
what is currently expected for RUN II.

More recently, a new \emph{sensitivity study} has
appeared~\cite{Babukhadia:2003zu}, where the low mass region only has
been revisited and new luminosity curves have been drawn.  Mainly
using the $q\bar{q}\rightarrow VH$ ($V\!=\!W^\pm,Z^0$) production
mode, it appears that new analyses techniques will allow to obtain
better results with less integrated luminosity. A $3\sigma$ discovery
of a SM Higgs boson with mass up to $125$~GeV will now require only
about 5~fb$^{-1}$, while 10~fb$^{-1}$ could allow a $5\sigma$
discovery of a SM Higgs boson with mass up to about $120$~GeV, right
at the LEP2 lower bound limit.

\begin{figure}
\centering
\begin{tabular}{l}
\includegraphics[scale=0.5]{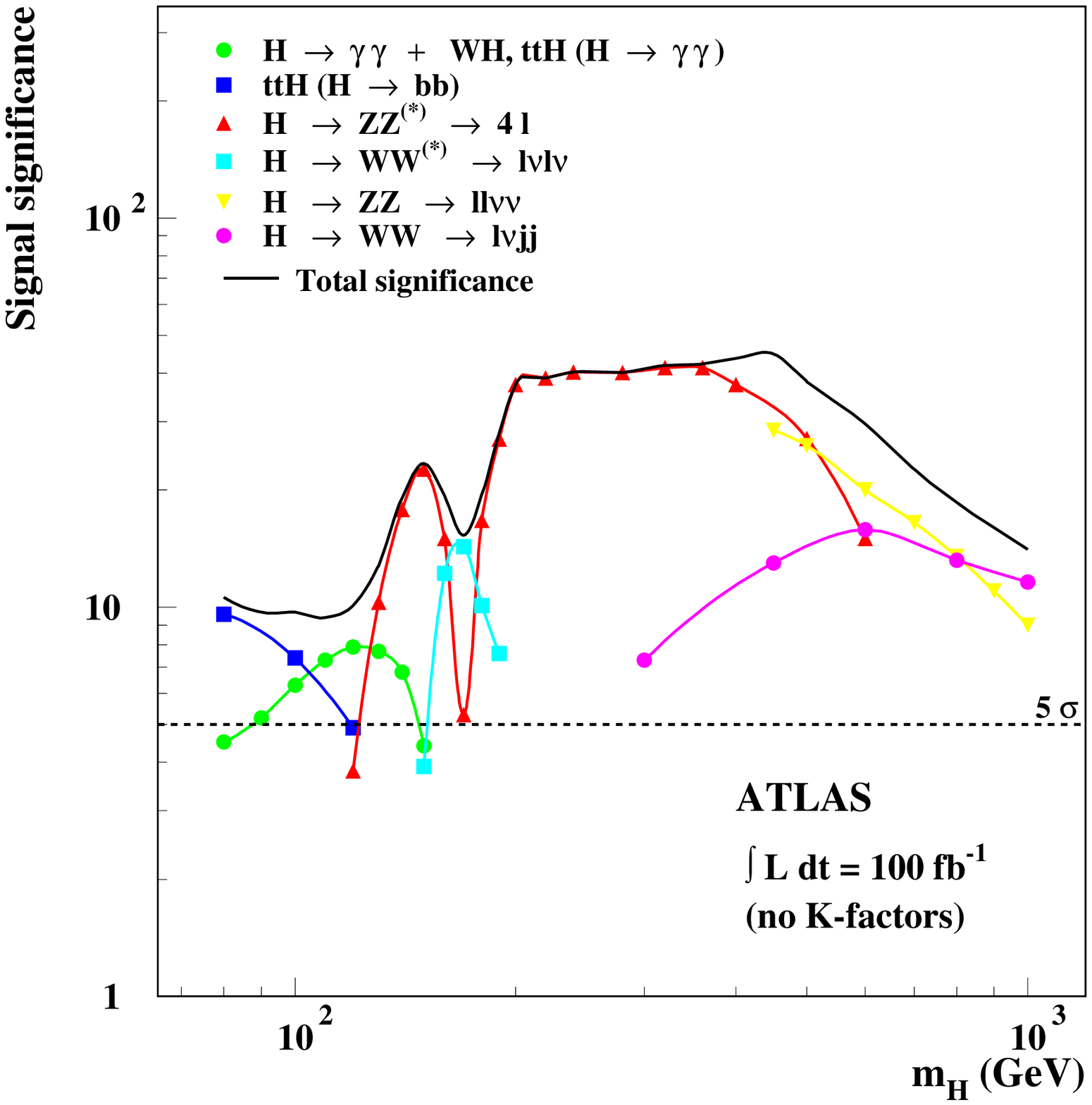}\\
\includegraphics[scale=0.5]{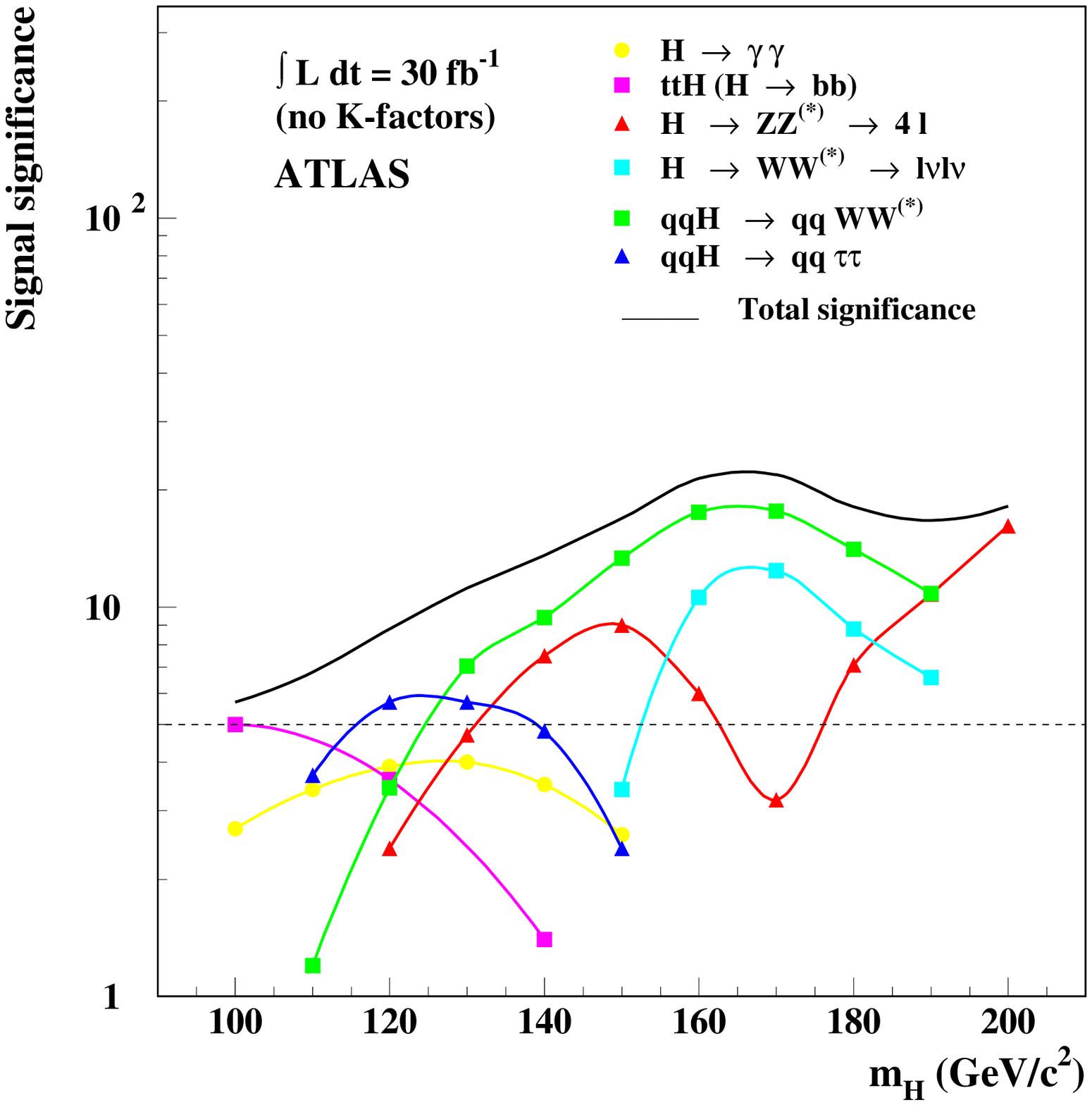}
\end{tabular}
\caption[]{Significance for the SM Higgs boson discovery in various
detection channels as a function of $M_H$. The upper plot is for
100~fb$^{-1}$ of data and with no $qqH$ channel included, the lower
plot for 30~fb$^{-1}$ and with the $qqH$ channel included over the
mass range $M_H\!<\!200$~GeV. Results are from the ATLAS
collaboration~\cite{atlas:asai_etal}.\label{fig:lhc_atlas_reach}}
\end{figure}
\begin{figure}
\centering
\includegraphics[scale=0.5]{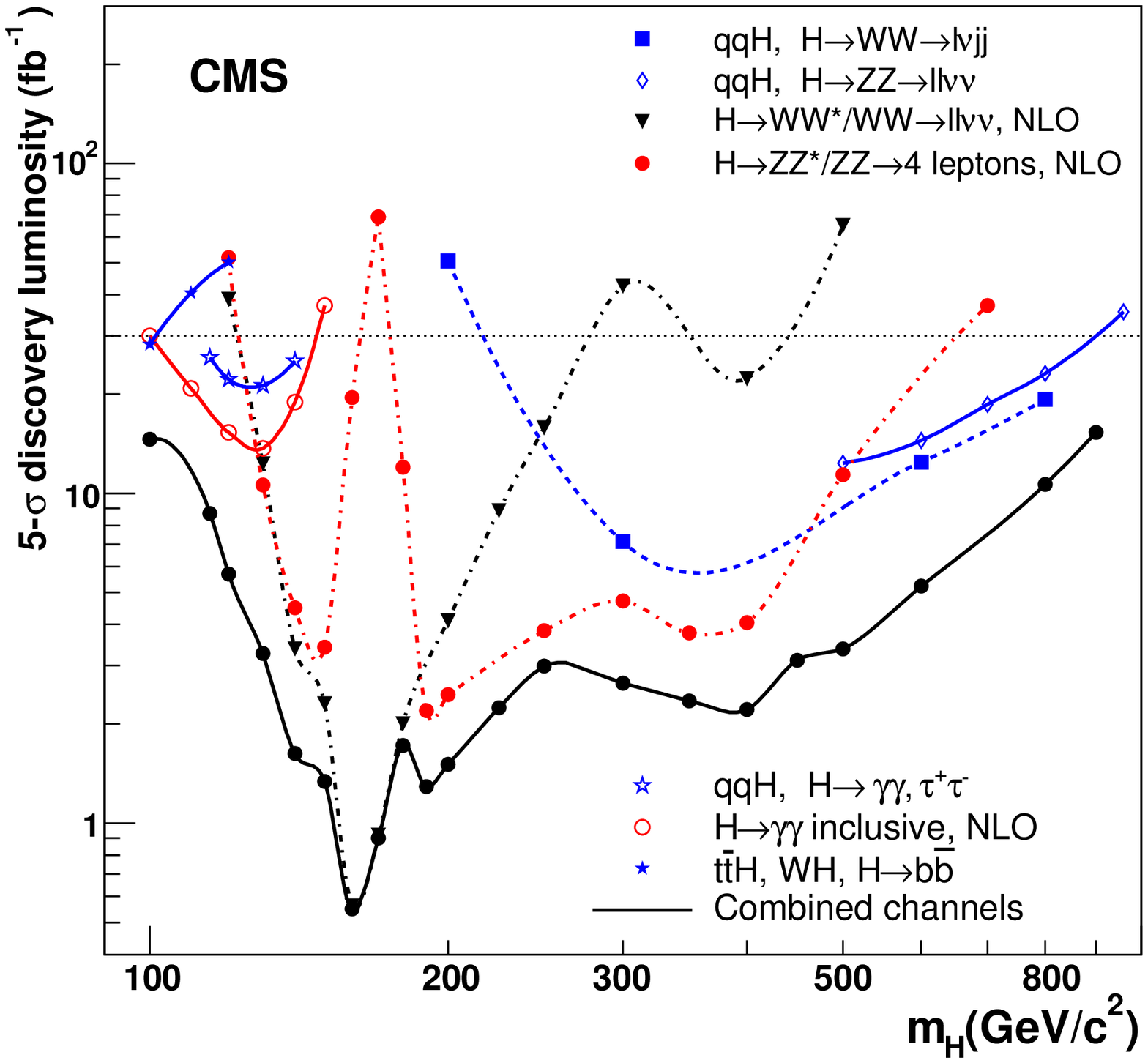}
\caption[]{Luminosity required to reach a $5\sigma$
discovery signal in CMS, using various detection channels, as a
function of $M_H$. From
Ref.~\cite{cms:abdullin_etal}.\label{fig:lhc_cms_reach}}
\end{figure}
At the LHC, all production modes will play an important role, thanks
to the higher statistics available. In particular, it is natural to
distinguish between a light ($M_H\!<\!130-140$~GeV) and heavy
($M_H>130-140$~GeV) mass region, as becomes evident by simultaneously
looking at both production cross sections (see
Fig.~\ref{fig:sm_higgs_lhc}) and decay branching ratios (see
Fig.~\ref{fig:sm_higgs_br}) over the entire $115-1000$~GeV SM Higgs
boson mass range. In the region of $M_H\!<\!130-140$~GeV the SM Higgs
boson at the LHC will be searched mainly in the following channels:
\begin{eqnarray}
\label{eq:higgs_boson_lhc_below_mh140}
&&gg\rightarrow H\,,\, H\rightarrow\gamma\gamma,W^+W^-,ZZ\,\,\,,\\
&&qq\rightarrow qqH\,,\,H\rightarrow\gamma\gamma,W^+W^-,ZZ,\tau^+\tau^-\,\,\,,
\nonumber\\
&&q\bar{q},gg\rightarrow t\bar{t}H\,,\,H\rightarrow b\bar{b},\tau^+\tau^-\,\,\,, 
\nonumber
\end{eqnarray}
while above that region, i.e. for $M_H\!>\!130-140$~GeV, the discovery
modes will be:
\begin{eqnarray}
\label{eq:higgs_boson_lhc_above_mh140}
&&gg\rightarrow H\,,\, H\rightarrow W^+W^-,ZZ\,\,\,,\\
&&qq\rightarrow qqH\,,\,H\rightarrow\gamma\gamma,W^+W^-,ZZ\,\,\,, 
\nonumber\\
&&q\bar{q},gg\rightarrow t\bar{t}H\,,\,H\rightarrow W^+W^-\,\,\,. \nonumber
\end{eqnarray}
These have been the modes used by both ATLAS and CMS to provide us
with the discovery reach illustrated in
Figs.~\ref{fig:lhc_atlas_reach} and \ref{fig:lhc_cms_reach}.
The ATLAS plots give the signal significance for a total integrated
luminosity of 100~fb$^{-1}$ (upper plot) and of 30~fb$^{-1}$ (lower
plot). The high luminosity (upper) plot belongs to the original ATLAS
technical design report~\cite{atlas:1999tdr}, and the weak boson
fusion channels had not been studied in detail at that time. The lower
luminosity (lower) plot is taken from a more updated
study~\cite{atlas:asai_etal}, and the weak boson fusion channels have
been included in the low mass region, up to about
$M_H\!\simeq\!200$~GeV, where they play an instrumental role towards
discovery. Other instrumental channels in the low mass region are the
inclusive Higgs production with $H\rightarrow\gamma\gamma$ and, below
$M_H\!=\!130$~GeV, $t\bar{t}H$ production with $H\rightarrow
b\bar{b}$. In the high mass region, the inclusive production with
$H\rightarrow ZZ,WW$ dominates, although CMS has found a substantial
contribution coming from weak gauge boson fusion with $H\rightarrow
ZZ,WW$.

\subsubsection{Studies of a SM Higgs boson}
\label{subsubsec:sm_higgs_properties}
If a Higgs boson signal is established, the LHC will have the capacity
of measuring several of its properties at some level of accuracy. In
particular, it will be able to measure its mass, width, and
couplings. At the same time, the charge and color quantum numbers of
the newly discovered particle will be established by detecting a single
production-decay channel; while a precise determination of its spin
and parity will probably require more statistics than available at the
LHC and will have to wait for a high energy Linear Collider to be
established (see Section~\ref{subsec:higgs_linear_collider}).

\begin{figure}
\begin{tabular}{cc}
\includegraphics[scale=0.4]{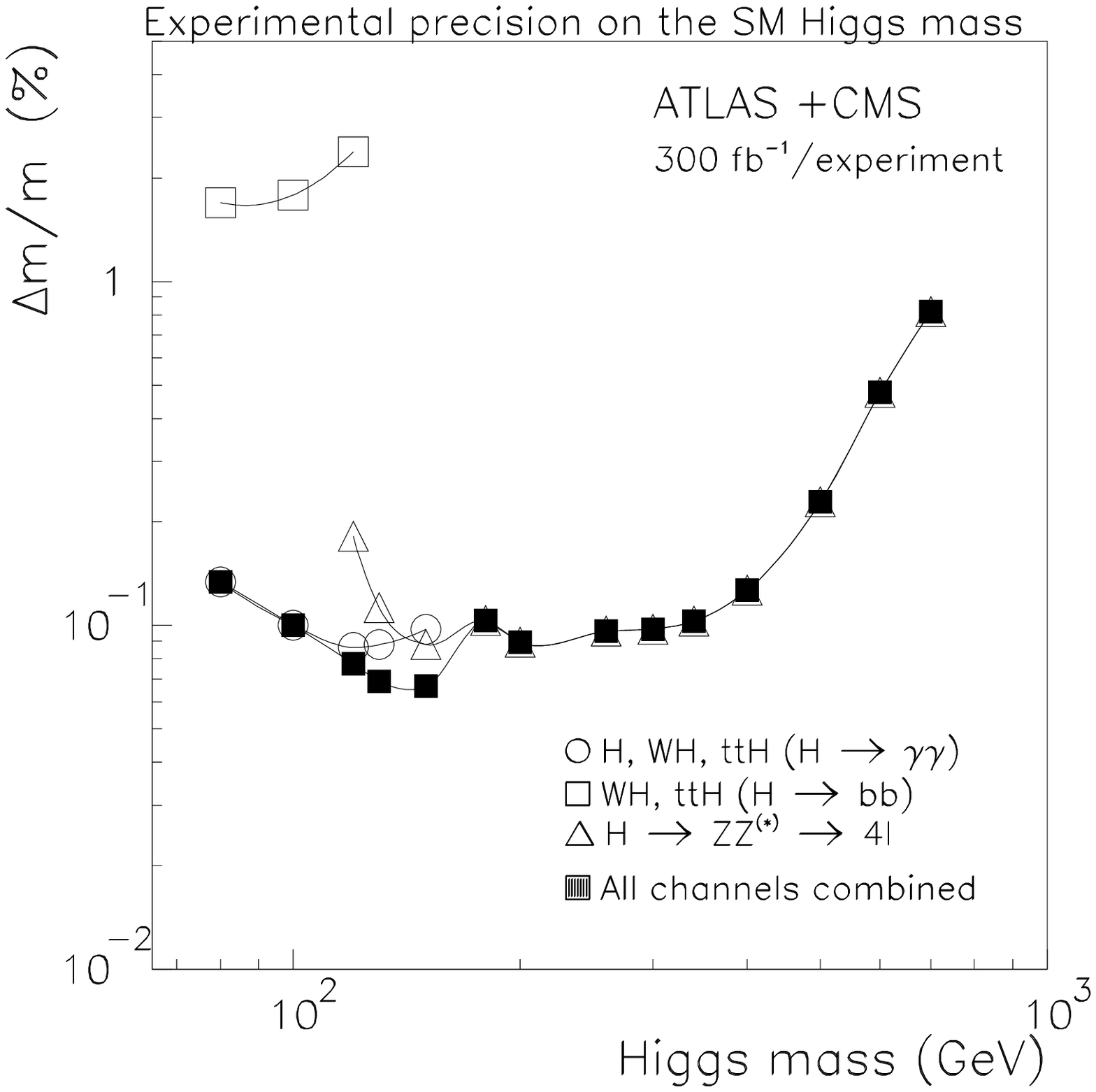}&
\includegraphics[scale=0.4]{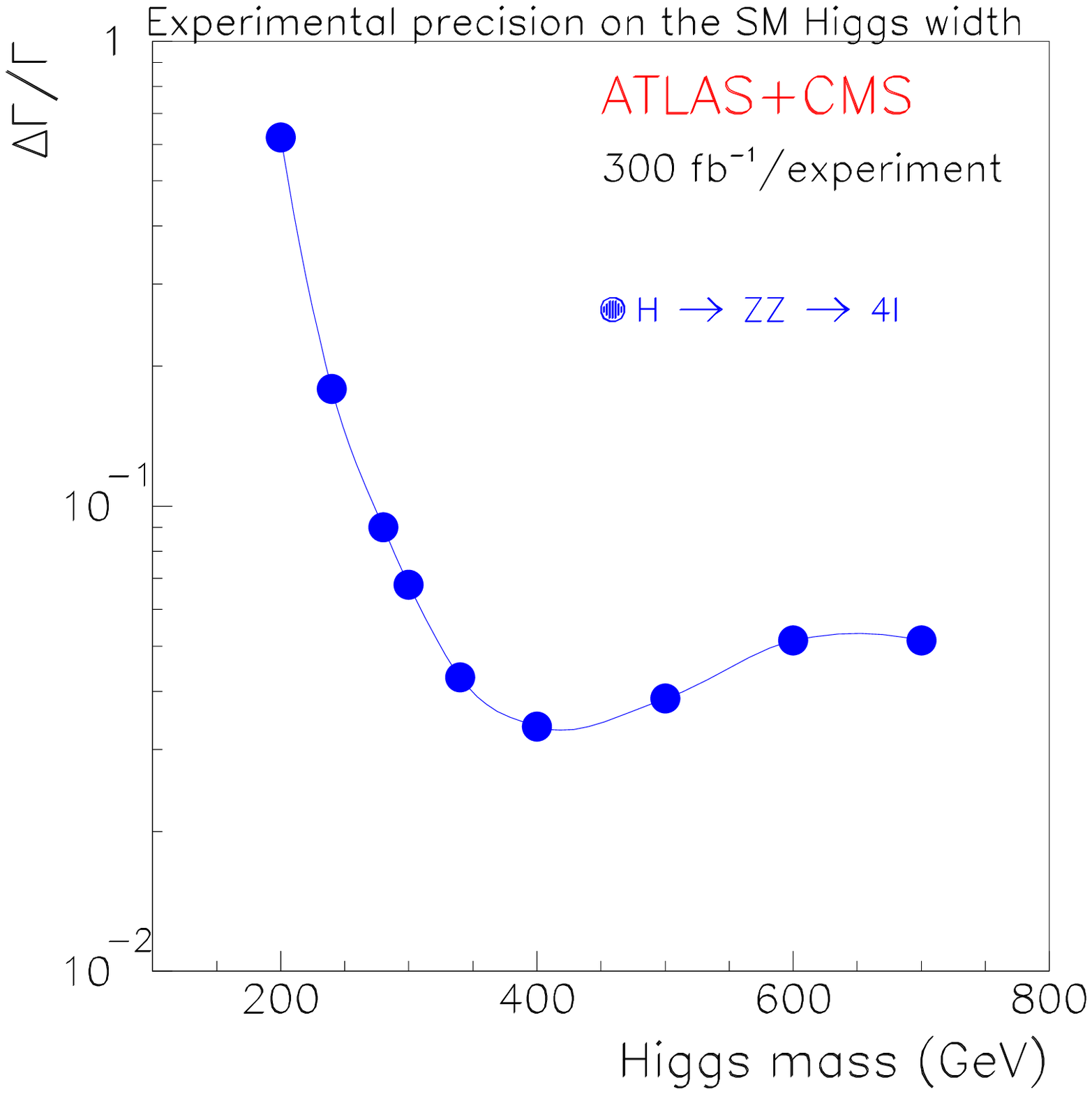}
\end{tabular}
\caption[]{Expected precision in the measurement of a SM-like Higgs
boson mass ($M_H$) and width ($\Gamma_H$) combining 300~fb$^{-1}$ of
data per experiment from ATLAS and CMS. From
Ref.~\cite{Gianotti:2000tz}.\label{fig:higgs_mass_width_lhc}}
\end{figure}
The two plots in Fig.~\ref{fig:higgs_mass_width_lhc} show the
precision with which both the mass and width of a SM-like Higgs boson
will be determined by ATLAS and CMS combining 300~fb$^{-1}$ of data
per experiment. We see that below $M_H\!\simeq\!400$~GeV the Higgs
mass can be determined with a precision of about 0.1\%, through
$H\rightarrow ZZ\rightarrow 4l$, complemented by $H,WH,t\bar{t}H
(H\rightarrow\gamma\gamma)$ and $t\bar{t}H (H\rightarrow b\bar{b})$ in
the low mass region. Above $M_H\!\simeq\!400$~GeV the accuracy
deteriorates for the smaller statistics available, although precisions
of the order of 1\% can still be obtained. We also see that the Higgs
width above $M_H\!\simeq\!200$~GeV will be entirely determined through
$H\rightarrow ZZ\rightarrow 4l$, while below $M_H\!\simeq\!200$~GeV
it is too small to be resolved experimentally and can only be
determined indirectly, as we will discuss in the following.

Finally, many studies in recent years have pointed to the fact that
the LHC, under minimal theoretical assumptions, will have the
potential to measure several Higgs boson couplings with an accuracy in
the 10-30\% range. The proposed strategy~\cite{Zeppenfeld:2000td}
consists of measuring the production-decay channels listed in
Eqs.~(\ref{eq:higgs_boson_lhc_below_mh140}) and
(\ref{eq:higgs_boson_lhc_above_mh140}) for a light
($M_H\!\le\!130-140$~GeV) or heavy ($M_H\!\ge\!130-140$~GeV) Higgs
boson respectively, and combine them to extract individual partial
widths or ratios of partial widths. Indeed, if a given
production-decay channel is observed, one can write that the
experimentally measured product of production cross section times
decay branching ratio corresponds, in the narrow width approximation,
to the following expression:
\begin{equation}
\label{eq:production_decay_ratio}
(\sigma_p(H)\mathrm{Br}(H\rightarrow dd))^{exp}=\frac{\sigma_p^{th}(H)}{\Gamma^{th}_p}
\frac{\Gamma_d\Gamma_p}{\Gamma}\,\,\,,
\end{equation}
where $\Gamma_p$ and $\Gamma_d$ are the partial widths associated with
the production and decay channels respectively, while $\Gamma$ is the
Higgs boson total width. The coefficient
$\sigma_p^{th}(H)/\Gamma^{th}_p$ can be calculated, while
$\Gamma_p$, $\Gamma_d$, and $\Gamma$ is what needs to be determined.
To each production-decay channel one can therefore associate a
measurable observable
\begin{equation}
\label{eq:z_p_d_observables}
Z^{(p)}_d=\frac{\Gamma_p\Gamma_d}{\Gamma}\,\,\,,
\end{equation}
where $p$ and $d$ label the production and decay channels
respectively. $Z^{(p)}_d$ is obtained from the experimental
measurement of $(\sigma_p(H)\mathrm{Br}(H\rightarrow dd))^{exp}$, normalized
by the theoretically calculable coefficient
$\sigma_p^{th}(H)/\Gamma^{th}_p$.  A signal in the $(p,d)$ channel
will measure $Z^{(p)}_d$, and therefore the product of Higgs couplings
$y_p^2y_d^2$, since $\Gamma_p\simeq y_p^2$ and $\Gamma_d\simeq y_d^2$.
Combining many different $(p,d)$ channels, a system of equations of
the form of Eq.~(\ref{eq:z_p_d_observables}) is obtained. Ratios of
partial widths $\Gamma_i/\Gamma_j$, and therefore ratios of Higgs
couplings, can then be derived in a model independent way, e.g.:
\begin{eqnarray}
\label{eq:ratios_gammai_gammaj}
\frac{\Gamma_b}{\Gamma_\tau}&=&\frac{Z^{(t)}_b}{Z^{(t)}_\tau}
\longrightarrow\frac{y_b}{y_\tau}\,\,\,,\\
\frac{\Gamma_t}{\Gamma_g}&=&\frac{Z^{(t)}_\tau Z^{(w)}_\gamma}
{Z^{(w)}_\tau Z^{(g)}_\gamma}
\longrightarrow\frac{y_t}{y_g}\,\,\,,\nonumber
\end{eqnarray}
while individual partial width can be obtained with the further
assumptions that: \emph{i)} the total width is the sum of all SM
partial widths, i.e. there is no new physics or invisible width
contributions, and \emph{ii)} the $y_W$ and $y_Z$ couplings are
related by the $SU(2)_L$ weak isospin symmetry. This is required by
the fact that, in $qq\rightarrow qqH$, the $W^+W^-\rightarrow H$ and
$ZZ\rightarrow H$ fusion processes cannot be distinguished
experimentally. 
\begin{figure}
\centering
\includegraphics[scale=0.5]{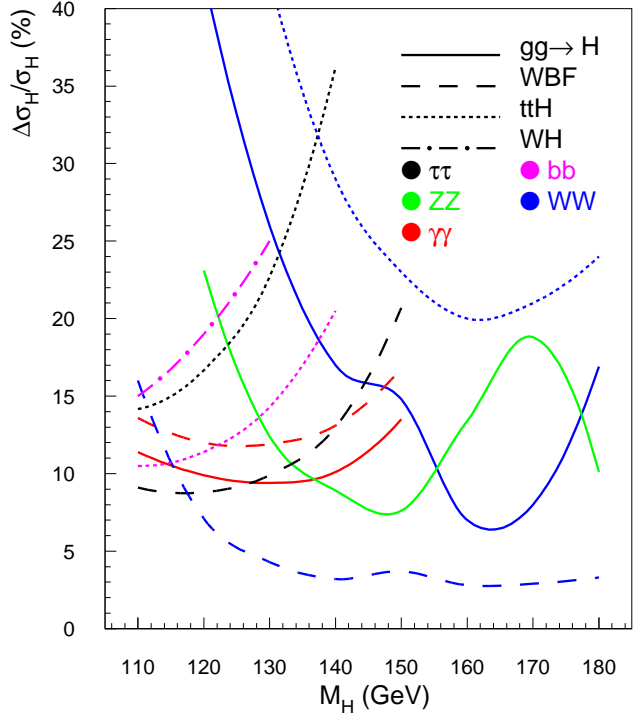}
\caption[]{Relative accuracies on the measurement of the cross section
of a scalar SM-like Higgs boson in the production+decay channels
listed in Eqs.~(\ref{eq:higgs_boson_lhc_below_mh140}) and
(\ref{eq:higgs_boson_lhc_above_mh140}). All channels have been
rescaled to a total integrated luminosity of 200~fb$^{-1}$, except
$pp\rightarrow t\bar{t}H,H\rightarrow W^+W^-$ and $q\bar{q}\rightarrow
WH,H\rightarrow b\bar{b}$ for which 300 ~fb$^{-1}$ have been used, and
$gg\rightarrow H,H\rightarrow W^+W^-$ that was studies with
30~fb$^{-1}$. From
Ref.~\cite{Belyaev:2002ua}. \label{fig:higgs_prod_decay_accuracies} }
\end{figure}
\begin{figure}
\centering
\includegraphics[scale=0.6]{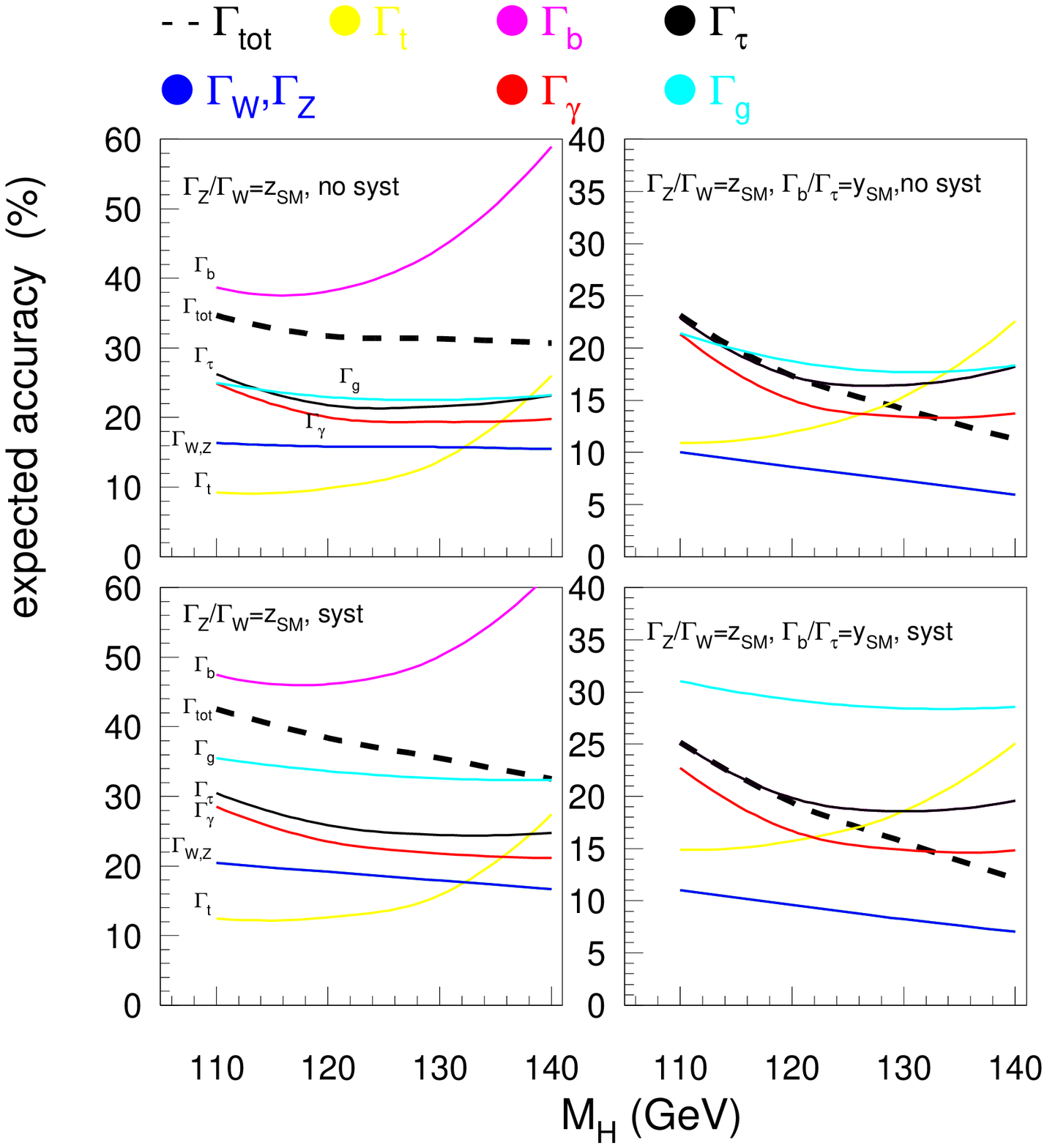}
\caption[]{Relative accuracy on the individual rates $\Gamma_i$
expected at the LHC, when the ratio $\Gamma_b/\Gamma_\tau$ is fixed to
its SM value (right plots) or not (left plots). The integrated
luminosity used is as explained in the caption of
Fig.~\ref{fig:higgs_prod_decay_accuracies}. The upper plots show the
accuracies obtained without including any theoretical systematic
error, while the lower plots include the systematic theoretical errors
explained in the text. From
Ref.~\cite{Belyaev:2002ua}. \label{fig:higgs_couplings_accuracies}}
\end{figure}

The accuracy with which both individual couplings and ratios of
couplings can be extracted is mainly determined by the experimental
error on the $Z^{(p)}_d$ measurements and by the theoretical
uncertainty on the prediction of $\sigma_p^{th}(H)$ in
Eq.~(\ref{eq:production_decay_ratio}).
Fig.~\ref{fig:higgs_prod_decay_accuracies} shows the estimated
relative accuracy with which various channel will be detected at the
LHC, assuming 200~fb$^{-1}$ of data available for most channels.  For
the purpose of illustration, Fig.~\ref{fig:higgs_couplings_accuracies}
shows the accuracy with which some of the SM Higgs partial width as
well as its total width could be determined at the LHC, when the
technique described above is implemented. The upper plots do not
include any theoretical systematic error, while the lower plots
include a theoretical systematic error of: 20\% for $gg\rightarrow H$,
5\% for $qq\rightarrow qqH$, and 10\% for $pp\rightarrow
t\bar{t}H$. At the same time, all plots assumes the $SU(2)$ gauge
induced relation between $y_W$ and $y_Z$, while the right hand plots
also assume a SM-like relation between $y_b$ and $y_\tau$. The more
the assumptions, the better the accuracy with which the considered
couplings can be determined, and the more model dependence is
introduced in the coupling determination. More sophisticated analyses
have appeared in recent
studies~\cite{Assamagan:2004mu,Duhrssen:2004cv}. Overall, we can
however conclude that the LHC has a great potential of giving a first
fairly precise indication of the nature of the couplings of a Higgs
boson candidate, although under some (well justified) model
assumptions. In particular, in the specific case of the top-quark
Yukawa coupling, the LHC will be for a long time the only machine to
be able to measure it with enough precision, since the measurement of
$y_t$ in $e^+e^-\rightarrow t\bar{t}H$ at a $\sqrt{s}\!=\!500$~GeV
Linear Collider is statistically very limited, as we will see in
Section~\ref{subsec:higgs_linear_collider}.

\subsubsection{Searching for a MSSM Higgs boson at the Tevatron and the
LHC}
\label{subsubsec:mssm_higgs_searches}
\begin{figure}
\centering
\begin{tabular}{cc}
\begin{minipage}{0.4\linewidth}
{\includegraphics[scale=0.6]{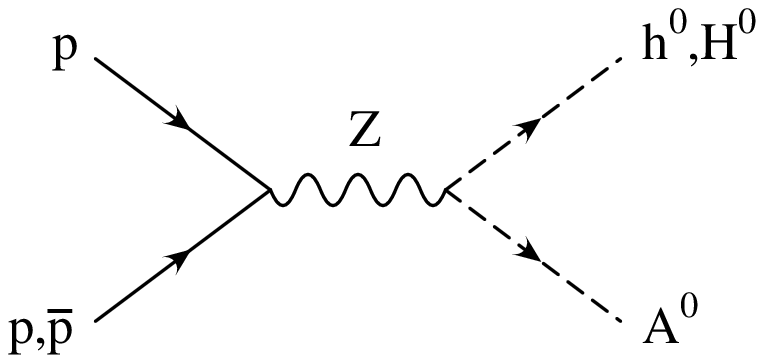}}
\end{minipage} &
\begin{minipage}{0.4\linewidth}
{\includegraphics[scale=0.6]{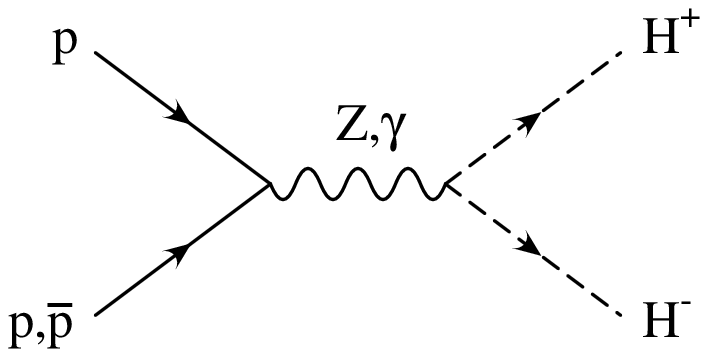}}
\end{minipage}
\end{tabular}
\caption[]{Associated production modes for MSSM Higgs bosons. 
\label{fig:mssm_higgs_associated_production}}
\end{figure}

\begin{figure}
\hspace{-0.5truecm}
\begin{tabular}{lr}
\begin{minipage}{0.5\linewidth}
{\includegraphics[scale=0.35,angle=-90]{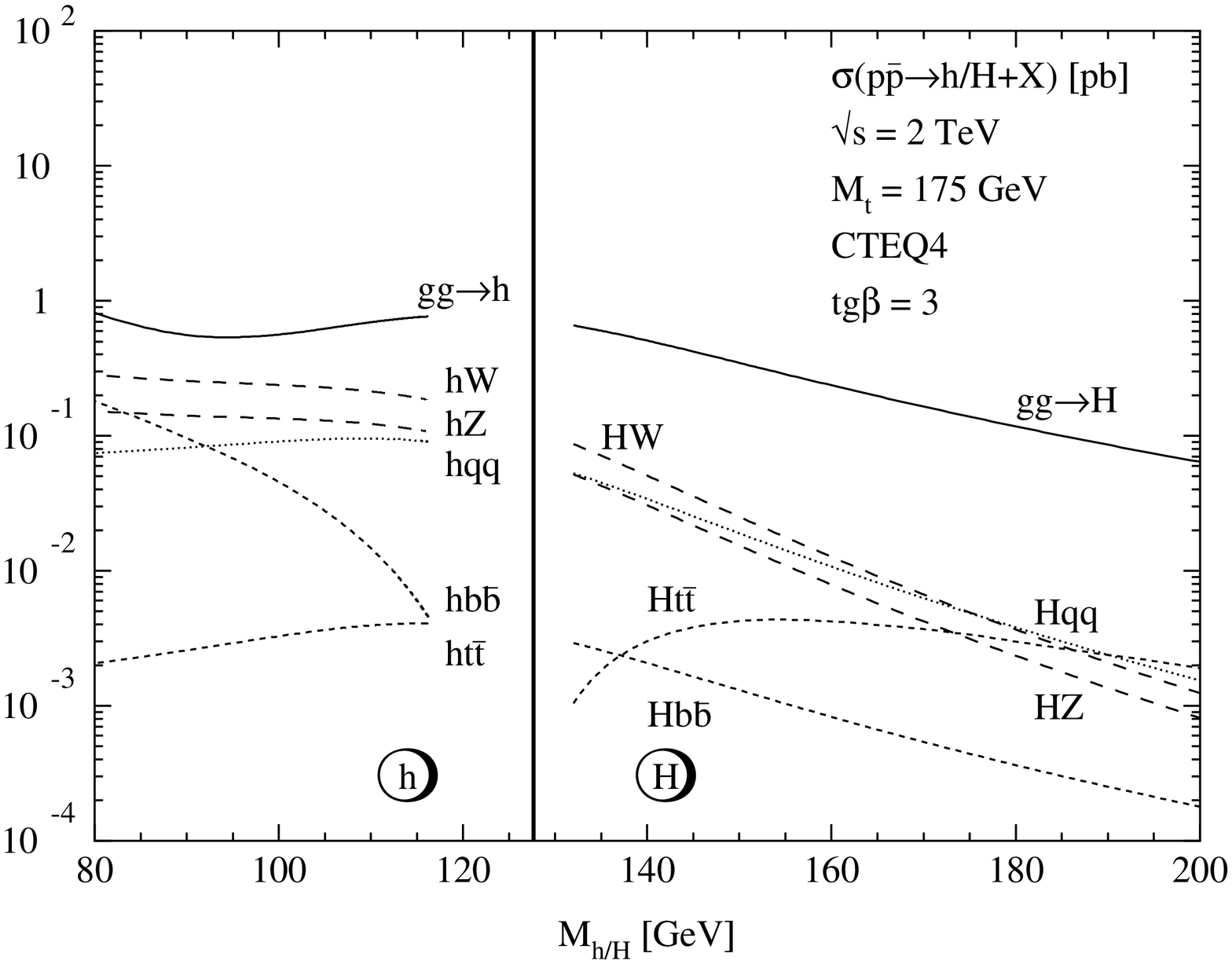}}
\end{minipage}&
\begin{minipage}{0.5\linewidth}
{\includegraphics[scale=0.35,angle=-90]{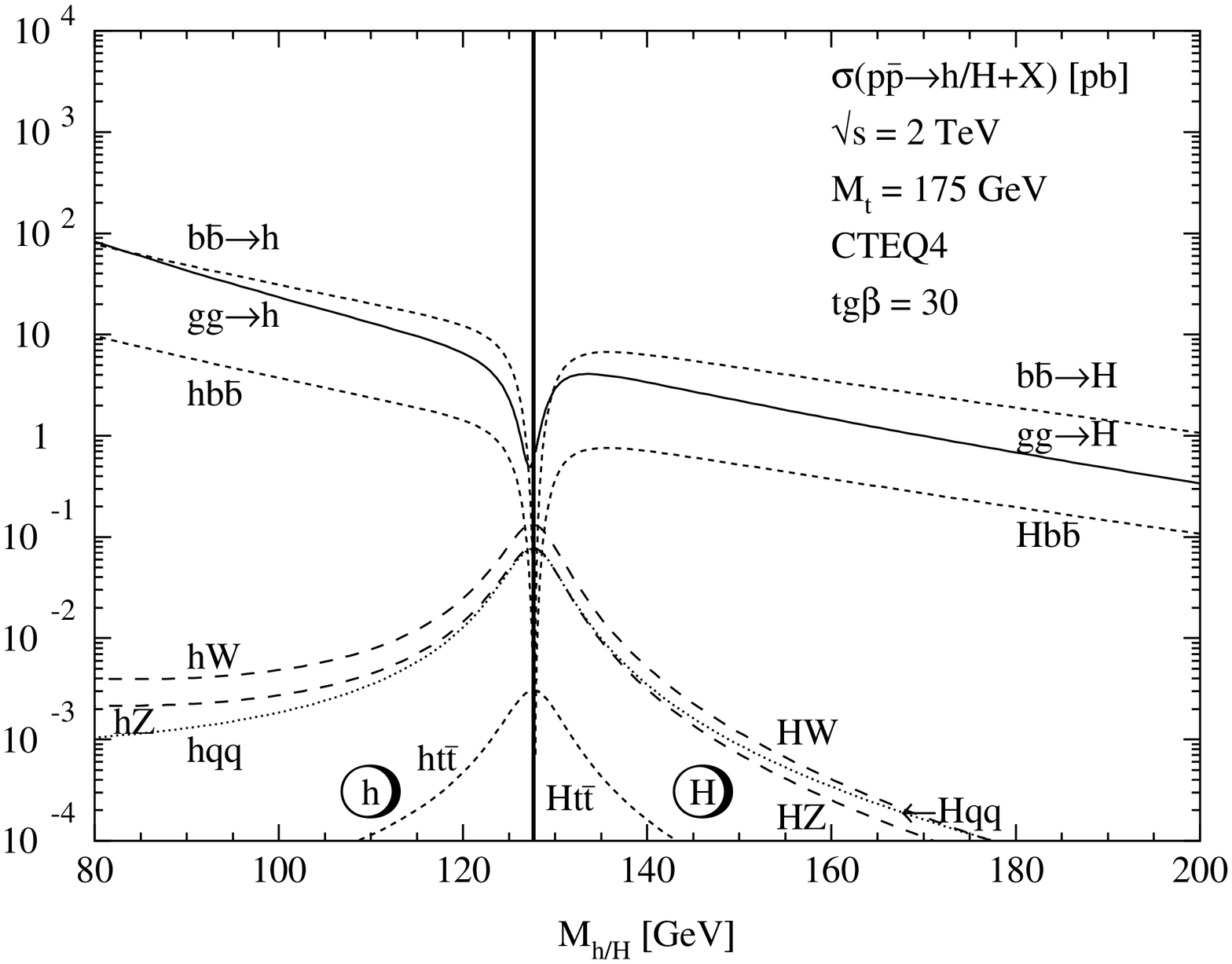}}
\end{minipage}\\
\begin{minipage}{0.5\linewidth}
{\vspace{0.4truecm}
\includegraphics[scale=0.35,angle=-90]{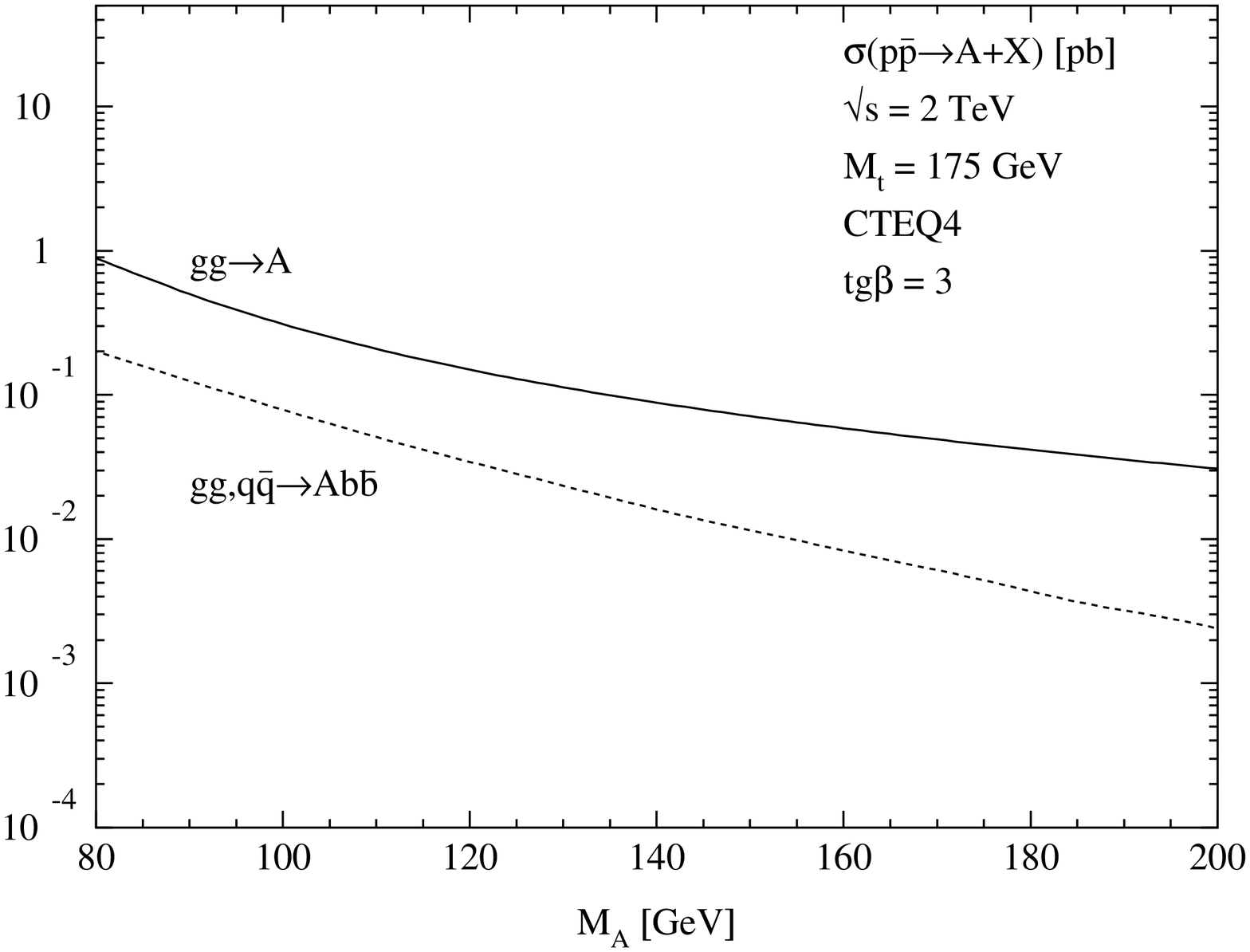}}
\end{minipage}&
\begin{minipage}{0.5\linewidth}
{\vspace{0.4truecm}
\includegraphics[scale=0.35,angle=-90]{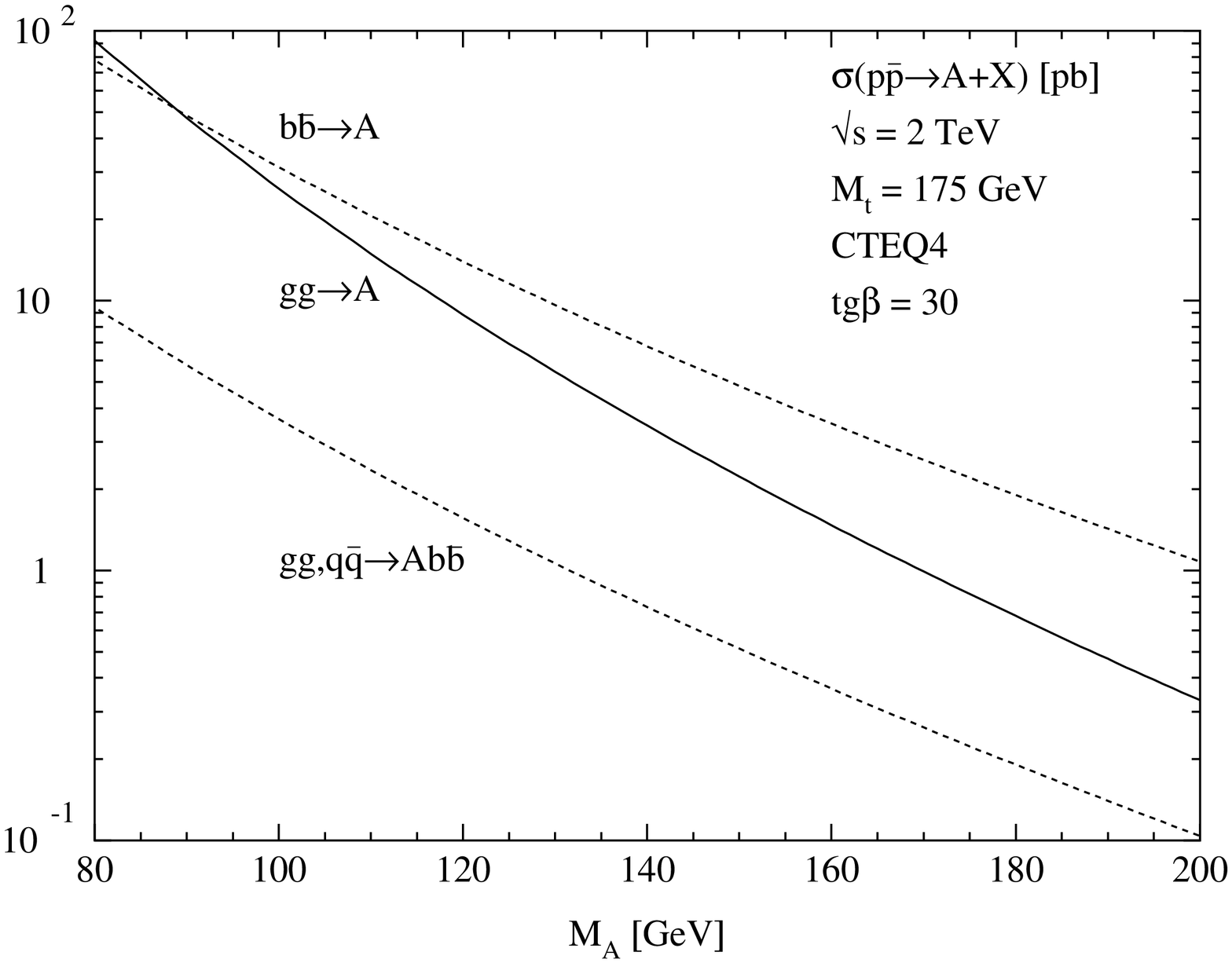}}
\end{minipage}
\end{tabular}
\caption[]{Neutral MSSM Higgs boson cross sections at the Tevatron
($\sqrt{s}\!=\!2$~TeV). The upper plots are for the neutral scalar
Higgs bosons, $h^0$ and $H^0$, while the lower plots are for the
neutral pseudoscalar Higgs boson, $A^0$. The left hand plots are for
$\tan\beta\!=\!3$ while the right hand plots are for
$\tan\beta\!=\!30$. From
Ref.~\cite{Carena:2002es}. \label{fig:mssm_higgs_tevatron}}
\end{figure}

\begin{figure}
\hspace{-1.truecm}
\begin{tabular}{lr}
\begin{minipage}{0.5\linewidth}
{\includegraphics[scale=0.6]{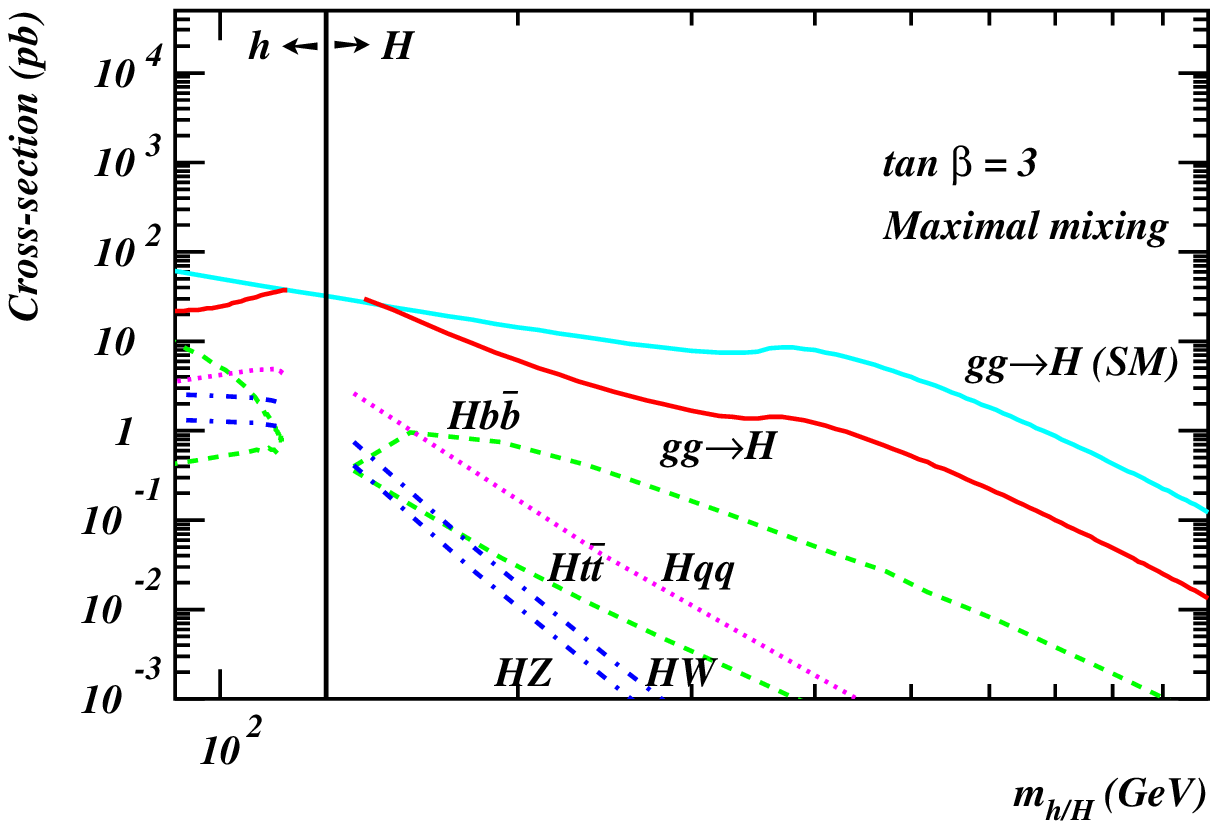}}
\end{minipage}&
\begin{minipage}{0.5\linewidth}
{\includegraphics[scale=0.6]{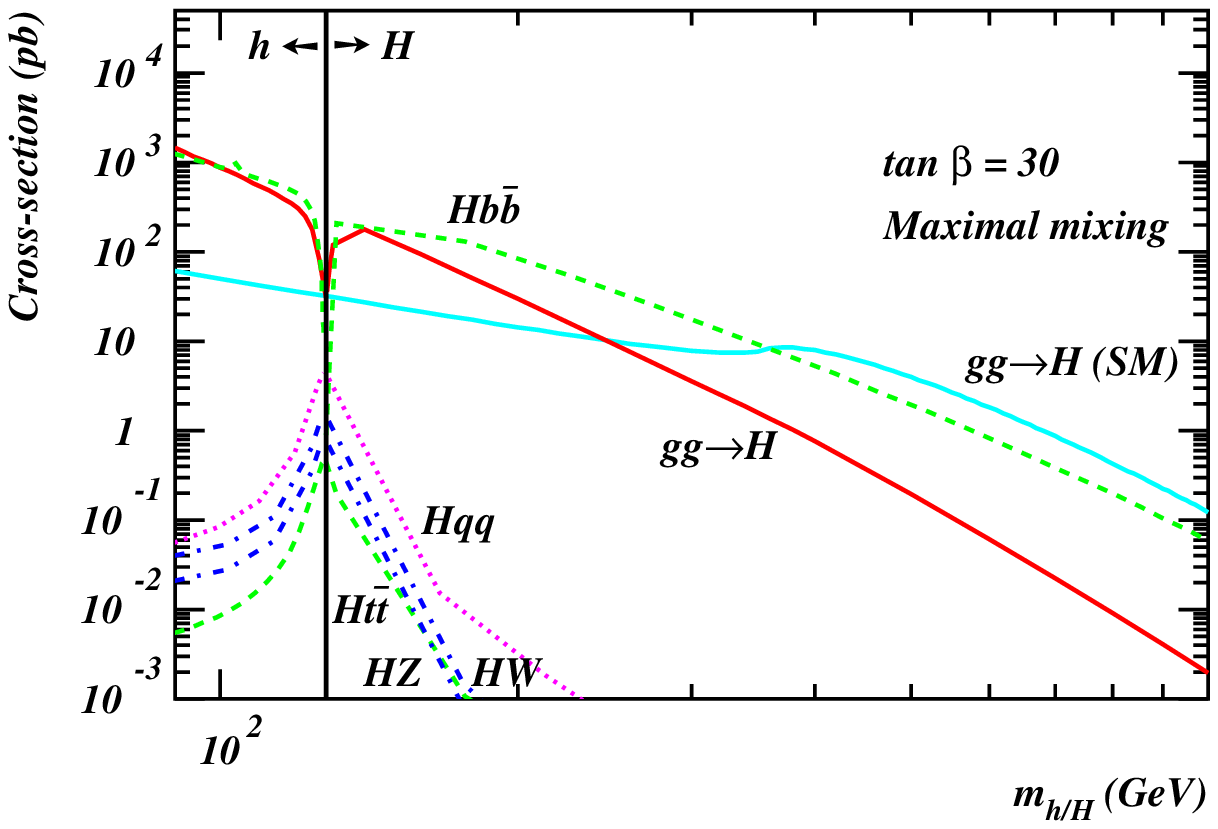}}
\end{minipage}\\
\begin{minipage}{0.5\linewidth}
{\includegraphics[scale=0.6]{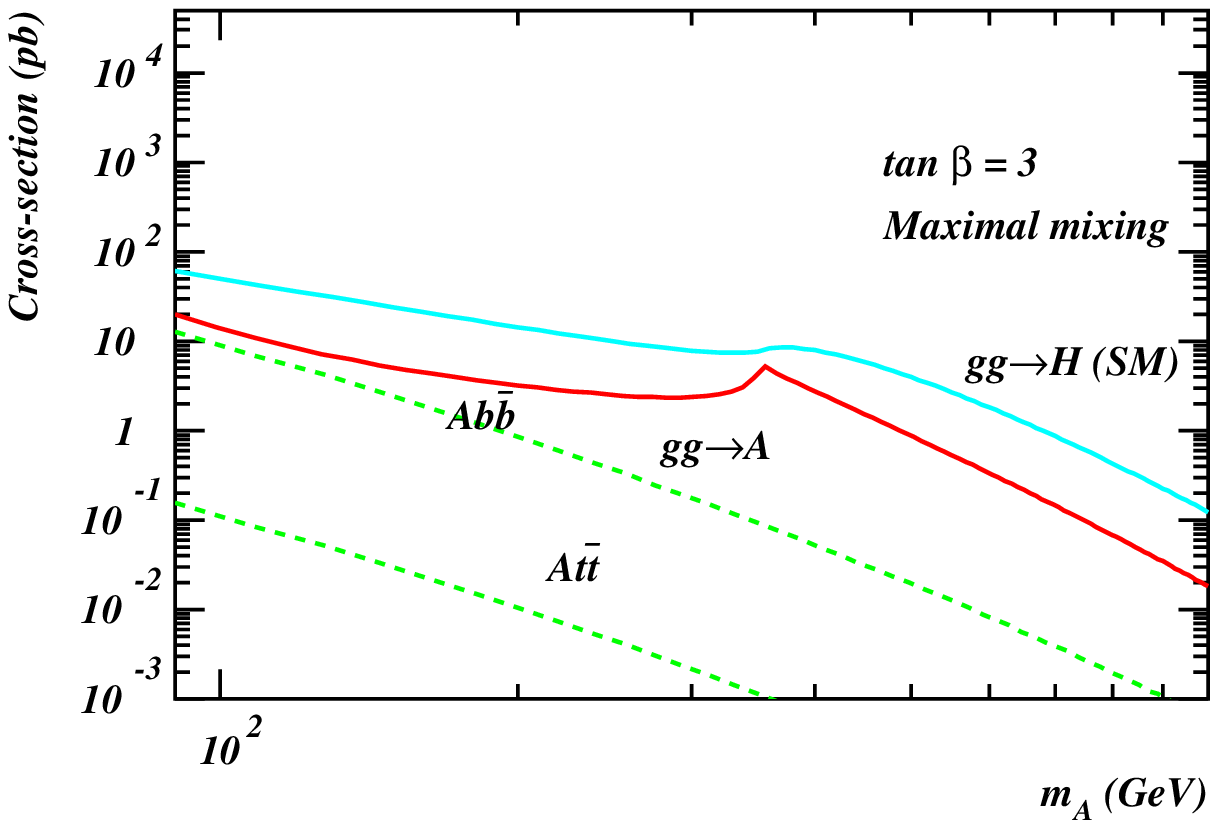}}
\end{minipage}&
\begin{minipage}{0.5\linewidth}
{\includegraphics[scale=0.6]{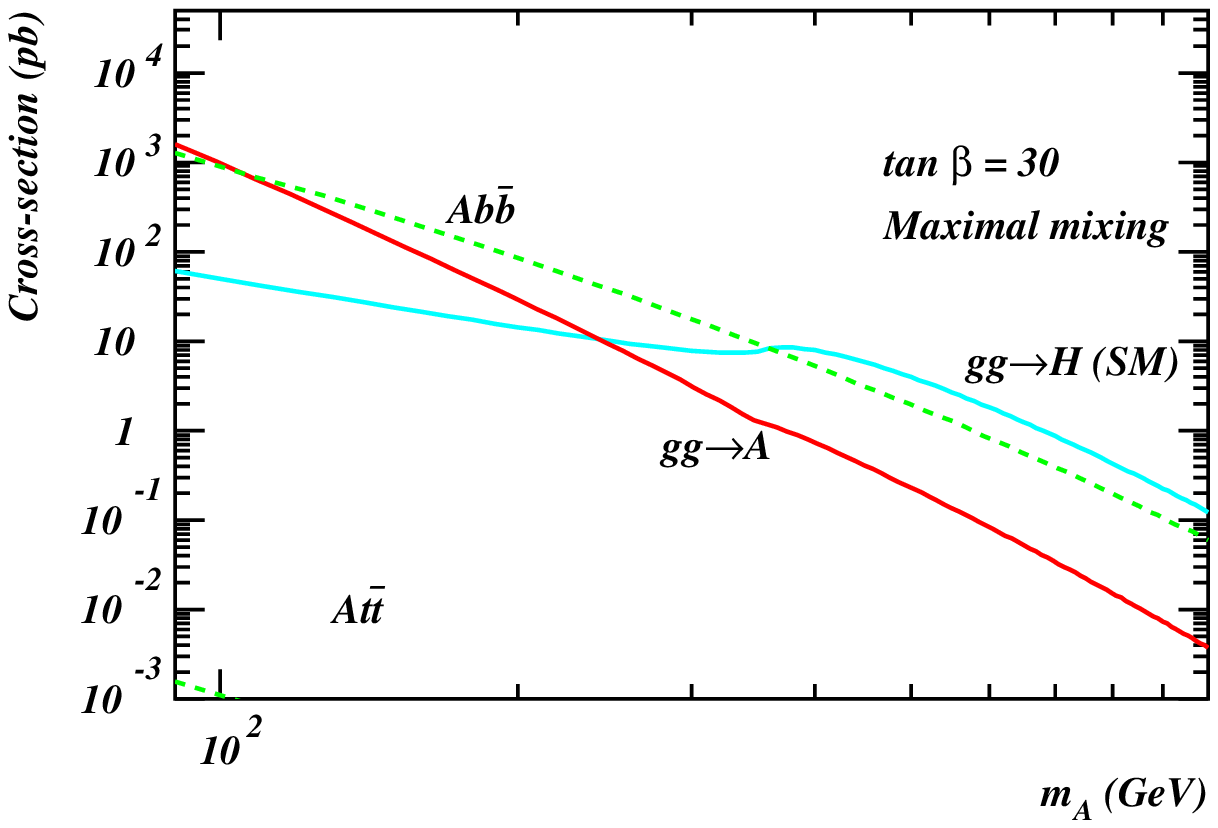}}
\end{minipage}
\end{tabular}
\caption[]{Neutral MSSM Higgs boson cross sections at the LHC
($\sqrt{s}\!=\!14$~TeV). The upper hand plots are for the neutral
scalar Higgs bosons, $h^0$ and $H^0$, while the lower plots are for
the neutral pseudoscalar Higgs boson, $A^0$. The left hand plots are
for $\tan\beta\!=\!3$ while the right hand plots are for
$\tan\beta\!=\!30$. From
Ref.~\cite{Carena:2002es}. \label{fig:mssm_higgs_lhc}}
\end{figure}

Most of the characteristics of the MSSM Higgs couplings that determine
the pattern of decays reviewed in
Section~\ref{subsec:mssm_higgs_branching_ratios} also affect the
mechanism of production of the MSSM Higgs bosons. In particular:
\begin{itemize}
\item for $M_A\gg M_Z$, the so called \emph{decoupling limit},
\begin{list}{$\longrightarrow$}
  {\setlength{\topsep}{0.truecm}\setlength{\parskip}{-0.8truecm}
    \setlength{\itemsep}{0.truecm}
    \setlength{\leftmargin}{1.5truecm}\setlength{\labelwidth}{1.5truecm}}
\item $h^0\longrightarrow H_{SM}$, while
\item $M_A\simeq M_H\,\,\,$ and $\,\,\,g_{(A,H)b\bar{b}}\gg
  g_{H_{SM}b\bar{b}}\,\,\,$, $\,\,\,g_{HVV}\ll g_{H_{SM}VV}$,
\end{list}
 \item while for $M_A\le M_Z$ and $\tan\beta\gg 1$:
\begin{list}{$\longrightarrow$}
  {\setlength{\topsep}{0.truecm}\setlength{\parskip}{-0.8truecm}
    \setlength{\itemsep}{0.truecm}
    \setlength{\leftmargin}{1.5truecm}\setlength{\labelwidth}{1.5truecm}}
\item $g_{HVV}\simeq g_{H_{SM}VV}$, while
\item $M_A\simeq M_h\,\,\,$ and $\,\,\,g_{(A,h)b\bar{b}}\gg
  g_{H_{SM}b\bar{b}}\,\,\,$, $\,\,\,g_{hVV}\ll g_{H_{SM}VV}$.
\end{list}
\end{itemize}
If we assume supersymmetric particles to be heavy enough that the
decay of a Higgs boson into supersymmetric particles as well as the
production of a Higgs boson through the decay of a supersymmetric
particle is forbidden, the available production processes are the SM
ones plus the associate production modes illustrated in
Fig.~\ref{fig:mssm_higgs_associated_production}, as well as the
production of a charged Higgs boson via the decay of a $t/\bar{t}$ quark.
A summary of the neutral MSSM Higgs boson production rates at the
Tevatron and at the LHC is given in
Figs.~\ref{fig:mssm_higgs_tevatron} and \ref{fig:mssm_higgs_lhc}, for
two values of $\tan\beta$, in the maximal mixing scenario (see
Section~\ref{subsubsec:higgs_mssm}). For the Tevatron the mass range
is limited to the range kinematically accessible while for the LHC the
entire Higgs mass range up to 1~TeV is covered. It is important to
notice how for large $\tan\beta$ ($\tan\beta\!=\!30$ in the plots of
Figs.~\ref{fig:mssm_higgs_tevatron} and \ref{fig:mssm_higgs_lhc}) the
production of both scalar and pseudoscalar neutral Higgs boson with
bottom quarks becomes dominant. In particular the inclusive production
(denoted in the Figs.~\ref{fig:mssm_higgs_tevatron} and
\ref{fig:mssm_higgs_lhc} as $b\bar{b}\rightarrow\phi^0$, for
$\phi^0\!=\!h^0,H^0,A^0$) becomes larger than the otherwise dominant
gluon-gluon fusion mode ($gg\rightarrow H$), while the exclusive
production (denoted in Figs.~\ref{fig:mssm_higgs_tevatron} and
\ref{fig:mssm_higgs_lhc} as $p\bar{p},pp\rightarrow
b\bar{b}\phi^0$) is right below gluon-gluon fusion but above all other
production modes. More details on exclusive vs inclusive production of
a Higgs boson with bottom quarks will be given in
Section~\ref{sec:theory}. We also notice the subleading role played by
vector boson fusion production ($qq\rightarrow qq\phi^0$) due to the
suppression (or absence in the case of $A^0$) of the $\phi^0VV$
couplings (for $V\!=\!W^\pm,Z$). Finally, the associated production
modes for neutral Higgs bosons (see
Fig.~\ref{fig:mssm_higgs_associated_production}) have in general very
small cross sections and are not considered in the plots of
Figs.~\ref{fig:mssm_higgs_tevatron} and
\ref{fig:mssm_higgs_lhc}.

\begin{figure}
\begin{tabular}{cc}
\begin{minipage}{0.5\linewidth}
{\vspace{0.3truecm}\includegraphics[scale=0.4]{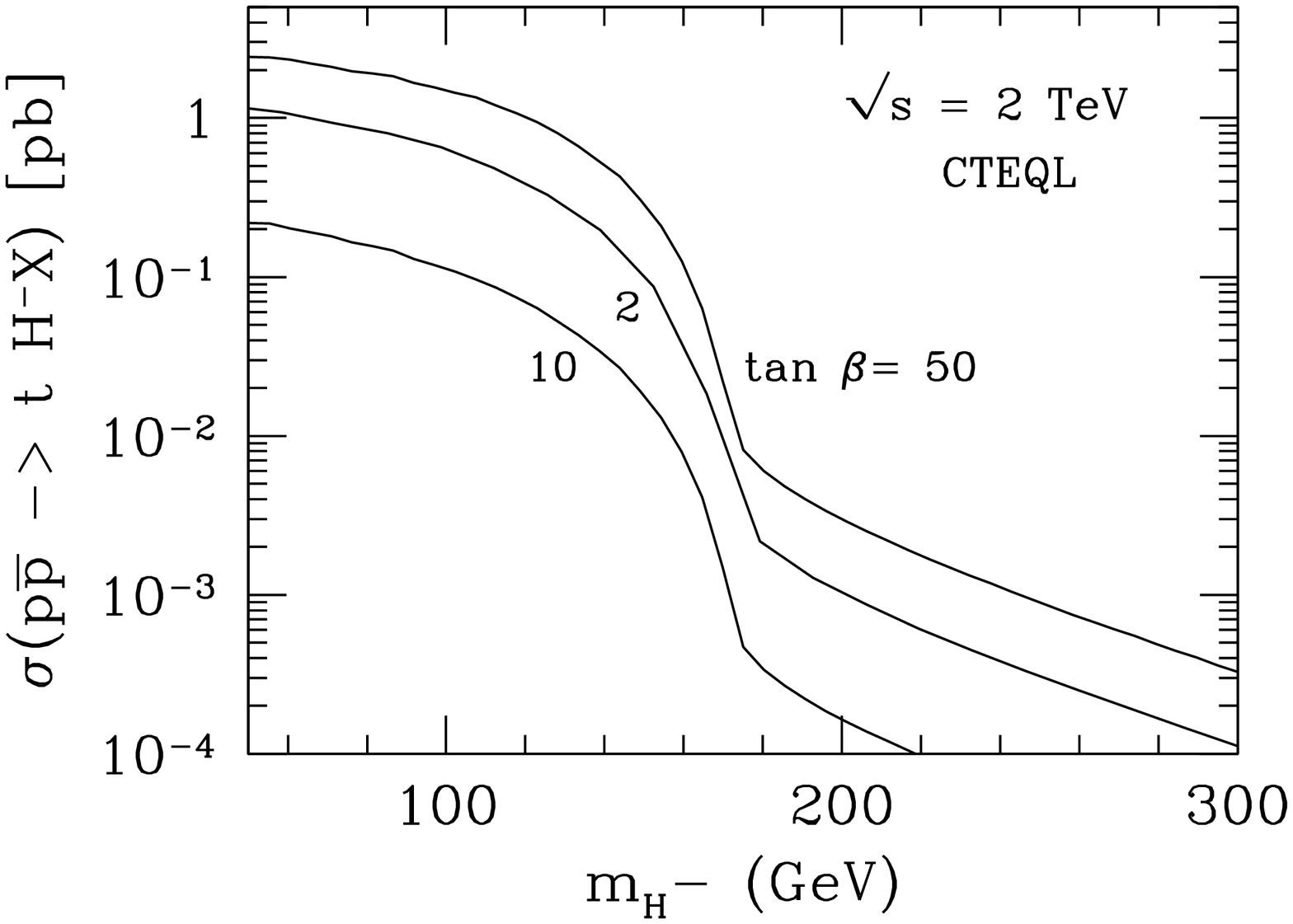}}
\end{minipage} &
\begin{minipage}{0.5\linewidth}
{\includegraphics[scale=0.42]{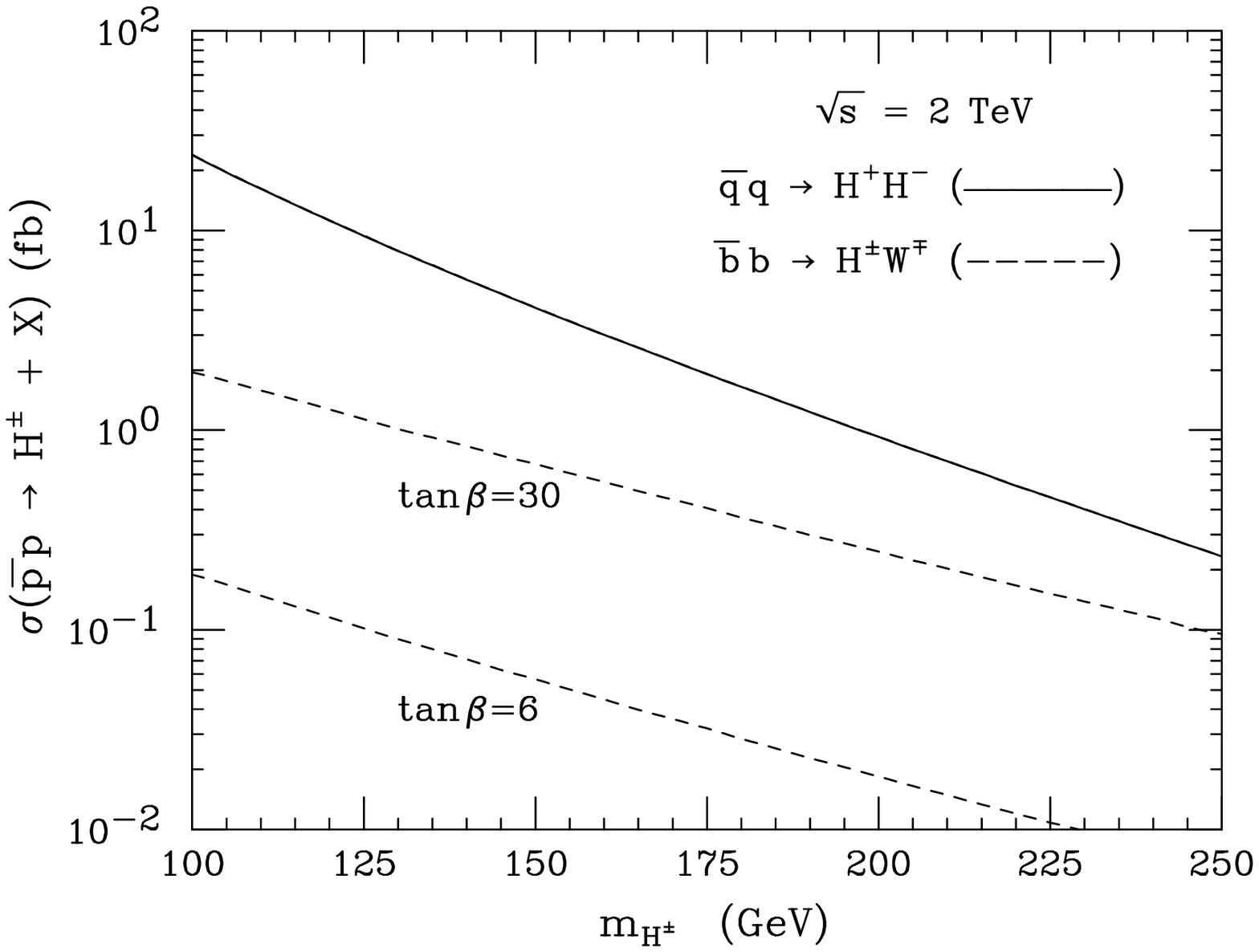}}
\end{minipage}
\end{tabular}
\caption[]{Charged Higgs boson cross sections at the Tevatron, as a function of
$M_{H^\pm}$, for different values of $\tan\beta$. From
Ref.~\cite{Carena:2002es}. \label{fig:mssm_charged_higgs_tevatron}}
\end{figure}

\begin{figure}
\begin{tabular}{cc}
\begin{minipage}{0.5\linewidth}
{\vspace{0.3truecm}\includegraphics[scale=0.4]{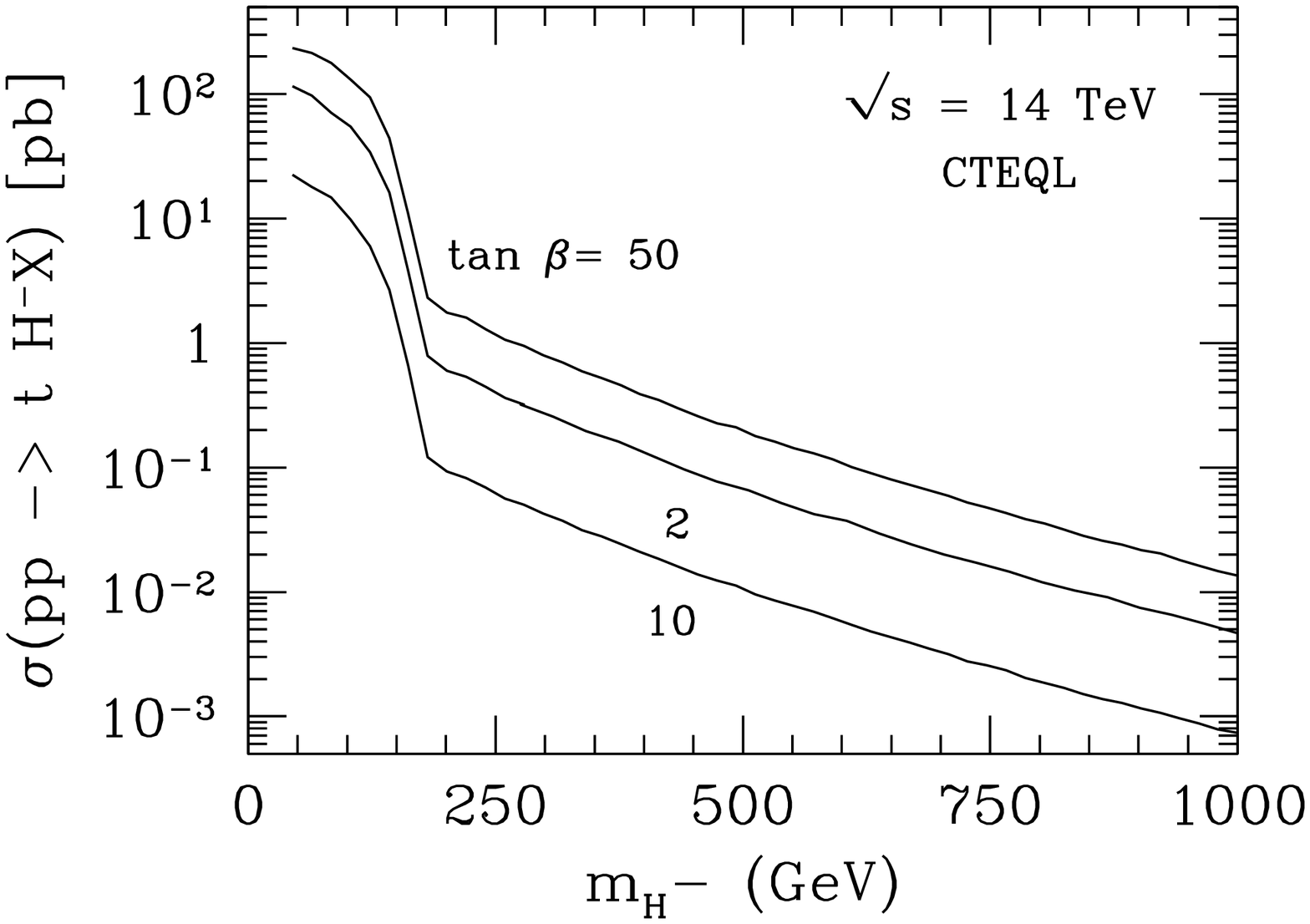}}
\end{minipage} &
\begin{minipage}{0.5\linewidth}
{\includegraphics[scale=0.42]{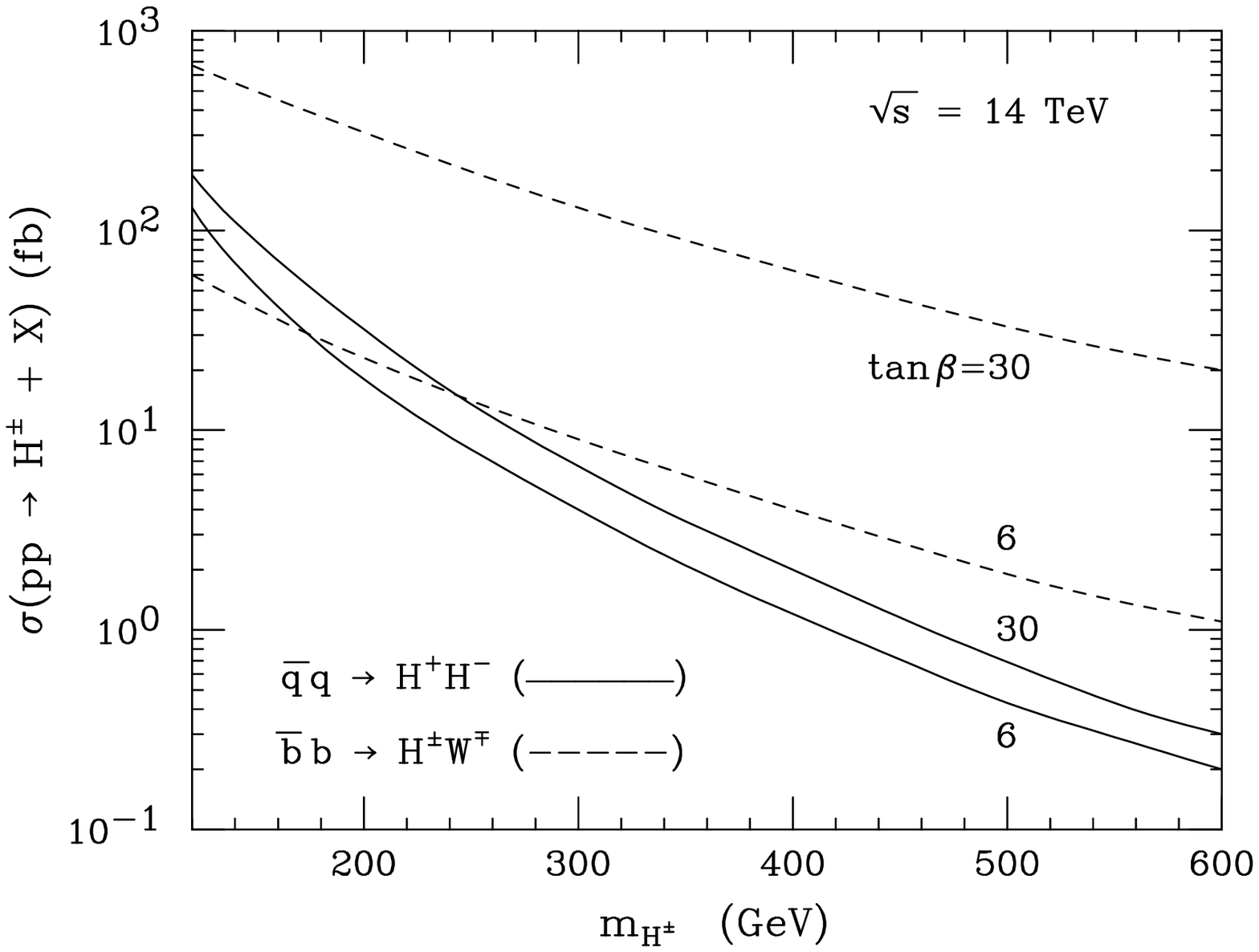}}
\end{minipage}
\end{tabular}
\caption[]{Charged Higgs boson cross sections at the LHC as functions of
$M_{H^\pm}$, for different values of $\tan\beta$. From
Ref.~\cite{Carena:2002es}. \label{fig:mssm_charged_higgs_lhc}}
\end{figure}

For the case of the charged MSSM Higgs boson, we need to distinguish
two cases: \emph{i)} $M_{H^\pm}<m_t-m_b$ and \emph{ii)
$M_{H^\pm}>m_t-m_b$}. When $M_{H^\pm}<m_t-m_b$, $H^\pm$ is mainly
produced via the decay of a top or anti-top in $t\bar{t}$ production,
i.e. via $p\bar{p},pp\rightarrow t\bar{t}$ with $t\rightarrow bH^+$ or
$\bar{t}\rightarrow \bar{b}H^-$. On the other hand, when
$M_{H^\pm}>m_t-m_b$, $H^\pm$ is mainly produced through $p\bar{p}\rightarrow
\bar{t}bH^+, t\bar{b}H^-$, as well as through the tree level 
production modes $q\bar{q}\rightarrow H^+H^-$ and $b\bar{b}\rightarrow
W^\pm H^\mp$, and at one loop through the associated modes
$gg\rightarrow H^+H^-$ and $gg\rightarrow W^\pm H^\mp$. The overall
cross section for the Tevatron and the LHC is illustrated in
Figs.~\ref{fig:mssm_charged_higgs_tevatron} and
\ref{fig:mssm_charged_higgs_lhc} as a function of the charged Higgs
boson mass, for different values of $\tan\beta$. The threshold
behavior at $M_H^\pm\simeq m_t-m_b$ is clearly visible.

\begin{figure}
\hspace{-1.truecm}
\begin{tabular}{cc}
\begin{minipage}{0.5\linewidth}
{\includegraphics[scale=0.4]{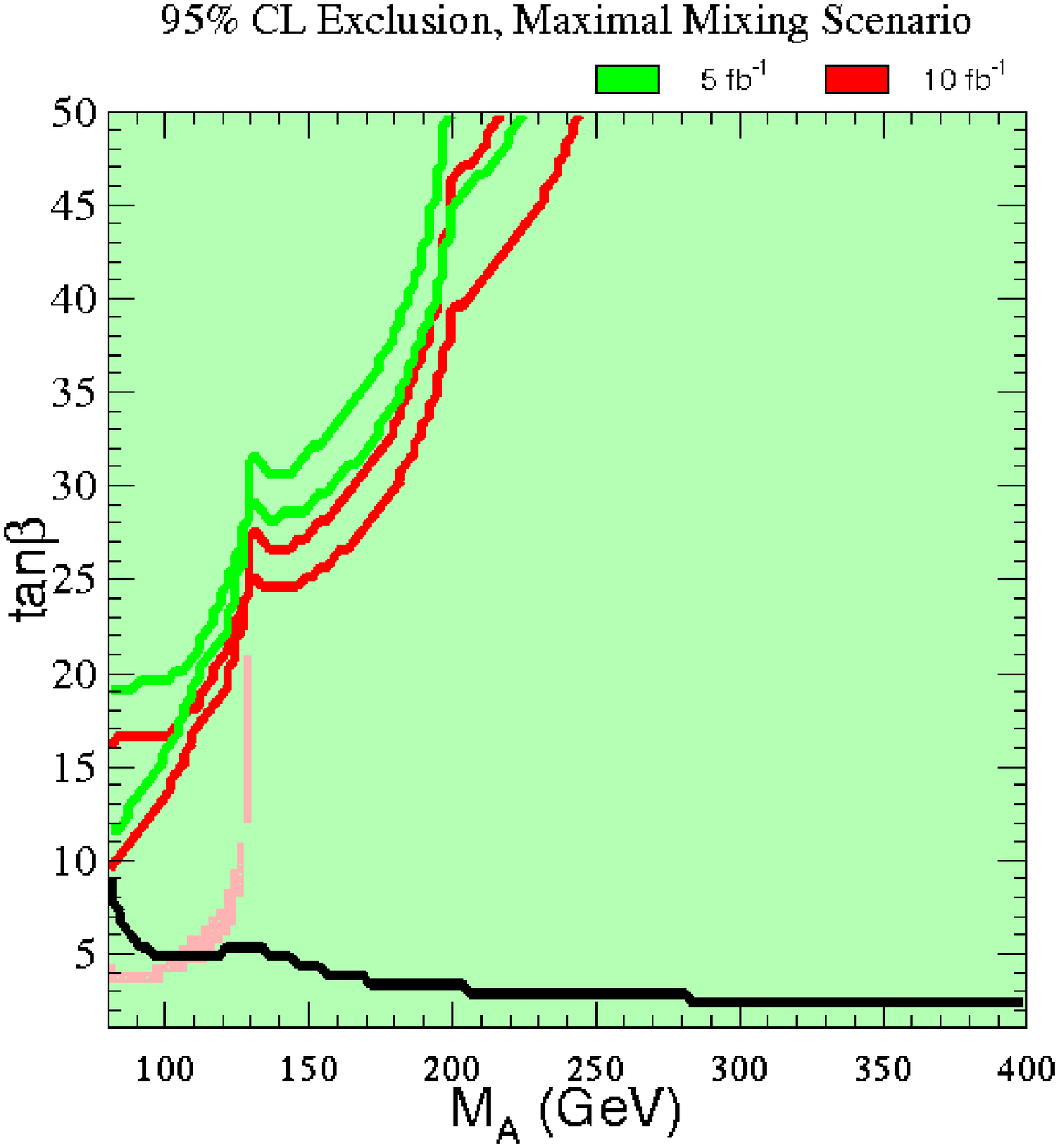}}
\end{minipage} &
\begin{minipage}{0.5\linewidth}
{\includegraphics[scale=0.4]{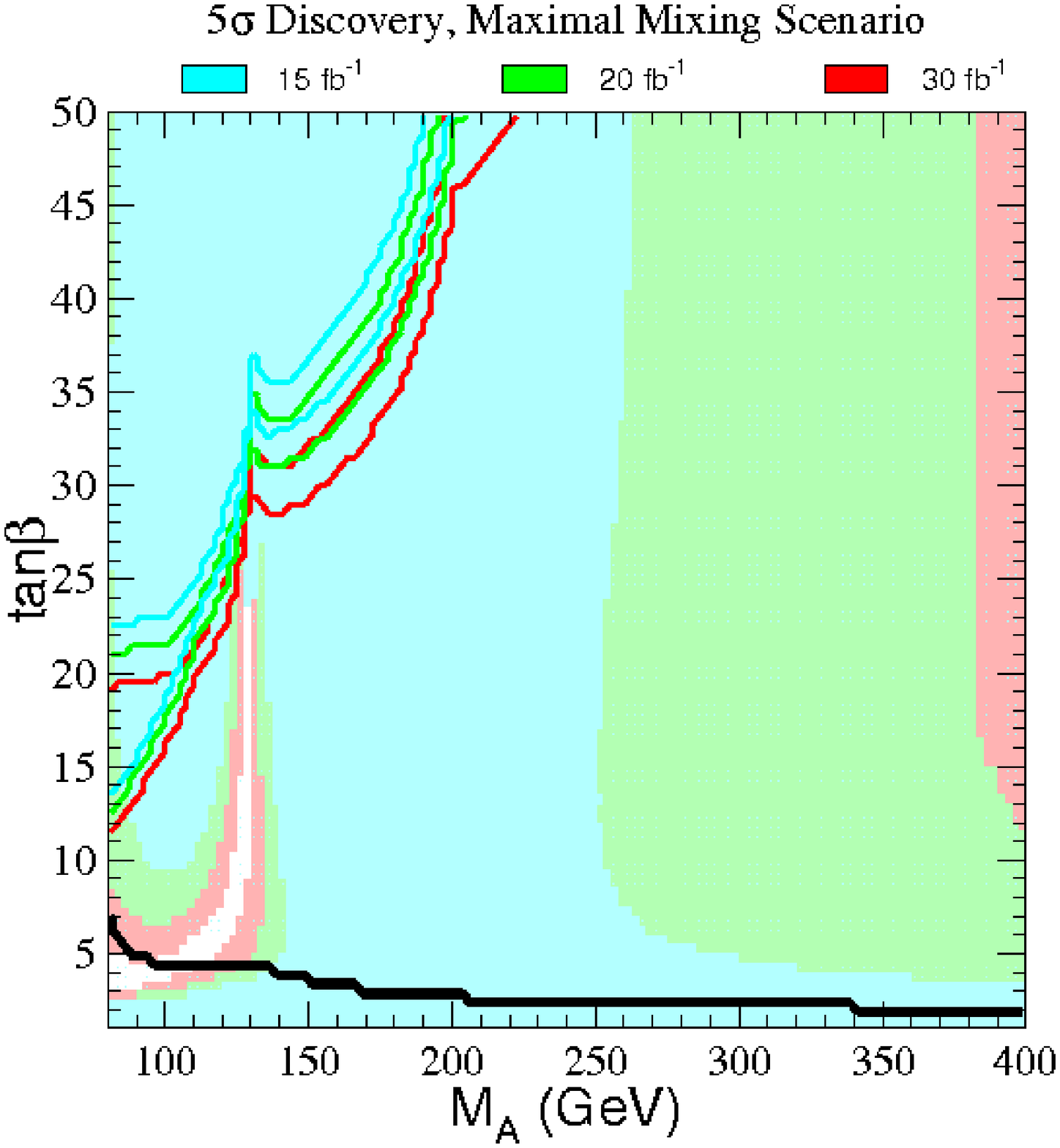}}
\end{minipage}
\end{tabular}
\caption[]{95\% C.L. exclusion region and $5\sigma$ discovery region
in the $(M_A,\tan\beta)$ plane, for the maximal mixing scenario. The
shaded regions correspond to the search channel $q\bar{q}\rightarrow
V\phi^0\,\,(\phi^0\rightarrow b\bar{b})$ (with $\phi^0=h^0,H^0$),
while the regions delimited by the colored solid lines correspond to
the search channel $gg,q\bar{q}\rightarrow
b\bar{b}\phi^0\,\,(\phi^0\rightarrow b\bar{b})$ (with
$\phi^0=h^0,H^0,A^0$). The two sets of lines correspond to simulations
from CDF and D$\emptyset$ respectively. The region below the black
line has been excluded by searches at LEP2. From
Ref.~\cite{Carena:2002es}. See also
Ref.~\cite{Carena:2000yx}.\label{fig:mssm_reach_tevatron}}
\end{figure}

The evidence or the absence of evidence for an MSSM Higgs boson at the
Tevatron or at the LHC will place definite bounds on the parameter
space of the MSSM. The reach of the Tevatron in the $(M_A,\tan\beta)$
plane is illustrated in Fig.~\ref{fig:mssm_reach_tevatron} for
different integrated luminosities. The shaded regions are spanned by
using only the $q\bar{q}\rightarrow V\phi^0\,\,(\phi^0\rightarrow
b\bar{b})$ channel (with $\phi^0=h^0,H^0$), while the region above the
solid curves are covered by using the $gg,q\bar{q}\rightarrow
b\bar{b}\phi^0\,\,(\phi^0\rightarrow b\bar{b})$ channel (with
$\phi^0=h^0,H^0,A^0$). The region below the black solid line is
excluded by searches at LEP2. Fig.~\ref{fig:mssm_reach_tevatron} shows
that, although discovery may require integrated luminosities that are
beyond the reach of RUN2 of the Tevatron, with 5~fb$^{-1}$ CDF and
D$\emptyset$ will be able to exclude (in the maximal mixing scenario)
almost all the parameter space of the MSSM at 95\% C.L., a pretty
impressive result by itself!
\begin{figure}
\centering
\includegraphics[scale=0.9]{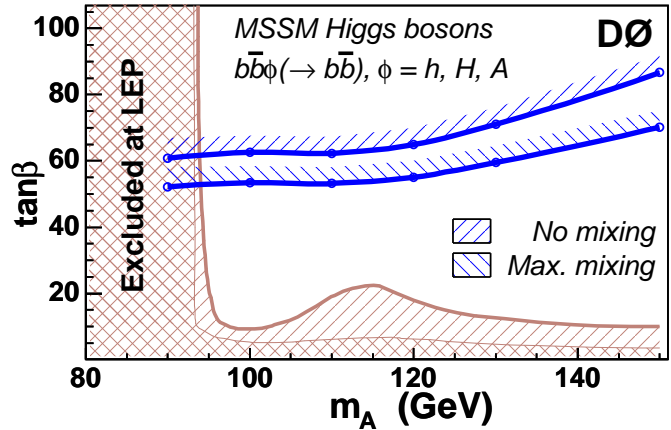}
\caption[]{The 95\% C.L. upper limit on $\tan\beta$ as a function of
$M_A$ for two scenarios of the MSSM, no mixing and maximal
mixing. Also shown are the limits obtained by the LEP experiments for
the same two scenarios of the MSSM. From
Ref.~\cite{Abazov:2005yr}, D$\emptyset$ analysis based
on 260~pb$^{-1}$ of data. \label{fig:search_bbH_d0}.}
\end{figure}
Both CDF and D$\emptyset$ have indeed already presented results from
searches conducted in the $p\bar{p}\rightarrow
b\bar{b}\phi^0\,\,(\phi^0\rightarrow b\bar{b})$ channel with three or
four $b$-quark jets tagged in the final state~\cite{Abazov:2005yr}.
The most recent results, from D$\emptyset$, are illustrated in
Fig.~\ref{fig:search_bbH_d0}, where we can see that, depending on
$M_A$, values of $\tan\beta$ as low as $\tan\beta\!=\!50$ have already
been excluded.
\begin{figure}
\begin{tabular}{cc}
\begin{minipage}{0.5\linewidth}
{\includegraphics[scale=0.45]{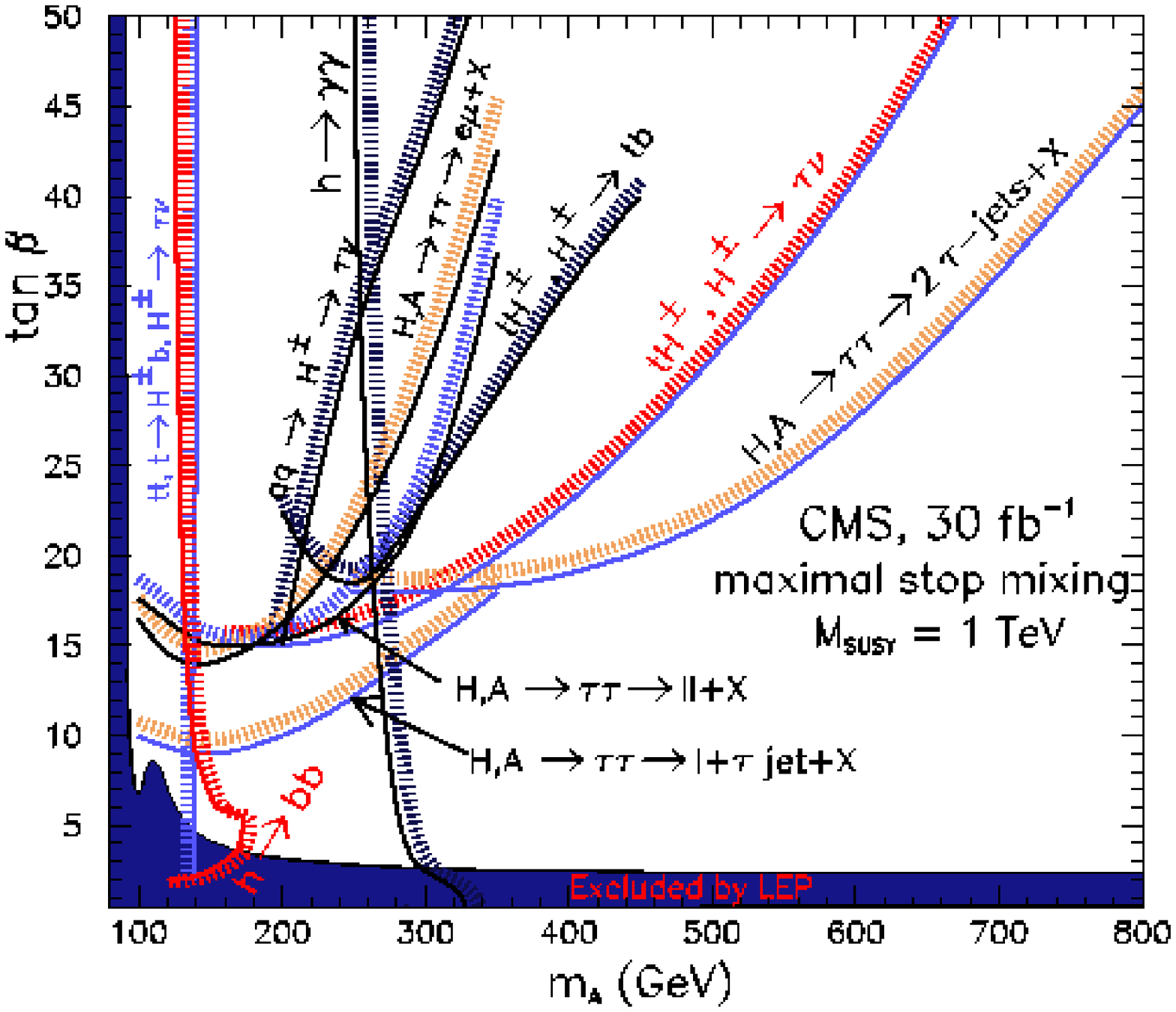}}
\end{minipage} &
\begin{minipage}{0.5\linewidth}
{\includegraphics[scale=0.4]{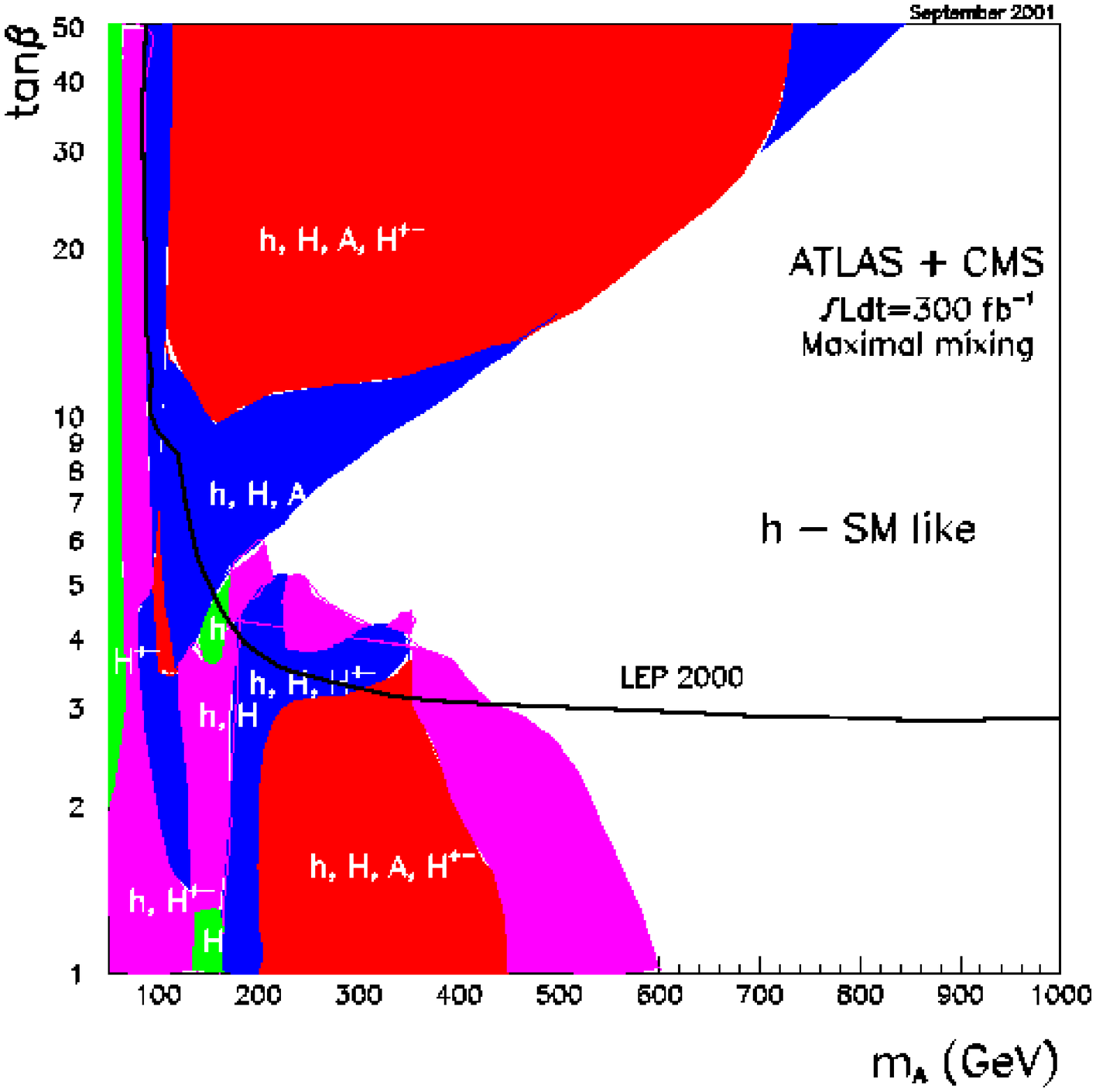}}
\end{minipage}
\end{tabular}
\caption[]{$5\sigma$ discovery contours for MSSM Higgs boson detection
in various channels, shown in the $(M_A,\tan\beta)$ parameter space,
assuming the maximal mixing scenario. The left hand plot is from the
CMS experiment and assumes 30~fb$^{-1}$ of integrated luminosity,
while the right hand plot combines ATLAS and CMS with a total
integrated luminosity of 300~fb$^{-1}$. 
\label{fig:mssm_reach_lhc}}
\end{figure}
The LHC $5\sigma$ discovery reach in the $(M_A,\tan\beta)$ parameter
space is illustrated in Fig.~\ref{fig:mssm_reach_lhc}. Thanks to the
high luminosity available and to the complementarity of various
production and decay modes, the entire $(M_A,\tan\beta)$ parameter
space can be covered, up to $M_A$ of the order of 1~TeV. This gives us
the exciting perspective that the LHC will be able to either discover
or completely rule out the existence of an MSSM Higgs boson!

\subsection{Higgs boson studies at a future $e^+e^-$ Linear Collider} 
\label{subsec:higgs_linear_collider}
As we all know, an $e^+e^-$ collider provides a very clean
environment, with relatively simple signatures and very favorable
signal to background ratios. Therefore, one of the most important
roles that a high energy $e^+e^-$ collider (to which we will refer as
International Linear Collider (ILC) or simply Linear Collider (LC))
will play is to unambiguously identify any new particle discovered at
the Tevatron or at the LHC, through a thorough program of precision
measurements. This is true in Higgs boson physics as well, where we
expect the mass, width, spin, and couplings of any Higgs boson
candidates to be determined at the few percent level. In this Section
I will mainly focus on a SM Higgs boson, since this is enough to
illustrate the role played by a LC in Higgs physics, and since the
true impact on the study of the MSSM parameter space will be
shaped by the discoveries occured by the time a LC is built.

\begin{figure}
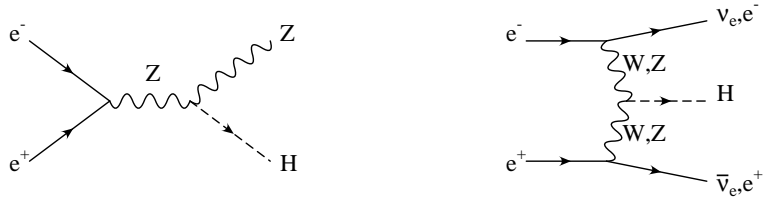

\begin{tabular}{cc}
\begin{minipage}{0.4\linewidth}
\includegraphics[scale=0.6]{epem_ZH}
\end{minipage} &
\begin{minipage}{0.4\linewidth}
\includegraphics[scale=0.6]{epem_Hnunu}
\end{minipage}
\end{tabular}
\caption[]{Higgs boson production via Higgs strahlung and $W^+W^-,ZZ$ fusion at
$e^+e^-$ colliders. \label{fig:epem_ZH_Hnunu}}
\end{figure}

\begin{figure}
\begin{minipage}{0.8\linewidth}
\includegraphics[scale=0.7]{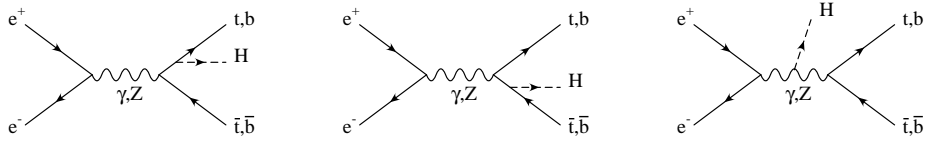}
\end{minipage}
\caption[]{Higgs boson production in association with $t\bar{t}$ pairs at
$e^+e^-$ colliders. \label{fig:epem_ttH}}
\end{figure}

\begin{figure}
\begin{tabular}{cc}
\begin{minipage}{0.4\linewidth}
\includegraphics[scale=0.6]{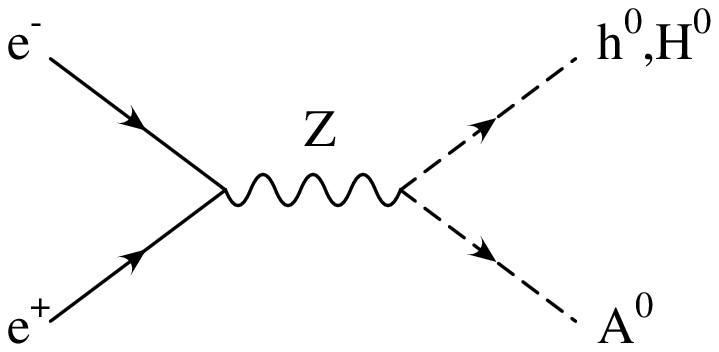}
\end{minipage} &
\begin{minipage}{0.4\linewidth}
\includegraphics[scale=0.6]{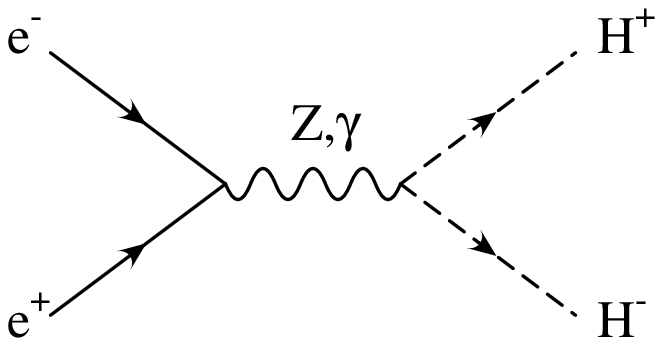}
\end{minipage}
\end{tabular}
\caption[]{MSSM Higgs pair production  at $e^+e^-$ colliders. \label{fig:epem_hA_HH}}
\end{figure}

\begin{figure}
\begin{tabular}{cc}
\begin{minipage}{0.5\linewidth}
\includegraphics[scale=0.65]{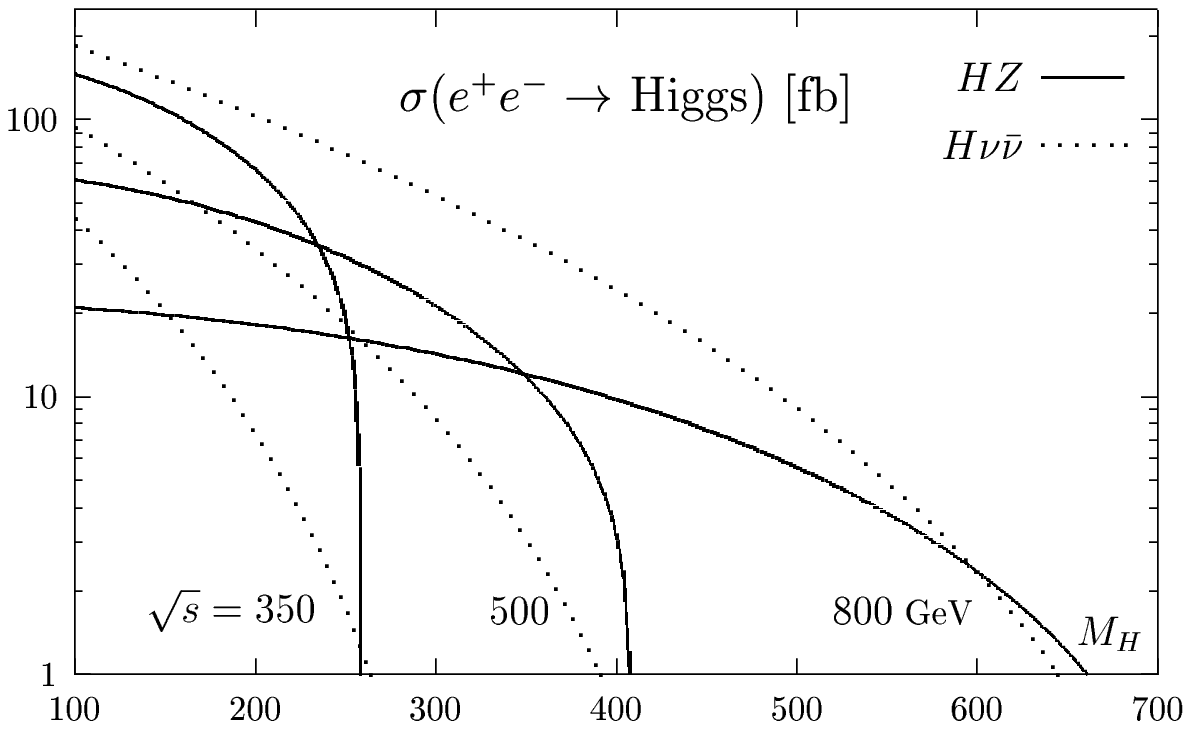}
\end{minipage} &
\begin{minipage}{0.5\linewidth}
\includegraphics[scale=0.35]{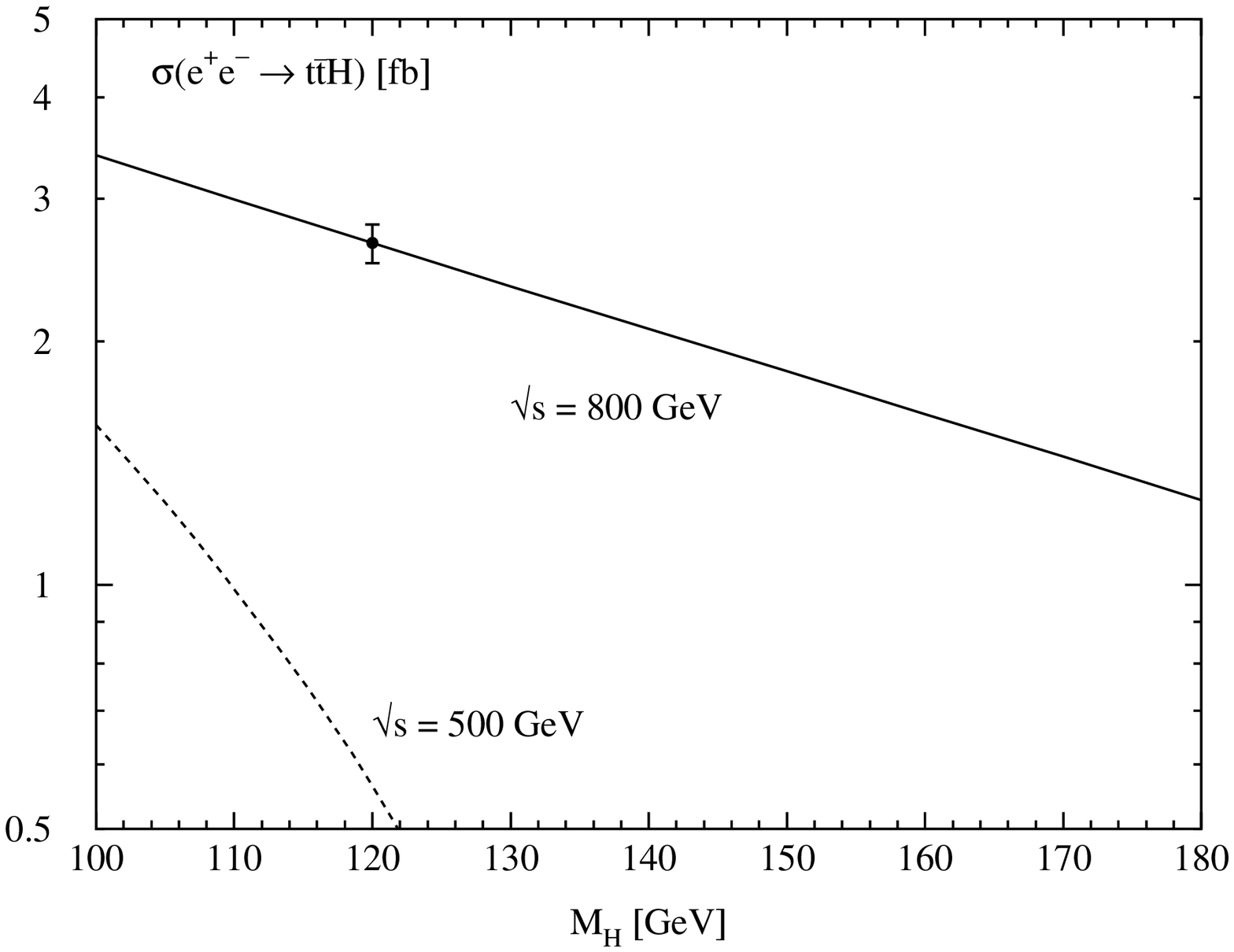}
\end{minipage}
\end{tabular}
\caption[]{SM Higgs boson production cross sections in $e^+e^-$
collisions, for various center of mass energies ($\sqrt{s}$). From
Ref.~\cite{Aguilar-Saavedra:2001rg}. \label{fig:sm_higgs_epem}}
\end{figure}

\begin{figure}
\centering
\includegraphics[scale=0.7]{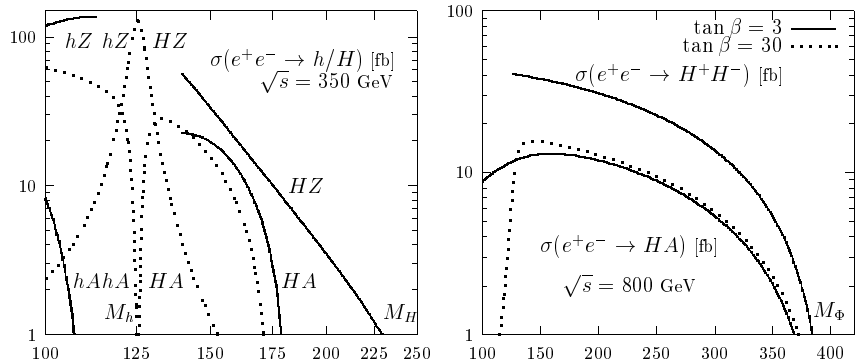}
\caption[]{MSSM Higgs boson production cross sections in $e^+e^-$
collisions, for two choices of $\tan\beta$: $\tan\beta\!=\!3$ (solid)
$\tan\beta\!=\!30$ (dotted), at $\sqrt{s}\!=\!350$~GeV, for $h^0$ and
$H^0$, and at $\sqrt{s}\!=\!800$~GeV, for $A^0$ and $H^\pm$.
From Ref.~\cite{Aguilar-Saavedra:2001rg}. \label{fig:mssm_higgs_epem}}
\end{figure}

First of all, the most important SM Higgs boson production processes in
$e^+e^-$ collisions are illustrated in Figs.~\ref{fig:epem_ZH_Hnunu}
and
\ref{fig:epem_ttH} and are: \emph{i)} $e^+e^-\rightarrow Z H$, the
Higgs strahlung or associated production with $Z$ gauge bosons,
\emph{ii)} $e^+e^-\rightarrow H\nu\bar{\nu}$, the $W^+W^-$ fusion
production, and $e^+e^-\rightarrow He^+e^-$, the $ZZ$ fusion
production, \emph{iii)} $e^+e^-\rightarrow t\bar{t}H$, the associated
production with a $t\bar{t}$ pair. In addition, in the MSSM we also
have:
\emph{iv)} $e^+e^-\rightarrow h^0A^0,H^0A^0$ and $e^+e^-\rightarrow
H^+H^-$, the pair production of two Higgs bosons, either neutral or
charged, as illustrated in Fig.~\ref{fig:epem_hA_HH}.
The cross sections of these processes as functions of the
corresponding Higgs masses are illustrated in
Figs.~\ref{fig:sm_higgs_epem} and \ref{fig:mssm_higgs_epem} for
various center of mass energies ($\sqrt{s}$). 
$e^+e^-\rightarrow Z H$ and $e^+e^-\rightarrow H\nu\bar{\nu}$ are the
leading production modes for a SM Higgs bosons. Their relative size
varies with the center of mass energy, since $\sigma(e^+e^-\rightarrow
Z H)$ scales as $1/s$ ($s$-channel process), while
$\sigma(e^+e^-\rightarrow H\nu\bar{\nu})$ scales as $\log(s)$
($t$-channel process).  $e^+e^-\rightarrow H\nu\bar{\nu}$ always
dominates over $e^+e^-\rightarrow He^+e^-$ by almost one order of
magnitude. In the MSSM, $e^+e^-\rightarrow H\nu\bar{\nu}$ plays a
lesser role, due to the suppression of the $VVH$ coupling
($V\!=\!W^\pm,Z$), but the $h^0A^0$, $H^0A^0$ and $H^+H^-$ pair
production modes become important. $t\bar{t}H$ production is always
very rare, in particular at center of mass energies around 500~GeV or
lower, but it plays a really important role at higher energies, around
800~GeV-1~TeV, for the determination of the top-quark Yukawa coupling,
as we will discuss later. Other rare production modes that could play
an important role in determining some Higgs boson properties are the
double Higgs boson production modes: $e^+e^-\rightarrow HHZ$ and
$e^+e^-\rightarrow HH\nu\bar{\nu}$

\begin{figure}
\centering
\includegraphics[scale=0.35]{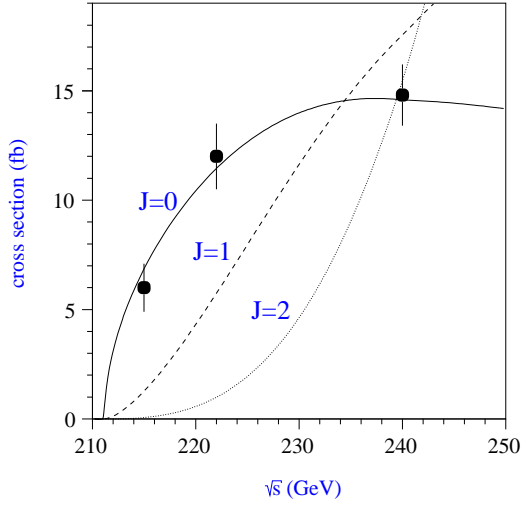}
\caption[]{The $e^+e^-\rightarrow ZH$ cross section energy dependence
near threshold for $M_H\!=\!120$~GeV and spin
$J^P\!=\!0^+,1^-,2^+$. From
Ref.~\cite{Aguilar-Saavedra:2001rg}. \label{fig:sm_higgs_spin}}
\end{figure}

With a LC running at energies between 350~GeV and 1~TeV, one or more
Higgs bosons can be observed over the entire mass spectrum and all
its properties can be precisely studied.  Reconstructing the recoiling
$l^+l^-$ mass (for $l\!=\!e,\mu$) in $e^+e^-\rightarrow HZ\rightarrow
Hl^+l^-$, where the $Z$ is mono-energetic, allows an excellent and
model independent determination of the Higgs boson mass. $Z\rightarrow
q\bar{q}$ decays can also be used and actually provide a very large
statistics. Accuracies of the order of 50-80~MeV can be obtained,
depending on the center of mass energy and the Higgs boson mass. The
spin and parity of the Higgs boson candidate (expected to be
$J^P\!=\!0^+$) can be determined in several ways, among others:
\emph{i)} from the onset of $\sigma(e^+e^-\rightarrow Z H)$, since the
energy dependence near threshold strongly depend on the $J^P$ quantum
number of the radiated $H$ (see Fig.~\ref{fig:sm_higgs_spin});
\emph{ii)} from the angular distribution of $H$ and $Z$ in
$e^+e^-\rightarrow ZH\rightarrow 4f $; \emph{iii)} from the
differential cross section in $e^+e^-\rightarrow t\bar{t}H$.

\begin{figure}
\centering
\includegraphics[scale=0.4]{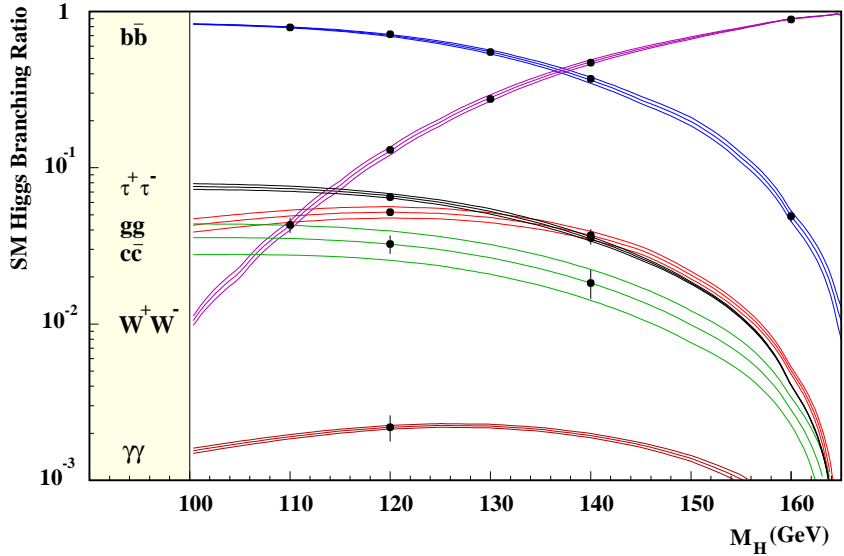}
\caption[]{The theoretical predictions (bands, due to the uncertainty
in the quark masses and $\alpha_s$) and experimental accuracies (dots
with error bars) for the SM Higgs branching ratios at
$\sqrt{s}\!=\!350$~GeV with 500~fb$^{-1}$ of data. From
Ref.~\cite{Aguilar-Saavedra:2001rg}. \label{fig:lc_higgs_coupling_accuracies}}
\end{figure}
Finally, a high energy LC will measure the Higgs boson couplings to
unprecedented precision and in a model independent way.  Thanks to the
precise knowledge of the initial state energy configuration (once the
initial state radiation, or beam strahlung, has been properly taken
into account), one can indeed measure both $\sigma(e^+e^-\rightarrow
HZ\rightarrow Hl^+l^-)$ and $\sigma(e^+e^-\rightarrow
W^*W^*\nu\bar{\nu}\rightarrow H\nu\bar{\nu})$, reconstructing the mass
recoiling against the $l^+l^-$ or $\nu\bar{\nu}$ pair, and from there
determine in a model independent way the isolated $HZZ$ and $HWW$
couplings\footnote{We notice that the two production modes
$e^+e^-\rightarrow HZ\rightarrow H\nu\bar{\nu}$ and $e^+e^-\rightarrow
W^*W^*\nu\bar{\nu}\rightarrow H\nu\bar{\nu}$ can be well isolated,
since their distribution in the invariant $\nu\bar{\nu}$ mass are very
distinctive. See Ref.~\cite{Aguilar-Saavedra:2001rg} for further
details.}. This is probably the most important intrinsic difference
between measuring the Higgs couplings at a lepton versus a hadron
collider. At a lepton collider the Higgs couplings to the weak gauge
bosons, i.e. the Higgs couplings associated to the production mode
($y_p$ of Section~\ref{subsubsec:sm_higgs_properties}), can be
isolated in a model independent way. Any other coupling can then be
also determined in a model independent way, measuring the individual
$Br(H\rightarrow XX)$ in $e^+e^-\rightarrow HZ$ followed by
$H\rightarrow XX$. Several recent studies have confirmed the
possibility of determining Higgs couplings to both gauge boson and
fermions within a few percent (2-5\%). For instance, for a Higgs boson
of $M_H\!=\!120$~GeV, the bottom-quark Yukawa coupling, $y_b$, could
be determined within 2\%, the $\tau$ one, $y_\tau$, within 5\%, and
the charm quark one, $y_c$ within 6\% due to the larger error on the
charm quark mass. This will test in a very stringent way the
proportionality of the Higgs couplings to the mass of the interacting
fermion.  Even the indirect coupling of the Higgs boson to a pair of
gluons, arising at the one-loop level (see
Section.~\ref{subsec:higgs_sm}), will be determined with a precision
of 4-5\%.  This will allow an indirect check of the top-quark Yukawa
coupling for a SM Higgs boson (when the top-quark loop dominates), or
will result in some anomalous coupling if new physics contributes in
the loop. In this second case, probably, other Higgs coupling will
show anomalous behaviors. The attainable precisions at a LC running at
$\sqrt{s}\!=\!350$~GeV, and with a 500~fb$^{-1}$ of data, are
summarized in Fig.~\ref{fig:lc_higgs_coupling_accuracies}.

\begin{figure}
\begin{tabular}{cc}
\begin{minipage}{0.5\linewidth}
{\includegraphics[scale=0.4]{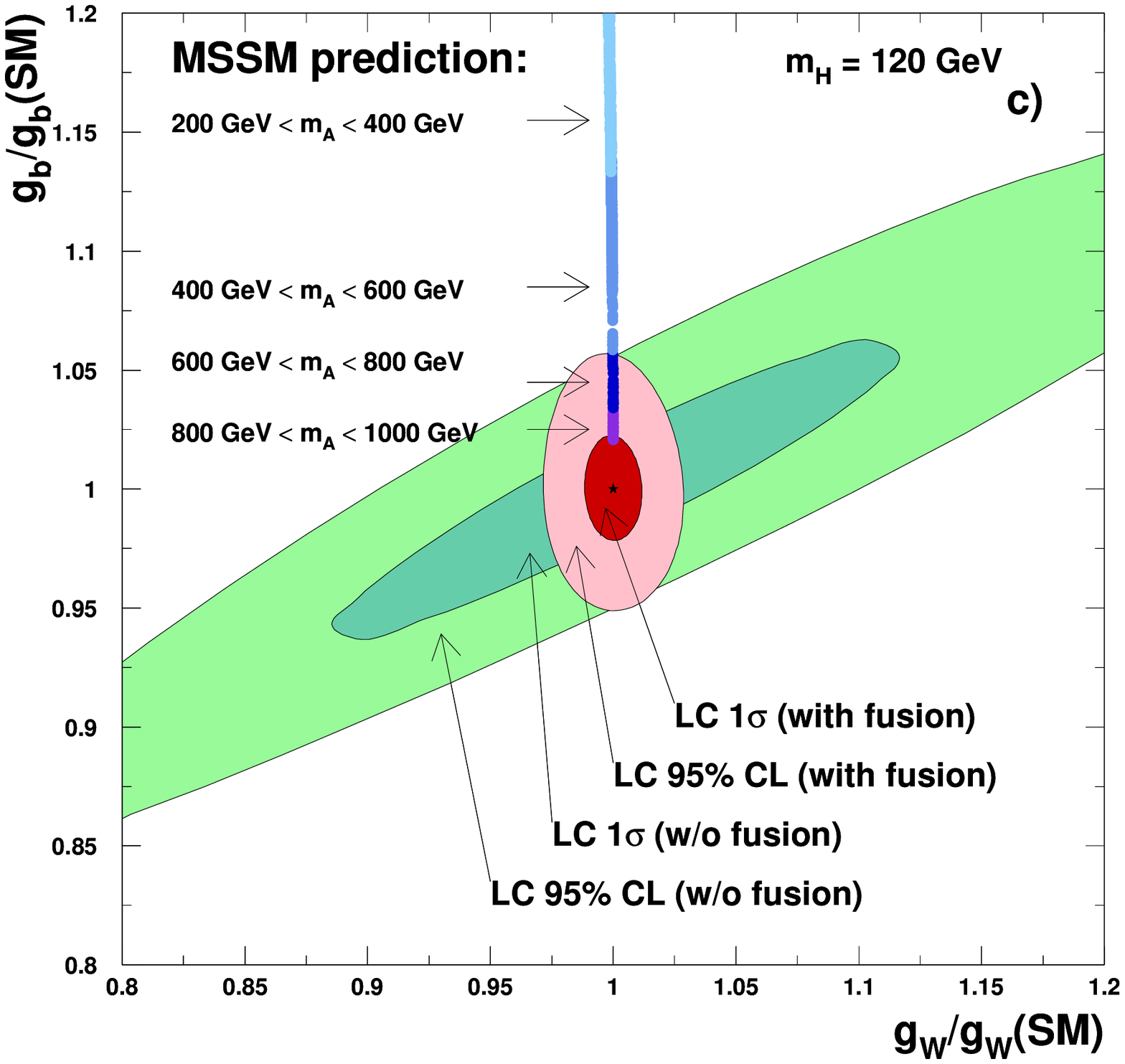}}
\end{minipage} &
\begin{minipage}{0.5\linewidth}
{\includegraphics[scale=0.4]{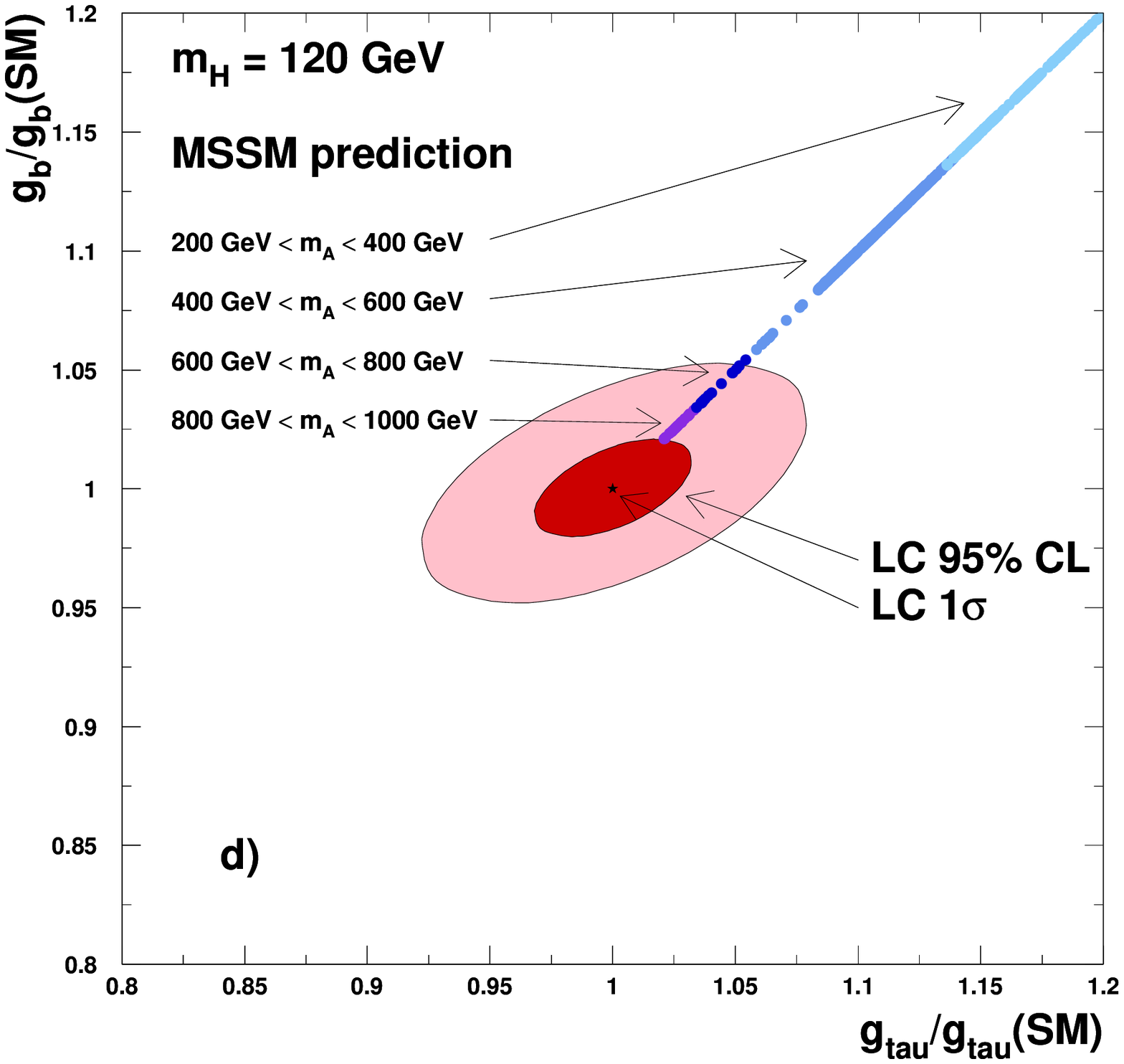}}
\end{minipage}
\end{tabular}
\caption[]{Higgs couplings determination at a LC (TESLA in this case) with 500~fb$^{-1}$ of
data, for $M_H\!=\!120$~GeV. The labels \emph{with fusion} and
\emph{w/o fusion} refers to $WW$-fusion inclusion/exclusion. From
Ref.~\cite{Aguilar-Saavedra:2001rg}. \label{fig:higgs_couplings_lc}}
\end{figure}
A LC will then be the ideal machine to discover small new physics
effects. For instance, it could play a fundamental role in
distinguishing the $h^0$ Higgs boson of the MSSM from the SM Higgs
boson, in regions of the MSSM parameter space close to the decoupling
limit. With this respect, Fig.~\ref{fig:higgs_couplings_lc} shows two
combined fits: the first one combines the ratio between the
bottom-quark Yukawa coupling to $h^0$ and its SM expected value and
the ratio between the $h^0$ coupling to $W$ bosons and its SM expected
value; the second one combines the same ratios for the bottom-quark
and $\tau$-lepton Yukawa couplings. Different contours illustrates the
precision with which the correlation between the combined ratios of
couplings can be measured. On the same plots we see what the ratios
would look like in the MSSM for different ranges of $M_A$. A definite
distinction between SM and MSSM Higgs bosons can clearly be
established.

\begin{figure}
\centering
\includegraphics[scale=0.5]{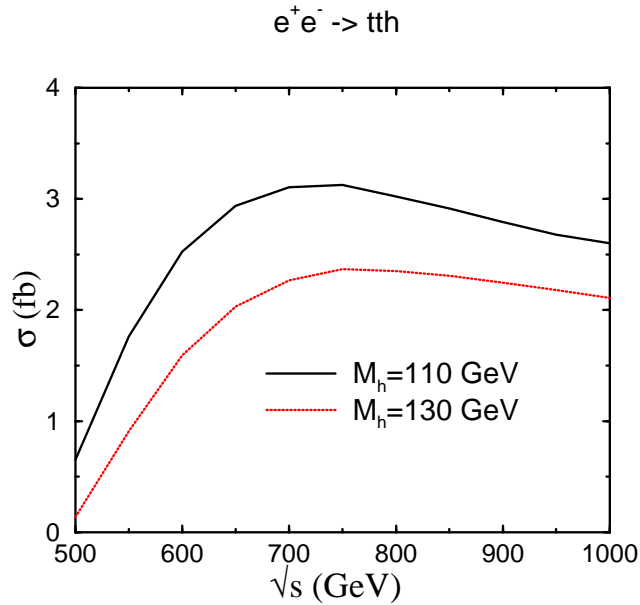}
\caption[]{Cross section for $e^+e^-\rightarrow t\bar{t}H$, in the SM,
as a function of the center of mass energy $\sqrt{s}$, for two
different values of the Higgs boson
mass. \label{fig:epem_tth_cross_section}}
\end{figure}
\begin{figure}
\centering
\includegraphics[scale=0.4]{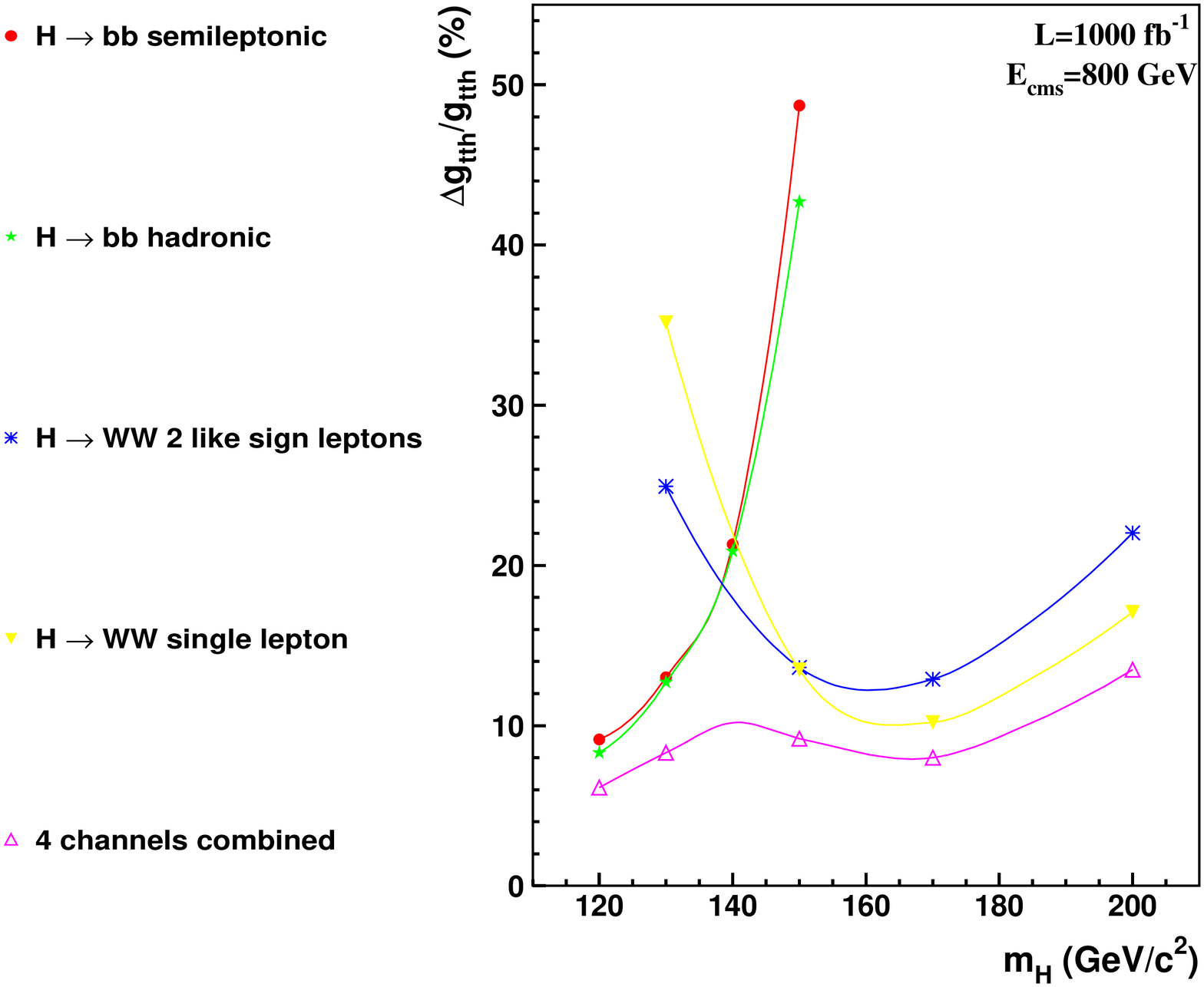}
\caption[]{Expected accuracies for the measurement of the top-quark Yukawa
coupling, $y_t$, in the process $e^+e^-\rightarrow t\bar{t}H$, as a
function of $M_H$, for $\sqrt{s}\!=\!800$~GeV and 1 ab$^{-1}$ of data
in various decay channels. From
Ref.~\cite{Djouadi:2005gi}.\label{fig:yt_combined}}
\end{figure}

The only problem in completing a full study of the Higgs boson
couplings is the determination of the top-quark Yukawa coupling, and
of the Higgs self couplings. The top-quark Yukawa coupling is
indirectly determined by measuring the $t\bar{t}H$ cross section, when
the $Z$ contribution is under control. This cross section is very
small at $\sqrt{s}\!=\!500$~GeV, and peaks around
$\sqrt{s}\!=\!800$~GeV (see Fig.~\ref{fig:epem_tth_cross_section})
Recent studies show that with a LC operating at $\sqrt{s}\!=\!500$~GeV
the top-quark Yukawa coupling, $y_t$, for a $M_H\!=\!120$~GeV Higgs
boson, will probably be determined only at the 25\% precision level,
while with a LC operating at $\sqrt{s}\!=\!800$~GeV precisions as high
as $5-6\%$ becomes available. An updated summary of the existing
studies can be seen in Fig.~\ref{fig:yt_combined}.  The initial phase
of a high energy LC will therefore not be able to give us probably the
most important Yukawa coupling, and with this respect the role of the
LHC becomes crucial, since, as we have seen in
Section~\ref{subsubsec:sm_higgs_properties}, the LHC can obtain $y_t$
within 10-15\% accuracy, although with some intrinsic model
dependence.

\begin{figure}
\centering
{\includegraphics[scale=0.5]{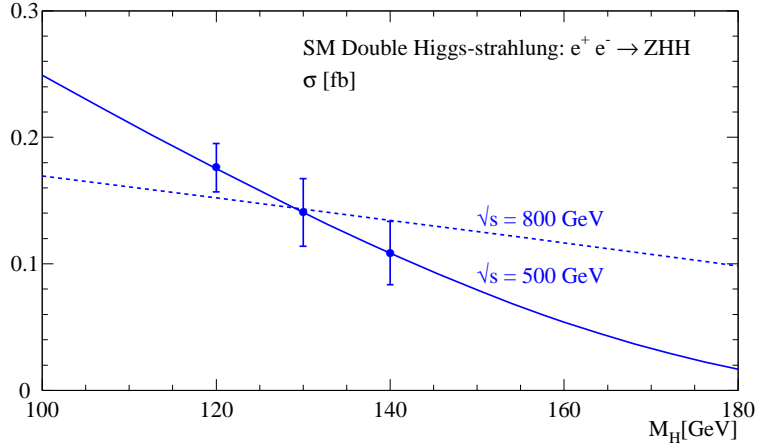}}
\caption[]{Cross section for $e^+e^-\rightarrow HHZ$, in the SM, as a
function of $M_H$, for two values of the center of mass energy
$\sqrt{s}$. The dots with error bars show the achievable experimental
accuracies for 1000~fb$^{-1}$ of data. From
Ref.~\cite{Aguilar-Saavedra:2001rg}. \label{fig:epem_HHZ_cross_section}}
\end{figure}

The production of Higgs boson pairs is also very rare (see
Fig.~\ref{fig:epem_HHZ_cross_section}), and the measurement of the
Higgs boson self-couplings will probably have to wait for a very high energy
Linear Collider, like the CLIC collider, a multi-TeV $e^+e^-$ machine
being studied at CERN.

Finally, the total width of a SM-like Higgs boson can be determined in
a model independent way by using any well measured branching
ratio. For example, one can use that $\Gamma\!=\!\Gamma(H\rightarrow
WW^*)/Br(H\rightarrow WW^*)$, where $Br(H\rightarrow WW^*)$ is
measured directly and $\Gamma(H\rightarrow WW^*)$ can be calculated
from the direct determination of the $HWW$ coupling.

\section{Highlights of theoretical calculations in Higgs boson physics}
\label{sec:theory}
I would like to conclude these lectures by reviewing some important
theoretical results that have recently been obtained in the
calculation of Higgs boson physics observables, i.e. total and differential
cross sections. Having to limit my discussion to a selection of
topics, I prefer to focus on hadron colliders, since the discovery of
a Higgs particle very much depends on our ability to provide precise
theoretical predictions for hadronic cross sections. In this context I
will concentrate on a few processes for which some outstanding progress
has been made recently.

The cross section for $p\bar{p}$ and $pp$ collisions to produce a
final state containing a Higgs bosons ($H$) can be schematically
written as:
\begin{equation}
\label{eq:hadronic_cross_section}
\sigma(pp,p\bar{p}\rightarrow H+X)=
\sum_{ij}\int dx_1 dx_2
f_i^p(x_1)f_j^{p,\bar{p}}(x_2)\hat{\sigma}(ij\rightarrow H+X)\,\,\,,
\end{equation}
where the partonic cross section $\hat\sigma(ij\rightarrow H+X)$ is
convoluted with the Parton Distribution Functions (PDF) of partons $i$
and $j$. $f_i^{p,\bar{p}}(x)$ denotes indeed the PDF of parton $i$
into a proton (or anti-proton) and can be interpreted as the
probability of finding parton $i$ into a proton (or anti-proton) with
a fraction $x$ of its longitudinal momentum. Both the partonic cross
section and the parton distribution functions are calculated perturbatively. At
hadron colliders, the most important effects arise from strong
interactions, and it is therefore mandatory to have the QCD
perturbative expansion of $\sigma(pp,p\bar{p}\rightarrow H+X)$ under
control. At each order in the perturbative expansion, the calculation
of both $\hat\sigma(ij\rightarrow H+X)$ and $f_i^{p,\bar{p}}(x)$
contains ultraviolet divergences that are subtracted through a
standard renormalization procedure. This, at each finite order, leaves
a dependence on the renormalization scale, $\mu_R$. In the same way,
when the PDF's are defined, a
factorization scale $\mu_F$ is introduced in the calculation of
$f_i^{p,\bar{p}}(x)$. The dependence on both $\mu_R$ and $\mu_F$ is
indicative of the residual theoretical uncertainty present at a given
perturbative order, and should improve the higher the order of QCD
corrections that are taken into account.

Indeed, it is well known that the theoretical predictions for most
Higgs production hadronic cross sections at lowest or leading order
(LO) are affected by a very large renormalization and factorization
scale dependence. In general, at least the next-to-leading order (NLO)
of corrections need to be calculated and this should stabilize or
improve the theoretical prediction for the cross section, making the
residual theoretical uncertainty comparable or smaller than the
corresponding experimental precision. In some cases, as we will see,
even next-to-next-to-leading order (NNLO) QCD corrections are
necessary to obtain reliable theoretical predictions.

\begin{figure}
\begin{tabular}{cc}
\begin{minipage}{0.5\linewidth}{
\includegraphics[bb=140 405 500 710,scale=0.6]{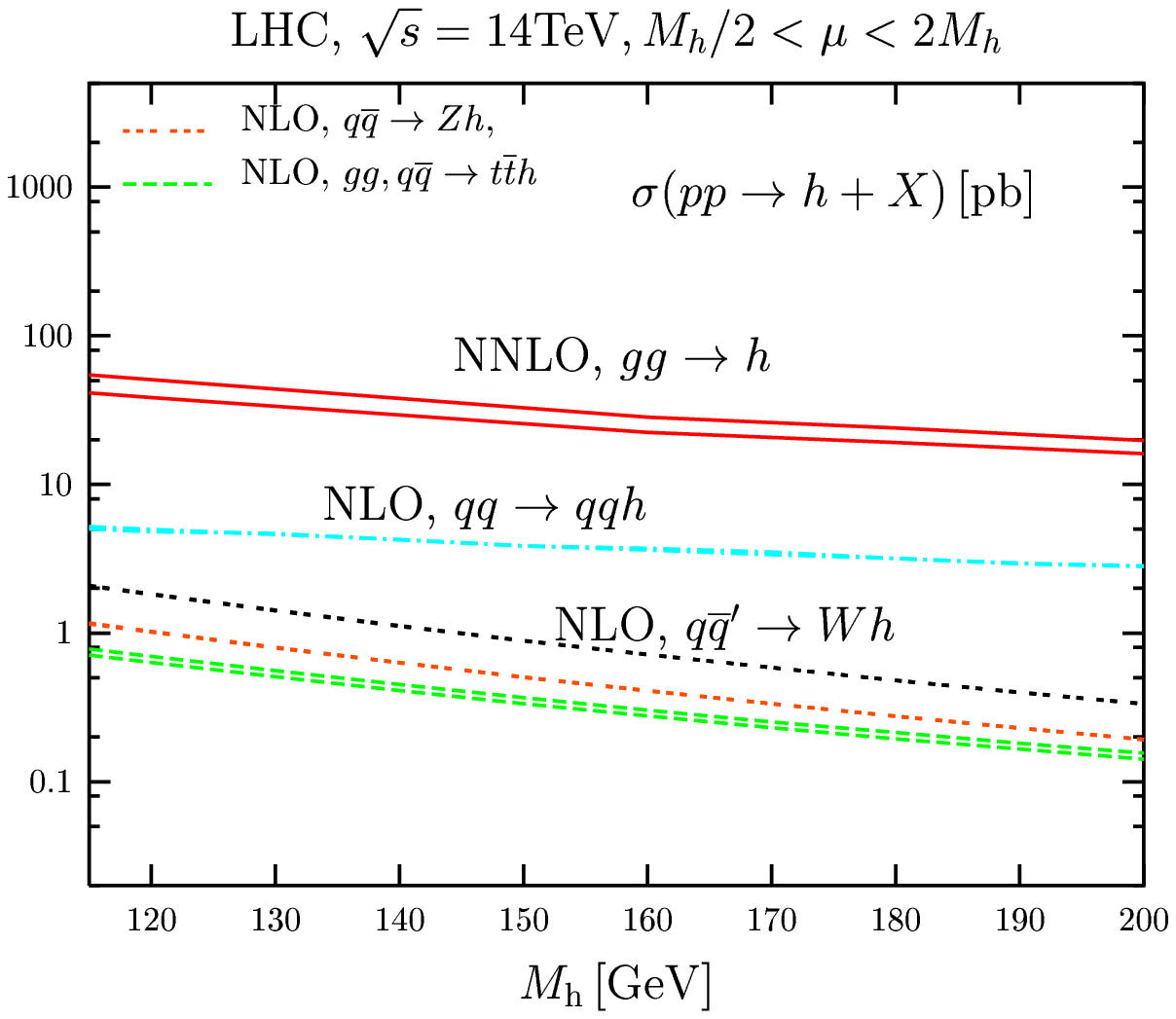}}
\end{minipage}&
\begin{minipage}{0.5\linewidth}{
\includegraphics[bb=140 405 500 710,scale=0.6]{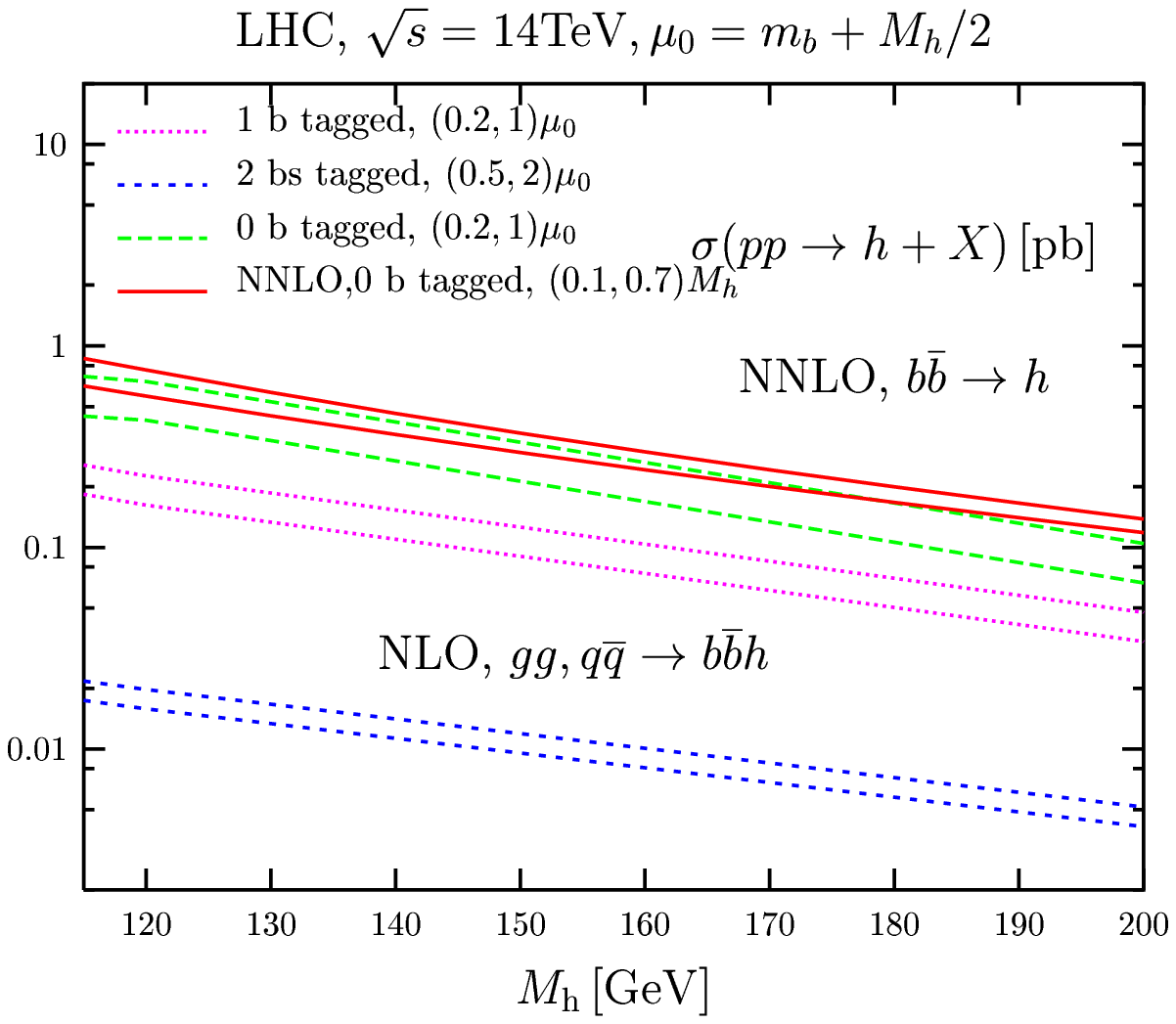}}
\end{minipage}
\end{tabular}
\caption[]{Residual renormalization and factorization
scale dependence ($\mu\!=\!\mu_R\!=\!=\!\mu_F$) of the SM Higgs boson
production cross section, when all available orders of QCD corrections
are included. The scale $\mu$ is varied according to the original work
present in the literature, and can therefore be slightly different
case by case. \label{fig:nlo_nnlo_residual_mu_dependence}}
\end{figure}

To be more specific, the NLO cross section for $pp,p\bar{p}\rightarrow
H+X$ can in full generality be written as:
\begin{equation}
\label{eq:sigma_nlo}
\sigma_{p\bar p,pp}^{NLO}=\sum_{i,j}\int dx_1dx_2
\mathcal{F}_i^{p}(x_1,\mu_R\mu_F\mathcal{F}_j^{\bar p,p}(x_2,\mu_R\mu_F
\hat{\sigma}_{ij}^{NLO}(x_1,x_2,\mu_R,\mu_F)\,\,\,,
\end{equation}
where we have made explicit the dependence on both renormalization and
factorization scale.
$\mathcal{F}_i^{p,\bar{p}}$ denote the NLO PDF's, while
$\hat{\sigma}_{ij}^{NLO}$ is the parton level cross section calculated
at NLO as:
\begin{equation}
\label{eq:sigma_hat_nlo}
\hat{\sigma}_{ij}^{NLO}=
\hat{\sigma}_{ij}^{LO}+ 
\frac{\alpha_s}{4\pi}\delta\hat{\sigma}_{ij}^{NLO}\,\,\,,
\end{equation}
where $\delta\hat\sigma_{ij}$ represents the $\mathcal{O}(\alpha_s)$
real and virtual corrections:
\begin{equation}
\label{eq:delta_sigma_nlo}
\delta\hat{\sigma}^{NLO}_{ij}=
\hat{\sigma}^{ij}_{virt}+\hat{\sigma}^{ij}_{real}\,\,\,.
\end{equation}
In a similar way, and with due differences, we could write
$\sigma_{p\bar p,pp}^{NNLO}$. A lot of theoretical effort has gone in
recent years in the calculation of $\delta\hat{\sigma}^{NLO}_{ij}$ and
$\delta\hat{\sigma}^{NNLO}_{ij}$ for several Higgs production
processes. I have collected in Table~\ref{tab:nlo_nnlo} all the
existing work on higher order QCD corrections to Higgs production
modes. My apologies for any omission!
\begin{table}
\caption[]{Existing QCD corrections for various SM Higgs production
processes. \label{tab:nlo_nnlo}}
{\begin{tabular}{c|l}
process & $\sigma_{NLO,NNLO}$ by\\
\hline
$gg\rightarrow H$ & 
\begin{minipage}{1.\linewidth}{
\scriptsize
\vspace{0.3truecm}
\begin{list}{}
  {\setlength{\topsep}{0.5truecm}\setlength{\parskip}{-0.8truecm}
    \setlength{\itemsep}{0.truecm}
    \setlength{\leftmargin}{0.0truecm}\setlength{\labelwidth}{1.5truecm}}
\item S.Dawson, NPB 359 (1991), 
      A.Djouadi, M.Spira, P.Zerwas, PLB 264 (1991)
\item C.J.Glosser \textit{et al.}, JHEP 0212 (2002); 
      V.Ravindran \textit{et al.}, NPB 634 (2002)
\item D. de Florian \textit{et al.}, PRL 82 (1999)
\item R.Harlander, W.Kilgore, PRL 88 (2002) (NNLO)
\item C.Anastasiou, K.Melnikov, NPB 646 (2002) (NNLO)
\item V.Ravindran \textit{et al.}, NPB 665 (2003) (NNLO)
\item S.Catani \textit{et al.} JHEP 0307 (2003) (NNLL),
G.Bozzi \textit{et al.}, PLB 564 (2003),hep-ph/0508068
\end{list}}
\end{minipage}\\
\hline
$q\bar{q}\rightarrow (W,Z)H$ & 
\begin{minipage}{1.\linewidth}{
\scriptsize
\vspace{0.3truecm}
\begin{list}{}
  {\setlength{\topsep}{0.5truecm}\setlength{\parskip}{-0.8truecm}
    \setlength{\itemsep}{0.truecm}
    \setlength{\leftmargin}{0.0truecm}\setlength{\labelwidth}{1.5truecm}}
\item T.Han, S.Willenbrock, PLB 273 (1991)
\item O.Brien, A.Djouadi, R.Harlander, PLB 579 (2004) (NNLO)
\end{list}}
\end{minipage}\\
\hline
$q\bar{q}\rightarrow q\bar{q}H$ & 
\begin{minipage}{1.\linewidth}{
\scriptsize
\vspace{0.3truecm}
\begin{list}{}
  {\setlength{\topsep}{0.5truecm}\setlength{\parskip}{-0.8truecm}
    \setlength{\itemsep}{0.truecm}
    \setlength{\leftmargin}{0.0truecm}\setlength{\labelwidth}{1.5truecm}}
\item T.Han, G.Valencia, S.Willenbrock, PRL 69 (1992)
\item T.Figy, C.Oleari, D.Zeppenfeld, PRD 68 (2003)
\end{list}}
\end{minipage}\\
\hline
$q\bar{q},gg\rightarrow t\bar{t}H$ & 
\begin{minipage}{1.\linewidth}{
\scriptsize
\vspace{0.3truecm}
\begin{list}{}
  {\setlength{\topsep}{0.5truecm}\setlength{\parskip}{-0.8truecm}
    \setlength{\itemsep}{0.truecm}
    \setlength{\leftmargin}{0.0truecm}\setlength{\labelwidth}{1.5truecm}}
\item W.Beenakker \textit{et al.}, PRL 87 (2001), NPB 653 (2003)
\item S.Dawson \textit{et al.}, PRL 87 (2001), PRD 65 (2002), PRD 67,68 (2003)
\end{list}}
\end{minipage}\\
\hline
$q\bar{q},gg\rightarrow b\bar{b}H$ & 
\begin{minipage}{1.\linewidth}{
\scriptsize
\vspace{0.3truecm}
\begin{list}{}
  {\setlength{\topsep}{0.5truecm}\setlength{\parskip}{-0.8truecm}
    \setlength{\itemsep}{0.truecm}
    \setlength{\leftmargin}{0.0truecm}\setlength{\labelwidth}{1.5truecm}}
\item S.Dittmaier, M.Kr\"amer, M.Spira, PRD 70 (2004)
\item S.Dawson \textit{et al.}, PRD 69 (2004), PRL 94 (2005)
\end{list}}
\end{minipage}\\
\hline
$gb(\bar{b})\rightarrow b(\bar{b})H$ & 
{\scriptsize J.Campbell \textit{et al.}, PRD 67 (2003)}\\
\hline
$b\bar{b}\rightarrow H$ & 
\begin{minipage}{1.\linewidth}{
\scriptsize
\vspace{0.3truecm}
\begin{list}{}
  {\setlength{\topsep}{0.5truecm}\setlength{\parskip}{-0.8truecm}
    \setlength{\itemsep}{0.truecm}
    \setlength{\leftmargin}{0.0truecm}\setlength{\labelwidth}{1.5truecm}}
\item D.A.Dicus \textit{et al.} PRD 59 (1999); 
      C.Balasz \textit{et al.}, PRD 60 (1999).
\item R.Harlander, W.Kilgore, PRD 68 (2003) (NNLO)
\end{list}}
\end{minipage}\\
\hline
\end{tabular}}
\end{table}
The result of this effort can be naively summarized by investigating
the residual renormalization ($\mu_R$) and factorization ($\mu_F$)
scale dependence in each SM Higgs boson production mode. This is
illustrated in Fig.~\ref{fig:nlo_nnlo_residual_mu_dependence} for the
case of the LHC. For each production mode the scale
$\mu\!=\!\mu_R\!=\!=\!\mu_F$ has been varied according to the
corresponding original literature as indicated in
Fig.~\ref{fig:nlo_nnlo_residual_mu_dependence}, such that the scale
interval can be
different case by case. Since different production modes have very
different cross sections, we have separated them into two plots,
containing the leading and sub-leading production modes respectively.
The bands in Fig.~\ref{fig:nlo_nnlo_residual_mu_dependence} are in no
way indicative of the overall theoretical error, since they do not
include systematic errors coming from PDFs and other input
parameters. Moreover the effect of setting $\mu_R\neq\mu_F$ needs and
has been investigated case by case, but it is not included in
Fig.~\ref{fig:nlo_nnlo_residual_mu_dependence}. Nevertheless
Fig.~\ref{fig:nlo_nnlo_residual_mu_dependence} gives us a qualitative
idea of the perturbative stability of the existing theoretical
predictions for the SM Higgs boson production cross sections. Overall
the existing theoretical predictions are in good control.

In a parallel series of lectures given at this
school~\cite{wacheroth:tasi04} you have been exposed to the complexity
of higher order QCD calculations, and to the variety of techniques
that have been developed to perform them. I will not then directly
proceed and comment about the results of some higher order calculations
in Higgs physics. In particular, I would like to report about:
\emph{i)} the calculation of $gg\rightarrow H$ at NNLO of QCD, 
a pioneer effort that has provided for the first time a
reliable theoretical results for the most important Higgs
production mode at hadron colliders; \emph{ii)} the calculation of
$pp,p\bar{p}\rightarrow t\bar{t}H,b\bar{b}H$ at NLO of QCD, 
a challenging task, due to the many massive degrees of
freedom involved, that has provided for the first time reliable 
theoretical predictions for the cross sections of these two physically 
very important production modes. Both \emph{i)} and \emph{ii)} rely on
the development of several innovative techniques that have allowed the
successful completion of both calculations. Given the degree of
technicalities involved, I will not review them in detail, but only
point to the kind of difficulties that had to be faced. The interested
reader can find all necessary technical details in the original
literature, listed in Table ~\ref{tab:nlo_nnlo} and in the bibliography.

\subsection{$gg\rightarrow H$ at NNLO}
\label{subsec:gg_to_H_nnlo}

\begin{figure}
\centering
\includegraphics[scale=0.7]{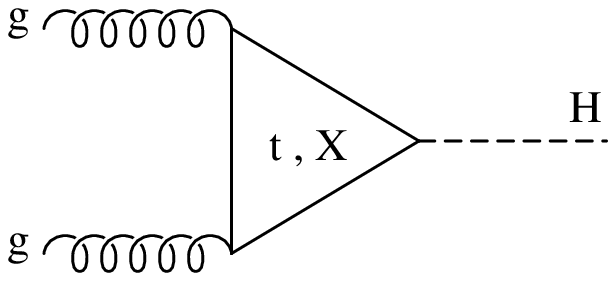}
\caption[]{The $gg\rightarrow H$ production process at lowest order.\label{fig:gg_to_H}}
\end{figure}
Most of the basic ideas that motivate the techniques used in the NNLO
calculation of the cross section for the $gg\rightarrow H$ production
process have been already introduced in
Section~\ref{subsubsec:sm_higgs_loop_decay}, where we discussed the
$H\rightarrow gg$ loop-induced decay.  In particular we know that in
the SM, the main contribution to $gg\rightarrow H$ comes form the
top-quark loop (see Fig.~\ref{fig:gg_to_H}) since:
\begin{equation}
\label{eq:gg_to_H_lo}
\sigma_{LO}=
\frac{G_F\alpha_s(\mu)^2}{288\sqrt{2}\pi}
\left|\sum_q A_q^H(\tau_q)\right|^2\,\,\,,
\end{equation}
where $\tau_q=4m_q^2/M_H^2$ and $A_q^H(\tau_q)\le 1$ with   
$A_q^H(\tau_q)\rightarrow 1$ for $\tau_q\rightarrow\infty$. 
\begin{figure}
\hspace{1.truecm}
\begin{minipage}{0.4\linewidth}
\centering
\includegraphics[scale=0.35]{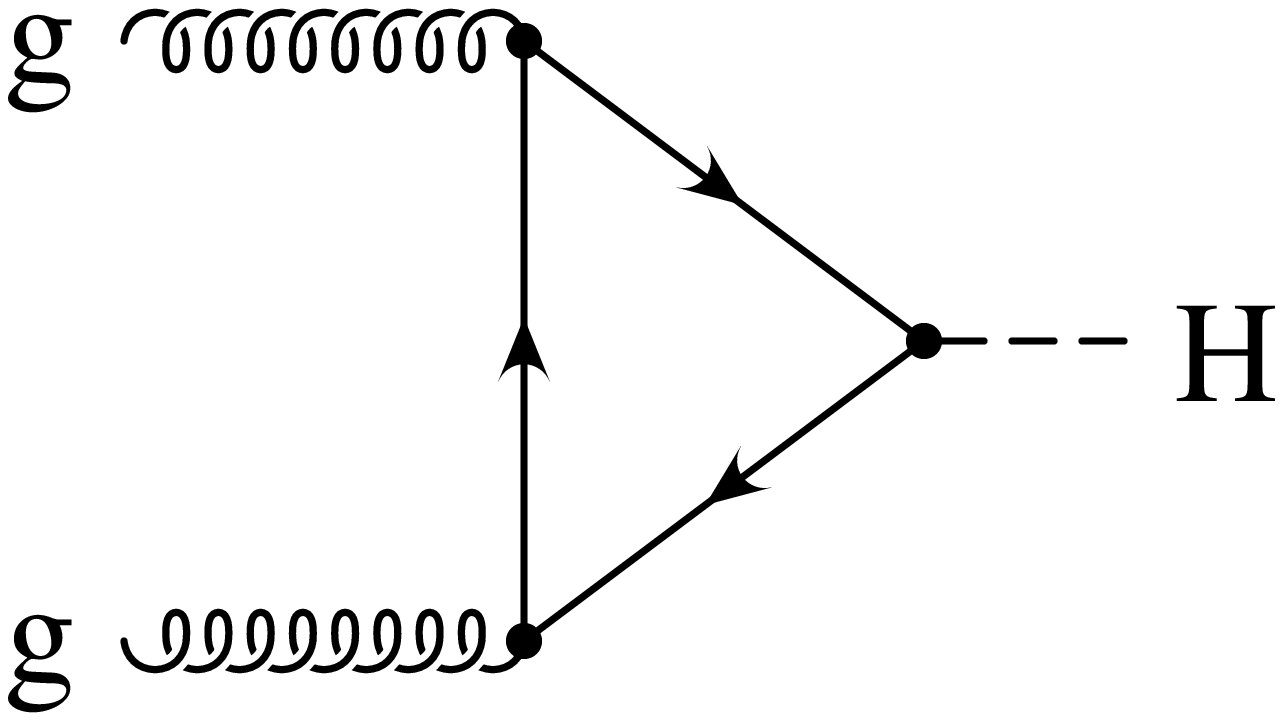}
\end{minipage}
$\longrightarrow$
\begin{minipage}{0.4\linewidth}
\centering
\includegraphics[scale=0.35]{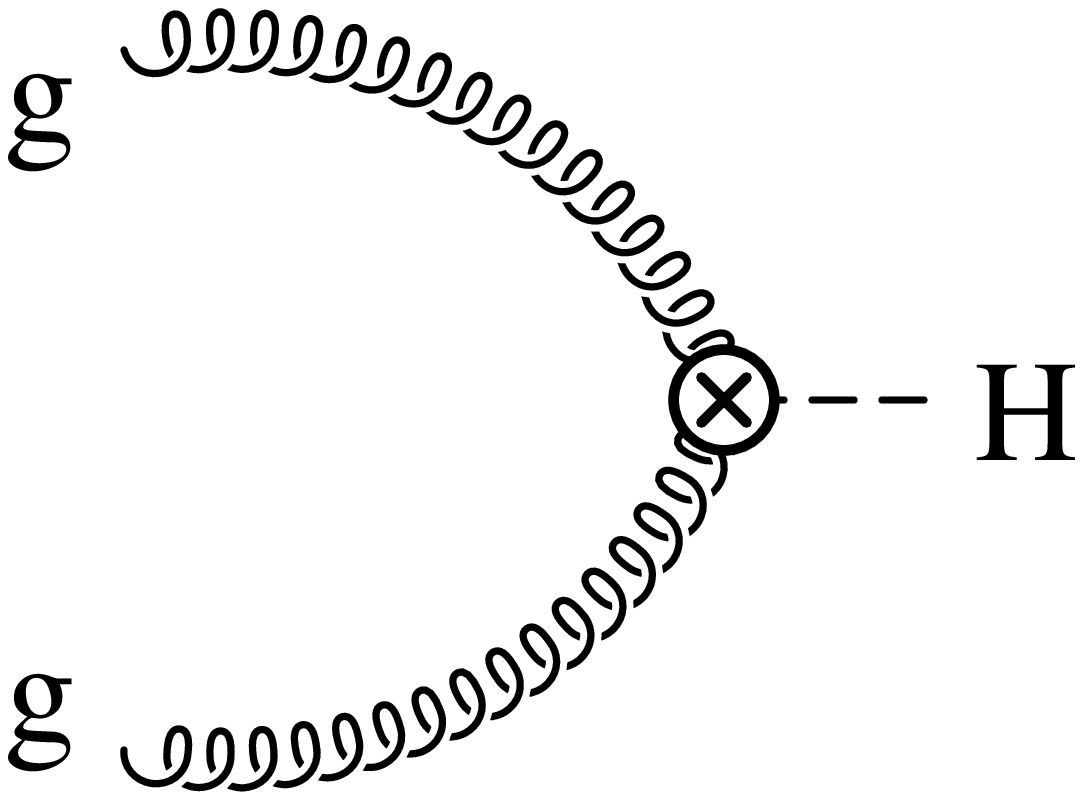}
\end{minipage}
\caption[]{The top-quark loop contribution to $gg\rightarrow H$ gives origin 
to a $ggH$ effective vertex in the $m_t\rightarrow\infty$ limit. 
\label{fig:gg_to_H_effective_vertex}}
\end{figure}

As we saw in Section~\ref{subsubsec:sm_higgs_loop_decay}, one can 
work in the infinite top-quark mass 
limit and reduce the one-loop $Hgg$ vertex to a tree level effective 
vertex, derived from an effective Lagrangian of the form:
\begin{equation}
\label{eq:gg_to_H_effective_lagrangian}
\mathcal{L}_{eff}=
\frac{H}{4v}C(\alpha_s)G^{a\mu\nu}G^a_{\mu\nu}\,\,\,,
\end{equation}
where the coefficient $C(\alpha_s)$, including NLO and NNLO QCD 
corrections, can be written as:
\begin{equation}
\label{eq:gg_to_H_wilson_coefficient}
C(\alpha_s)=\frac{1}{3}\frac{\alpha_s}{\pi}\left[1+c_1\frac{\alpha_s}{\pi}+
c_2\left(\frac{\alpha_s}{\pi}\right)^2+\cdots\right]\,\,\,.
\end{equation}
NLO and NNLO QCD corrections to $gg\rightarrow H$ can then be
calculated as corrections to the effective $Hgg$ vertex, and the complexity
of the calculation is reduced by one order of loops. 

\begin{figure}
\centering
\includegraphics[scale=0.5]{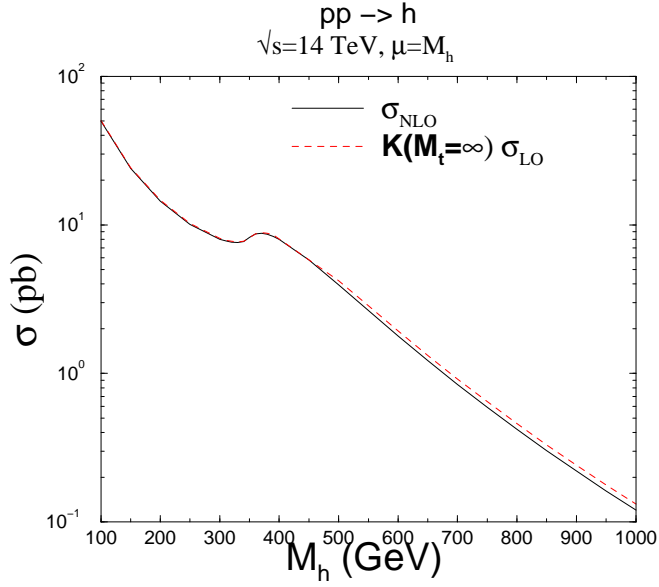}
\caption[]{The NLO cross section for 
  $gg\rightarrow H$ as a function of $M_H$. The two curves represent
  the results of the exact calculation (solid) and of the infinite
  top-quark mass limit calculation (dashed), where the NLO cross
  section has been obtained as the product of the $K$-factor
  ($K\!=\!\sigma_{NLO}/\sigma_{LO}$) calculated in the
  $m_t\!\rightarrow\!\infty$ limit times the LO cross section. From
  Ref.~\cite{Dawson:1998yi}.  \label{fig:gg_to_H_Kfactor}}
\end{figure}

The NLO order of QCD corrections has actually been calculated both
with and without taking the infinite top-quark mass limit. The
comparison between the exact and approximate calculation shows an
impressive agreement at the level of the total cross section, and, in
particular, at the level of the $K$-factor, i.e. the ratio between NLO
and LO total cross sections ($K\!=\!\sigma_{NLO}/\sigma_{LO}$), as
illustrated in Fig.~\ref{fig:gg_to_H_Kfactor}.
It is indeed expected that methods like the infinite top quark mass
limit may not reproduce the correct kinematic distributions of a given
process at higher order in QCD, but are very reliable at the level of
the total cross section, in particular when the cross section receives
large momentum independent contribution at the first order of QCD
corrections. As for the $H\rightarrow gg$ decay process, the NLO
corrections to $gg\rightarrow H$ are very large, changing the LO cross
section by more than 50\%. Since the $gg\rightarrow H$ is the leading
Higgs boson production mode at hadron colliders, it has been clear for
quite a while that a NNLO calculation was needed in order to
understand the behavior of the perturbatively calculated cross
section, and if possible, in order to stabilize its theoretical
prediction.

\begin{figure}
\begin{minipage}{0.3\linewidth}{
\includegraphics[scale=0.3]{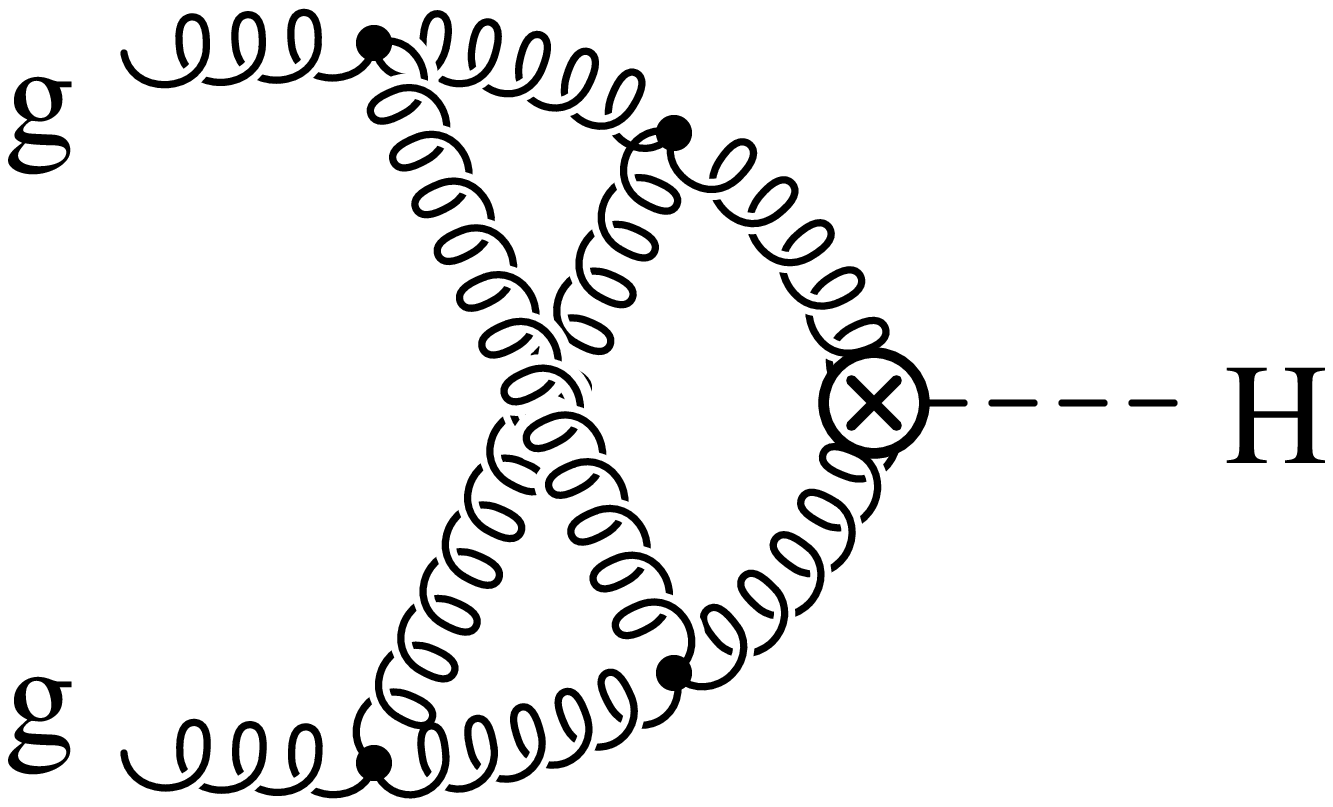}}
\end{minipage}
\begin{minipage}{0.3\linewidth}{
\includegraphics[scale=0.3]{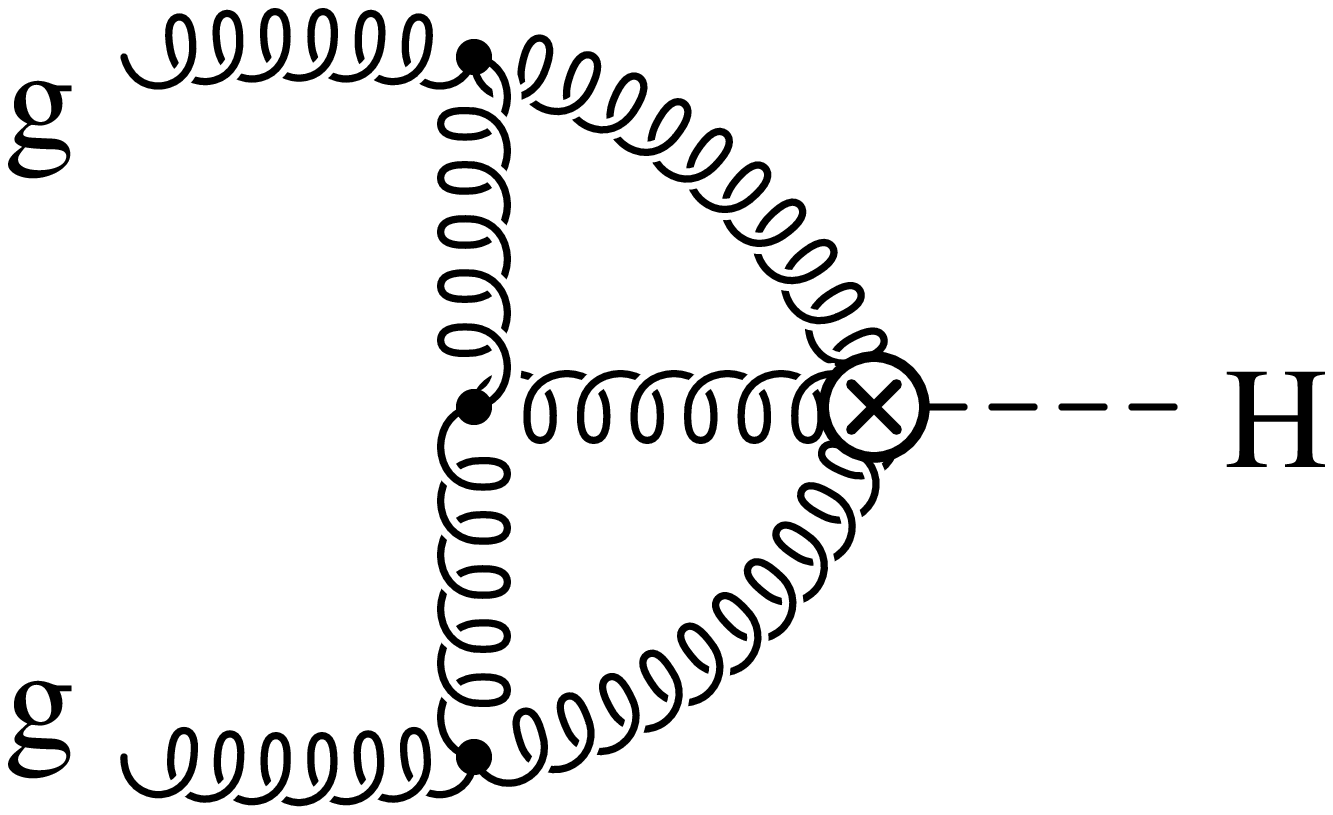}}
\end{minipage}
\caption[]{Two-loop diagrams that enter the NNLO QCD corrections to
  $gg\rightarrow H$. \label{fig:gg_to_H_nnlo}}
\end{figure}

Recently the NNLO corrections to the total cross section have been
calculated using the infinite top-quark mass limit (see
Table~\ref{tab:nlo_nnlo}). The calculation of the NNLO QCD corrections
involves then 2-loop diagrams like the ones shown in 
Fig.~\ref{fig:gg_to_H_nnlo}, instead of the original
3-loop diagrams (a quite formidable task!).
\begin{figure}
\centering
\includegraphics[scale=0.65]{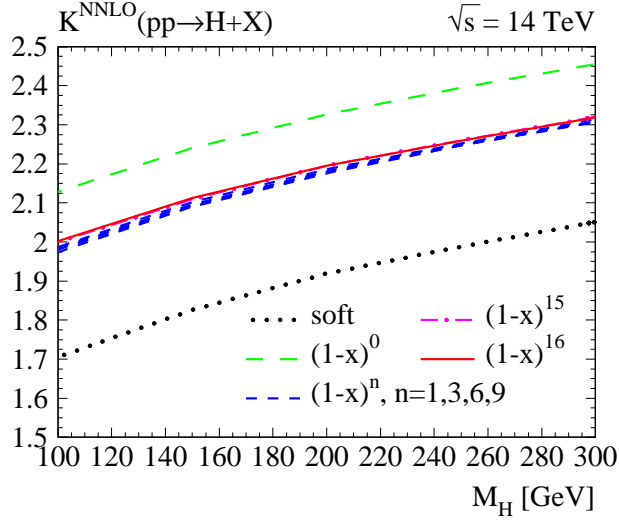}
\caption[]{$K$-factor for $gg\rightarrow H$ at the LHC
  ($\sqrt{s}\!=\!14$~TeV), calculated adding progressively more terms
  in the expansion of Eq.~(\ref{eq:gg_to_H_sigma_hat_ij_nth_term}).
  From Harlander and Kilgore as given in Table~\ref{tab:nlo_nnlo}.
  \label{fig:gg_to_H_soft_expansion_convergence}}
\end{figure}
Moreover, thanks to the $2\rightarrow 1$ kinematic of the
$gg\rightarrow H$ process, the cross section has in one case be calculate in
the so called \emph{soft limit}, i.e. as an expansion in the parameter
$x\!=\!M_H^2/\hat{s}$ about $x\!=\!1$, where $\hat{s}$ is the partonic
center of mass energy (see paper by Harlander and Kilgore in 
Table~\ref{tab:nlo_nnlo}). The $n$-th term in the expansion of 
$\hat\sigma_{ij}$ of Eq.~(\ref{eq:hadronic_cross_section}):
\begin{equation}
\label{eq:gg_to_H_sigma_hat_ij_exp}
\hat{\sigma}_{ij}=\sum_{n\ge 0}\left(\frac{\alpha_s}{\pi}\right)^n
\hat{\sigma}^{(n)}_{ij}\,\,\,,
\end{equation}
can then be written in the soft limit ($x\rightarrow 1$) as follows:
\begin{equation}
\label{eq:gg_to_H_sigma_hat_ij_nth_term}
\hat{\sigma}^{(n)}_{ij}= \underbrace{
  a^{(n)}\delta(1-x)+\sum_{k=0}^{2n-1}b_k^{(n)}
  \left[\frac{\ln^k(1-x)}{1-x}\right]_+}_{\mbox{purely soft terms}} +
\underbrace{
\sum_{l=0}^{\infty}\sum_{k=0}^{2n-1}c_{lk}^{(n)}(1-x)^l\ln^k(1-x)}_
{\mbox{collinear+hard terms}}
\end{equation}
where we have made explicit the origin of different terms in the
expansion. The NNLO cross section is then obtained by calculating the
coefficients $a^{(2)}$, $b^{(2)}_k$ , and $c^{(2)}_{lk}$,
 for $l\ge 0$ and $k\!=\!0,\ldots,3$.
\begin{figure}
\centering
\includegraphics[scale=0.6]{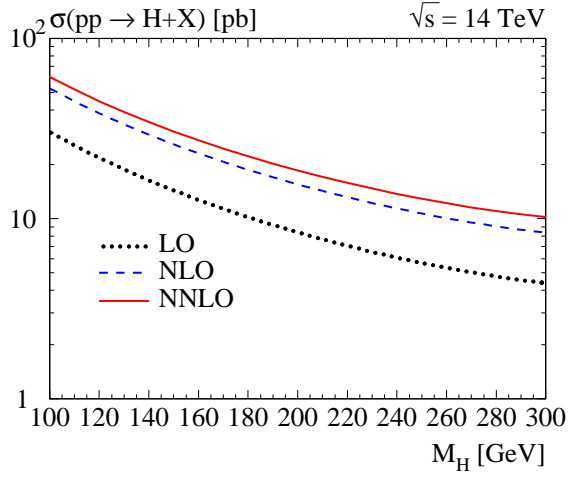}
\caption[]{Cross section for $gg\rightarrow H$ at the LHC 
($\sqrt{s}\!=\!14$~TeV), calculated at LO, NLO and NNLO of
  QCD corrections, as a function of $M_H$, for
  $\mu_F\!=\!\mu_R\!=\!M_H/4$. From Harlander and Kilgore in
  Table~\ref{tab:nlo_nnlo}. \label{fig:gg_to_H_nnlo_nlo_lo}}
\end{figure}
In Fig.~\ref{fig:gg_to_H_soft_expansion_convergence} we see
the convergence behavior of the expansion in
Eq.~(\ref{eq:gg_to_H_sigma_hat_ij_nth_term}). Just adding the first few
terms provides a remarkably stable $K$-factor. The results shown in
Fig.~\ref{fig:gg_to_H_soft_expansion_convergence} have been indeed
confirmed by a full calculation~\cite{Anastasiou:2002yz}, 
where no soft approximation has been used. 

\begin{figure}
\begin{tabular}{cc}
\begin{minipage}{0.44\linewidth}{
\includegraphics[bb=100 265 470 565,scale=0.6]{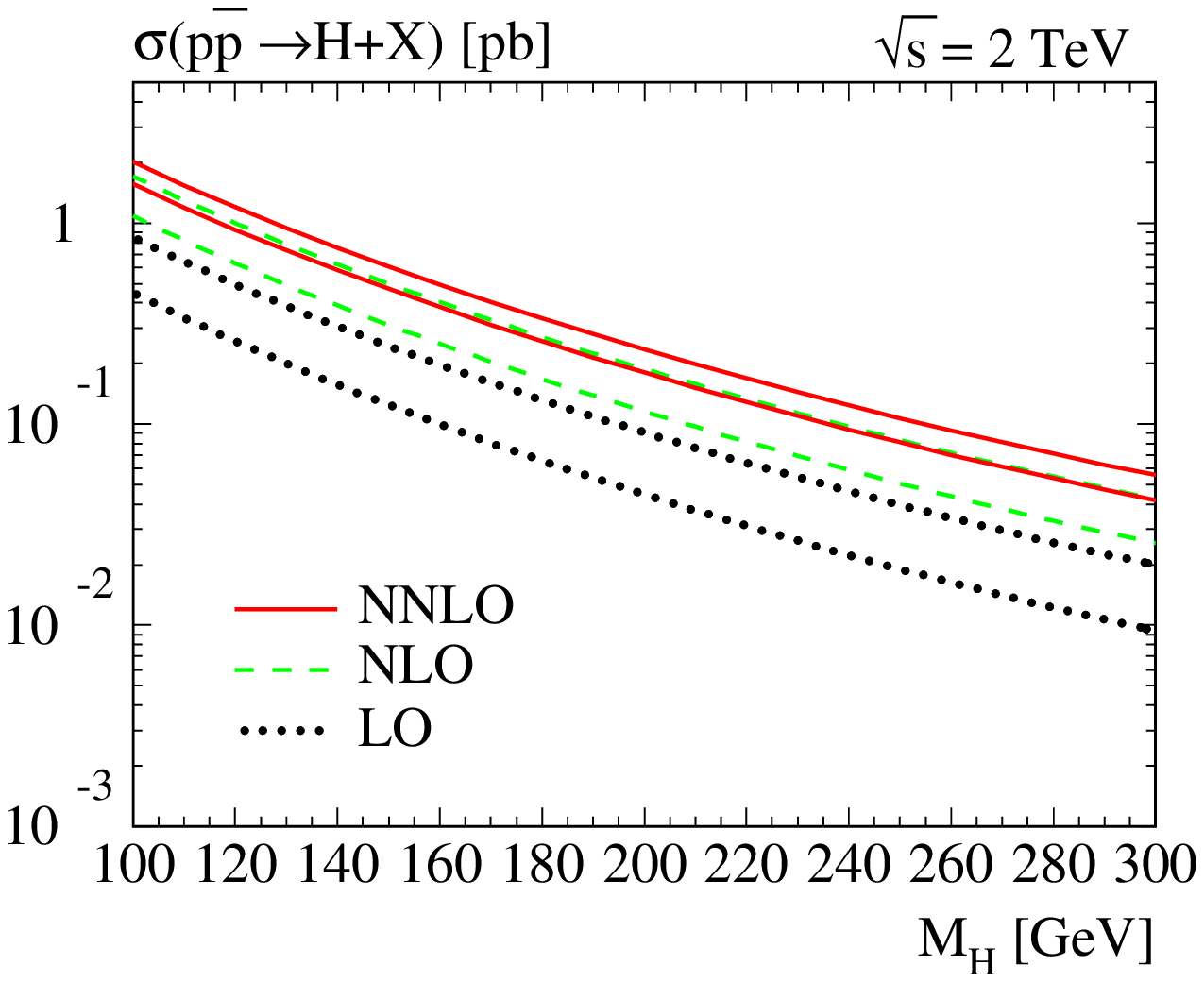}}
\end{minipage}&
\hspace{1.truecm}
\begin{minipage}{0.44\linewidth}{
\includegraphics[bb=100 265 470 565,scale=0.6]{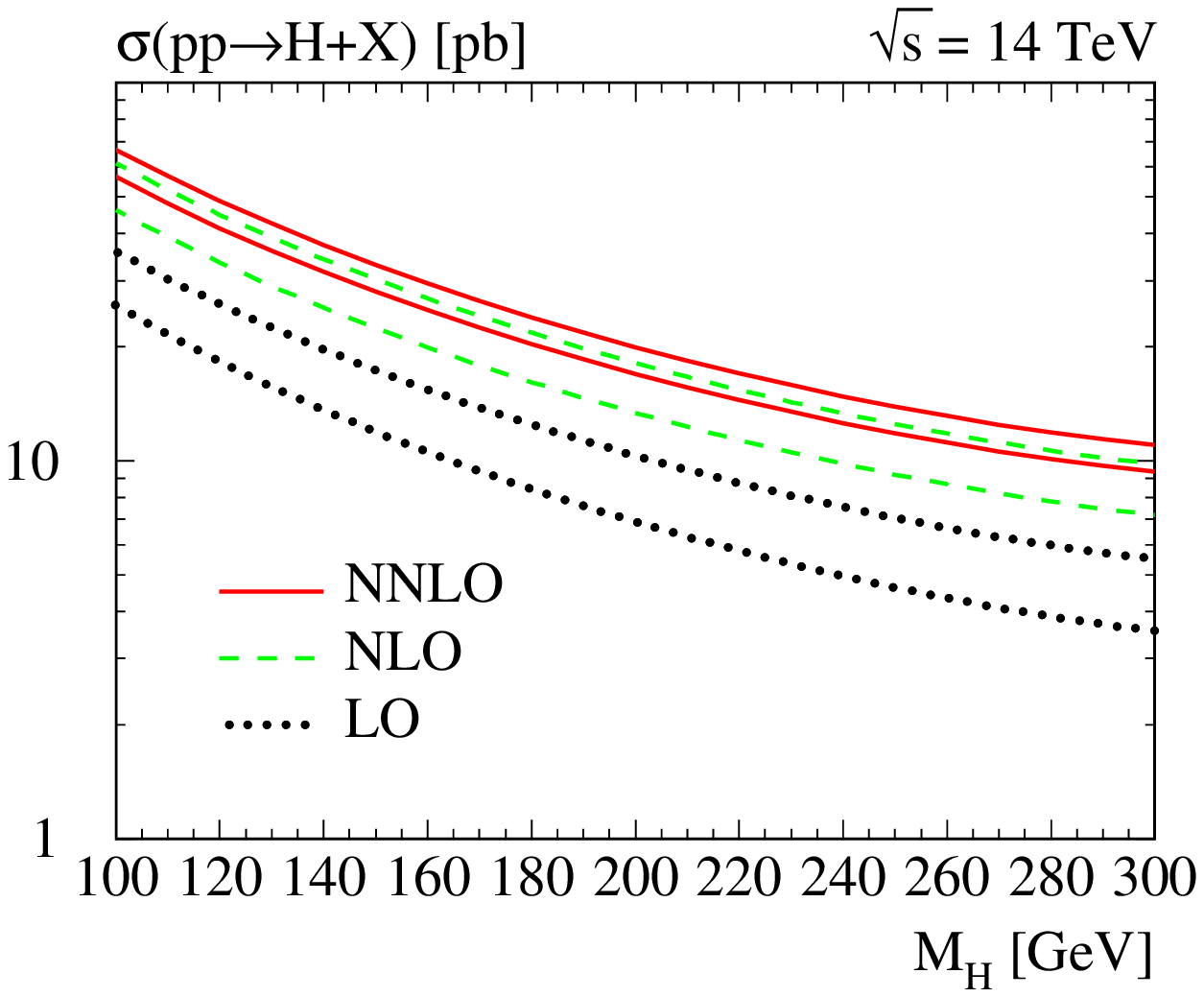}}
\end{minipage}
\end{tabular}
\caption[]{Residual renormalization/factorization scale dependence of
  the LO, NLO, and NNLO cross section for $gg\rightarrow H$, at the
  Tevatron ($\sqrt{s}\!=\!2$~TeV) and at the LHC
  ($\sqrt{s}\!=\!14$~TeV), as a function of $M_H$. The bands are
  obtained by varying $\mu_R\!=\!\mu_F$ by a factor of 2 about the
  central value $\mu_F\!=\!\mu_R\!=\!M_H/4$. 
 From Ref.~\cite{Harlander:2002wh}. \label{fig:gg_to_H_nnlo_nlo_lo_mu_dep}}
\end{figure}

The results of the NNLO calculation~\cite{Harlander:2002wh,Anastasiou:2002yz}
are illustrated in Figs.~\ref{fig:gg_to_H_nnlo_nlo_lo} and
\ref{fig:gg_to_H_nnlo_nlo_lo_mu_dep}.
\begin{figure}
\begin{tabular}{cc}
\begin{minipage}{0.45\linewidth}{\hspace{-1.truecm}
\includegraphics[scale=0.65]{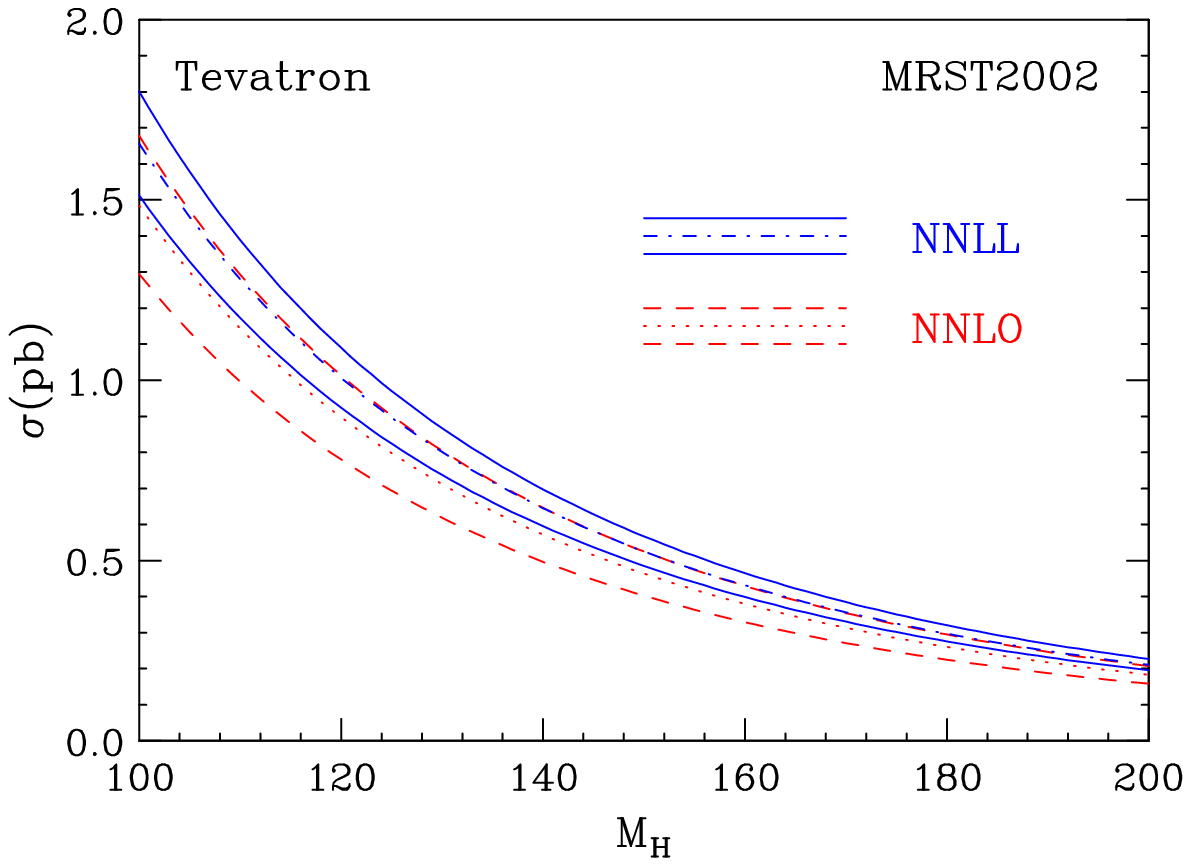}}
\end{minipage}&
\begin{minipage}{0.45\linewidth}{
\includegraphics[scale=0.65]{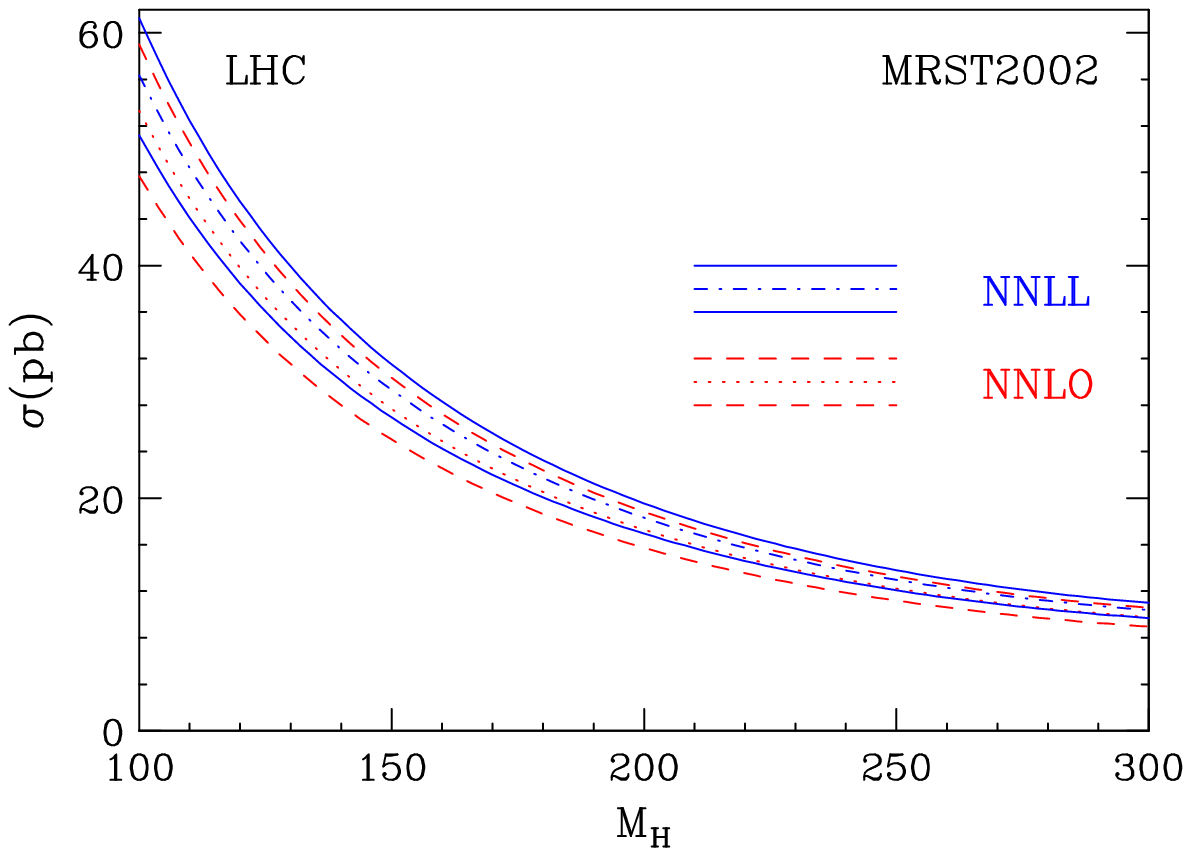}}
\end{minipage}
\end{tabular}
\caption[]{NNLL and NNLO cross sections for Higgs boson production via
gluon-gluon fusion at both the Tevatron and the LHC. From
Ref.~\cite{Catani:2003zt}\label{fig:gg_to_H_nnll_total_xsc}}
\end{figure}
In Fig.~\ref{fig:gg_to_H_nnlo_nlo_lo} we can observe the convergence
of the perturbative calculation of $\sigma(gg\rightarrow H$), since
the difference between NLO and NNLO is much smaller than the original
difference between LO and NLO. This is further confirmed in
Fig.~\ref{fig:gg_to_H_nnlo_nlo_lo_mu_dep}, where we see that the
uncertainty band of the NNLO cross section overlaps with the
corresponding NLO band. Therefore the NNLO term in the perturbative
expansion only modify the NLO cross section within its NLO theoretical
uncertainty. This is precisely what one would expect from a good
convergence behavior. Moreover, the narrower NNLO bands in
Fig.~\ref{fig:gg_to_H_nnlo_nlo_lo_mu_dep} shows that the NNLO result
is pretty stable with respect to the variation of both renormalization
and factorization scales. This has actually been checked thoroughly in
the original papers, by varying both $\mu_R$ and $\mu_F$ independently
over a range broader than the one used in
Fig.~\ref{fig:gg_to_H_nnlo_nlo_lo_mu_dep}.

\begin{figure}
\centering
\begin{tabular}{c}
\begin{minipage}{0.8\linewidth}{
\includegraphics[scale=0.6]{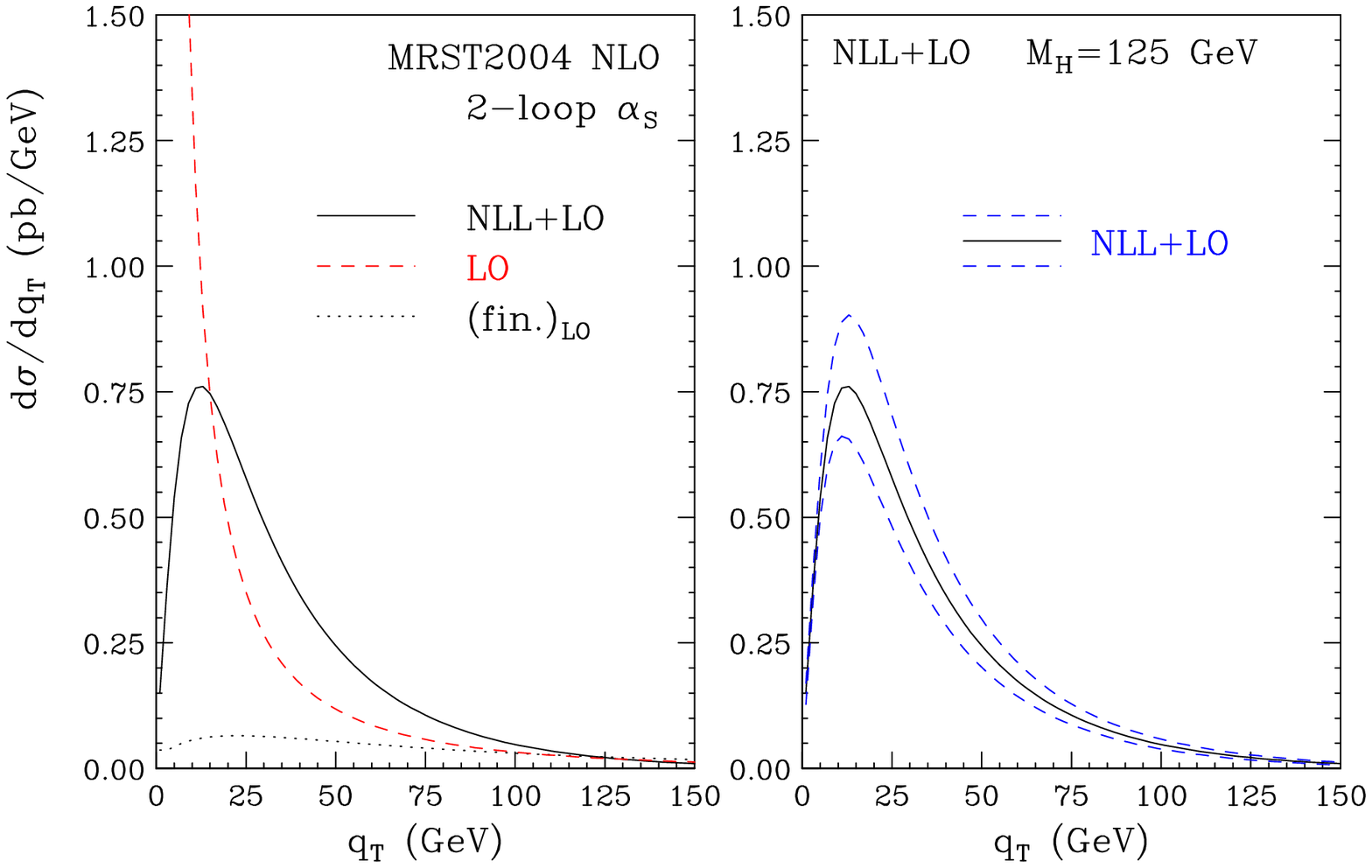}}
\end{minipage} \\
\begin{minipage}{0.8\linewidth}{
\includegraphics[scale=0.6]{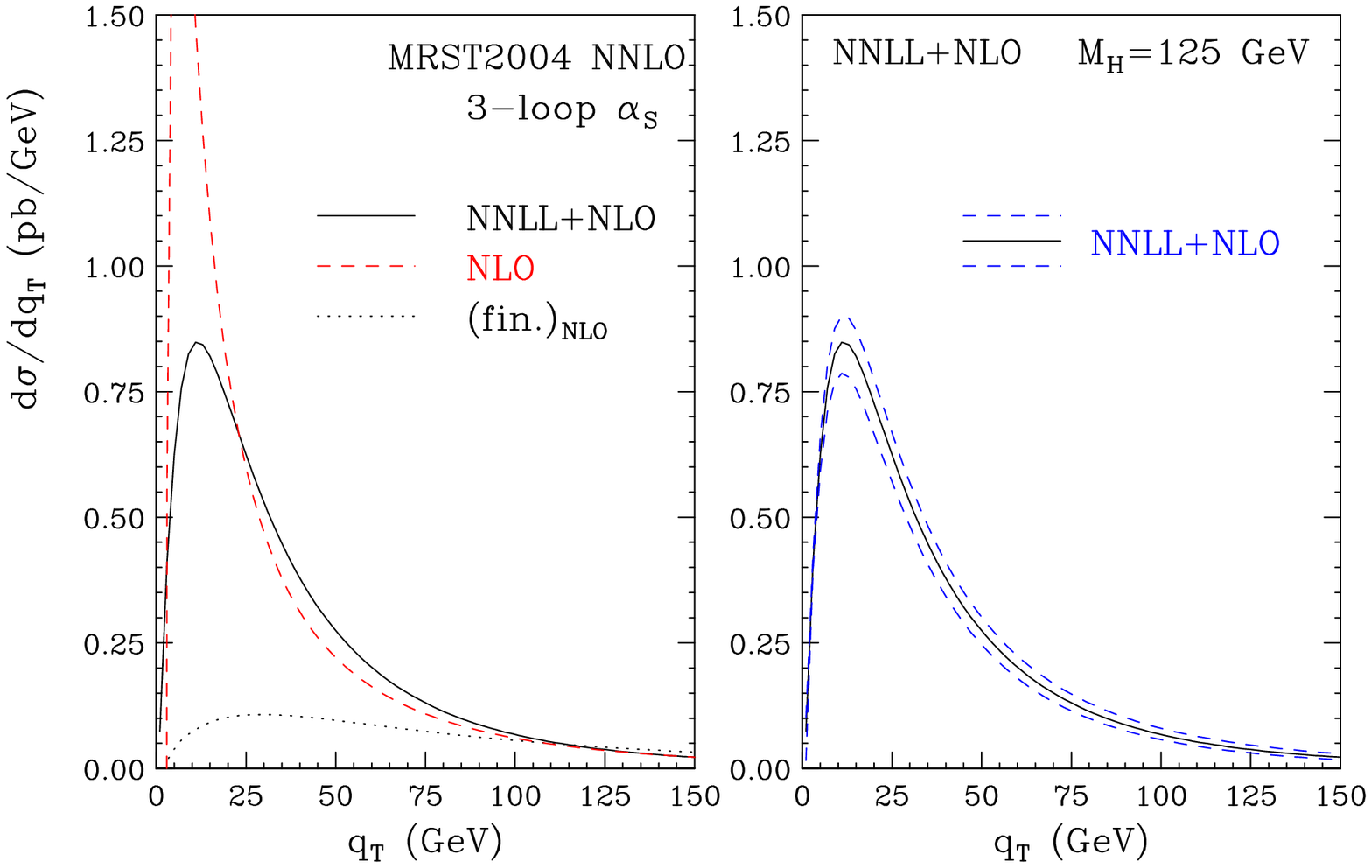}}
\end{minipage}
\end{tabular}
\caption[]{The $q_T$ spectrum  at the LHC with $M_H\!=\!125$~GeV.
The upper plots show: (left) setting $\mu_R\!=\!\mu_F\!=\!Q\!=\!M_H$, the results
at NNL+LO accuracy compared with the LO spectrum and the finite
component of the LO spectrum; (right) the uncertainty band from the
variation of the scales $\mu_R$ and $\mu_F$ at NLL+LO accuracy.
The lower plots show the same at NNLL+NLO accuracy.
From
Ref.~\cite{Bozzi:2005wk}.\label{fig:gg_to_H_nnll_partial_xsc}}
\end{figure}
The NNLO cross section for $gg\rightarrow H$ has been further improved
by Catani et al.~\cite{Catani:2003zt} by resumming up to the
next-to-next-to leading order of soft logarithms. Using the techniques
explained in their papers, they have been able to obtain the
theoretical results shown in Figs.~\ref{fig:gg_to_H_nnll_total_xsc}
and \ref{fig:gg_to_H_nnll_partial_xsc} for the total and differential
cross sections respectively. In particular, we see from
Fig.~\ref{fig:gg_to_H_nnll_total_xsc} that the NNLO and NNLL results
nicely overlap within their uncertainty bands, obtained from the
residual renormalization and factorization scale dependence. The
residual theoretical uncertainty of the NNLO+NNLL results has been
estimated to be 10\% from perturbative origin plus 10\% from the use
of NLO PDF's instead of NNLO PDF's. Moreover, in
Fig.~\ref{fig:gg_to_H_nnll_partial_xsc} we see how the resummation of
NNL crucially modify the shape of the Higgs boson transverse momentum
distribution at low transverse momentum ($q_T$), where the soft
$\ln(M_H^2/q_T^2)$ are large and change the behavior of the
perturbative expansion in $\alpha_s$~\cite{Bozzi:2005wk}.

\subsection{$pp,p\bar{p}\rightarrow t\bar{t}H$  and
$pp,p\bar{p}\rightarrow b\bar{b}H$ at NLO}
\label{subsec:tth_bbh_nlo}
\begin{figure}
\begin{tabular}{cc}
\begin{minipage}{0.5\linewidth}
\includegraphics[scale=0.45]{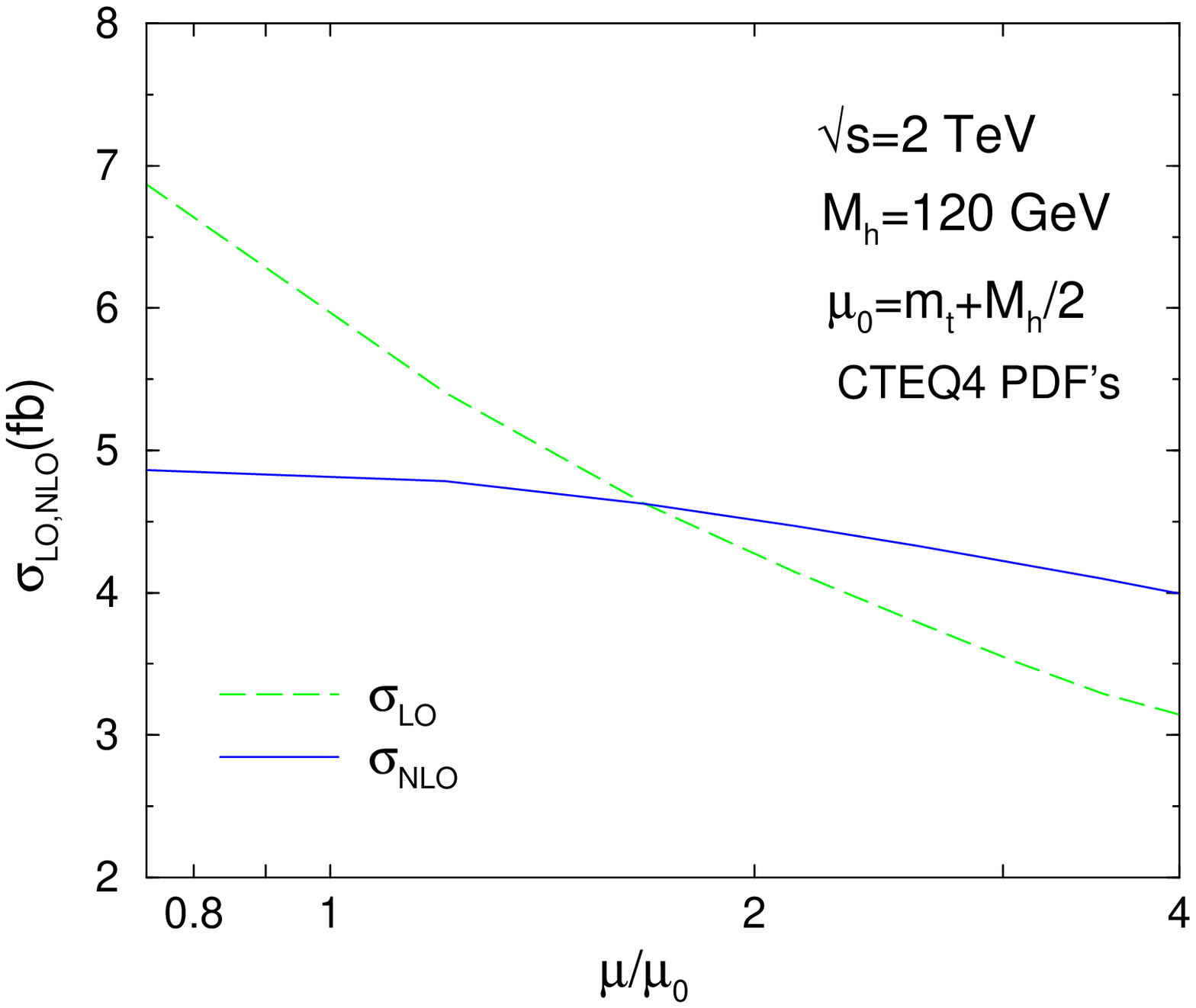}
\end{minipage}&
\begin{minipage}{0.5\linewidth}
\includegraphics[scale=0.45]{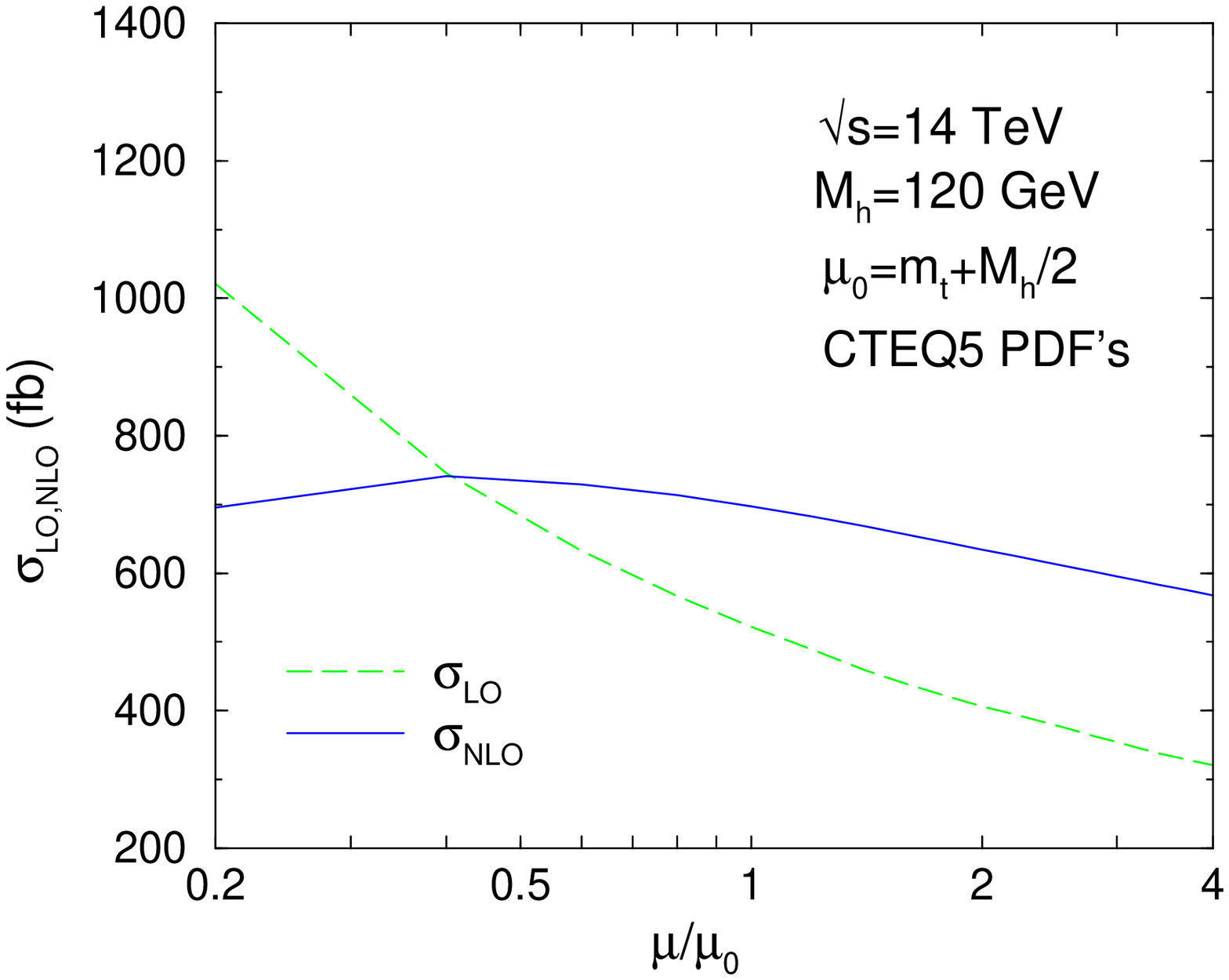}
\end{minipage}
\end{tabular}
\caption[]{Dependence of $\sigma_{LO,NLO}(p\bar{p},pp\rightarrow
t\bar{t}H)$ on the renormalization/factorization scale $\mu$, at both
the Tevatron and the LHC, for $M_H\!=\!120$~GeV. The reference scale
$\mu_0$ is taken to be $\mu_0\!=\!m_t+M_H/2$. From
Refs.~\cite{Reina:2001bc} and
\cite{Dawson:2003zu}.\label{fig:tth_nlo_mudep}}
\end{figure}

\begin{figure}
\begin{center}
\begin{tabular}{cc}
\begin{minipage}{0.5\linewidth}
\includegraphics[scale=0.45]{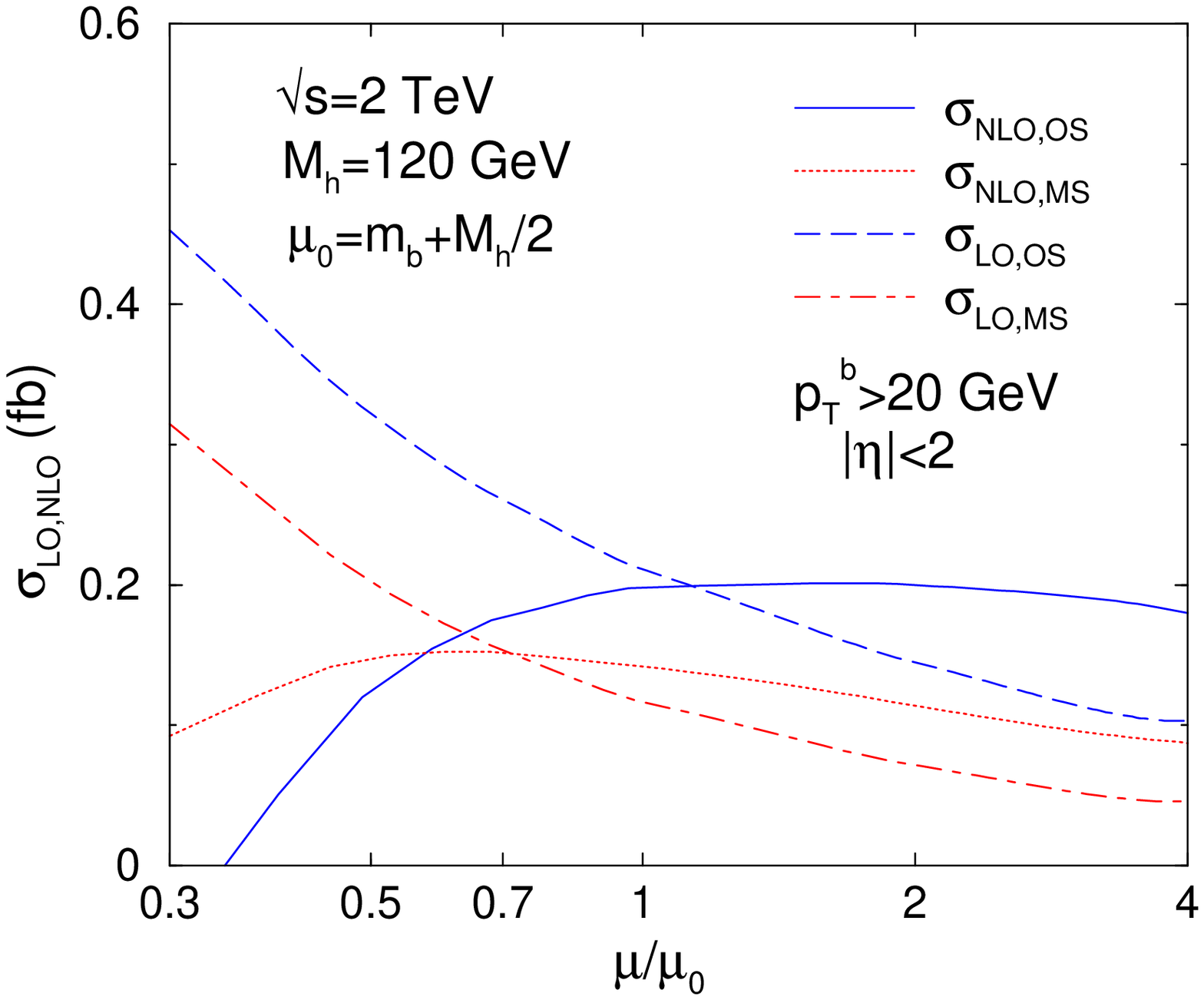}
\end{minipage}&
\begin{minipage}{0.5\linewidth}
\includegraphics[scale=0.45]{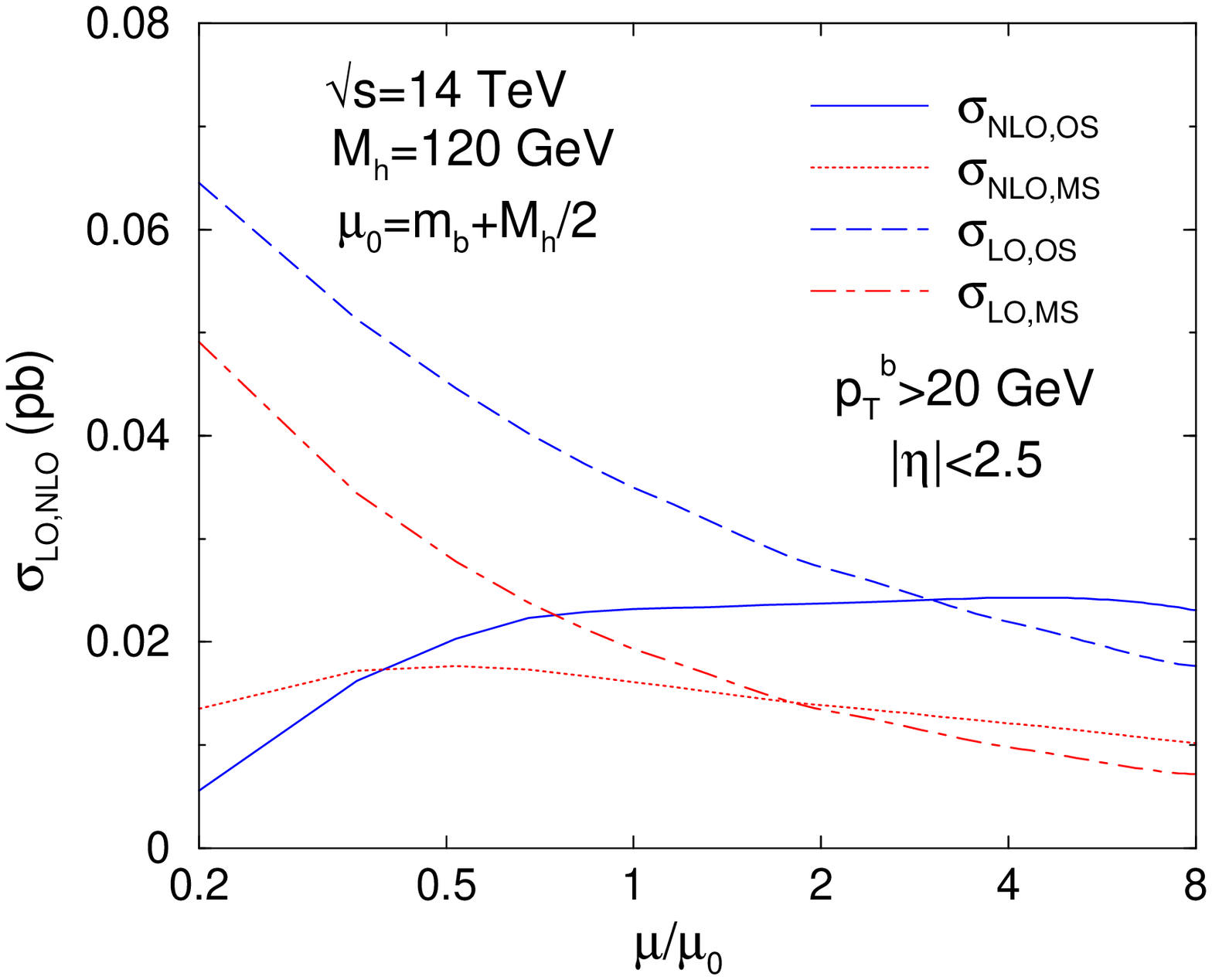}
\end{minipage}
\end{tabular}
\end{center}
\caption[]{Scale dependence of the LO and NLO cross sections for
$p\bar{p}\rightarrow b\bar{b}H$ (Tevatron) and $pp\rightarrow
b\bar{b}H$ (LHC), for $M_H\!=\!120$~GeV. The curves labeled
$\sigma_{LO,OS}$ and $\sigma_{NLO,OS}$ use the $OS$ renormalization
scheme, while the curves labeled $\sigma_{LO,MS}$ and
$\sigma_{NLO,MS}$ use the $\overline{MS}$ renormalization
scheme. From Ref.~\cite{Dawson:2003kb}.\label{fig:bbh_2b_mudep}}
\end{figure}

The associated production of a Higgs boson with heavy quark pairs,
$pp,p\bar{p}\rightarrow t\bar{t}H$ and $b\bar{b}H$, has been for a
while the only Higgs production process for which the NLO of QCD
corrections had not been calculated. Given the relevance of both
production modes to Higgs physics (see discussion in
Section~\ref{subsec:higgs_tevatron_lhc}) and the large renormalization
and factorization scale dependence of the LO cross sections (see,
e.g., the LO curves in Figs.~\ref{fig:tth_nlo_mudep} and
\ref{fig:bbh_2b_mudep}), a full NLO calculation was mandatory. This
has been completed in the papers by Beenakker et
al.~\cite{Beenakker:2001rj,Beenakker:2002nc} and Dawson et
al.~\cite{Reina:2001sf,Reina:2001bc,Dawson:2002tg,Dawson:2003zu}
listed in Table~\ref{tab:nlo_nnlo} and we will briefly report about
their most important results in this Section.

\begin{figure}
\begin{tabular}{l}
\includegraphics[scale=0.6]{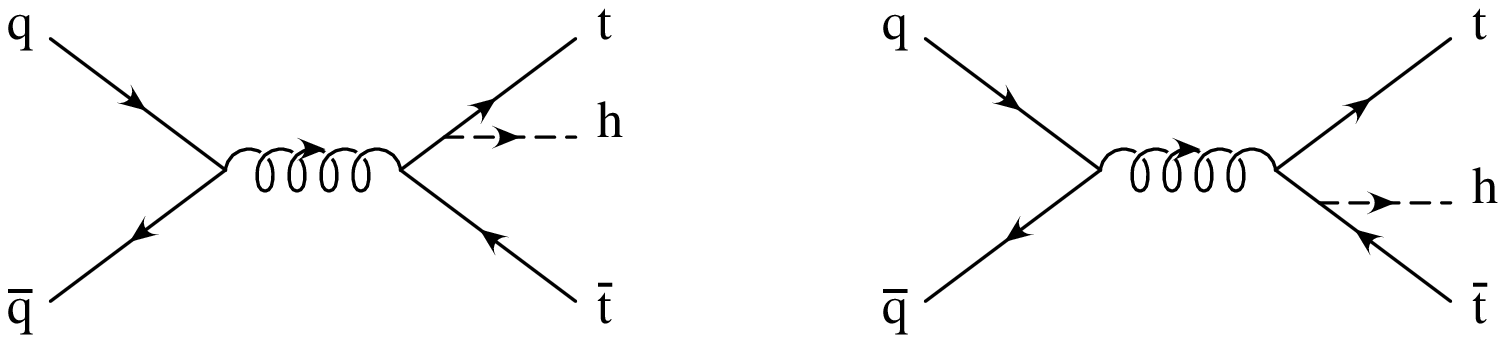}
\vspace{1.truecm}\\
\includegraphics[scale=0.8]{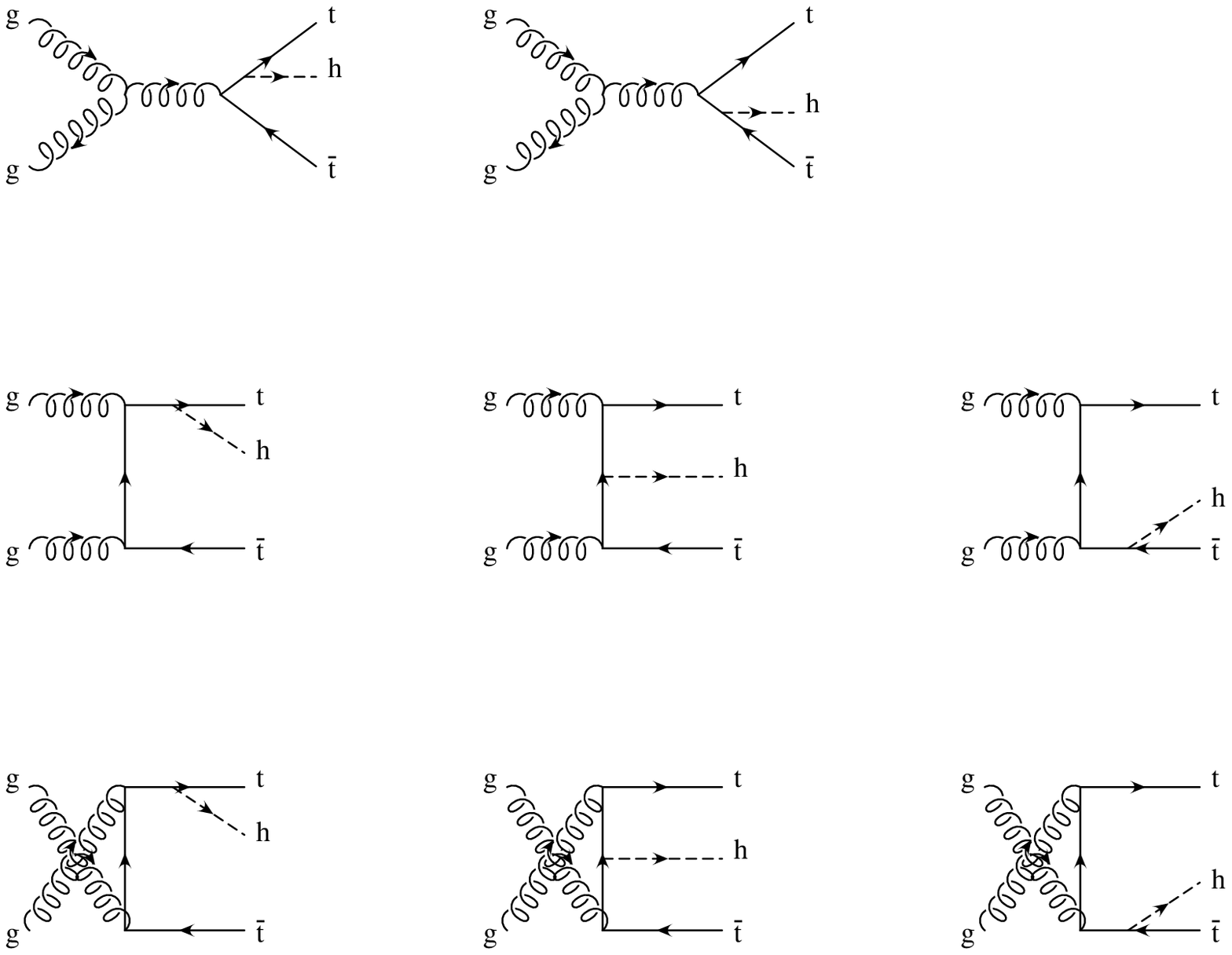}
\end{tabular}
\caption[]{Tree level Feynman diagrams for $q\bar{q}\rightarrow
t\bar{t}H$ and $gg\rightarrow t\bar{t}H$. Analogous diagrams, with
$t\rightarrow b$, contribute to the $q\bar{q}\rightarrow b\bar{b}H$
and $gg\rightarrow b\bar{b}H$ processes.\label{fig:qqgg_QQh}}
\end{figure}
\begin{figure}
\includegraphics[scale=0.75]{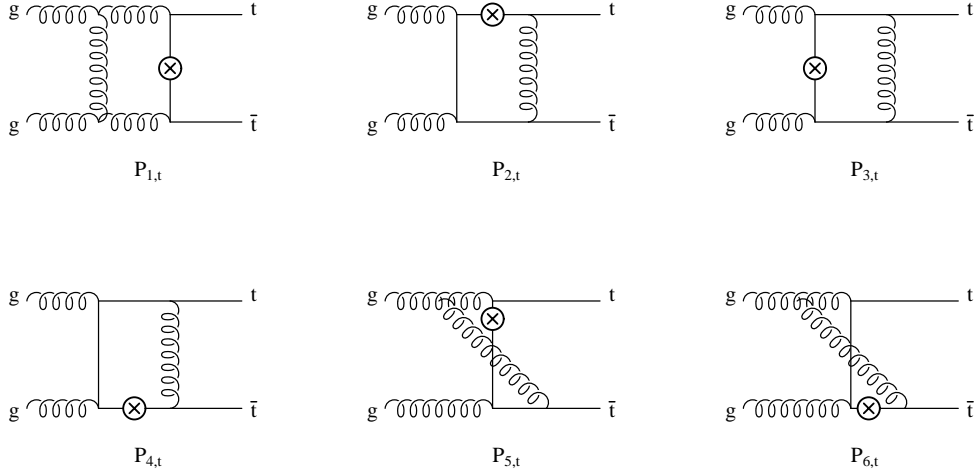}
\caption[]{Example of $\mathcal{O}(\alpha_s)$ virtual corrections to
$gg\rightarrow t\bar{t}H$: pentagon diagrams. The circled crosses
denote all possible insertion of the final Higgs boson leg. All
$t$-channel diagrams have corresponding $u$ channel diagrams, where
the two initial state gluon legs are crossed. Analogous diagrams with
$t\rightarrow b$ contribute to the $\mathcal{O}(\alpha_s)$ virtual
corrections to $gg\rightarrow b\bar{b}H$.
\label{fig:tth_virtual}}
\end{figure}
\begin{figure}
\includegraphics[scale=0.75]{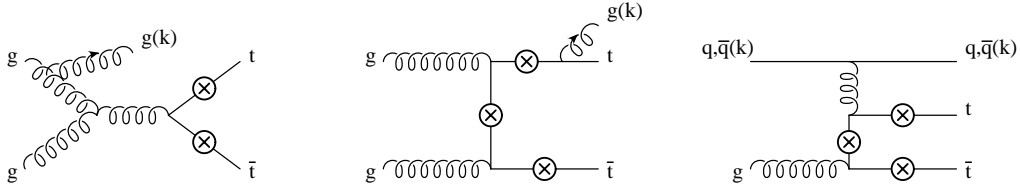}
\caption[]{Examples of $\mathcal{O}(\alpha_s)$ real corrections to
$gg\rightarrow t\bar{t}H$. The circled crosses denote all possible
insertion of the final Higgs boson leg. Analogous diagrams with
$t\rightarrow b$ contribute to the $\mathcal{O}(\alpha_s)$ real
corrections to $gg\rightarrow b\bar{b}H$.\label{fig:tth_real}}
\end{figure}

The NLO calculation of $pp,p\bar{p}\rightarrow Q\bar{Q}H$ (for
$Q\!=\!b,t$) presents several challenges, since it has to deal with a
$2\rightarrow 3$ process that involves all massive particles in the
final state. At tree level, $Q\bar{Q}H$ proceeds through the
$q\bar{q},gg\rightarrow Q\bar{Q}H$ parton level processes illustrated
in Fig.~\ref{fig:qqgg_QQh} for $Q\!=\!t$.  As expected, the
$q\bar{q}\rightarrow Q\bar{Q}H$ dominates for large fraction of the
parton longitudinal momentum $x$, while $gg\rightarrow Q\bar{Q}H$
dominates at small $x$. This translate into the fact that the parton
level cross section for $t\bar{t}H$ production is dominated by
$q\bar{q}\rightarrow t\bar{t}H$ at the Tevatron, and by $gg\rightarrow
t\bar{t}H$ at the LHC, while $b\bar{b}H$ production is always
dominated by $gg\rightarrow b\bar{b}H$.  The $\mathcal{O}(\alpha_s)$
virtual corrections include up to pentagon diagrams, such that the
problem of calculating scalar and tensor integrals with up to five
denominators, several of which massive, has to be faced. Most
integrals have to be calculated analytically since both ultraviolet
(UV) and infrared (IR) singularities have to be extracted.  The
$\mathcal{O}(\alpha_s)$ real corrections involve factoring out IR
(soft and collinear) divergences from a $2\rightarrow 4$ phase space
with several massive particles. Samples of Feynman diagrams
corresponding to the $\mathcal{O}(\alpha_s)$ virtual and real
corrections are illustrated in Figs.~\ref{fig:tth_virtual} and
\ref{fig:tth_real}.
Several new methods and algorithms have been used by the two
collaborations that have calculated the NLO cross section for
$pp,p\bar{p}\rightarrow Q\bar{Q}H$, and we refer to their papers for
all technical details. In the following we will comment separately on
$t\bar{t}H$ and $b\bar{b}H$ NLO results.

For $t\bar{t}H$ production, the most important outcome of the full NLO
calculation is illustrated in Fig.~\ref{fig:tth_nlo_mudep}, where the
renormalization ($\mu_R$) and factorization scale ($\mu_F$) dependence
of the LO and NLO total inclusive cross section is presented, for a SM
Higgs boson mass of $M_H\!=\!120$~GeV, at both the Tevatron and the
LHC. We note that the factorization and renormalization scales have
been set equal in the plots of Fig.~\ref{fig:tth_nlo_mudep},
$\mu_R=\mu_F=\mu$, while in the original work both scales have been
first varied independently to verify that $\mu_R=\mu_F=\mu$ is not a
particular point at which both scale dependences accidentally mutually
cancel.  It is evident from the plots of Fig.~\ref{fig:tth_nlo_mudep}
that the NLO total cross section sensitivity to $\mu_R$ and $\mu_F$ is
drastically reduced with respect to the LO cross section. Indeed, the
residual systematic error coming from scale dependence is at NLO
reduced to about 15\%, as opposed to more than 100\% of the LO cross
section. The NLO predictions for $t\bar{t}H$ production can now be
confidently used to interface with experimental analyses.

Let us now consider $b\bar{b}H$ production\footnote{For an updated
review see Ref.~\cite{Dawson:2005vi}}. Naively, one would expect the
calculation of $b\bar{b}H$ production at NLO to follow that of
$t\bar{t}H$, with the universal replacement of the top-quark mass with
the bottom-quark mass, $m_t \leftrightarrow m_b$.  However, the
theoretical prediction of $b\bar{b}H$ production at hadron colliders
involves a few subtle issues not encountered in the calculation of
$t\bar{t}H$ production. Indeed, both from the experimental and
theoretical standpoint, it is important to distinguish between {\it
inclusive} and {\it exclusive} $b\bar{b}H$ production.  More
specifically, the production of a Higgs boson with a pair of $b$
quarks can be detected via: ({\it i}) a fully {\it exclusive}
measurement, where both $b$ jets are observed, ({\it ii}) a fully {\it
inclusive} measurement, where no $b$ jet is observed, or ({\it iii}) a
{\it semi-inclusive} measurement, where at least one $b$ jet is
observed.

Experimentally, $b$ quarks are identified or {\it tagged} by imposing
selection cuts on their \emph{transverse momentum} and their angular
direction with respect to the beam axis or \emph{pseudorapidity}.
Inclusive modes have larger cross sections, but also larger
background, such that more exclusive modes are often preferred
experimentally.  Moreover, only the exclusive and semi-inclusive modes
are unambiguously proportional to the bottom-quark Yukawa coupling.

Theoretically, different  approaches may be adopted
depending on the fact that a final state $b$ quark is either treated
inclusively (untagged) or exclusively (tagged).  Indeed, when a final
state $b$ quark is not identified through some selection cuts, the
corresponding integration over its phase space, in particular over its
transverse momentum, gives rise to logarithms of the form:
\begin{equation}
\label{eq:lambda_b}
\Lambda_b = \log\biggl(\frac{\mu_H^2}{m_b^2}\biggr)\,\,,
\end{equation}
where $m_b$ and $\mu_H$ represent the lower and upper bounds of the
integration over the transverse momentum of the final state $b$ quark.
$\mu_H$ is typically of $\mathcal{O}(M_H)$ and therefore, due to the
smallness of the bottom-quark mass, these logarithms can be quite
large.  Additionally, the same logarithms appear at every order in the
perturbative expansion of the cross section in $\alpha_s$, due to
recursive gluon emission from internal bottom-quark lines.  If the
logarithms are large, the convergence of the perturbative expansion of
the cross section could be severely hindered and it can be advisable
to reorganize the expansion in powers of $\alpha_s^n \Lambda_b^m$,
further resumming various orders of logarithms via renormalization
group techniques\footnote{The logarithms mentioned here also appear in
the $t\bar{t}H$ calculation but, since $\mu_H$ is typically of the
order of $m_t$, the logarithms are small and the convergence of the
perturbative expansion in $\alpha_s$ is preserved.}.

\begin{figure}[hbt]
\begin{center}
\includegraphics[scale=0.9]{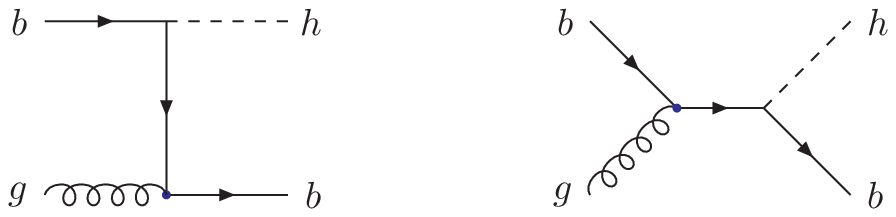}
\vspace*{-0.2cm}
\caption[]{Tree level Feynman diagram for $bg\to bH$ in the 5FNS.\label{fig:bghb_feyn}}
\end{center}
\end{figure}

\begin{figure}[hbt]
\begin{center}
\includegraphics[scale=0.9]{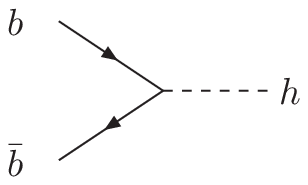}
\vspace*{-0.2cm}
\caption[]{Tree level Feynman diagram for $b\bar{b}\to H$ in the 5FNS.\label{fig:bbh_feyn}}
\end{center}
\end{figure}

Currently, there are two approaches to calculating the inclusive and
semi-inclusive cross sections for Higgs production with bottom quarks.
Working under certain kinematic approximations, and adopting the
so-called {\it five-flavor-number scheme} (5FNS), the collinear
logarithms, $\Lambda_b$, can be factored out and resummed by
introducing a bottom-quark Parton Distribution Function
(PDF)~\cite{Barnett:1987jw,Olness:1987ep,Dicus:1988cx}.  This approach
restructures the calculation to be an expansion in both $\alpha_s$ and
$\Lambda_{b}^{-1}$. At tree level, the semi-inclusive production is
then described by the process $bg\rightarrow bH$ illustrated in
Fig.~\ref{fig:bghb_feyn}, while the fully inclusive production process
becomes $b\bar{b}\rightarrow H$, illustrated in
Fig.~\ref{fig:bbh_feyn}. Alternatively, working with no kinematic
approximations, and adopting the so-called {\it four-flavor-number
scheme} (4FNS), one can compute the cross section for $p\bar{p},pp \to
b\bar{b}H$ at fixed order in QCD with no special treatment of the
collinear logarithms, considering just the parton level processes
$q\bar{q},gg \to b\bar{b}H$ illustrated in Fig.~\ref{fig:qqgg_QQh}
(with $t\rightarrow b$) and their radiative corrections.

The fully exclusive $b\bar{b}H$ production cross section can only be
computed in the 4FNS framework.  As far as the inclusive and
semi-inclusive production cross sections go, the comparison between
the 4FNS and 5FNS needs to consider QCD corrections beyond the LO, in
order to work with stable results. Indeed, the two calculation
schemes represent different perturbative expansion of the same
physical cross section, and therefore should agree at sufficiently high
order.  The discussion to follow is based on the NLO calculation of
$q\bar{q},gg \to b\bar{b}H$~\cite{Dittmaier:2003ej,Dawson:2003kb} and
$bg\rightarrow bH$~\cite{Campbell:2002zm}, and on the NNLO calculation
of $b\bar{b}\rightarrow H$~\cite{Harlander:2003ai}. 

It should be noted that our discussion for the production of a
{\it{scalar}} Higgs boson with bottom quarks applies equally well to
the production of a {\it{pseudoscalar}} Higgs boson.  In fact, if one
neglects the bottom-quark mass in the calculation of the NLO
corrections, the predictions for $b\bar{b}A^0$ is identical to those
for $b\bar{b}h^0(H^0)$ upon rescaling of the Yukawa couplings (see
Section~\ref{subsubsec:mssm_higgs_couplings_fermions}).  On the other
hand, for massive $b$ quarks, the situation becomes more complicated
due to the $\gamma_5$ matrix appearing in the $b\bar{b}A^0$ Yukawa
coupling.  The $\gamma_5$ Dirac matrix is intrinsically a
four-dimensional object and care must be taken in its treatment when
regularizing the calculation in dimensional regularization ($d \neq
4$).  However, bottom-quark mass effects are expected to be small,
$\mathcal{O}(\frac{m_b^2}{M_{h}^2})$, and predictions for
$b\bar{b}h^0$, upon rescaling of the Yukawa coupling, provide good
estimates of $b\bar{b}A^0$ production even in the massive $b$-quark
case. In the following we will present results for the exclusive,
semi-inclusive, and inclusive cross sections separately.

\begin{figure}
\begin{center}
\begin{tabular}{cc}
\begin{minipage}{0.5\linewidth}
\includegraphics[scale=0.45]{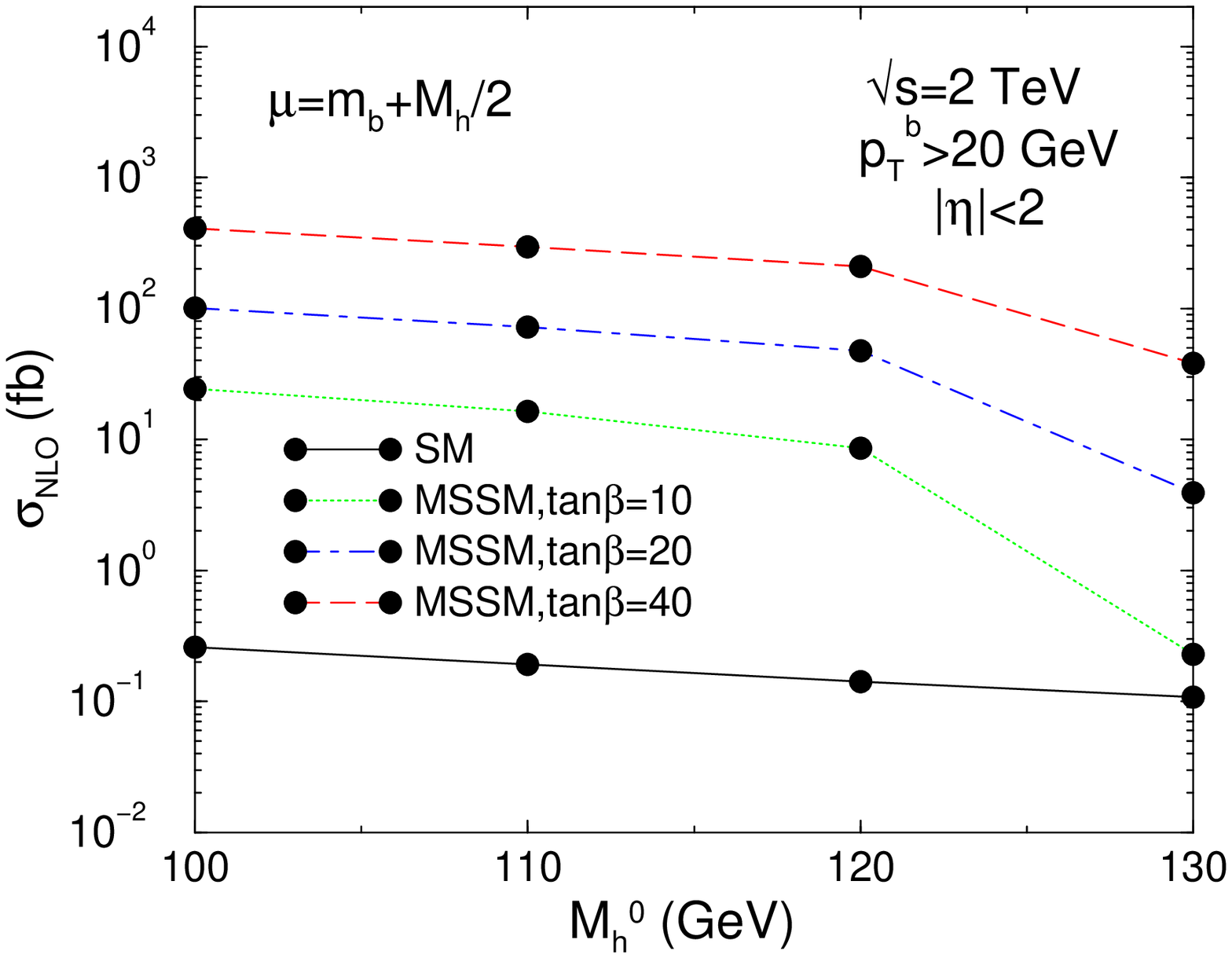}
\end{minipage}&
\begin{minipage}{0.5\linewidth}
\includegraphics[scale=0.45]{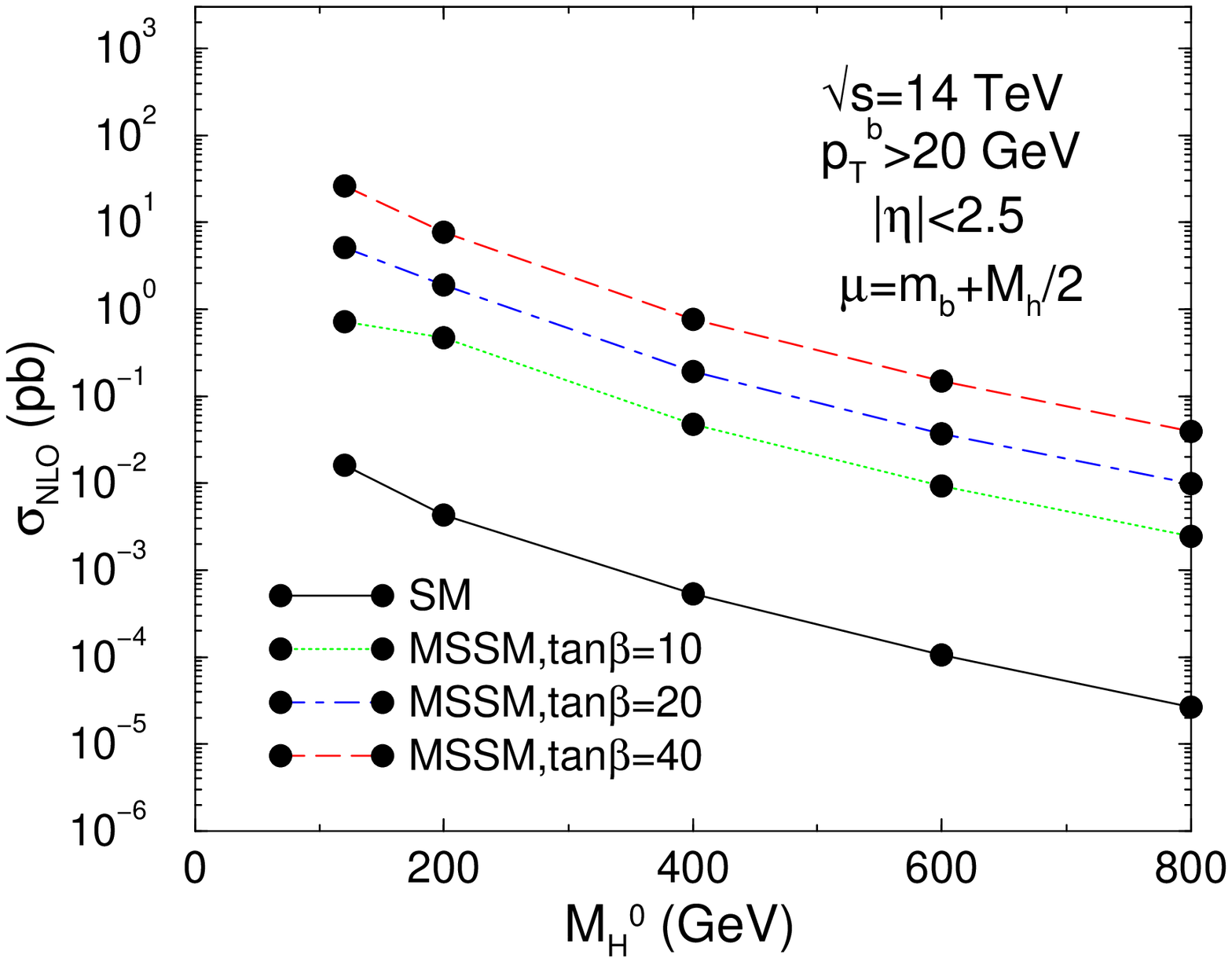}
\end{minipage}
\end{tabular}
\end{center}
\caption[]{NLO cross section for $p\bar{p}\rightarrow b\bar{b}H$
(Tevatron) and $pp\rightarrow b\bar{b}H$ (LHC) as a function of $M_H$,
in the SM and in the MSSM with $\tan\beta\!=\!10,20$ and $40$. For the
Tevatron the production process considered is $p\bar{p}\to
b\bar{b}h^0$ with $M_{h^0}\!=\!100,110,120$, and 130~GeV, while for
the LHC it is $pp\to b\bar{b}H^0$ with $M_{H^0}\!=\!120,200,400,600$,
and 800~GeV. From
Ref.~\cite{Dawson:2003kb}.\label{fig:bbh_2b_mhdep_sm_mssm}}
\end{figure}

The \emph{fully exclusive} $b\bar{b}H$ NLO total cross section is
illustrated in Figs.~\ref{fig:bbh_2b_mudep} and
\ref{fig:bbh_2b_mhdep_sm_mssm}, both for the Tevatron and for the
LHC. Both $b$ quarks in the final state are identified by the cuts
explicitly given in Fig.~\ref{fig:bbh_2b_mudep}, which have been
chosen to closely mimic experimental searches.  The curves in
Fig.~\ref{fig:bbh_2b_mudep} show the dependence of the LO and NLO
exclusive $b\bar{b}H$ cross section from both renormalization and
factorization scales (set equal in these plots,
i.e. $\mu_R\!=\!\mu_F\!=\!\mu$). The two sets of curves represent the
case in which the bottom-quark mass in the bottom-quark Yukawa
coupling is renormalized in the on-shell scheme ($OS$, blue curves) or
in the modified Minimal Subtraction scheme ($\overline{MS}$, red
curves). As we have already observed in
Section~\ref{subsubsec:sm_higgs_to_fermions} when we considered the
decays of a Higgs boson into quark pairs, the bottom-quark
renormalized mass varies substantially when the renormalization scale
$\mu$ is varied from scales of the order of $m_b$ to scales of the
order of $M_H$, and therefore it is important to know how the large
logarithms that determine the running of $m_b$ are treated in the
different renormalization schemes. This is particularly true for the
factor $m_b$ that appears in the bottom-quark Yukawa coupling,
$y_b\!=\!m_b/v$, since the cross section depends quadratically on
$y_b$. It is much less relevant for the cross section kinematic
dependence on $m_b$, coming from the amplitude square or from the
integration over the final state phase space. For this reason, the
$OS$ and $\overline{MS}$ labels in Fig.~\ref{fig:bbh_2b_mudep} refer
to the cases in which only the bottom-quark mass in $y_b$ is
renormalized one way or the other. The kinematic $b$-quark mass is
always taken to be the pole mass.
Looking at the curves in Fig.~\ref{fig:bbh_2b_mudep} we learn that, as
expected, the NLO set of QCD corrections stabilizes the total cross
section, drastically reducing the scale dependence of the LO cross
section. Moreover we see that there is a non negligible dependence on
the bottom-quark mass renormalization scheme. This dependence is
intrinsic of any perturbative calculation, and should decrease the
more orders are added in the perturbative expansion of a given
physical observable.  However, from the behavior of the residual scale
dependence, it is possible to estimate which renormalization scheme
provides a better perturbative expansion. In the case illustrated in
Fig.~\ref{fig:bbh_2b_mudep}, both $OS$ and $\overline{MS}$ NLO cross
sections show a well defined plateau region where the cross section is
very mildly dependent on the scale $\mu$. Nevertheless, the
$\overline{MS}$ cross section overall performs better, since it has a
more regular behavior also at small scales. As a result,
$\overline{MS}$ is often the preferred choice in all processes that
depend on the bottom-quark Yukawa coupling.  Finally,
Fig.~\ref{fig:bbh_2b_mhdep_sm_mssm} shows the dependence of the total
cross section from the Higgs boson mass, in both the SM (solid black
curve) and the MSSM (dashed colored curves, corresponding to different
values of $\tan\beta$), over a significant $M_H$ range for both the
Tevatron and the LHC. The quadratic growth of the MSSM cross sections
with $\tan\beta$ is evident, and this graphically confirms the
possibility of finding evidence of new physics already at the
Tevatron, if an MSSM-like 2HDM with large $\tan\beta$ is realized in
nature (see Sec.~\ref{subsubsec:mssm_higgs_searches}).

\begin{figure}
\centering
\includegraphics[scale=0.9]{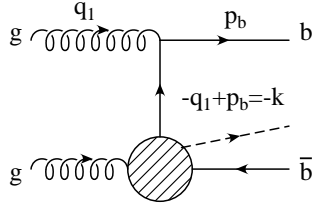}
\caption[]{Tree level Feynman diagrams for $gg\rightarrow b\bar{b}H$
illustrating the almost collinear emission of a bottom quark (upper
leg).\label{fig:bgbh_blob}}
\end{figure}

Let us now turn to the \emph{semi-inclusive} and \emph{inclusive}
$b\bar{b}H$ cross sections. For both production modes, much effort has
been spent recently in understanding the difference between 4FNS and
5FNS in order to assess the reliability of the existing theoretical
results\cite{Campbell:2004pu,Dittmaier:2003ej,Kramer:2004ie,Dawson:2004sh}. First
of all, let us briefly review the idea behind the introduction of a
bottom-quark PDF, which naturally leads to the 5FNS framework. For the
purpose of illustration, consider the prototype case depicted in
Fig.~\ref{fig:bgbh_blob}: one of the final state bottom quarks is
directly originating from the $g\rightarrow b\bar{b}$ splitting of one
of the initial state gluons, while the shaded blob represents all the
possible non collinear configurations of the remaining particles. In
the $m_b\rightarrow 0$ limit, the $g\rightarrow b\bar{b}$
configuration gives origin to collinear singularities, when the two
$b$ quarks are emitted in the same direction of the splitting
gluon. In the case of $b\bar{b}H$ production, the singularities appear
in the $p_T^b\rightarrow 0$ phase space region, where $p_T^b$ is the
transverse momentum of the upper leg bottom quark in
Fig.~\ref{fig:bgbh_blob}. If we take $m_b\neq 0$, these singularities
are regulated by the non zero $b$-quark mass, leaving behind
\emph{collinear} logarithms $\Lambda_b$ of the form given in 
Eq.~(\ref{eq:lambda_b}).
The contribution to the total partonic cross section from this diagram
can be written as:
\begin{eqnarray} 
\label{eq:dsigma_ggbbh}
d\hat{\sigma}_{gg\rightarrow b\bar{b}H} &=& 
\frac{1}{(2)2E_1 2E_2}
\frac{d^3 p_b}{(2\pi)^3}\frac{1}{2E_b}
\frac{d^3 p_{\bar{b}}}{(2\pi)^3}\frac{1}{2E_{\bar{b}}}
\frac{d^3 p_H}{(2\pi)^3}\frac{1}{2E_H}
\overline{\sum}|\mathcal{M}_{gg \rightarrow b\bar{b}H}|^2\nonumber\\
\,\,\,&&\cdot(2\pi)^4\delta^{(4)}(q_1+q_2-p_b-p_{\bar{b}}-p_H)\,\,\,,
\end{eqnarray}
where we have used the momentum notation of Fig.~\ref{fig:bgbh_blob},
such that $E_{1,2}$ are the energies of the initial gluons while
$E_{b,\bar{b},H}$ are the energies of the final state particles. The
amplitude for this process is denoted by $\mathcal{M}_{gg \rightarrow
b\bar{b}H}$. Parameterizing the $b$-quark propagator momentum as:
\begin{equation}
k^{\mu} = zq^{\mu}_{1} + \beta n^{\mu} + k^{\mu}_{\perp}\,\,\,,
\end{equation}
where $z$ is the fraction of the initial gluon momentum carried by the
$b$ quark, and $k^{\mu}_{\perp}$ is transverse to both $q^{\mu}_{1}$
and $n^{\mu}$, for an arbitrary vector $n^\mu$, with
$k^{\mu}_{\perp}k_{\perp\mu} = -k_{T}^2$, one can explicitly show
(after some Dirac algebra manipulations) that:
\begin{equation}
\label{eq:m2_ggbbh_red}
|\mathcal{M}_{gg \rightarrow b\bar{b}H}|^{2} \simeq  g_s^2 C_F
\frac{4}{k_T^2} \left(\frac{1-z}{z}\right) P_{qg}(z)
|\mathcal{M}_{g\bar{b} \rightarrow \bar{b}H}|^{2}\,\,\,,
\end{equation}
where $P_{qg}(z)$ is the Altarelli-Parisi splitting function for
$g\rightarrow q\bar{q}$:
\begin{equation}
\label{eq:p_qg}
P_{qg} = \frac{1}{2}[z^2 + (1-z)^2]\,\,\,,
\end{equation}
$C_F=(N^2-1)/2/N$ (for $N\!=\!3$ colors), and 
$\mathcal{M}_{g\bar{b} \rightarrow \bar{b}h}$ is the amplitude for
$g\bar{b}\rightarrow \bar{b}H$. The previous equation is approximate, i.e. it has
been obtained by neglecting higher powers of $k_T^2$, keeping only the
terms that would give the most singular or leading contribution upon
integration over the \emph{collinear} $b$-quark phase space:
\begin{equation}
\label{eq:ps_b}
\frac{d^{3}p_{b}}{(2\pi)^{3}}\frac{1}{2E_{b}} \simeq
\frac{1}{16\pi^{2}}\frac{dzdk_{T}^{2}}{(1-z)}\,\,\,,
\end{equation}
where we have used that in the small $k_T$ limit the transverse part
of the four momentum of the outgoing quark, $p_b^T$ coincides with
$k_T$.  Inserting Eqs.~(\ref{eq:m2_ggbbh_red}) and (\ref{eq:ps_b})
into Eq.~(\ref{eq:dsigma_ggbbh}) one finds:
\begin{equation}
\label{eq:dsigma_ggbbh_bgbh}
d\hat{\sigma}_{gg \rightarrow b\bar{b}H} \simeq
 \frac{dk_{T}^{2}}{k_{T}^{2}}dz
\frac{\alpha_{s}}{2\pi}P_{qg}(z)
d\hat{\sigma}_{g\bar{b} \rightarrow \bar{b}H}\,\,\,.
\end{equation}
The integration over $k_T^2$, with lower bound $m_b^2$ and upper bound
$\mu_H^2$, gives origin to the \emph{collinear} logarithm $\Lambda_b$
introduced in Eq.~(\ref{eq:lambda_b}). Moreover, when one convolutes
with the gluon PDFs, $g(x_i,\mu)$ (for $i\!=\!1,2$), of the two
initial gluons to obtain the hadronic cross section, a bottom-quark
PDF of the form:
\begin{equation}
\label{eq:b_pdf}
b(x,\mu_F)=\frac{\alpha_s(\mu_F)}{2 \pi}\Lambda_b \int_x^1
\frac{dy}{y} P_{qg}\left(\frac{x}{y}\right)g(y,\mu_F)\,\,\,
\end{equation}
naturally appears. In Eq.~(\ref{eq:b_pdf}) $\mu_F$ represents the
factorization scale. The collinear $\Lambda_b$ logarithms are factored
out and then resummed in the bottom-quark PDF when the factorization
scale is set to $\mu_F\!=\!\mu_H$. Indeed, Eq.~(\ref{eq:b_pdf}) gives
the bottom-quark parton density at the lowest order in $\alpha_s$,
while the leading $\alpha_s^n\Lambda_b^n$ logarithms are resummed via
the DGLAP equation upon evolution:
\begin{equation}
\label{eq:DGLAP}
\frac{d}{d\log\mu}\,b(x,\mu)=\frac{\alpha_s(\mu)}{\pi} \int_x^1
\frac{dy}{y} P_{qg}\left(\frac{x}{y}\right)\,g(y,\mu)\,\,\,.
\end{equation}

The 5FNS approach is therefore based on the approximation that the
outgoing $b$ quarks are at small transverse momentum, since this is
the region of phase space that is emphasized by the $k_T$
expansion. The incoming $b$ partons are given zero momentum at leading
order, and acquire transverse momentum at higher order.  With the use
of a $b$-quark PDF, the 5FNS effectively reorders the perturbative
expansion to be one in $\alpha_s$ and $\Lambda_b^{-1}$.  To see how
this works, let us consider the perturbative expansion of the
inclusive process $b\bar{b}\rightarrow H$ (Fig.~\ref{fig:bbh_feyn}) which,
according to what we just saw, is intrinsically of order
$\alpha_s^2\Lambda_b^2$.  At NLO, the virtual and real corrections to
the tree level process make contributions of $\mathcal{O}(\alpha_s^3
\Lambda_b^2)$.  However, at NLO, we must also consider the
contribution from $bg\rightarrow bH$ where the final state $b$ is at high
transverse momentum.  This process makes a contribution of order $\alpha_s^2
\Lambda_b$ and is, thus, a correction of $\mathcal{O}(\Lambda_b^{-1})$
to the tree level cross section.  Similarly, at NNLO, besides the
myriad of radiative corrections of $\mathcal{O}(\alpha_s^4
\Lambda_b^2)$, we must also include the contribution from the process
$gg\rightarrow b\bar{b}H$, where both $b$ and $\bar{b}$ are at high
$p_T$.  The contribution from these diagrams are of order
$\alpha_s^2$, and are, thus, $\mathcal{O}(\Lambda_b^{-2})$ (or NNLO)
corrections to the tree level process $b\bar{b}\rightarrow
H$~\cite{Dicus:1988cx,Dicus:1998hs}.  The above discussion for
$b\bar{b} \to h$ also applies to the perturbative expansion of
$bg\rightarrow bH$.  In this case, the tree level process is of order
$\alpha_s^2 \Lambda_b$ and the contribution from $gg\rightarrow
b\bar{b}H$ is a NLO correction of $\mathcal{O}(\Lambda_b^{-1})$
~\cite{Campbell:2002zm}.

\begin{figure}[hbtp!]
\begin{center}
\begin{tabular}{lr}
\includegraphics[bb=150 500 430 700,scale=0.80]{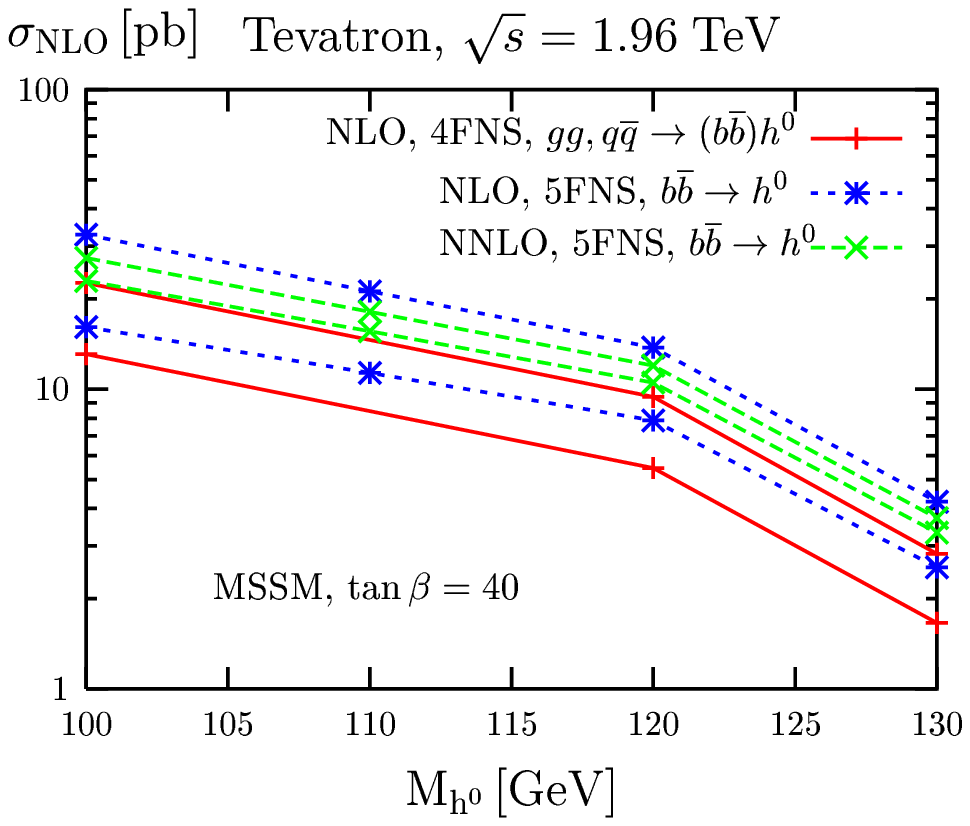} & 
\includegraphics[bb=150 500 430 700,scale=0.80]{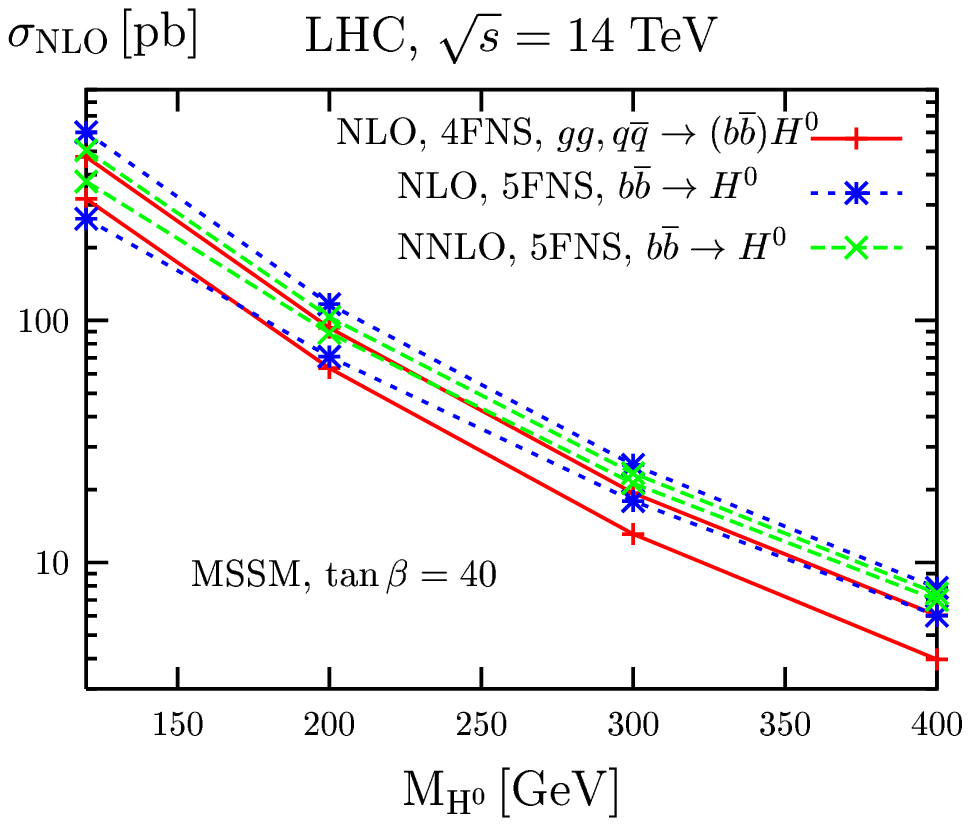}
\end{tabular}
\vspace*{8pt}
\caption[]{Total cross sections for $pp,p\bar p\rightarrow (b\bar{b})h$
  ($h=h^0,H^0$) in the MSSM with no bottom-quark jet identified in the
  final state in the 4FNS (at NLO) and 5FNS (at NLO and NNLO) as a
  function of the light and heavy MSSM Higgs boson masses, at both the
  Tevatron and the LHC.  The error bands have been obtained by varying
  the renormalization ($\mu_r$) and factorization ($\mu_f$) scales
  separately between $\mu_0/4$ and $\mu_0$ (with
  $\mu_0\!=\!m_b+M_H/2$) in the 4FNS, while keeping $\mu_r\!=\!M_h$
  and varying $\mu_f$ between $0.1M_H$ and $0.7M_H$ in the 5FNS (see
  Ref.~\cite{Harlander:2003ai} for details). From
  Ref.~\cite{Dawson:2005vi}.\label{fig:inclusive_4fns_5fns}}
\end{center}
\end{figure}

\begin{figure}[hbtp!]
\begin{center}
\begin{tabular}{lr}
\includegraphics[bb=150 500 430 700,scale=0.80]{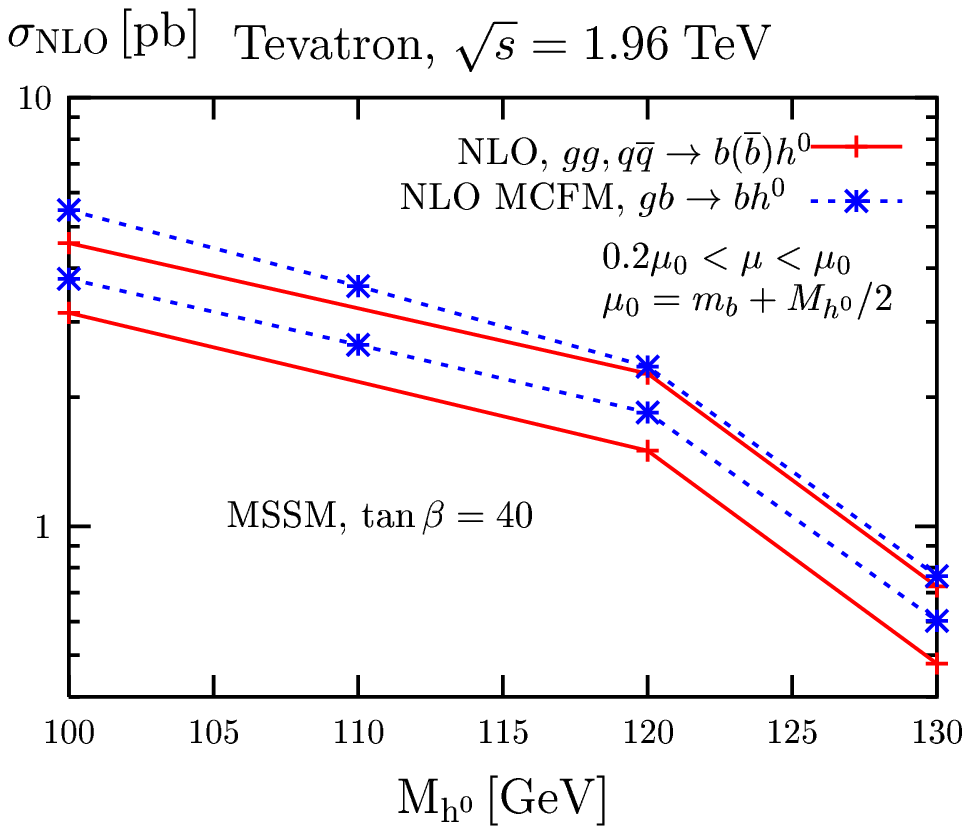} & 
\includegraphics[bb=150 500 430 700,scale=0.80]{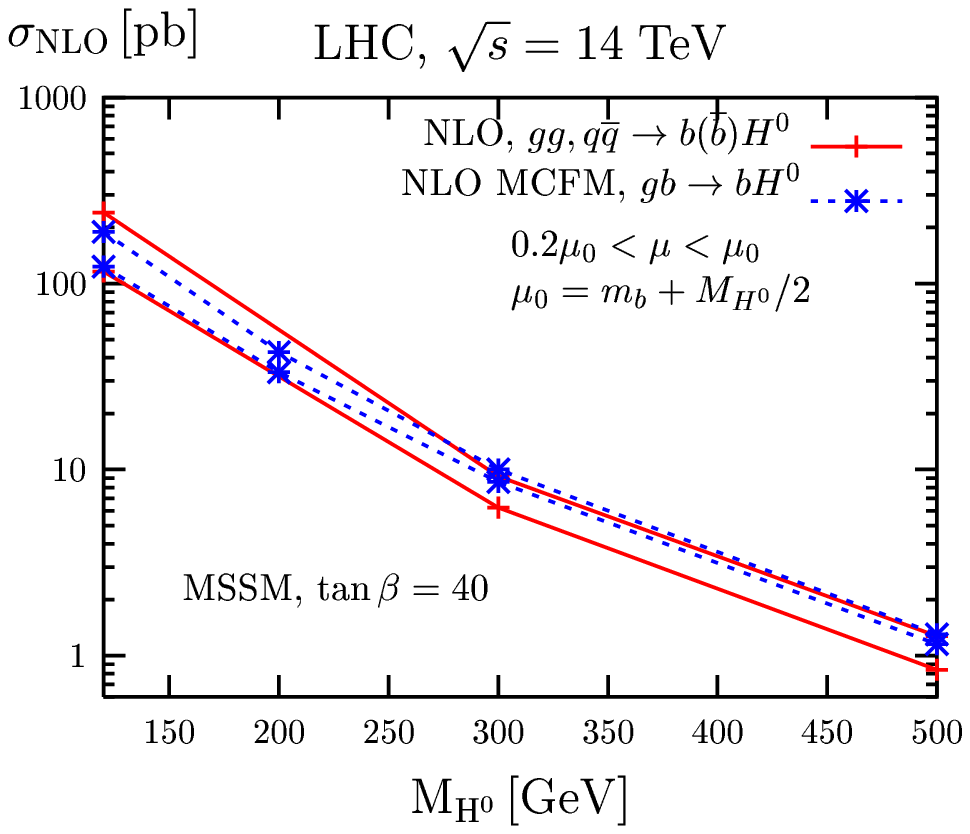}
\end{tabular}
\vspace*{8pt}
\caption[]{Total NLO cross section in the MSSM for $pp,p\bar{p}\rightarrow
  b(\bar{b}) h$ production at the Tevatron and the LHC as a function
  of $M_{h^0,H^0}$. We varied $\mu_r$ and $\mu_f$ independently to
  obtain the uncertainty bands, as explained in the text.  The solid
  curves correspond to the 4FNS, the dashed curves to the 5FNS. The
  error bands have been obtained by varying the renormalization
  ($\mu_r$) and factorization ($\mu_f$) scales separately between
  $0.2\mu_0$ and $\mu_0$ (with $\mu_0\!=\!m_b+M_H/2$). From
  Ref.~\cite{Dawson:2005vi}.\label{fig:semi_inclusive_4fns_5fns}}
\end{center}
\end{figure}

The comparison between 4FNS and 5FNS has been initially performed in
the
SM~\cite{Campbell:2004pu,Dittmaier:2003ej,Kramer:2004ie,Dawson:2004sh}.
This has been crucial to understand several important issues. However,
since the production of a Higgs boson with bottom quarks will only be
physically interesting if the bottom-quark Yukawa coupling is enhanced
beyond its SM value, we will present the 4FNS vs 5FNS comparison in
the MSSM, with $\tan\beta=40$~\cite{Dawson:2005vi}.
Figs.~\ref{fig:inclusive_4fns_5fns} and
\ref{fig:semi_inclusive_4fns_5fns} show the comparison 
for the inclusive and semi-inclusive total cross sections
respectively, at both the Tevatron and the LHC. They both give the
cross section as a function of the Higgs boson mass, chosen to be
$h^0$ at the Tevatron and $H^0$ at the LHC. In the inclusive case,
Fig.~\ref{fig:inclusive_4fns_5fns} illustrates both the NLO and NNLO
predictions. Both at LO and NLO (NNLO) the band are obtained by
varying the renormalization and factorization scales independently as
explained in the figure captions.  In both the inclusive and
semi-inclusive case there is good agreement between the 4FNS and 5FNS
results within their respective scale uncertainties, although the 5FNS
tends to always give slightly higher results.

It is actually very satisfactory that the NLO (and NNLO) calculations
for the semi-inclusive and inclusive $b\bar{b}H$ production agree
within their systematic errors. Indeed, if all perturbative orders
were to be considered, the two approaches would produce the very same
outcome. The truncation of the perturbative series at a given order
gives origin to discrepancies, because the 4FNS and 5FNS perturbative
series are differently ordered. However, we see that considering the
first (second) order of corrections already brings the agreement
between the two schemes within the respective theoretical
uncertainties.  The results shown in
Figs.~\ref{fig:semi_inclusive_4fns_5fns} and
\ref{fig:inclusive_4fns_5fns} represent indeed a major advancement,
since the comparisons existing in the literature before
Refs.~\cite{Campbell:2004pu,Dittmaier:2003ej,Kramer:2004ie,Dawson:2004sh}
showed a pronounced disagreement.  This was mainly due to the absence
of the (now available) NLO results for $q\bar{q},gg\rightarrow
b\bar{b}H$.

\section*{Acknowledgments}
I would like to thank the organizers of TASI 2004 for inviting me to
lecture and for providing such a stimulating atmosphere for both
students and lecturers. I am most thankful to Chris~B.~Jackson and
Fernando Febres-Cordero for carefully reading this manuscript and
providing me with several valuable comments. This work was supported
in part by the U.S.  Department of Energy under grant
DE-FG02-97ER41022.

\end{document}